\documentclass[12pt, letterpaper]{article}
\pdfoutput=1
\usepackage{float}
\usepackage{amsfonts}
\usepackage{amsmath}
\usepackage{bm}
\usepackage{url}
\usepackage{cite}
\usepackage{hyperref}
\usepackage{slashed}
\usepackage{tikz}
\usepackage{graphicx}
\usepackage{epstopdf}
\usepackage{subfigure}
\usepackage{mciteplus}
\usepackage{amssymb}
\usepackage{color}
\usepackage{mathrsfs}
\usepackage{fancybox}
\usepackage{lipsum}
\usepackage{eurosym}
\usepackage{tcolorbox}
\usepackage{tensor}

\newcommand{\Tr}{\operatorname{Tr}} 

\edef\restoreparindent{\parindent=\the\parindent\relax}
\usepackage{parskip}
\restoreparindent

\numberwithin{equation}{section}
\allowdisplaybreaks

\usepackage{silence}
\newcommand*\centermathcell[1]{\omit\hfil$\displaystyle#1$\hfil\ignorespaces}

\newcommand{\slr}{\text{SL}^+(2, \mathbb{R})}
\newcommand{\sltr}{\text{SL}(2, \mathbb{R})}

\newcommand{\osp}{\text{OSp}(1|2)}
\newcommand{\ospp}{\text{OSp}^+(1|2)}
\newcommand{\posp}{\text{OSp}'(1|2)}

\newcommand{\B}{\scriptscriptstyle \text{B}}
\newcommand{\T}{\scriptscriptstyle \text{T}}
\newcommand{\F}{\scriptscriptstyle \text{F}}
\newcommand{\+}{\scalebox{.4}{$+$}} 
\newcommand{\m}{\scalebox{.4}{$-$}} 

\newcommand{\NS}{\text{NS}}
\newcommand{\R}{\text{R}}
	
\title{Supergroup Structure of Jackiw-Teitelboim Supergravity}

\begin{document}

\begin{titlepage}

\setcounter{page}{0} \baselineskip=15.5pt \thispagestyle{empty}

\hfill UTTG-05-21

\vspace{1cm}

\begin{center}

\def\thefootnote{\fnsymbol{footnote}}
\begin{changemargin}{0.05cm}{0.05cm}
\begin{center}
{\Large \bf Supergroup Structure of \\[10 pt] Jackiw-Teitelboim Supergravity}
\end{center} 
\end{changemargin}

~\\[1cm]
{Yale Fan${}^{\rm a}$\footnote{\href{mailto:yalefan@gmail.com}{\protect\path{yalefan@gmail.com}}} and Thomas G. Mertens${}^{\rm b}$\footnote{\href{mailto:thomas.mertens@ugent.be}{\protect\path{thomas.mertens@ugent.be}}}}
\\[0.3cm]
\vspace{0.7cm}
{\normalsize{\sl ${}^{\rm a}$Theory Group, Department of Physics,
\\[1mm]
University of Texas at Austin, Austin, TX 78712, USA}} \\[3mm]
{\normalsize{\sl ${}^{\rm b}$Department of Physics and Astronomy,
\\[1mm]
Ghent University, Krijgslaan, 281-S9, 9000 Gent, Belgium}} \\
\vspace{0.5cm}

\end{center}

\vspace{0.2cm}
\begin{changemargin}{1cm}{1cm}
{\small \noindent
\begin{center}
\textbf{Abstract}
\end{center}}
We develop the gauge theory formulation of $\mathcal{N}=1$ Jackiw-Teitelboim supergravity in terms of the underlying $\text{OSp}(1|2, \mathbb{R})$ supergroup, focusing on boundary dynamics and the exact structure of gravitational amplitudes. We prove that the BF description reduces to a super-Schwarzian quantum mechanics on the holographic boundary, where boundary-anchored Wilson lines map to bilocal operators in the super-Schwarzian theory. A classification of defects in terms of monodromies of $\text{OSp}(1|2, \mathbb{R})$ is carried out and interpreted in terms of character insertions in the bulk. From a mathematical perspective, we construct the principal series representations of $\text{OSp}(1|2, \mathbb{R})$ and show that whereas the corresponding Plancherel measure does \emph{not} match the density of states of $\mathcal{N}=1$ JT supergravity, a restriction to the positive subsemigroup $\text{OSp}^+(1|2, \mathbb{R})$ yields the correct density of states, mirroring the analogous results for bosonic JT gravity. We illustrate these results with several gravitational applications, in particular computing the late-time complexity growth in JT supergravity.
\end{changemargin}
\vspace{0.3cm}

\vfil

\end{titlepage}

\newpage

\tableofcontents

\setcounter{tocdepth}{2}
\setcounter{footnote}{0}
\addtolength{\baselineskip}{0.1mm}
\addtolength{\abovedisplayskip}{1mm}
\addtolength{\belowdisplayskip}{1mm}


\section{Introduction and Overview}

The simplicity of the gravitational path integral in low dimensions, particularly in Jackiw-Teitelboim (JT) gravity \cite{Jackiw:1984je, Teitelboim:1983ux, Almheiri:2014cka, Jensen:2016pah, Maldacena:2016upp, Engelsoy:2016xyb}, has yielded important insights such as the paradigm of ensemble duality for effective theories of gravity \cite{Saad:2019lba, Giddings:1988wv, Marolf:2020xie, Blommaert:2019wfy, Belin:2020hea, Giddings:2020yes, Anous:2020lka, Saad:2021rcu, Altland:2021rqn}, an improved understanding of topological effects on boundary correlation functions \cite{Blommaert:2019hjr, Saad:2019pqd, Blommaert:2020seb, Stanford:2020wkf}, an explicit calculational scheme for local quantum gravitational observables \cite{Mertens:2019bvy, Blommaert:2020yeo, Lin:2019qwu}, and concrete applications of new gravitational entropy formulas (reviewed in \cite{Almheiri:2020cfm}).  The tractability of lower-dimensional models of gravity stems from the fact that such theories are perturbatively equivalent to topological gauge theories.  In the particular case of JT gravity or its $\mathcal{N}=1$ supersymmetric counterpart, this gauge theory is a BF model based on the bosonic group $\sltr$ \cite{Fukuyama:1985gg, Isler:1989hq, Chamseddine:1989yz, Jackiw_1992} or the supergroup $\text{OSp}(1|2, \mathbb{R})$ \cite{Astorino:2002bj}.  Similar constructions (e.g., \cite{Witten:1988hc, Achucarro:1989gm, Coussaert:1995zp, Kim:2015qoa, Cotler:2018zff}) have elucidated many aspects of 3d gravity over the years.  As recent progress has demonstrated, a careful understanding of the gauge theory description has much to teach us about the gravitational path integral.

Nonetheless, there exist profound structural differences between gravity and ordinary gauge theory that must be reconciled to effectively bring gauge-theoretic tools to bear on problems in holography and nonperturbative quantum gravity.  These differences have been well-appreciated in the context of 3d gravity \cite{Witten:2007kt}.  We recapitulate them here and adapt the discussion to 2d. Namely, to upgrade gauge theory to gravity, one must do the following:
\begin{itemize}
\item[\textbf{I.}] \textbf{Restrict the integration space to smooth metrics only.} The gauge theory formulation of quantum gravity, whenever it exists, admits gauge field configurations that correspond to singular geometries. The canonical example is the classical solution $\mathbf{A}=0$ in 3d Chern-Simons gravity, which corresponds to a non-invertible metric. Hence the gauge formulation contains ``too much'' information compared to gravity, which raises the question of how to naturally exclude these extraneous configurations. Luckily, in 2d $\sltr$ BF theory, this problem can be formulated very precisely. The path integral of this theory computes the volume of the moduli space of flat $\sltr$ connections ($\mathbf{F}=0$) on any given Riemann surface $\Sigma$, which we denote by $\mathcal{M}(G, \Sigma)$ for $G = \sltr$. This moduli space has a connected ``hyperbolic'' component, called Teichm\"uller space $\mathcal{T}(\Sigma) \subset \mathcal{M}(G, \Sigma) $, that parametrizes smooth geometries.\footnote{More precisely, the moduli space of flat $\text{PSL}(2, \mathbb{R})$ connections on a Riemann surface $\Sigma$ of genus $g$ and $n$ boundaries with $-\chi(\Sigma) = 2g + n - 2 > 0$ has $1 - 2\chi(\Sigma)$ connected components labeled by all integer values of the Euler class $e$ satisfying $|e|\leq -\chi(\Sigma)$ \cite{Goldman1988} (when $n > 0$, one assumes hyperbolic monodromies for each boundary component, and $e$ refers to the corresponding relative Euler class). The components with $|e| = -\chi(\Sigma)$ correspond to smooth hyperbolic metrics. The other components contain hyperbolic metrics with isolated conical singularities whose angular excesses are multiples of $2\pi$ \cite{Goldman88geometricstructures}.} The main technical question that remains is how to accomplish the restriction to Teichm\"uller space within a concrete amplitude calculation.
\item[\textbf{II.}] \textbf{Quotient by large diffeomorphisms.} Gravity contains large diffeomorphisms that are invisible from the gauge theory perspective. Again, we can be very explicit in 2d: Teichm\"uller space $\mathcal{T}(\Sigma)$ contains surfaces that are considered equivalent in gravity by virtue of large diffeomorphisms. The prototypical example of such diffeomorphisms is the modular group SL$(2,\mathbb{Z})$ acting on the modulus of a torus surface. The generalization to any Riemann surface is that there exists a discrete group of large diffeomorphisms, the mapping class group $\operatorname{MCG}(\Sigma)$, by which we must quotient the space of smooth geometries (modulo small diffeomorphisms) to reach the true integration space of inequivalent geometries: the moduli space of 2d Riemann surfaces $\mathcal{M}(\Sigma) \simeq \mathcal{T}(\Sigma)/\operatorname{MCG}(\Sigma)$.
\item[\textbf{III.}] \textbf{Sum over topologies.} The gravitational path integral naturally contains a summation over different topologies consistent with the prescribed boundaries. By contrast, gauge theory is defined on a fixed spacetime manifold and does not automatically include any such summation. In 2d, we can again be explicit. For a 2d Riemann surface $\Sigma$, the Gauss-Bonnet theorem states that
\begin{equation}
\chi(\Sigma) = 2-2g-n = \frac{1}{4\pi} \int_\Sigma R + \frac{1}{2\pi}\oint_{\partial \Sigma} K,
\label{GB}
\end{equation}
where $g$ is the genus, $n$ is the number of boundaries, and $K$ is the trace of the extrinsic curvature on each of the boundaries. For a given number of boundaries $n$, restricting to a fixed genus $g$ hence defines a constrained gravitational path integral in which the metric tensor is required to satisfy \eqref{GB}. While the resulting constrained path integral is not ill-defined, it may fail to capture important physical effects (such as the downward part of the Page curve \cite{Almheiri:2019qdq, Penington:2019kki}, or the late-time ramp and plateau in boundary correlation functions \cite{Saad:2019lba, Blommaert:2019hjr, Saad:2019pqd}). To accommodate such effects within the gauge formulation, a summation over different topologies must be introduced by hand. This issue is an inherent limitation of any gauge theory description, and we will not concern ourselves with it further in this work.
\end{itemize}

Our motivation in this work is to explore precisely those global aspects of gauge theory that manifest themselves in a gravitational description.  In particular, our goal is to elaborate on the structural link between the geometry of 2d gravity and the algebraic framework of group theory and representation theory.  Within JT gravity, which has played a central role in recent advances due to its exact solubility, the questions that we ask include: What is the detailed structure of the JT gravity path integral?  Can one compute refined observables (correlation functions) beyond those of \cite{Saad:2019lba}?  Our specific focus is on understanding a single feature---supersymmetry---that leads to richer physics and improved UV behavior, and that may be present in top-down constructions of such models.

Supersymmetry aside, the $\mathfrak{sl}(2, \mathbb{R})$ BF theory presentation of ordinary JT gravity provides a convenient language for computing diffeomorphism-invariant observables (boundary correlation functions).  While the natural home of gauge theory is a fixed topology (particularly the disk), disk correlators are the foundation of correlators in arbitrary genus.  In gauge theory language, the known diagrammatic rules for disk amplitudes \cite{Mertens:2017mtv} take as their basic ingredient a certain momentum space integration measure $d\mu(k)$ (or density of states, via $E\sim k^2$).  It has been argued that this integration measure follows directly from the Plancherel measure on the space of continuous irreps of a modification of $\sltr$, namely the semigroup $\slr$ \cite{Blommaert:2018oro, Blommaert:2018iqz}.

In this paper, we apply these lessons to supergravity.  We undertake a detailed study of the $\osp$ (or $\mathfrak{osp}(1|2)$) supergroup gauge theory formulation of $\mathcal{N} = 1$ JT supergravity, emphasizing the $\ospp$ supersemigroup structure.  Both the exact solution for the partition function \cite{Maldacena:2016upp, Cotler:2016fpe, Stanford:2017thb} and the dual matrix ensemble of JT gravity \cite{Saad:2019lba} have been generalized to JT supergravity \cite{Stanford:2019vob}.  In this paper, we likewise generalize both the exact group-theoretic computation of correlators and the semigroup structure in JT gravity \cite{Blommaert:2018oro,Blommaert:2018iqz} to JT supergravity.  In past work, the boundary correlators have been obtained for $\mathcal{N} = 1$ JT supergravity by exploiting its relation to 2d Liouville CFT \cite{Mertens:2017mtv, Mertens:2018fds, Lam:2018pvp}.  Other approaches, possibly also amenable to supersymmetrization, include direct 1d path integral calculations \cite{Bagrets:2016cdf, Bagrets:2017pwq, Belokurov:2019els} as well as methods relating JT dynamics to that of a particle on AdS$_2$ in an infinite magnetic field and the universal cover of $\sltr$ \cite{Kitaev:2018wpr, Yang:2018gdb, Iliesiu:2019xuh}. As compared to these other approaches, the conceptual unity and simplicity of the group-theoretic approach allows for a clean generalization not only to JT supergravity, but also potentially to other theories of dilaton gravity. We return to this perspective in the concluding section.

There are both conceptual and technical reasons for tackling the problem of JT supergravity.  Conceptually, adding supersymmetry allows us to address questions such as: How robust is the semigroup structure of 2d dilaton gravity?  Does it persist in more complicated theories of gravity? Moreover, there exist further links to be made with minimal string theory and Liouville gravity, first suggested for the disk partition function in \cite{Saad:2019lba, StanfordSeiberg} and worked out for several amplitudes in \cite{Mertens:2020hbs, Mertens:2020pfe}. Taking the point of view that 2d string theory gives a more microscopic definition of such models (in which the worldsheet expansion becomes a universe expansion), JT supergravity is a natural setting for using tools from minimal (super)string theory to understand quantum gravity \cite{Johnson:2019eik, Johnson:2020heh, Johnson:2020exp, Johnson:2021owr, Okuyama:2020qpm, Mertens:2020hbs, Mertens:2020pfe}.  On the technical side, we develop various elements of supergroup representation theory from scratch, many of which may be of independent interest.

A more detailed summary of our results is as follows.

In \textbf{Section \ref{supersch}}, we begin by examining the path integral of JT supergravity in BF language and deriving the super-Schwarzian quantum mechanics that governs its boundary dynamics.  We further classify the various possible super-Schwarzian models in terms of coadjoint orbits of the $\mathcal{N} = 1$ super-Virasoro algebra.

In \textbf{Section \ref{bilocal}}, we enrich this correspondence with operator insertions.  We demonstrate that the first-order formulation of $\mathcal{N}=1$ JT supergravity with boundary-anchored gravitational Wilson line insertions is equivalent to the $\mathcal{N}=1$ super-Schwarzian theory with bilocal operator insertions:
\begin{equation}
\int \left[\mathcal{D}\mathbf{B}\right]\left[\mathcal{D}\mathbf{A}\right] \mathcal{W}_j(\tau_1, \tau_2)_{IJ} \cdots e^{-S_\text{JT}^{\mathcal{N} = 1}} = \int [\mathcal{D} F]\, [\mathcal{D}\eta]\, \mathcal{O}_h(\tau_1,\theta_1,\tau_2,\theta_2)_{m} \cdots e^{-S_\text{Sch}^{\mathcal{N} = 1}}.
\end{equation}
The bulk action $S_\text{JT}^{\mathcal{N} = 1}$ is a functional of a dilaton supermultiplet $\mathbf{B}$ and a superconnection $\mathbf{A}$, both valued in $\mathfrak{osp}(1|2)$, while the boundary action $S_\text{Sch}^{\mathcal{N} = 1}$ is a functional of a bosonic reparametrization mode $F(\tau)$ and its superpartner $\eta(\tau)$.  $(\tau, \theta)$ are 1d superspace coordinates.  The boundary-anchored Wilson line is given by
\begin{equation}
\mathcal{W}_j(\tau_1,\tau_2) = \mathcal{P}\exp\left[-\int_{\tau_1}^{\tau_2} R_j(\mathbf{A})\right].
\end{equation}
It forms a $\dim R\times \dim R$ matrix, from which we pick a certain element $IJ$. The super-Schwarzian bilocal operator takes the form
\begin{equation}
\label{defbilocal}
\mathcal{O}_{h}(\tau_1,\theta_1,\tau_2,\theta_2) = \left(\frac{D\theta'_1 D\theta'_2}{\tau'_1 - \tau'_2 - \theta'_1\theta'_2}\right)^{2h}, \quad D\equiv \partial_\theta + \theta\partial_\tau,
\end{equation}
where $(\tau', \theta')$ is a superconformal transformation of $(\tau, \theta)$ dictated by $F, \eta$.  It contains four components $m=1,2,3,4$ when expanded in the Grassmann variables $\theta_{1,2}$. Each of these four components can be uniquely mapped to a pair of indices $IJ$ on the left-hand side. The representation $j$ of the Wilson line is related to the weight $h$ of the bilocal operator on the right-hand side by $j=-h$.

In \textbf{Section \ref{semigroup}}, we show that the full structure of JT supergravity amplitudes suggests that the aforementioned restriction to smooth geometries can be naturally implemented by restricting the full group to its positive subsemigroup. We provide arguments in favor of this scheme, and then explicitly compute the measure on the set of continuous representations of the subsemigroup and demonstrate agreement with the density of states of the gravitational system. This result shows how gravity and gauge theory match at lowest genus $g=0$, where only item \textbf{I} of the complications listed above is present.

In \textbf{Section \ref{applications}}, we present a few physical applications of our results by explicitly computing several gravitational amplitudes: the boundary two-point function, the Wheeler-DeWitt wavefunction, and defect insertions. Using these defect insertions, one can glue surfaces together to reach different topologies. It is here that item \textbf{II} in the above list of complications makes an appearance. As a last example, we compute the late-time complexity growth in this model and exhibit a similar physical result as in the bosonic case: the linear growth in complexity persists even after classical gravity ceases to hold.

In \textbf{Section \ref{discussion}}, we conclude by commenting on several outstanding problems and intriguing extensions whose full treatment is postponed to future work.

In the interest of conveying our main ideas as clearly as possible, many of their technical foundations are left to extensive appendices.  \textbf{Appendix \ref{app:supernumbers}} summarizes our conventions for supernumbers.  \textbf{Appendix \ref{app:bosonic}} serves both to review bosonic JT gravity and to present some new results using techniques that we apply also to JT supergravity. \textbf{Appendix \ref{firstorderformalism}} reviews the relation between the first- and second-order formulations of JT supergravity. \textbf{Appendix \ref{app:details}} provides some further details on coadjoint orbits of the super-Virasoro group and on super-Schwarzian bilocal operators. The results of \textbf{Appendix \ref{osprep}} form the technical core of this paper.  Here, we aim to provide a comprehensive overview of the representation theory of $\osp$, which could be of interest on its own to some readers.  In particular, we compute the Plancherel measure for $\osp$ in Section \ref{app:whit}.  Finally, \textbf{Appendix \ref{ospprep}} provides some technical proofs for the positive subsemigroup of $\osp$.

\section{Super-Schwarzian and Defect Classification}
\label{supersch}

In this section, we discuss the kinematics of JT supergravity as a supergroup BF theory, focusing in particular on the boundary dynamics.  Specifically, we show that the boundary action of a constrained particle on the $\osp$ group manifold reduces to the $\mathcal{N}=1$ super-Schwarzian. Moreover, we show how to classify defect insertions in terms of monodromies of the super-Schwarzian system.

The procedure is to implement the Brown-Henneaux gravitational boundary conditions \cite{Brown:1986nw} on the BF model in the bulk. Solving them boils down to solving the supersymmetric Hill's equation, which can be done in terms of reparametrization functions of the supercircle $S^{1|1}$.

\subsection{Gravitational Boundary Action} \label{gravboundact}

The first-order action of $\mathcal{N} = 1$ JT supergravity in Euclidean signature can be written as a BF theory with gauge algebra $\mathfrak{osp}(1|2)$:\footnote{The supertrace of a supermatrix $M = \left[\begin{array}{c|c} A & B \\ \hline C & D \end{array}\right]$ is defined as $\operatorname{STr}M = \Tr A - \Tr D$.}
\begin{equation}
\label{sjtac}
S_\text{JT}^{\mathcal{N} = 1} = \int_\mathcal{M} \operatorname{STr}(\mathbf{B}\mathbf{F}).
\end{equation}
We have introduced the $\mathfrak{osp}(1|2)$-valued fields
\begin{equation}
\mathbf{B} = \phi^a J_a - \phi J_2 + \lambda^\alpha Q_\alpha, \qquad \mathbf{A} = e^a J_a + \omega J_2 + \psi^\alpha Q_\alpha,
\label{sjtfields}
\end{equation}
where $\mathbf{B}$ is a zero-form and $\mathbf{A}$ is a one-form connection with field strength $\mathbf{F} = d\mathbf{A} + \mathbf{A}\wedge \mathbf{A}$.  We have implicitly chosen an imaginary contour of integration for $\mathbf{B}$ in the path integral.  The component fields consist of scalar Lagrange multipliers $\phi^a$, a dilaton $\phi$, a dilatino $\lambda^\alpha$, the zweibein $e^a$, the spin connection $\omega$, and the gravitino $\psi^\alpha$.  The indices $a\in \{0, 1\}$ and $\alpha\in \{+, -\}$ denote doublets of $\mathfrak{so}(2)$, while the bosonic components of $\mathbf{B}$ and the bosonic components of $\mathbf{A}$ each combine into $\mathfrak{sl}(2, \mathbb{R})$ triplets.  The $\mathfrak{osp}(1|2)$ generators $J_{0, 1, 2}$ and $Q_\pm$ are described below.  

The supergroup $\osp$ is defined as the subgroup of $\text{GL}(1|2,\mathbb{R})$ matrices
\begin{equation}
\label{osppar}
g = \left[\begin{array}{cc|c}
a & b & \alpha \\
c & d & \gamma \\ \hline
\beta & \delta & e
\end{array}\right]
\end{equation}
consisting of five bosonic variables $a$, $b$, $c$, $d$, $e$ and four fermionic (Grassmann) variables $\alpha$, $\beta$, $\gamma$, $\delta$ that satisfy the relations
\begin{equation}
\label{definingcond}
\alpha = \pm(a\delta - b \beta), \qquad \gamma = \pm(c \delta - d \beta), \qquad e = \pm(1 + \beta \delta), \qquad ad-bc = 1 +\delta \beta,
\end{equation}
for either choice of sign $\pm$.  Restricting to a single one of the signs leads to the projective supergroup denoted by $\posp = \osp/\mathbb{Z}_2$ in \cite{Stanford:2019vob}.  See Appendix \ref{osprep} for details.

We denote the Cartan-Weyl generators in the above defining representation by \cite{Frappat:1996pb}
\begin{gather}
\label{defgen}
H = \left[\begin{array}{cc|c} 
1/2 & 0 & 0 \\
0 & -1/2 & 0 \\
\hline
0 & 0 & 0
\end{array} \right], \quad E^- = \left[\begin{array}{cc|c} 
0 & 0 & 0 \\
1 & 0 & 0 \\
\hline
0 & 0 & 0
\end{array} \right], \quad E^+ = \left[\begin{array}{cc|c} 
0 & 1 & 0 \\
0 & 0 & 0 \\
\hline
0 & 0 & 0
\end{array} \right], \\[5 pt]
\label{defgen2}
F^- = \left[\begin{array}{cc|c} 
0 & 0 & 0 \\
0 & 0 & -1/2 \\
\hline
1/2 & 0 & 0
\end{array} \right], \quad F^+ = \left[\begin{array}{cc|c} 
0 & 0 & 1/2 \\
0 & 0 & 0 \\
\hline
0 & 1/2 & 0
\end{array} \right],
\end{gather}
which satisfy the $\mathfrak{osp}(1|2)$ Lie superalgebra:
\begin{alignat}{2}
[H, E^\pm] &= \pm E^\pm, \quad & [E^+, E^-] &= 2H, \nonumber \\
[H, F^\pm] &= \pm\frac{1}{2}F^\pm, \quad & [E^\pm, F^\mp] &= -F^\pm, \label{ospline} \\
\{F^+, F^-\} &= \frac{1}{2}H, \quad & \{F^\pm, F^\pm\} &= \pm\frac{1}{2}E^\pm. \nonumber
\end{alignat}
In \eqref{sjtfields}, we have defined the following linear combinations of $\mathfrak{osp}(1|2)$ generators:
\begin{equation}
J_0 = -H, \quad J_1 = \frac{1}{2}(E^- + E^+), \quad J_2 = \frac{1}{2}(E^- - E^+), \quad Q_- = -F^-, \quad Q_+ = F^+.
\end{equation} 
Therefore, in matrix form, we have:
\begin{equation}
\mathbf{B} = \frac{1}{2}\left[\begin{array}{cc|c} 
-\phi^0 & \phi^1 + \phi & \lambda^+ \\
\phi^1 - \phi & \phi^0 & \lambda^- \\
\hline
-\lambda^- & \lambda^+ & 0
\end{array} \right], \qquad
\mathbf{A} = \frac{1}{2}\left[\begin{array}{cc|c} 
-e^0 & e^1 - \omega & \psi^+ \\
e^1 + \omega & e^0 & \psi^- \\
\hline
-\psi^- & \psi^+ & 0
\end{array} \right].
\end{equation}
We summarize the details of the derivation of \eqref{sjtac} from superspace and the relation to the metric formulation in Appendix \ref{firstorderformalism}.

For a manifold with boundary, the BF action \eqref{sjtac} gets augmented by a boundary term:
\begin{equation}
\label{BFbdy}
S_\text{JT}^{\mathcal{N} = 1} = \int_\mathcal{M} \operatorname{STr}(\mathbf{B}\mathbf{F}) - \frac{1}{2} \oint_{\partial \mathcal{M}} d\tau \operatorname{STr}(\mathbf{B}\mathbf{A}_\tau),
\end{equation}
where the Euclidean coordinate $\tau$ is tangent to the boundary $\partial \mathcal{M}$. It will play the role of time coordinate further on. We choose the mixed boundary condition
\begin{equation}
\left.\mathbf{B}\right|_{\partial \mathcal{M}} = \left.\mathbf{A}_\tau\right|_{\partial \mathcal{M}}.
\label{mixedBC}
\end{equation}
The above boundary action and condition can be found in several ways \cite{Mertens:2018fds}. One is to simply demand a good variational principle for the BF action on $\mathcal{M}$. Another is to follow the usual relation between Chern-Simons theory in 3d and the boundary WZW action. Dimensionally reducing that setup automatically generates this boundary term in the action, along with this specific boundary condition.

Starting with the action \eqref{BFbdy}, the solution of the gravitational path integral proceeds along familiar lines. We first path-integrate over the $\mathbf{B}$ fields, which figure as Lagrange multipliers in the action. The resulting dynamics then reduces to a pure boundary contribution from flat connections:
\begin{equation}
\label{pi}
\int_{\mathbf{F}=0} \left[\mathcal{D} \mathbf{A}_\tau\right] \exp\left[\frac{1}{2} \oint_{\partial \mathcal{M}} d\tau \operatorname{STr}(\mathbf{A}_\tau^2)\right].
\end{equation}
Within the first-order formulation of an OSp$(1|2)$ gauge connection, one can impose the gravitational (or Brown-Henneaux) boundary conditions as \cite{Henneaux:1999ib}:
\begin{equation}
\label{bhs}
\left.\mathbf{A}_\tau\right|_{\partial \mathcal{M}} = \left[\begin{array}{cc|c} 
0 & T_{\B}(\tau) & T_{\F}(\tau) \\
1 & 0 & 0 \\
\hline
0 & T_{\F}(\tau) & 0
\end{array} \right],
\end{equation}
where the boundary degrees of freedom are parametrized by a bosonic function $T_{\B}(\tau)$, interpretable as the energy, and a fermionic function $T_{\F}(\tau)$, interpretable as the supercharge. This leads to the boundary action
\begin{equation}
\frac{1}{2}\oint_{\partial \mathcal{M}} d\tau \operatorname{STr}(\mathbf{A}_\tau^2) = \oint_{\partial \mathcal{M}} d\tau\, T_{\B}(\tau).
\end{equation}
The components $T_{\B}(\tau)$ and $T_{\F}(\tau)$ can be packaged into a single fermionic superfield
\begin{equation}
\mathcal{V}(\tau,\theta) \equiv T_{\F}(\tau) + \theta T_{\B}(\tau),
\label{fermsuperfield}
\end{equation}
where we introduced the Grassmann coordinate $\theta$ as the fermionic partner of the bosonic boundary coordinate $\tau$. It is a general fact that we can write any fermionic superfield $\mathcal{V}(\tau,\theta)$ as a super-Schwarzian derivative of two new superfields $\tau'(\tau,\theta)$ and $\theta'(\tau,\theta)$:
\begin{equation}
\mathcal{V}(\tau,\theta) = - \frac{D^4\theta'}{D\theta'} + \frac{2D^3\theta' D^2 \theta'}{(D\theta')^2} \equiv -\operatorname{Sch}(\tau', \theta'; \tau, \theta),
\label{SuS}
\end{equation}
where $\tau'$ is bosonic and $\theta'$ is fermionic, satisfying $D\tau' = \theta' D\theta'$. This constraint can further be solved explicitly as
\begin{equation}
\tau' = F(\tau + \theta\eta(\tau)), \qquad \theta' = \sqrt{\partial_\tau F(\tau)}\left(\theta + \eta(\tau) + \frac{1}{2}\theta\eta(\tau)\partial_\tau \eta(\tau)\right),
\label{SSsolution}
\end{equation}
in terms of a bosonic function $F(\tau)$ and its fermionic superpartner $\eta(\tau)$. In terms of these functions, the stress tensor and its superpartner can be written as
\begin{align}
\label{susyact}
T_{\B}(\tau) &= \frac{1}{2}\left(\{F,\tau\} + \eta \partial_\tau^3\eta + 3 \partial_\tau\eta\partial_\tau^2\eta - \{F,\tau\}\eta\partial_\tau\eta\right), \\
T_{\F}(\tau) &= \partial_\tau^2\eta + \frac{1}{2}\eta\partial_\tau\eta\partial_\tau^2\eta + \frac{1}{2}\eta \{F,\tau\}. \label{susyactother}
\end{align}
We view this change of variables as a field redefinition in the path integral:
\begin{equation}
(T_{\B},T_{\F}) \, \to \, (F,\eta).
\end{equation}
The new fields $(F,\eta)$ are not in one-to-one correspondence with the components of the stress tensor, since the solutions to the super-Schwarzian differential equation \eqref{SuS} are subject to a super-M\"obius ambiguity:
\begin{equation}
\label{ospredun}
\tau'\to \frac{a\tau'-c - \beta \theta'}{-b\tau'+d+\delta \theta'}, \qquad \theta'\to \frac{\alpha \tau'- \gamma +e \theta'}{-b\tau'+d+\delta \theta'},
\end{equation}
where the entries are taken from the projective group $\text{OSp}'(1|2,\mathbb{R}) = \text{OSp}(1|2,\mathbb{R})/\mathbb{Z}_2$ \eqref{osppar}. This means that one should identify field configurations differing only by such transformations.

We consider the resulting model on a supercircle $S^{1|1}$ with $T_{\B}(\tau+\beta)= T_{\B}(\tau)$ and $T_{\F}(\tau+\beta) = \pm T_{\F}(\tau)$. In the end, the path integral \eqref{pi} becomes
\begin{equation}
Z = \int_{\operatorname{SDiff}_{\mathcal{N} = 1}(S^{1|1})/H} [\mathcal{D} F]\, [\mathcal{D}\eta]\, e^{-S_\text{Sch}^{\mathcal{N} = 1}[F, \eta]}, \quad S_\text{Sch}^{\mathcal{N} = 1} = -\oint_{\partial \mathcal{M}}d\tau\, T_{\B}(\tau),
\end{equation}
where the Lagrangian is expressed in terms of the fields $F(\tau)$ and $\eta(\tau)$ (the ``su\-per\-re\-pa\-ram\-e\-triza\-tion modes'') by substituting \eqref{susyact} for $T_{\B}(\tau)$. The stabilizer $H$ is the subgroup of OSp$(1|2,\mathbb{R})$ \eqref{ospredun} that respects the periodicity constraints of $\theta'$ and $\tau'$, as we will work out more explicitly below. It contains information on the precise functional form of $F \circ_H f$ in terms of a new variable $f(\tau)$, as well as $\eta(\tau)$. These models are all $\mathcal{N}=1$ super-Schwarzian theories that have different geometric interpretations and uses.

\subsection{Super-Hill's Equation, Monodromies, and Defects}
\label{sect:hill}

For the sake of analyzing the different possible super-Schwarzian models in detail, we reformulate the gravitational boundary conditions in terms of the supersymmetric Hill's equation.\footnote{Some aspects of this analysis appeared in \cite{Cardenas:2018krd}. We will have need of a more extensive treatment to prepare for the calculation of boundary-anchored Wilson lines in Section \ref{bilocal}.} By the flatness condition on any off-shell bulk connection, we have
\begin{equation}
\left.\mathbf{A}_\tau\right|_{\partial \mathcal{M}} = g\partial_\tau g^{-1},
\end{equation}
and we can rewrite the gravitational boundary conditions in terms of constraints on the boundary group element $g \in \osp$.  This group element $g$ is generically multi-valued and can have nontrivial monodromy when encircling the boundary circle; the gauge connection $\mathbf{A}_\tau$, however, has fixed periodicity constraints. Depending on the sector (NS or R), we have:
\begin{equation}
\label{perA}
\left.\mathbf{A}_\tau(\tau+\beta)\right|_{\partial \mathcal{M}} = \begin{cases}
(-)^F \left.\mathbf{A}_\tau(\tau)\right|_{\partial \mathcal{M}} (-)^F \quad & \text{(NS)}, \\
\left.\mathbf{A}_\tau(\tau)\right|_{\partial \mathcal{M}} \quad & \text{(R)},
\end{cases}
\end{equation}
where the presence of the ``sCasimir'' operator $(-)^F \equiv \text{diag}(+1,+1,-1)$ ensures that the fermionic pieces (i.e., $T_{\F}(\tau)$ in \eqref{bhs}) flip sign upon traversing the boundary circle.

Parametrizing
\begin{equation}
g^{-1} = \left[\begin{array}{cc|c} 
A & B & A\delta - B \beta \\
C & D & C\delta - D \beta \\
\hline
\beta & \delta & 1+\beta \delta
\end{array} \right], \quad
g = \left[\begin{array}{cc|c} 
D & -B & -\delta \\
-C & A & \beta \\
\hline
C\delta-D\beta & B\beta-A\delta & 1+\beta\delta \\
\end{array} \right],
\end{equation}
the boundary condition \eqref{bhs} is written in full as
\begin{equation}
\left[\begin{array}{cc|c} 
A & B & A\delta - B \beta \\
C & D & C\delta - D \beta \\
\hline
\beta & \delta & 1+\beta \delta
\end{array} \right] 
\left[\begin{array}{cc|c} 
0 & T_{\B} & T_{\F} \\
1 & 0 & 0 \\
\hline
0 & T_{\F} & 0
\end{array} \right] = \left[\begin{array}{cc|c} 
A' & B' & (A\delta - B \beta)' \\
C' & D' & (C\delta - D \beta)' \\
\hline
\beta' & \delta' & (1+\beta \delta)'
\end{array} \right],
\end{equation}
leading to the coupled differential equations
\begin{alignat}{3}
\label{eqns}
B' &= A T_{\B} + (A\delta-B\beta)T_{\F}, \qquad & A' &= B, \qquad & (A\delta-B\beta)' &= AT_{\F}, \nonumber \\
D' &= C T_{\B} + (C\delta-D\beta)T_{\F}, \qquad & C' &= D, \qquad & (C\delta-D\beta)' &= CT_{\F}, \\
\delta' &= \beta T_{\B} + (1+\beta\delta)T_{\F}, \qquad & \beta' &= \delta, \qquad & (\beta\delta)' &= \beta T_{\F}. \nonumber
\end{alignat}
We would like to recast these equations as the supersymmetric Hill's equation, which takes the form
\begin{equation}
\label{sHill}
(D^3 - \mathcal{V}) \psi = 0,
\end{equation}
with $D\equiv \partial_\theta + \theta\partial_\tau$ and $\mathcal{V}$ defined in \eqref{fermsuperfield}. Writing $\psi(\tau,\theta) = \psi_{\text{bot}}(\tau) + \theta \psi_{\text{top}}(\tau)$ (where we make no assumptions about the Grassmann parity of $\psi$), this equation becomes the coupled system
\begin{align}
\label{hillcom}
\psi_{\text{bot}}'' - T_{\B}(\tau) \psi_{\text{bot}} + T_{\F}(\tau) \psi_{\text{top}} &=0, \qquad \psi_{\text{top}}' - T_{\F}(\tau) \psi_{\text{bot}} =0
\end{align}
for the bottom and top components of $\psi$.

The general solution to the supersymmetric Hill's equation \eqref{sHill} is known. Writing the superfield $\mathcal{V}$ as a super-Schwarzian derivative as before,
\begin{equation}
\mathcal{V}(\tau, \theta) = - \frac{D^4\alpha}{D\alpha} + \frac{2D^3\alpha D^2 \alpha}{(D\alpha)^2} = -\operatorname{Sch}(A, \alpha; \tau, \theta),
\label{VasSS}
\end{equation}
it can be established that up to super-M\"obius transformations, the solutions of \eqref{sHill} consist of two bosonic superfields and one fermionic superfield \cite{Arvis:1982tq}:
\begin{align}
\psi_1 &= (D\alpha)^{-1}, \nonumber \\
\psi_2 &= A (D\alpha)^{-1}, \label{solhill} \\
\psi_3 &= -\alpha (D\alpha)^{-1}, \nonumber
\end{align}
written in terms of a bosonic superfield $A(\tau, \theta)$ and a fermionic superfield $\alpha(\tau, \theta)$ constrained by $DA = \alpha D\alpha$. Writing each solution as $\psi_i \equiv \psi_{i,\text{bot}} + \theta \psi_{i, \text{top}}$,\footnote{\label{fn4} Writing $\alpha=\alpha_{\F} + \theta \alpha_{\B}$ and $A = A_{\B} + \theta A_{\F}$, we have explicitly that
\begin{alignat}{2}
\psi_{1, \text{bot}} &= \frac{1}{\alpha_{\B}}, \qquad & \psi_{1, \text{top}} &= -\frac{\alpha_{\F}'}{\alpha_{\B}^2}, \\
\psi_{2, \text{bot}} &= \frac{A_{\B}}{\alpha_{\B}}, \qquad & \psi_{2, \text{top}} &= \frac{A_{\F}}{\alpha_{\B}} - \frac{A_{\B}\alpha_{\F}'}{\alpha_{\B}^2}, \\
\psi_{3, \text{bot}} &= -\frac{\alpha_{\F}}{\alpha_{\B}}, \qquad & \psi_{3, \text{top}} &= -1 - \frac{\alpha_{\F}\alpha_{\F}'}{\alpha_{\B}^2},
\end{alignat}
as well as
\begin{equation}
A_{\B}' = \alpha_{\B}^2 - \alpha_{\F}\alpha_{\F}', \qquad A_{\F} = \alpha_{\B}\alpha_{\F}.
\end{equation}
}
one can check (using, for instance, $D^2 = \partial_\tau$) that the three solutions \eqref{solhill} satisfy the interrelations
\begin{align}
\psi_1 \partial_\tau \psi_2 - \psi_2 \partial_\tau \psi_1 &= 1 - \psi_3 \partial_\tau \psi_3, \label{relions} \\
\psi_{1, \text{bot}} \partial_\tau\psi_{3, \text{bot}} - \partial_\tau\psi_{1, \text{bot}} \psi_{3, \text{bot}} &= \psi_{1, \text{top}}, \\
\psi_{2, \text{bot}} \partial_\tau\psi_{3, \text{bot}} - \partial_\tau\psi_{2, \text{bot}} \psi_{3, \text{bot}} &= \psi_{2, \text{top}}, \\
1 + \psi_{3,\text{bot}} \partial_\tau \psi_{3,\text{bot}} &= -\psi_{3,\text{top}}. \label{relionslast}
\end{align}
These relations form the analogue of the Wronskian condition for the $\osp$ system.

Comparing the structure of the equations \eqref{hillcom} to the equations \eqref{eqns}, we identify the boundary group element as
\begin{equation}
g^{-1} = \left[\begin{array}{cc|c} 
\psi_{1, \text{bot}} & \psi_{1, \text{bot}}' & \psi_{1, \text{top}} \\
\psi_{2, \text{bot}} & \psi_{2, \text{bot}}' & \psi_{2, \text{top}} \\
\hline
\psi_{3, \text{bot}} &\psi_{3, \text{bot}}' & -\psi_{3, \text{top}}
\end{array} \right], \quad
g = \left[\begin{array}{cc|c} 
\psi_{2, \text{bot}}' & -\psi_{1, \text{bot}}' & -\psi_{3, \text{bot}}' \\
-\psi_{2, \text{bot}} & \psi_{1, \text{bot}} & \psi_{3, \text{bot}} \\
\hline
\psi_{2, \text{top}} & -\psi_{1, \text{top}} & -\psi_{3, \text{top}}
\end{array} \right],
\label{boundaryg}
\end{equation}
where the relations \eqref{relions}--\eqref{relionslast} indeed implement the $\osp$ (more precisely, $\posp$) restrictions \eqref{definingcond} on the supermatrix.  The most general solution of the supersymmetric Hill's equation, written in matrix form as
\begin{equation}
g^{-1}\left[\begin{array}{cc|c} 
0 & T_{\B} & T_{\F} \\
1 & 0 & 0 \\
\hline
0 & T_{\F} & 0
\end{array} \right] = \partial_\tau g^{-1},
\end{equation}
is obtained by taking $g^{-1}\to S^{-1}g^{-1}$ for arbitrary $S^{-1}\in \osp$, or equivalently,
\begin{equation}
\left[\begin{array}{c} \psi_1 \\ \psi_2 \\ \hline \psi_3 \end{array}\right]\to S^{-1}\left[\begin{array}{c} \psi_1 \\ \psi_2 \\ \hline \psi_3 \end{array}\right], \qquad S^{-1} = \left[\begin{array}{cc|c}
d & -b & -\delta \\
-c & a & \beta \\ \hline
\gamma & -\alpha & e
\end{array}\right]\in \osp.
\end{equation}
This implements a super-M\"obius transformation on $A(\tau, \theta) = \psi_2/\psi_1$ and $\alpha(\tau, \theta) = -\psi_3/\psi_1$ of precisely the form \eqref{ospredun}.  The superfield $\mathcal{V}$ in \eqref{sHill} and \eqref{VasSS} is invariant under such transformations.

The classification of solutions to the supersymmetric Hill's equation (i.e., of equivalence classes of solutions related by super-M\"obius transformations) leads naturally to a classification of defects in JT supergravity.  Such a classification is equivalent to the classification of conjugacy classes of $\osp$.

Depending on $T_{\B}$ and $T_{\F}$, the solutions of the supersymmetric Hill's equation can have nontrivial monodromies:
\begin{equation}
\label{mono}
g(\tau + \beta) =
\begin{cases}
(-)^F g(\tau) M \quad & \text{(NS)}, \\
g(\tau)M \quad & \text{(R)}.
\end{cases}
\end{equation}
Within the NS sector, the factor of $(-)^F$ ensures that $\left.\mathbf{A}_\tau\right|_{\partial \mathcal{M}}$ has the correct periodicity as in \eqref{perA}.\footnote{It is instructive to work out the simplest example, in which $g(\tau+\beta) = (-)^F g(\tau) (-)^F$. This leads to the periodicity conditions where $\psi_{1,2,\text{bot}}$ and $\psi_{3,\text{top}}$ are periodic and the other components are antiperiodic. These are indeed solutions to \eqref{hillcom} with $T_{\B}(\tau+\beta)=T_{\B}(\tau)$ and $T_{\F}(\tau+\beta)=-T_{\F}(\tau)$.}

By the equivalence relation
\begin{equation}
\label{Sdep}
g\sim gS, \qquad S \in \osp,
\end{equation}
the monodromies are parametrized by conjugacy classes of group elements:
\begin{equation}
\label{conjc}
M\sim S M S^{-1}.
\end{equation}
Conjugacy classes of $\osp$ are discussed in \cite{Stanford:2019vob}, particularly Section 3.5.4 and Appendix A.3.  Each conjugacy class can be thought of as associated with a spin structure, corresponding to the holonomy of a flat $\osp$ connection around a circle with spin structure of Neveu-Schwarz (NS; antiperiodic) or Ramond (R; periodic) type.  Working within $\posp$ ($\osp$ modulo the action of the scalar matrices $\pm 1$), where all elements have Berezinian $+1$,\footnote{The Berezinian (superdeterminant) is defined for an invertible supermatrix $M = \left[\begin{array}{c|c} A & B \\ \hline C & D \end{array}\right]$ (one for which both bosonic blocks $A$ and $D$ are invertible) as
\begin{equation}
\operatorname{Ber}(M) = \det(A - BD^{-1}C)\det(D)^{-1} = \det(A)\det(D - CA^{-1}B)^{-1}.
\end{equation}} the NS-type conjugacy classes are obtained by multiplying the R-type conjugacy classes by $\operatorname{diag}(+1, +1, -1)$ (which commutes with purely bosonic group elements).\footnote{Such an element, which commutes with bosonic generators and anticommutes with fermionic generators, belongs to the ``scentre'' of the universal enveloping algebra of $\mathfrak{osp}(1|2)$ \cite{Arnaudon:1996qe}.}
The different monodromies $M$ and their stabilizers $H$ are shown in the first two columns of Table \ref{monodromies}.

\begin{table}[!htbp]
\centering
\begin{equation*}
\begin{array}{c||c|c|c}
\text{Class} & \text{Monodromy $M$} & \text{Stabilizer $H$} & \text{sVirasoro}
\\
\hline 
\hline & & 
\\
\begin{array}{c} \textbf{Hyperbolic} \\
\text{(R or NS)} 
\end{array} & \arraycolsep=3pt \left[\begin{array}{cc|c} 
\cosh \pi \Lambda & \sinh \pi \Lambda & 0 \\
\sinh \pi \Lambda & \cosh \pi \Lambda & 0 \\
\hline
0 & 0 & \pm 1
\end{array} \right] & \arraycolsep=3pt
\left[\begin{array}{cc|c} 
\cosh \pi t & \sinh \pi t & 0 \\
\sinh \pi t & \cosh \pi t & 0 \\
\hline
0 & 0 & \pm' 1
\end{array} \right] & L_0 \\ & &
\\
\hline & &
\\
\textbf{Elliptic} & \arraycolsep=3pt \left[\begin{array}{cc|c} 
\cos \pi \Theta & -\sin \pi \Theta & 0 \\
\sin \pi \Theta & \cos \pi \Theta & 0 \\
\hline
0 & 0 & 1
\end{array} \right] & \arraycolsep=3pt
\left[\begin{array}{cc|c} 
\cos\pi\phi & -\sin \pi\phi & 0 \\
\sin \pi\phi & \cos \pi\phi & 0 \\
\hline
0 & 0 & 1
\end{array} \right] & L_0 \\ & &
\\
\hline & &
\\
\begin{array}{c} \textbf{Special} \\ \textbf{Elliptic I} \\
\text{(R or NS)} 
\end{array} & \left[\begin{array}{cc|c} 
1 & 0 & 0 \\
0 & 1 & 0 \\
\hline
0 & 0 & 1
\end{array} \right] & \begin{array}{c} H_{\R} = \text{OSp}(1|2,\mathbb{R}) \rule[-1.6 em]{0 pt}{0.5 em} \\ \hline \rule[1.6 em]{0 pt}{0.5 em}
H_{\NS} = \text{SL}(2,\mathbb{R})\times \mathbb{Z}_2 \end{array} & \begin{array}{c} L_0, L_{\pm n}, G_{\pm \frac{n}{2}} \\ \text{\scriptsize ($n$ even)} \rule[-1 em]{0 pt}{0.5 em} \\ \hline \rule[1 em]{0 pt}{0.5 em}
L_0, L_{\pm n} \\ \text{\scriptsize ($n$ even)} \end{array} \\ & & \\
\hline & &
\\
\begin{array}{c} \textbf{Special} \\ \textbf{Elliptic II} \\
\text{(R or NS)} 
\end{array}& 
 \left[\begin{array}{cc|c} 
1 & 0 & 0 \\
0 & 1 & 0 \\
\hline
0 & 0 & -1
\end{array} \right] & 
\begin{array}{c}
H_{\R} = \text{SL}(2,\mathbb{R})\times \mathbb{Z}_2 \rule[-1.6 em]{0 pt}{0.5 em} \\ \hline \rule[1.6 em]{0 pt}{0.5 em}
H_{\NS} = \text{OSp}(1|2,\mathbb{R})
\end{array} & \begin{array}{c} L_0, L_{\pm n} \\ \text{\scriptsize ($n$ odd)} \rule[-1 em]{0 pt}{0.5 em} \\ \hline \rule[1 em]{0 pt}{0.5 em}
L_0, L_{\pm n}, G_{\pm \frac{n}{2}} \\ \text{\scriptsize ($n$ odd)} \end{array} \\ & &
\\
\hline & &
\\
\begin{array}{c} \textbf{Parabolic} \\
\text{(NS)} 
\end{array} &  \left[\begin{array}{cc|c} 
1 & 1 & 0 \\
0 & 1 & 0 \\
\hline
0 & 0 & -1
\end{array} \right] & 
\left[\begin{array}{cc|c} 
1 & b & 0 \\
0 & 1 & 0 \\
\hline
0 & 0 & \pm 1
\end{array} \right] & L_0 \\ & &
\\
\hline & &
\\
\begin{array}{c} \textbf{Parabolic} \\
\text{(R)} 
\end{array} & 
 \left[\begin{array}{cc|c} 
1 & 1 & 0 \\
0 & 1 & 0 \\
\hline
0 & 0 & 1
\end{array} \right] & 
\left[\begin{array}{cc|c} 
1 & b & \pm \delta \\
0 & 1 & 0 \\
\hline
0 & \delta & \pm 1
\end{array} \right] & L_0, G_0
\end{array}
\end{equation*}
\caption{Inequivalent monodromy matrices $M$ identified with (constant-representative) $\mathcal{N}=1$ Virasoro coadjoint orbits. We list the stabilizer subgroups $H = \{S\in \text{OSp}(1|2, \mathbb{R}) \,|\, MS = SM\}$ preserving these monodromy matrices. These subgroups are identified with the $\mathcal{N}=1$ Virasoro subalgebras preserving the value of the super-Schwarzian derivative. The NS-type special elliptic orbits are the only exceptions to this rule. In these cases, the correct stabilizer is denoted by $H_{\NS}$, and it preserves $M(-)^F$: $M(-)^F S = SM (-)^F$.}
\label{monodromies}
\end{table}

Before providing a more in-depth discussion of these different classes, we first transfer this structure of the monodromy matrix to the actual fields $(F,\eta)$ within the group element $g^{-1}$. In all cases shown in Table \ref{monodromies}, the (inverse) monodromy matrix is bosonic block diagonal, and it acts as:
\begin{equation}
M^{-1}g^{-1}(\tau) = \left[\begin{array}{cc|c} 
M_{11} & M_{12} & 0 \\
M_{21} & M_{22} & 0 \\
\hline
0 & 0 & M_{33}
\end{array} \right]  \left[\begin{array}{cc|c} 
\psi_{1,\text{bot}} & \psi_{1,\text{bot}}' & \psi_{1,\text{top}} \\
\psi_{2,\text{bot}} & \psi_{2,\text{bot}}' & \psi_{2,\text{top}} \\
\hline
\psi_{3,\text{bot}} &\psi_{3,\text{bot}}' & -\psi_{3,\text{top}}
\end{array} \right]
\label{monoact}
\end{equation}
where $\left[\begin{array}{cc} 
M_{11} & M_{12} \\
M_{21} & M_{22}\end{array} \right]$ is an (inverse) SL$(2,\mathbb{R})$ monodromy matrix and $M_{33} = \pm 1$. In light of \eqref{monoact}, \eqref{mono} can be decomposed into the component relations\footnote{Care has to be exercised here since our parametrization in footnote \ref{fn4} was for only one component of the OSp group. One can accommodate both components by having $\psi_3$ everywhere with a $\pm$ symbol in front of its expression.}
\begin{alignat}{2}
A_{\B}(\tau+ \beta) &= \frac{M_{21} + M_{22}A_{\B}(\tau)}{M_{11} + M_{12} A_{\B}(\tau)}, \qquad & A_{\F}(\tau+\beta ) &= \frac{A_{\F}(\tau)}{(M_{11} + M_{12} A_{\B}(\tau))^2}, \\
 \alpha_{\F}(\tau+\beta) &= \frac{M_{33} \alpha_{\F}(\tau)}{M_{11} + M_{12} A_{\B}(\tau)}, \qquad & \alpha_{\B}(\tau + \beta) &= \frac{M_{33} \alpha_{\B}(\tau)}{M_{11} + M_{12} A_{\B}(\tau)}. 
\end{alignat}
These monodromy relations are indeed realized by the reparametrization solution\footnote{To match these expressions, we should let $\theta \to - \theta$ as $\tau\to \tau+\beta$ in the sector where $\eta(\tau+\beta)=-\eta(\tau)$.}
\begin{align}
\tau'(\tau, \theta) &\equiv A(\tau, \theta) = F(\tau + \theta \eta(\tau)), \label{Areparam} \\
\theta'(\tau, \theta) &\equiv \alpha(\tau, \theta) = \sqrt{\partial_\tau F(\tau)} \left(\theta + \eta(\tau) + \frac{1}{2}\theta \eta(\tau) \partial_\tau \eta(\tau) \right) \label{alphareparam}
\end{align}
(compare to \eqref{SSsolution}) upon writing
\begin{equation}
F(\tau) = \begin{cases}
\tan \frac{\pi}{\beta} \Theta f(\tau) & \text{(elliptic)}, \\
\tanh \frac{\pi}{\beta} \Lambda f(\tau) & \text{(hyperbolic)}, \end{cases} \quad f(\tau + \beta) = f(\tau) +\beta, \quad \eta (\tau + \beta) = \pm \eta(\tau), 
\end{equation}
where periodicity for $\eta(\tau)$ gives the Ramond sector, and antiperiodicity realizes thermal fermionic boundary conditions, corresponding to the Neveu-Schwarz sector. For the par\-a\-bol\-ic defects (punctures), we instead have the relations
\begin{equation}
F(\tau+\beta) = F(\tau) + \beta, \qquad \eta(\tau+\beta) = \pm \eta(\tau).
\end{equation}
These relations can be summarized by the monodromy relations:
\begin{equation}
\label{perorbit}
F(\tau+ \beta) = M \cdot F(\tau), \qquad \eta(\tau+\beta)= \pm \eta(\tau),
\end{equation}
where $M$ is a (P)SL$(2,\mathbb{R})$ monodromy matrix.

This classification is closely related to the classification of coadjoint orbits of the super-Vi\-ra\-so\-ro group \cite{Bakas:1988mq, Delius:1990pt}; see \cite{Cotler:2018zff, Cardenas:2018krd} for recent partial treatments. The analogous case of bosonic JT gravity and coadjoint orbits of the Vi\-ra\-so\-ro group was discussed in Section 3 of \cite{Mertens:2019tcm} (see also Appendix F of \cite{Mefford:2020vde} for a review). We list in the final column of Table \ref{monodromies} the $\mathcal{N}=1$ Virasoro subalgebra that preserves the super-Schwarzian derivative \eqref{SuS}. This is the stabilizer of the corresponding super-Virasoro coadjoint orbit. This list hence identifies these Virasoro orbits directly with the solution classes of the super-Hill's equation. By definition, coadjoint orbits contain the pair $(T_{\B}(\tau),T_{\F}(\tau))$ by acting with the full super-Virasoro group on a single fixed element defining the specific orbit. In most cases, there exists an element within the orbit that has a constant value of $(T_{\B},T_{\F})$. Coadjoint orbits \emph{without} such a constant representative admit no solutions to the equation
\begin{equation}
\partial_\tau \operatorname{Sch}(\tau',\theta';\tau,\theta)=0.
\end{equation}
However, this equation is also the saddle equation of any super-Schwarzian model. Hence if we restrict to defects for which there is a classical (saddle) interpretation, then we care only about the constant-representative orbits, and we can restrict to the class of orbits catalogued in Table \ref{monodromies}. Assuming this restriction, and the periodicity conditions \eqref{perorbit}, we give the explicit derivation of the final column in Appendix \ref{app:coadj}.

We next discuss the different orbits from Table \ref{monodromies} in more detail.

The \textbf{hyperbolic orbit} has stabilizer $U(1) \times \mathbb{Z}_2$, which has two connected components denoted by the $\pm'$ sign choice in the table.

The \textbf{elliptic orbits} have stabilizer $U(1)$ (in $\posp$, the set of elliptic monodromy matrices and the corresponding stabilizers have a single component since one can set $\Theta \to \pi - \Theta$ or $\phi \to \pi - \phi$ to map the two would-be components into each other). When $\Theta \in \mathbb{N}$, the stabilizer gets enhanced and we reach the special elliptic orbits. For $\Theta = n$ even, we obtain the type I special elliptic orbits, and for $\Theta = n$ odd, we obtain the type II special elliptic orbits.

The \textbf{special elliptic orbits} have the largest stabilizer. The stabilizer $H_{\R}$, defined as the set of matrices satisfying $MS = SM$, is the full group OSp$(1|2,\mathbb{R})$ for type I, but is reduced to SL$(2,\mathbb{R}) \times \mathbb{Z}_2$ for type II. This corresponds to the orbits relevant for the Ramond sector. For the Neveu-Schwarz special elliptic orbit, the relevant stabilizer $H_{\NS}$ is different than $H_{\R}$ due to the nontrivial fermionic periodicity conditions. It is instructive to work this out a bit more explicitly within the super-Schwarzian orbit language. We present the analysis in Appendix \ref{app:detNS}. The upshot is that for $n$ odd, the fermionic variables in the stabilizer group $\osp$ flip sign when going once around the thermal circle, $\tau \to \tau + \beta$. This corresponds to how supersymmetry is implemented for the NS vacuum: the fermionic charges carry half-integer spin under rotations around the thermal circle \cite{Fu:2016vas}. In our language, this sign flip is implemented by giving the matrix $S$ in \eqref{Sdep} (which parametrizes precisely the $\osp$ redundancy) a weak $\tau$-dependence that ensures that the fermionic parameters flip sign as we rotate around the thermal circle: $S(\tau+\beta)= (-)^F S(\tau) (-)^F$. Combining this condition with \eqref{mono}, we get instead of \eqref{conjc} the equivalence relation $M(-)^F \sim SM (-)^F S^{-1}$, which defines the stabilizer $H_{\NS}$:
\begin{equation}
H_{\NS} = \left\{S \in \text{OSp}(1|2,\mathbb{R})\, | \, M(-)^F = SM (-)^F S^{-1}\right\}.
\end{equation} 
In the end, the presence of the $(-)^F$ effectively maps the analysis to the same one as in the Ramond sector but swapping the roles of type I and type II, leading to an OSp$(1|2,\mathbb{R})$ stabilizer $H_{\NS}$ for type II and a reduced $\sltr \times \mathbb{Z}_2$ stabilizer for type I.

Finally, the \textbf{parabolic orbit} has stabilizer $\mathbb{R}\times \mathbb{Z}_2$ (the noncompact version of U$(1)\times \mathbb{Z}_2$) for the NS puncture. In the Ramond case, the actual stabilizer supergroup is the (noncompact) subgroup of OSp$(1|2,\mathbb{R})$ generated by the (commuting) parabolic generators $E^+$ and $F^+$. We denote this subgroup by $\mathbb{R}^{1|1}\times \mathbb{Z}_2$, by analogy with the notation $\mathbb{R}$ for the noncompact version of U$(1)$. This enhancement of the stabilizer for the R punctures was noticed and studied in several works \cite{Penner_2019, Ip_2019, Stanford:2019vob}. It can be matched to the Ramond vacuum, where two zero modes exist with generators $G_0$ and $L_0 = G_0^2$.

Since the NS parabolic orbit can be obtained by taking the formal limit $n \to 0$ of the special elliptic orbits, the explicit analysis in Appendix \ref{app:detNS} demonstrates that the fermionic generators are periodic and do not pick up a minus sign upon rotation.\footnote{The sign function in \eqref{signproblem} always evaluates to $+1$ in this limiting case.} One can visualize this by realizing that for this orbit, moving along the thermal boundary is a translation instead of a rotation, which hence does nothing to spinors.

Within amplitudes, these different orbits can be accounted for by suitable \emph{defect} insertions. From a gauge-theoretic perspective, these insertions can be interpreted as characters of the principal series representations of OSp$(1|2,\mathbb{R})$. From the orbit perspective, they have an interpretation in terms of classical limits of super-Virasoro modular S-matrices. This is in complete parallel to the bosonic case \cite{Mertens:2019tcm}. Explanations of these statements, and explicit expressions for these defects, will be discussed later on in Section \ref{s:defglue}. We next provide a geometric interpretation of these different orbits/defects.

\subsection{Bulk Interpretation of Orbits (or Defects)}
\label{s:defint}

In order to achieve a bulk gravitational interpretation of these defects, we first briefly discuss the metric formulation of JT supergravity, referring to Appendix \ref{firstorderformalism} for more technical background on this formulation and its equivalence to the first-order formalism. We write the JT supergravity action in superspace as\footnote{We have reinstated Newton's constant: the action \eqref{N1action} differs from \eqref{BFbdy} by a factor of $-1/4\pi G$.}
\begin{equation}
S_\text{JT}^{\mathcal{N} = 1} = -\frac{1}{16\pi G}\left[\int_\mathcal{M} d^2 z\, d^2\theta\, E\Phi(R_{+-} + 2) + 2\int_{\partial \mathcal{M}} d\tau\, d\theta\, \Phi_b K\right].
\label{N1action}
\end{equation}
In superconformal gauge, we have
\begin{equation}
R_{+-} = 2e^{-\Sigma}D\bar{D}\Sigma
\end{equation}
for some bosonic superfield $\Sigma$, where $D\equiv \partial_\theta + \theta\partial_z$ and $\bar{D}\equiv \partial_{\bar{\theta}} + \bar{\theta}\partial_{\bar{z}}$.  The $\Phi$ equation of motion $R_{+-} = -2$ is then equivalent to the super-Liouville equation
\begin{equation}
D\bar{D}\Sigma + e^\Sigma = 0.
\end{equation}
The solution can be written as
\begin{equation}
e^\Sigma = \frac{D\theta'\bar{D}\bar{\theta}'}{z' - \bar{z}' - \theta'\bar{\theta}'},
\label{bulksuper}
\end{equation}
in terms of (anti)holomorphic bosonic superfields $z'(z, \theta)$, $\bar{z}'(\bar{z}, \bar{\theta})$ and fermionic superfields $\theta'(z, \theta)$, $\bar{\theta}'(\bar{z}, \bar{\theta})$ satisfying the constraints
\begin{equation}
Dz' = \theta'D\theta', \qquad \bar{D}\bar{z}' = \bar{\theta}'\bar{D}\bar{\theta}'.
\end{equation}
These constraints imply that $(z', \theta')$ and $(\bar{z}', \bar{\theta}')$ are (anti)holomorphic superconformal transformations of $(z, \theta)$ and $(\bar{z}, \bar{\theta})$.  They can be solved explicitly into
\begin{equation}
\label{bulkpar}
z' = F(z + \theta\eta(z)), \qquad \theta' = \sqrt{\partial F(z)}\left(\theta + \eta(z) + \frac{1}{2}\theta\eta(z)\partial\eta(z)\right)
\end{equation}
in terms of a bosonic function $F(z)$ and a fermionic function $\eta(z)$, and similarly for $\bar{z}'$ and $\bar{\theta}'$ in terms of $\bar{F}(\bar{z})$ and $\bar{\eta}(\bar{z})$ \cite{Fu:2016vas}.  We refer to the bulk supergeometry \eqref{bulksuper} as super-AdS$_2$, for any choice of $F, \eta$. The subset of superconformal transformations that act as isometries of the solution \eqref{bulksuper} comprises the super-M\"obius group $\posp$ \cite{Arvis:1982tq}.  The length element in superspace is defined as $d\mathbf{z} \equiv dz + \theta d\theta$ and transforms as $d\mathbf{z}' = (D\theta')^2 d\mathbf{z}$. The supermetric of the Poincar\'e super upper half-plane (SUHP) in superconformal gauge can then be written in several ways:
\begin{equation}
\label{supmetric}
ds^2 = g_{MN}dZ^M dZ^N = e^{2\Sigma} d\mathbf{z} \otimes d\bar{\mathbf{z}} = \frac{(D\theta')^2(\bar{D}\bar{\theta}')^2}{|z' - \bar{z}' - \theta'\bar{\theta}'|^2}|dz + \theta d\theta|^2 = \frac{|dz' + \theta' d\theta'|^2}{|z' - \bar{z}' - \theta'\bar{\theta}'|^2},
\end{equation}
where the primed coordinates are the super-Poincar\'e coordinates and the unprimed coordinates are some preferred (or proper) coordinates.

To state the bulk interpretation of defects in a natural fashion, it is convenient to first relate the super-Schwarzian dynamical time reparametrizations $(F,\eta)$ to the bulk gravitational description. This can be done in terms of the dynamics of the holographic ``wiggly'' boundary. Our discussion roughly follows the treatment of \cite{Forste:2017kwy}, which is in turn based on that for bosonic JT gravity \cite{Jensen:2016pah, Maldacena:2016upp, Engelsoy:2016xyb}.

The super-Poincar\'e boundary lies at $z' = \bar{z}'$, $\theta' = \bar{\theta}'$ and is of codimension $1|1$. We write $z'=\tau'+iy'$ and $\bar{z}' = \tau'-iy'$, where $\tau'$ and $y'$ are the super-Poincar\'e time and radial coordinates, respectively. We regularize the holographic boundary by moving it inward. Its location is specified in \emph{preferred} coordinates to be
\begin{equation}
\label{regbdy}
y = \epsilon, \qquad \theta=\bar{\theta},
\end{equation}
where we have defined $y \equiv \frac{z - \bar{z}}{2i}$. From the form of the supermetric, one immediately sees that preserving these asymptotics requires (compare to \cite{Forste:2017kwy})
\begin{equation}
\label{as}
z' - \bar{z}' - \theta' \bar{\theta}' = 2i\epsilon D\theta' \bar{D}\bar{\theta}' + \mathcal{O}(\epsilon^2),
\end{equation}
which is a single bosonic constraint on the bulk super-Poincar\'e coordinates. In fact, this choice of regularized boundary \eqref{regbdy} imposes both a bosonic and a fermionic constraint, which when combined imply the asymptotic relation \eqref{as}. Namely, the location \eqref{regbdy} in proper coordinates can be translated into a location in the Poincar\'e SUHP coordinates in terms of a wiggly curve specified by the functions
\begin{align}
\tau'(\tau,\theta), \qquad y'(\tau,\theta), \qquad \theta'(\tau,\theta), \qquad \bar{\theta}'(\tau,\theta).
\end{align}
These functions are given explicitly by:
\begin{alignat}{2}
\label{tauloc}
\tau' &={}& \centermathcell{\textstyle \frac{1}{2}(A(\tau + i\epsilon, \theta) + A(\tau - i\epsilon, \theta))} &= F(\tau + \theta\eta(\tau)) + \mathcal{O}(\epsilon^2), \\
y' &={}& \centermathcell{\textstyle \frac{1}{2i}(A(\tau + i\epsilon, \theta) - A(\tau - i\epsilon, \theta))} &= \epsilon ( (D\alpha)^2 - \alpha D^2\alpha ) + \mathcal{O}(\epsilon^2), \label{yloc} \\
\theta' &={}& \centermathcell{\alpha(\tau + i\epsilon, \theta)} &= \alpha + i\epsilon D^2\alpha + \mathcal{O}(\epsilon^2), \label{thloc} \\
\bar{\theta}' &={}& \centermathcell{\alpha(\tau - i\epsilon, \theta)} &= \alpha - i\epsilon D^2\alpha + \mathcal{O}(\epsilon^2), \label{thbloc}
\end{alignat}
where the functions on the right are given in \eqref{Areparam} and \eqref{alphareparam}, and we have written $\alpha \equiv \alpha(\tau, \theta)$ for brevity.  In particular, we use $A\equiv \tau'|_\partial$ and $\alpha\equiv \theta'|_\partial$ to distinguish between the reparametrized boundary coordinates and their bulk counterparts. Inserting \eqref{yloc}, \eqref{thloc}, and \eqref{thbloc} into the left-hand side of \eqref{as}, one indeed verifies the relation \eqref{as}.

\begin{figure}[!htb]
\centering
\includegraphics[width=0.25\textwidth]{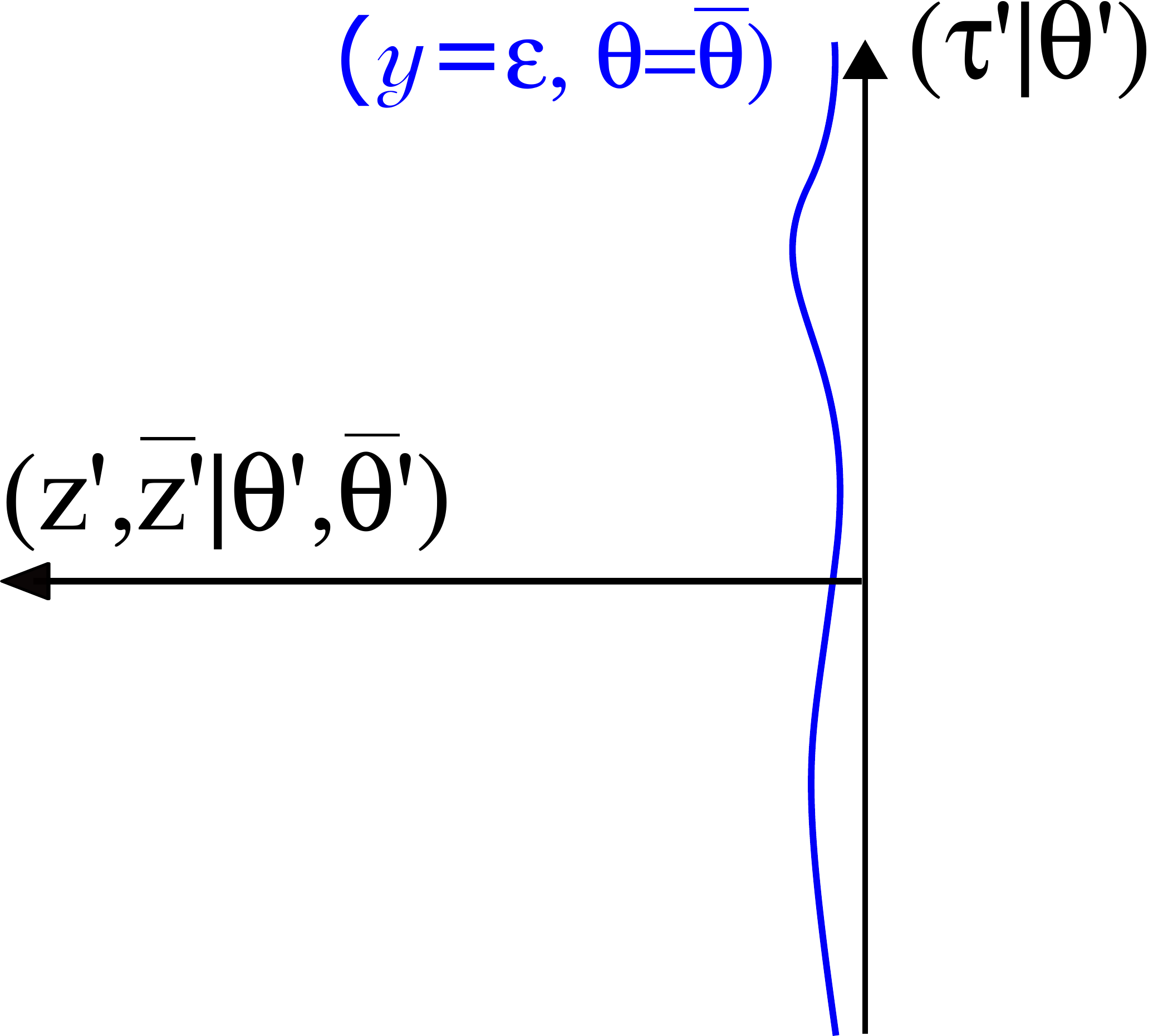}
\caption{Regularized holographic boundary at proper location \eqref{regbdy}. The actual shape of this $1|1$-dimensional boundary curve in super-Poincar\'e coordinates is described by the dynamical superreparametrization \eqref{tauloc}--\eqref{thbloc}, and is drawn as a wiggly curve.}
\label{superwiggly}
\end{figure}

Thus we obtain a $1|1$-dimensional curve embedded in the gravitational bulk (Figure \ref{superwiggly}). It was shown in \cite{Forste:2017kwy} that the dynamics governed by the boundary curve with these definitions is precisely the $\mathcal{N}=1$ super-Schwarzian.

\begin{figure}[!htb]
\centering
\includegraphics[width=0.55\textwidth]{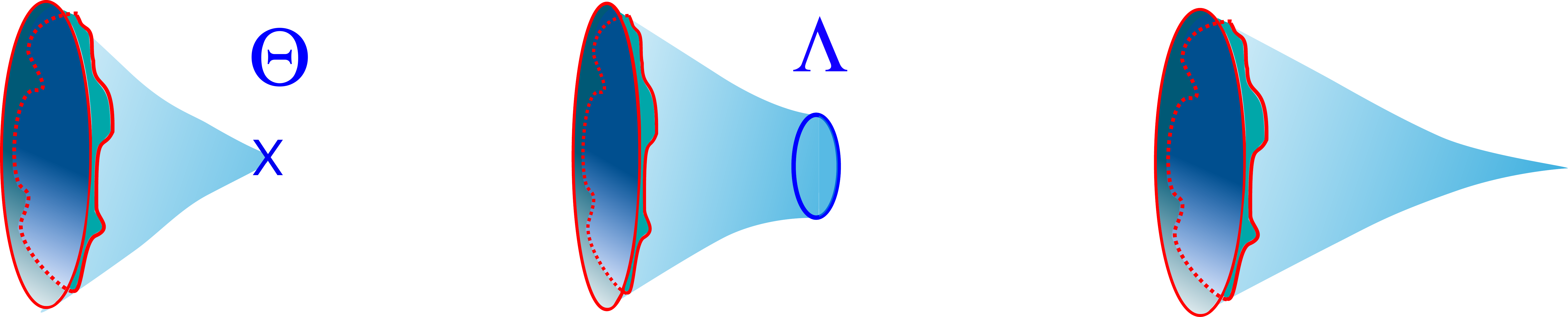}
\caption{Geometric interpretation of the different orbits. Left: elliptic orbit with a conical defect. Middle: hyperbolic orbit with a geodesic neck. Right: parabolic orbit with a cusp.}
\label{SUSYorbits}
\end{figure}

Given a certain off-shell boundary time reparametrization $\left(F(\tau), \eta(\tau) \right)$, one can naturally choose a bulk superframe that smoothly extrapolates this boundary frame into the bulk by using \eqref{bulkpar} and its antiholomorphic counterpart. Doing so leads to an off-shell bulk supergeometry
\begin{equation}
ds^2 = \frac{(D\theta')^2(\bar{D}\bar{\theta}')^2}{|z' - \bar{z}' - \theta'\bar{\theta}'|^2}|dz + \theta d\theta|^2
\end{equation}
with bosonic metric
\begin{equation}
ds^2 = \frac{\partial F(z) \bar{\partial}F(\bar{z})}{(F(z)-F(\bar{z}))^2} dz d\bar{z}.
\end{equation}
For the different monodromy classes, this bosonic submetric matches with the metric in bosonic JT gravity. Hence the interpretation there \cite{Mertens:2019tcm} immediately applies here as well:
\begin{itemize}
\item \textbf{Elliptic monodromies} with parameter $\Theta$ correspond to conical singularities with periodic identification $2\pi \Theta$. For integer $\Theta =n$, these correspond to replicated geometries. Unlike in bosonic JT gravity, we will need to make a distinction between even and odd values of $n$ when computing physical amplitudes since the stabilizer is not the same in these two cases.
\item \textbf{Hyperbolic monodromies} with parameter $\Lambda$ correspond to geometries with a wormhole of geodesic neck length $2\pi \Lambda$.
\item \textbf{Parabolic monodromies} correspond to geometries with a cusp at infinity. The periodic identification leads to a thermal AdS$_2$ geometry.
\end{itemize}
This classification is augmented by the fermionic boundary condition $\eta(\tau+\beta) = \pm \eta(\tau)$ (periodic or antiperiodic) for each class. The resulting geometries are illustrated in Figure \ref{SUSYorbits}.

\section{Bilocal Operators as Wilson Lines} \label{bilocal}

In this section, we utilize the explicit analysis of the gravitational boundary conditions in terms of super-Hill's equation, as presented in Section \ref{sect:hill}, to identify the super-Schwarzian bilocal operators \eqref{defbilocal} directly as boundary-anchored Wilson lines in the OSp$(1|2)$ formulation of JT supergravity. We augment this analysis by an explicit worldline path integral description of the Wilson line as a massive particle moving on the supermanifold.

First recall the identification of Wilson lines with bilocal operators purely within BF theory, starting with the disk for simplicity.  A Wilson line in the representation $R_j$ with boundary endpoints at $\tau_1$ and $\tau_2$ is given by
\begin{equation}
\mathcal{W}_j(\tau_1, \tau_2) = \mathcal{P}\exp\left[-\int_{\tau_1}^{\tau_2} R_j(\mathbf{A})\right].
\end{equation}
After integrating out the bulk scalar, which enforces the flatness of $\mathbf{A}$, we may freely deform the integration contour while preserving the endpoints to see that any such Wilson line is the unique solution to the one-dimensional initial value problem
\begin{equation}
\frac{d}{d\tau_2}\mathcal{W}_j(\tau_1, \tau_2) = -R_j(\mathbf{A}_\tau(\tau_2))\mathcal{W}_j(\tau_1, \tau_2), \quad \mathcal{W}_j(\tau_1, \tau_1) = 1.
\end{equation}
But for $\mathbf{A}$ flat (pure gauge on a disk), we have $\mathbf{A}_\tau = -\partial_\tau gg^{-1}$, so the bilocal operator
\begin{equation}
\mathcal{O}_{h=-j}(\tau_1, \tau_2) = R_j(g(\tau_2)g^{-1}(\tau_1))
\end{equation}
is also a solution to the same initial value problem.  Hence $\mathcal{W}_{j}(\tau_1, \tau_2) = \mathcal{O}_{h=-j}(\tau_1, \tau_2)$.  Sim\-i\-lar reasoning can be used to reduce a Wilson line on a more complicated topology (such as a Wilson line with endpoints on different boundary components) to a similar form, as long as $\mathbf{A}$ can be written as $-dgg^{-1}$ for a single-valued function $g$ along the support of the Wilson line and the boundary components---in other words, as long as the contour does not encircle a handle or a defect.  Otherwise, the topological class of the line becomes important.

In our case, $g|_\partial$ is further constrained by gravitational (super-Schwarzian) boundary conditions, and such bilocal operators can be viewed as boundary-to-boundary propagators of a bulk matter field coupled to JT gravity.

\subsection{Warmup: Finite Representations} \label{finitereps}

Let us first work out this interpretation for a Wilson line in the defining $j=1/2$ representation. This is a $2|1$-dimensional representation that has both lowest- and highest-weight states:
\begin{equation}
|{\textstyle \frac{1}{2}}\rangle = |\text{h.w.}\rangle, \qquad |0\rangle, \qquad |{\textstyle -\frac{1}{2}}\rangle = |\text{l.w.}\rangle,
\end{equation}
where
\begin{equation}
E^{-}|\text{l.w.}\rangle = F^- |\text{l.w.}\rangle = 0, \qquad E^{+}|\text{h.w.}\rangle = F^+ |\text{h.w.}\rangle = 0.
\end{equation}
In vector notation,
\begin{equation}
|\text{l.w.}\rangle = \left(\begin{array}{c} 
0 \\ 1 \\ \hline 0
\end{array} \right) 
\substack{\xrightarrow{2F^+} \\ \xleftarrow{-2F^-}} \left(\begin{array}{c} 
0 \\ 0 \\ \hline 1
\end{array} \right) \substack{\xrightarrow{2F^+} \\ \xleftarrow{2F^-}} \left(\begin{array}{c} 
1 \\ 0 \\ \hline 0
\end{array} \right) = |\text{h.w.}\rangle.
\end{equation}
Then the Wilson line in group theory language, by virtue of the identification \eqref{boundaryg}, can be written as the matrix element
\begin{equation}
\langle\text{l.w.}|g(\tau_2)g^{-1}(\tau_1)|\text{h.w.}\rangle = \psi_{1, \text{bot}}(\tau_2) \psi_{2, \text{bot}}(\tau_1) - \psi_{2, \text{bot}}(\tau_2)\psi_{1, \text{bot}}(\tau_1) + \psi_{3, \text{bot}}(\tau_2)\psi_{3, \text{bot}}(\tau_1).
\label{fiducialspin12osp}
\end{equation}
In fact, anticommuting the Grassmann parameters carefully shows that the matrix element between the states $(1 - 2\theta_1 F^-)|\text{h.w.}\rangle$ and $(1 - 2\theta_2 F^+)|\text{l.w.}\rangle$ results in a superspace bilocal operator:
\begin{align}
&\langle\text{l.w.}|(1 - 2(F^+)^\dag\theta_2)g(\tau_2)g^{-1}(\tau_1)(1 - 2\theta_1 F^-)|\text{h.w.}\rangle \\
&= \psi_1(\tau_2, \theta_2)\psi_2(\tau_1, \theta_1) - \psi_2(\tau_2, \theta_2)\psi_1(\tau_1, \theta_1) + \psi_3(\tau_2, \theta_2)\psi_3(\tau_1, \theta_1) = \frac{\tau'_1 - \tau'_2 - \theta'_1\theta'_2}{D_1\theta'_1 D_2\theta'_2}, 
\nonumber
\end{align}
where $\psi_j(\tau_i, \theta_i) = \psi_{j, \text{bot}}(\tau_i) + \theta_i\psi_{j, \text{top}}(\tau_i)$, $\tau'_i\equiv A(\tau_i, \theta_i)$, and $\theta'_i\equiv \alpha(\tau_i, \theta_i)$.

Thus Wilson lines between lowest- and highest-weight states yield standard Schwarzian or super-Schwarzian bilocal operators.  Other matrix elements yield bilocal operators that are more complicated in the Schwarzian language and that can be constructed from derivatives of the standard bilocal operators, as well as the Schwarzian derivative factors $T_{\B}(\tau)$ and $T_{\F}(\tau)$.  For example, defining $\Delta_\tau\equiv T_{\F}^{-1}(\tau)(\partial_\tau^2 - T_{\B}(\tau))$, we obtain the following result for $j = 1/2$:\footnote{We can identify bilocal operators with matrix elements of suitable group elements in the \emph{hyperbolic} basis (i.e., in a basis of eigenstates of the $\mathfrak{osp}(1|2)$ generator $H$).  Indeed, for the finite-dimensional bilocal operators considered here, $E^\pm$ are not diagonalizable. The eigenvalues of the generators, as well as properties like self-adjointness and diagonalizability, are representation-dependent. For instance, $E^\pm$ are nilpotent in finite-dimensional representations, but not necessarily so in infinite-dimensional representations.}
\begin{equation}
R_{1/2}(g(\tau_2)g^{-1}(\tau_1)) = \left[\begin{array}{ccc}
-\partial_{\tau_2} & -\partial_{\tau_2}\Delta_{\tau_1} & -\partial_{\tau_2}\partial_{\tau_1} \\
\Delta_{\tau_2} & \Delta_{\tau_2}\Delta_{\tau_1} & \Delta_{\tau_2}\partial_{\tau_1} \\
1 & \Delta_{\tau_1} & \partial_{\tau_1}
\end{array}\right]\langle{\textstyle -\frac{1}{2}}|g(\tau_2)g^{-1}(\tau_1)|{\textstyle \frac{1}{2}}\rangle
\end{equation}
in terms of the fiducial matrix element computed in \eqref{fiducialspin12osp}, where $R_{1/2}(g(\tau_2)g^{-1}(\tau_1))_{mm'} = \langle m|g(\tau_2)g^{-1}(\tau_1)|m'\rangle$ for $m, m' = \frac{1}{2}, 0, -\frac{1}{2}$.  The superspace bilocal operator is then the matrix element between the states $|{\textstyle \frac{1}{2}}\rangle - \theta_1|0\rangle$ and $|{\textstyle -\frac{1}{2}}\rangle - \theta_2|0\rangle$.  A quick proof of these relations, based on exploiting the supersymmetric Hill's equation, is presented in Appendix \ref{app:bilocal}.

The generalization to spin-$j$ representations is readily worked out, with the details again left to Appendix \ref{app:bilocal}. For example, one obtains for the mixed lowest/highest-weight matrix element:
\begin{equation}
\langle-j|g(\tau_2)g^{-1}(\tau_1)|j\rangle = [\psi_{1, \text{bot}}(\tau_2) \psi_{2, \text{bot}}(\tau_1) - \psi_{2, \text{bot}}(\tau_2)\psi_{1, \text{bot}}(\tau_1) + \psi_{3, \text{bot}}(\tau_2)\psi_{3, \text{bot}}(\tau_1)]^{2j},
\end{equation}
which is simply the appropriate power of \eqref{fiducialspin12osp}.  

Such operator insertions where $j\in \mathbb{N}$ are structurally unique in the super-Schwarzian model: they correspond to degenerate Virasoro representations, and their correlation functions are simpler than the other ones. Moreover, when coming from the minimal superstring, these operators correspond to the boundary tachyon vertex operators \cite{Mertens:2020pfe}. However, from a gravitational perspective, these operator insertions are somewhat unphysical, and a much more important role is played by the infinite-dimensional representations.

\subsection{Discrete Series Representations}

Our main interest lies in the infinite-dimensional lowest/highest-weight representations, which fall into a discrete series.  We call them the discrete representations.  Such representations are conveniently described in terms of a carrier space of functions on $\mathbb{R}^{1|1}$, with the group acting by super-M\"obius transformations.  We present the details in Appendix \ref{monomialreps}. The generators are written as differential operators acting on functions on $\mathbb{R}^{1|1}$:
\begin{gather}
\hat{H} = x\partial_x + \frac{1}{2}\vartheta\partial_\vartheta - j, \qquad \hat{E}^- = \partial_x, \qquad \hat{E}^+ = -x^2\partial_x - x\vartheta\partial_\vartheta + 2jx, \\
\hat{F}^- = \frac{1}{2}(\partial_\vartheta + \vartheta\partial_x), \qquad \hat{F}^+ = -\frac{1}{2}x\partial_\vartheta - \frac{1}{2}x\vartheta\partial_x + j\vartheta. \label{borelweilwil}
\end{gather}
We will come across this realization of the $\mathfrak{osp}(1|2)$ superalgebra several more times.

For a discrete representation, the bra and ket wavefunctions on the superline $\mathbb{R}^{1|1}$ are
\begin{equation}
\langle x, \vartheta|\text{h.w.}\rangle_j = x^{2j} = \frac{1}{x^{2h}}, \qquad \langle\text{l.w.}|x, \vartheta\rangle_j = \vartheta\delta(x) = \delta(x, \vartheta),
\end{equation}
where $j=-h<0$ and $h$ is the conformal weight of the local boundary operators \cite{Fitzpatrick:2016mtp, Hikida:2017ehf, Hikida:2018eih}. We thus have the Wilson line
\begin{equation}
\label{tocom}
\langle\text{l.w.}|g(\tau_2)g^{-1}(\tau_1)|\text{h.w.}\rangle = \int dx\, d\vartheta\, \delta(x, \vartheta) g(\tau_2)g^{-1}(\tau_1) x^{2j}.
\end{equation}
The group element itself is conveniently written in the Gauss-Euler form as
\begin{equation}
\label{GE}
g^{-1}(\phi, \gamma_{\m}, \gamma_{\+} | \theta_{\m}, \theta_{\+}) = e^{2\theta_{\m} \hat{F}^-}e^{\gamma_{\m} \hat{E}^-}e^{2\phi \hat{H}}e^{\gamma_{\+} \hat{E}^+}e^{2\theta_{\+} \hat{F}^+},
\end{equation}
where we identify from \eqref{boundaryg} the parameters
\begin{equation}
\label{paride}
e^\phi = \psi_{1, \text{bot}}, \quad \gamma_{\m} = \frac{\psi_{2, \text{bot}}}{\psi_{1, \text{bot}}}, \quad \gamma_{\+} = \frac{\psi_{1, \text{bot}}'}{\psi_{1, \text{bot}}}, \quad \theta_{\m} = \frac{\psi_{3, \text{bot}}}{\psi_{1, \text{bot}}}, \quad \theta_{\+} = \frac{\psi_{1, \text{top}}}{\psi_{1, \text{bot}}}.
\end{equation}
Using \eqref{GE} with the parameters \eqref{paride} and the Borel-Weil generators \eqref{borelweilwil}, we compute the successive applications of group elements as follows:
\begin{align}
x^{2j} &\xrightarrow{g^{-1}(\tau_1)} (\psi_{1, \text{bot}}(\tau_1)x + \psi_{2, \text{bot}}(\tau_1) + \psi_{3, \text{bot}}(\tau_1)\vartheta)^{2j} \\*
&\xrightarrow{g(\tau_2)} ((-\psi_{1, \text{bot}}'(\tau_2)\psi_{2, \text{bot}}(\tau_1) + \psi_{2, \text{bot}}'(\tau_2)\psi_{1, \text{bot}}(\tau_1) - \psi_{3, \text{bot}}'(\tau_2)\psi_{3, \text{bot}}(\tau_1))x \nonumber \\*
&\hspace{1.5 cm} + \psi_{1, \text{bot}}(\tau_2)\psi_{2, \text{bot}}(\tau_1) - \psi_{2, \text{bot}}(\tau_2)\psi_{1, \text{bot}}(\tau_1) + \psi_{3, \text{bot}}(\tau_2)\psi_{3, \text{bot}}(\tau_1) \nonumber \\*
&\hspace{1.5 cm} + (-\psi_{1, \text{top}}(\tau_2)\psi_{2, \text{bot}}(\tau_1) + \psi_{2, \text{top}}(\tau_2)\psi_{1, \text{bot}}(\tau_1) - \psi_{3, \text{top}}(\tau_2)\psi_{3, \text{bot}}(\tau_1))\vartheta)^{2j} \\
&\xrightarrow{\text{$x = 0$, $\vartheta = 0$}} (\psi_1(\tau_2, \theta_2)\psi_2(\tau_1, \theta_1) - \psi_2(\tau_2, \theta_2)\psi_1(\tau_1, \theta_1) + \psi_3(\tau_2, \theta_2)\psi_3(\tau_1, \theta_1))^{2j}|_\text{bot} \\
&= \left(\frac{\tau'_1 - \tau'_2 - \theta'_1\theta'_2}{D_1\theta'_1 D_2\theta'_2}\right)^{2j}\bigg|_\text{bot},
\end{align}
where in the end, we set $x=0$ and $\vartheta=0$ as imposed by the $\delta(x, \vartheta)$ bra wavefunction in \eqref{tocom}.  The steps are analogous to those in Appendix I of \cite{Blommaert:2018iqz}.  The recipe for computing more general matrix elements in the $(x, \vartheta)$ basis is described in Appendix \ref{monomialreps}. Since these more general matrix elements have not been systematically studied even in bosonic JT gravity, we also present the bosonic results in Appendix \ref{app:boscor}.

When $j=-h < 0$, we directly reproduce the bilocal operators in the super-Schwarzian theory. Hence the boundary-anchored Wilson lines with suitable representation indices for the bra and ket labels (as explained above) correspond to the components of the superspace bilocal operator \eqref{defbilocal}.

In the next subsection, we supplement this description with an intuitive first-quantized picture of the Wilson line. Unlike the current treatment, in which we compute the different Grassmann components of the bilocal operator separately, the procedure discussed next will immediately give the superspace description of the Wilson line.

\subsection{Gravitational Wilson Lines and Geodesics}
\label{s:gravwilson}

We claim that the following Euclidean worldline description in superspace constructs the $\mathcal{N}=1$ su\-per-Schwarzian bilocal operator:
\begin{equation}
\label{worldlinesuper}
\int \left[\mathcal{D}Z\right] e^{- m \int_{\tau_1}^{\tau_2} ds\, (g_{MN} \dot{Z}^M\dot{Z}^N)^{1/2}} = \left(\frac{D\theta'_1 D\theta'_2}{\tau'_1 - \tau'_2 - \theta'_1\theta'_2}\right)^{2h},
\end{equation}
where $m^2 = h(h - 1/2)$, $Z^M = z, \bar{z}, \theta, \bar{\theta}$ are coordinates spanning the $2|2$-dimensional supermanifold with metric \eqref{supmetric}, and $D\equiv \partial_\theta + \theta\partial_\tau$ (whether $D$ refers to the 1d superderivative or to the holomorphic 2d superderivative should be clear from context). Notice that $m^2 = j(j+1/2)$ equals the Casimir operator in the spin-$j$ representation.\footnote{This identification is not surprising if one thinks of it as a consequence of the massive Klein-Gordon equation on the $2|2$-dimensional Poincar\'e SUHP, which is the homogeneous space $\text{OSp}(1|2, \mathbb{R})/\text{U}(1)$.  The Casimir is the eigenvalue of the Laplacian on the $3|2$-dimensional group $\text{OSp}(1|2, \mathbb{R})$, and equals the parameter $m^2$ in the Klein-Gordon equation.} 

The worldline path integral in \eqref{worldlinesuper} is taken along all trajectories superdiffeomorphic to the boundary segment between both endpoints $(\tau_1, \theta_1)$ and $(\tau_2, \theta_2)$ in superspace, and the primed coordinates on the left are understood in the sense of \eqref{Areparam} and \eqref{alphareparam}. The proof of \eqref{worldlinesuper} is a direct generalization of the construction of \cite{Iliesiu:2019xuh}, and is based on earlier accounts in 3d pure gravity \cite{Ammon:2013hba, Castro:2018srf}. The details are presented in Appendix \ref{app:orbits}.

Here, we instead choose to present a more physical discussion by comparing both sides in the limit of large weight $h$, where the geodesic approximation holds.

\begin{figure}[!htb]
\centering
\includegraphics[width=0.3\textwidth]{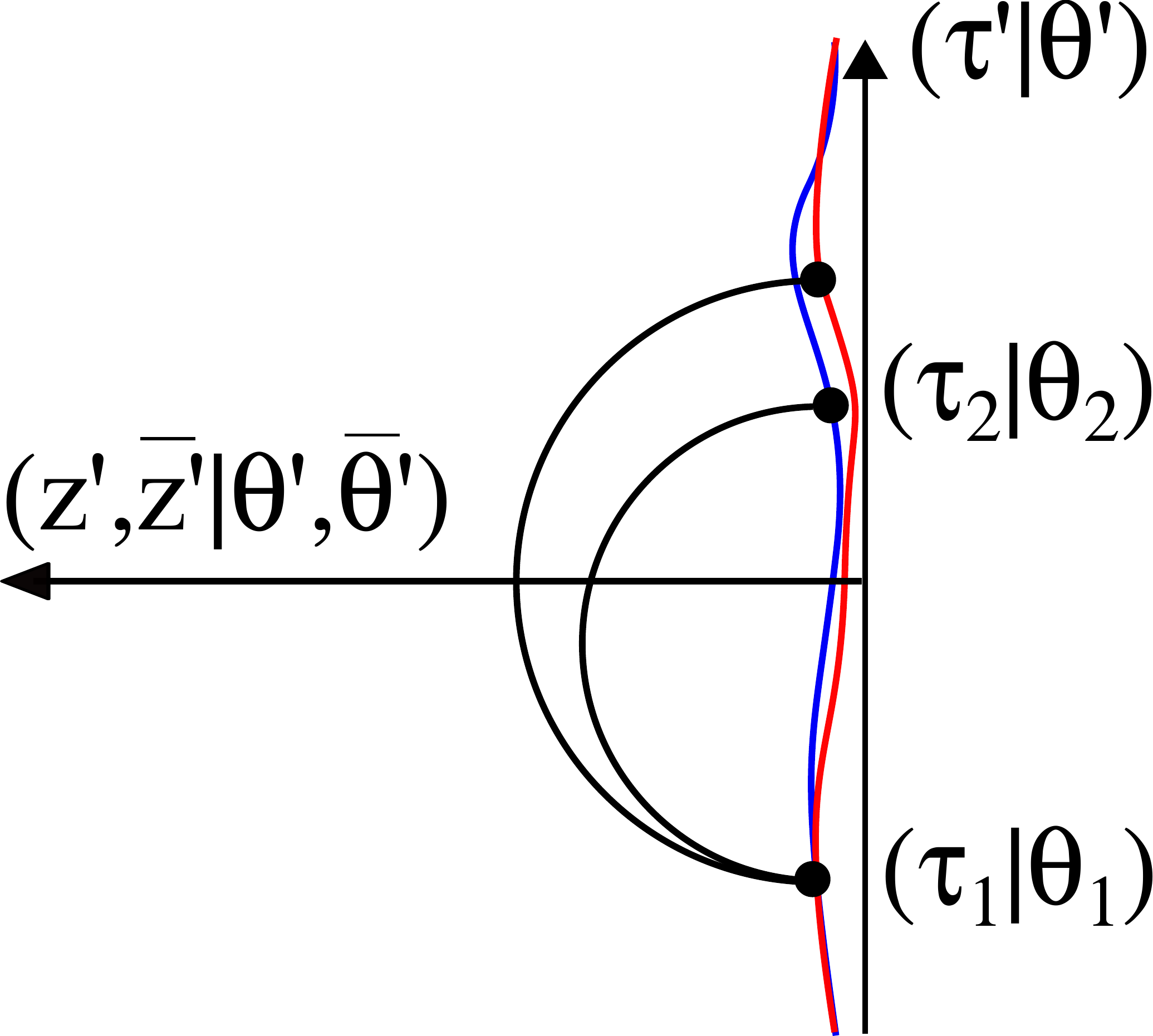}
\caption{Geodesics between two boundary endpoints in superspace.  The coordinates with primes are Poincar\'e SUHP coordinates.  The coordinates without primes are ``proper'' or preferred coordinates on the boundary superclock. The boundary is $1|1$-dimensional, while the bulk is $2|2$-dimensional. The bulk geodesics are $1|0$-dimensional.}
\label{geodistancesuper}
\end{figure}

A bulk geodesic in superspace is a curve of dimension $1|0$, and hence of codimension $1|2$ on a $2|2$-dimensional super-Riemann surface. It describes a trajectory $(z(s), \bar{z}(s), \theta(s), \bar{\theta}(s))$.  One can think of the wiggly boundary curve defined in Section \eqref{s:defint}, as infinitesimally ``thickened'' in the Grassmann direction $\theta$, whereas the bulk geodesics have no such thickening.\footnote{When using geodesic boundaries, one imagines these to be thickened into $1|2$ curves using the leaves from $D$ and $\bar{D}$, as explained in \cite{Stanford:2019vob}.}  This is illustrated in Figure \ref{geodistancesuper}.

For two endpoints $(\tau_1, \theta_1)$ and $(\tau_2, \theta_2)$ on the holographic wiggly boundary, for which according to \eqref{regbdy} $\theta_i = \bar{\theta}_i$ and $z_i = \bar{z}_i + 2i\epsilon$, one can compute the geodesic distance $d$ in the supermetric \eqref{supmetric} to be \cite{Matsumoto:1989hc}
\begin{equation}
\cosh d = 1 + \frac{|\tau'_1 - \tau'_2 - \theta'_1\theta'_2|^2}{2(z'_1 - \bar{z}'_1 - \theta'_1\bar{\theta}'_1)(z'_2 - \bar{z}'_2 - \theta'_2\bar{\theta}'_2)} = 1 + \frac{|\tau'_1 - \tau'_2 - \theta'_1\theta'_2|^2}{2\epsilon^2(D\theta'_1)^2 (D\theta'_2)^2},
\end{equation}
which is approximated by the formula
\begin{equation}
\label{aprogeod}
d \approx  \ln \frac{|\tau'_1 - \tau'_2 - \theta'_1\theta'_2|^2}{(D\theta'_1)^2 (D\theta'_2)^2} - \ln \epsilon^2
\end{equation}
for small $\epsilon$.  After subtracting the divergent term, we see that inserting \eqref{aprogeod} into the saddle-point approximation for the right-hand side of \eqref{worldlinesuper} indeed reproduces the left-hand side of \eqref{worldlinesuper} for large values of $h$ where $h\approx m$.  Note that the one-loop exactness of the worldline path integral would suggest that the finite correction to $m^2$ that results in the Casimir $h(h - 1/2)$ comes from evaluating the one-loop determinant.

Equation \eqref{aprogeod} is an expression in superspace, which can be expanded in the Grassmann variables. The bottom component $d_{\text{bot}}$ is interpretable as the geodesic distance in the bosonic submanifold, and this interpretation will be put to use further on in Section \ref{sect:complex} to compute the boundary-to-boundary wormhole length, including quantum gravitational corrections.

\section{Gravity as a Gauge Theory: Semigroup Structure}
\label{semigroup}

As discussed up to this point, the supergroup OSp$(1|2,\mathbb{R})$ suffices to describe the ``local'' dynamics of $\mathcal{N}=1$ JT supergravity.  In this section, we provide evidence that understanding the full quantum dynamics (and in particular, the precise form of the amplitudes) requires a refinement of this group-theoretic structure. The ultimate reason for this refinement is the discrepancy between gauge theory and gravity, as discussed in the introduction. We will argue that a natural way to implement this transfer is the proposal that gravity is in fact described by the semigroup OSp$^+(1|2,\mathbb{R})$ (which we will define in Section \ref{s:subsemi}).

\subsection{Motivation: Bosonic Winding Sectors on the Disk}

To motivate the discrepancy between full-fledged BF gauge theory and gravity, we present an insightful argument in the case of bosonic JT gravity. For completeness, we review the bosonic story in Appendix \ref{app:bosonic}, with the relevant group theory for $\sltr$ and the positive subsemigroup $\slr$ discussed in Appendices \ref{app:ha} and \ref{app:slr}, respectively.

The bosonic Schwarzian path integral that emerges from the $\sltr$ group structure is given by the Euclidean action
\begin{equation}
S = - \frac{1}{2}\int_{0}^{\beta}d\tau \left\{F,\tau\right\},
\end{equation}
integrated over all functions $F$ satisfying the monodromy constraint $F(\tau+\beta) = M \cdot F(\tau)$. Imposing trivial monodromy $M = \mathbf{1}$, we can reparametrize $F$ in a one-to-one fashion by writing $F(\tau) = \tan \smash{\frac{\pi}{\beta}}f(\tau)$ and allowing $f(\tau+\beta) = f(\tau) + n\beta$ for $n$ ranging over all positive integers.\footnote{The integer $n$ must be positive to satisfy the condition $F'\geq 0$ coming from $e^{2\phi} = F'$, where $f' \geq 0$ without loss of generality.} To isolate the actual vacuum orbit of bosonic JT gravity, we would further fix $n$ to a single value ($n=1$) and consider that particular theory. This further constraint defines gravity. Here, we drop this constraint, instead insisting on remaining one-to-one with the group-theoretic $\sltr$ structure.

Hence we are led to consider the following partition function, which has the asymptotic (gravitational) boundary conditions that produce the Schwarzian action, but not the $n=1$ winding constraint:
\begin{equation}
Z(\beta) = \sum_{n=1}^\infty\int_{f(\tau+\beta)=f(\tau)+\beta} \left[\mathcal{D}f\right] e^{\frac{1}{2} \int_{0}^{\beta}d\tau \left\{\tan \frac{\pi}{\beta}n f,\tau\right\}}.
\end{equation}
Each of the terms is one-loop exact, but care has to be taken for relative minus signs. At the one-loop level, upon plugging in $f(\tau) = \tau + \epsilon(\tau)$ and expanding in $\epsilon$, one finds $2(n-1)$ negative modes. Since the one-loop determinant is given by $\left(\det\mathcal{O}\right)^{-1/2}$ in terms of the quadratic operator $\mathcal{O}$, each pair of negative modes gives a factor of $-1$, leading to a total factor of $(-)^{n-1}$. Incorporating this minus sign into the Schwarzian answer of \cite{Mertens:2019tcm}, we find:
\begin{equation}
Z(\beta) = \sum_{n=1}^{\infty} (-)^{n-1}n \left(\frac{\pi}{\beta}\right)^{3/2}e^{\frac{\pi^2}{\beta}n^2} = \int_0^\infty dk \left(2\sum_{n=1}^{\infty} (-)^{n-1} k \sinh(2\pi n k)\right) e^{- \beta k^2}.
\end{equation}
The quantity in parentheses diverges, but we can give meaning to it by using the limit $q\to 0^+$ of the regularized expression\footnote{This same formula was written down in \cite{Yang:2018gdb} and interpreted as well in terms of multi-wound particle trajectories. We will see that the significance of the multi-wound paths is precisely that they encode the distinction between gauge theory and gravity.}
\begin{equation}
2\sum_{n=1}^{\infty} (-)^{n-1} e^{-2\pi q n } k \sinh(2\pi n k) = \frac{k \sinh 2 \pi k }{\cosh 2 \pi q + \cosh 2 \pi k}, \qquad k < q.
\end{equation}
For $q > 0$, the left-hand side converges in a nonempty strip, allowing it to be analytically continued to arbitrary $k$ via the right-hand side.  We then obtain\footnote{We can obtain the other principal series representations of $\sltr$ by dropping the minus signs for the negative modes and then using the related identity
\begin{equation}
2\sum_{n=1}^{\infty} e^{-2\pi q n } k \sinh(2\pi n k) = \frac{k \sinh 2 \pi k }{\cosh 2 \pi q - \cosh 2 \pi k}, \qquad k < q.
\end{equation}
This instead leads to the expression $Z(\beta) = \int_0^\infty dk \left( k \coth\pi k \right) e^{- \beta k^2}$.}
\begin{equation}
Z(\beta) = \int_0^\infty dk \left( k \tanh\pi k \right) e^{- \beta k^2},
\end{equation}
which is the partition function for the gravitational coset of $\sltr$ with measure $d\mu(k)\equiv dk\, k \tanh \pi k$.\footnote{The meaning of the word ``coset'' here is that we have implemented the gravitational boundary conditions at the boundary of the disk, reducing the boundary dynamics to the Schwarzian model rather than that of a particle on the $\sltr$ group manifold. We give more details on this implementation in the supersymmetric case in Section \ref{applications}.} As reviewed in Appendix \ref{app:ha}, this measure is indeed the Plancherel measure for the principal series representations of $\sltr$, and this is \emph{not} JT gravity: it would imply a density of states $\rho(E)\sim \tanh(\pi\sqrt{E})$ ($E > 0$). The bulk disk geometries corresponding to this summation have conical identifications of $2\pi n$ (they are replicated geometries), all of which save for $n=1$ carry conical singularities. Restricting this by hand to the \emph{smooth} hyperbolic component of the moduli space of configurations, one would land solely on $n=1$, and obtain the measure $d\mu(k)\equiv dk\, k \sinh(2\pi k)$ and hence the $\rho(E)\sim \sinh(2\pi\sqrt{E})$ density of states of JT gravity.

What this argument teaches us is that an additional constraint must be imposed on the BF theory to make contact with smooth gravity. In \cite{Blommaert:2018iqz}, evidence was provided that a particularly natural way to accomplish this is to restrict the group to the positive semigroup $\slr$. The calculation of the Plancherel measure for the principal series representations of $\slr$ is reviewed in Appendix \ref{app:slr}, leading to $d\mu(k) = dk\, k \sinh(2\pi k)$. In the next section, we will provide evidence that a similar construction works for supergravity in terms of the positive semigroup OSp$^+(1|2)$.\footnote{In \cite{Blommaert:2018iqz}, the term ``subsemigroup'' is used to emphasize that the semigroup $\slr$ is a subset of $\sltr$.  The corresponding term here would be ``subsupersemigroup,'' but we will often opt for ``subsemigroup'' to reduce verbiage.} 

It would be interesting to understand the application of the above winding argument directly for OSp$(1|2,\mathbb{R})$, which we postpone to future work.

\subsection{\texorpdfstring{$\text{OSp}^+(1|2, \mathbb{R})$}{OSp+(1|2, R)} Subsemigroup}
\label{s:subsemi}

It is known that $\mathcal{N}=1$ JT supergravity amplitudes contain the density of states \cite{Stanford:2017thb}
\begin{equation}
\label{JTdos}
\rho(E) \sim \frac{1}{\sqrt{E}}\cosh 2 \pi \sqrt{E}.
\end{equation}
This profile is constrained by physical arguments in the following way. First, it has the large-$E$ Bekenstein-Hawking growth $\rho(E) \approx e^{2\pi \sqrt{E}}$, matching the semiclassical black hole first law in JT supergravity: $S(E) = 2 \pi \sqrt{E}$. This is precisely the same first law as in the bosonic JT model \cite{Almheiri:2014cka}, because the fermions are turned off in the classical black hole solution. Second, it has a pole $\rho(E) \sim 1/\sqrt{E}$ as $E \to 0$. This is as expected, since the corresponding supercharge density $\rho(Q) \sim 1$, with $E = Q^2$, is then regular as $E \to 0$. This is also the same pole as the ``hard wall'' in the random matrix ensembles describing the very low-energy spectral statistics of $\mathcal{N}=1$ supergravity models.\footnote{The ``Bessel model'' plays the same role here as the Airy model does for bosonic JT gravity: it is an exactly solvable matrix model in the suitable universality class that describes the leading behavior of JT (super)gravity very close to the spectral edge $E=0$.}

Now, if JT supergravity were indeed described globally by OSp$(1|2,\mathbb{R})$ BF theory, then the above density of states \eqref{JTdos} would match precisely with the Plancherel measure on the space of irreps appearing in the Plancherel/Peter-Weyl decomposition of functions on the group manifold. Since the above density of states is continuous, it would need to match the Plancherel measure on the principal series representations of OSp$(1|2,\mathbb{R})$.\footnote{More explicitly, upon introducing the OSp$(1|2,\mathbb{R})$ spin label $j=-1/4+ik/2$ and the momentum variable $k\in \mathbb{R}^+$, the spacetime energy in BF models is identified with the Casimir eigenvalue $E = j(j+1/2)-1/16 = k^2/4$, where we chose to shift away the zero-point energy $1/16$.} However, this is not the case. Since the Plancherel measure on the principal series representations of OSp$(1|2,\mathbb{R})$ seems to be unavailable in the mathematics literature, we set out to find it in Appendix \ref{osprep}. In particular, we construct the principal series representations from first principles using parabolic induction in \ref{app:ps}, and compute the corresponding Plancherel measure in \ref{app:whit}. We obtain the result
\begin{equation}
\rho(E) \sim  \frac{1}{\sqrt{E}}\frac{\cosh 2 \pi \sqrt{E}}{1 + \cosh 2 \pi \sqrt{E}},
\end{equation}
which does not match the JT supergravity answer \eqref{JTdos}. In particular, we find the large-$E$ power-law asymptotics $\rho(E) \sim 1/\sqrt{E}$. In fact, we will argue in Section \ref{discussion} that the large-argument behavior of the Plancherel measure of any semisimple Lie (super)group takes the following form:
\begin{equation}
\rho(k) \sim k^{\left|\Delta_B^+\right| - \left|\Delta_F^+\right|},
\end{equation}
where the exponent is the number of positive bosonic roots minus the number of positive fermionic roots. This behavior immediately rules out the Plancherel measure on the space of principal series representations of \emph{any} Lie (super)group as a candidate for the physical density of states of black holes.

To find the correct structure, we take guidance from how the bosonic JT gravity model is related to SL$(2,\mathbb{R})$. In \cite{Blommaert:2018iqz}, it was argued that the bulk theory should be regarded as a BF theory of a subsemigroup of SL$(2,\mathbb{R})$.\footnote{In \cite{Iliesiu:2019xuh}, a different proposal was made in terms of a parametric limit of the universal cover of $\sltr$. It would be interesting to develop the superanalogue of that story as well, and to compare the two approaches in the supersymmetric case.} This is the subset of SL$(2,\mathbb{R})$ matrices for which all entries are positive in the defining representation:
\begin{equation}
\slr\equiv \left\{\left(\begin{array}{cc} a & b \\ c & d \end{array}\right), \quad ad - bc = 1, \quad a, b, c, d > 0\right\}.
\end{equation}
This subset is closed under multiplication, but not under taking inverses. It hence defines a semigroup that is a subset of a group, hence the name subsemigroup. It was shown in \cite{Blommaert:2018iqz} that if one subscribes to this structure, then one can find the correct density of states. For convenience, the argument is repeated in Appendix \ref{app:slr}.

A key motivation for following this approach is its deep relation with the theory of quantum groups. Somewhat surprisingly, the latter has been studied in much more depth than its classical limit. Let us review the argument. Structurally, the subsemigroup appears due its nice representation-theoretic properties.  In particular, the principal series representations $P_k$ are the only ones appearing in the Plancherel decomposition:
\begin{equation}
\label{pldecom}
\mathfrak{L}^2\left( \text{SL}^+(2,\mathbb{R})\right) = \int_{\oplus} d\mu(k) \, P_k \otimes P_k, \qquad d\mu(k) = dk\, k \sinh(2\pi k).
\end{equation}
This formal equation can be derived by taking a classical $q \to 1$ limit of the results of Ponsot and Teschner in terms of the set of so-called self-dual representations of the Faddeev modular double of U$_q (\mathfrak{sl}(2,\mathbb{R}))$ \cite{Ponsot:1999uf, Ponsot:2000mt}. When writing $q= e^{\pi i b^2}$, self-duality implies that the representation is simultaneously a representation of the dual quantum group with $b\to 1/b$. The $q$-deformed version of this statement was later rigorously derived in \cite{Ip}, and moreover conjectured to hold for the $q$-deformed positive subsemigroup of any simple Lie group \cite{Ip_2020}.

The story for JT gravity on its own still deserves much more investigation, but for the moment, we will accept it and attempt to see whether something similar could be true for supergravity.

We now define the analogous group-theoretic structure for the supersymmetric situation of interest in this work. The subsupersemigroup is defined in the defining representation of OSp$(1|2,\mathbb{R})$ by having $a,b,c,d > 0$ and no restriction on the Grassmann variables. The quantities $a,b,c,d$ are supernumbers, and their positivity properties are defined in Appendix \ref{app:supernumbers}. In particular, a supernumber is positive iff its body is positive.\footnote{This leaves the set of pure soul supernumbers undetermined in terms of positivity. Since this set is of measure zero in the set of all supernumbers, we will not care what positivity means in this case.} The intuition behind this definition is that Grassmann combinations should be thought of as infinitesimal compared to the purely bosonic variables. Under composition of semigroup elements $g_1 \cdot g_2$, we find that the new entries $a,b,c,d$ again all have positive bodies, and hence this positivity restriction indeed defines a semigroup:
\begin{equation}
\text{OSp}^+(1|2, \mathbb{R}) \equiv \left\{\left[\begin{array}{cc|c}
a & b & \alpha \\
c & d & \gamma \\ \hline
\beta & \delta & e
\end{array}\right]\in \text{OSp}(1|2, \mathbb{R}), \quad a,b,c,d > 0 \right\}.
\end{equation}
In this light, we make a conjecture similar to \eqref{pldecom} above that for the supergroup case,
\begin{equation}
\label{PlOsp}
\mathfrak{L}^2\left( \text{OSp}^+(1|2,\mathbb{R})\right) = \int_{\oplus} d\mu(k) \, P_k \otimes P_k, \qquad d\mu(k) = dk \cosh (\pi k),
\end{equation}
where only the principal series representations $P_k$ appear in the direct integral, and the measure reflects the correct gravitational density of states \eqref{JTdos}. We will use the semigroup approach in Section \ref{sect:plmeasu} to derive this Plancherel measure for $\text{OSp}^+(1|2,\mathbb{R})$, identifiable as the gravitational density of states $\rho(E) \sim \frac{1}{\sqrt{E}}\cosh 2 \pi \sqrt{E}$.

We first present several pieces of evidence in favor of the above conjecture \eqref{PlOsp}.

Firstly, it was shown in \cite{Hadasz:2013bwa} that the class of representations $P_k$ is self-dual in the setting of quantum supergroups, mirroring the statement in the bosonic case.

Secondly, we can solve the Casimir eigenvalue problem in the relevant subsector of the supergroup manifold for the subsemigroup $\text{OSp}^+(1|2,\mathbb{R})$. This is done in Appendix \ref{app:regu}, and in particular in \eqref{Cassubsemi}, where one can prove that only the principal series representations appear. The discrete representations of OSp (which figure in the Plancherel decomposition of the full supergroup $\text{OSp}(1|2,\mathbb{R})$) come from a different sector, beyond the subsemigroup.

\begin{figure}[!htb]
\centering
\includegraphics[width=0.8\textwidth]{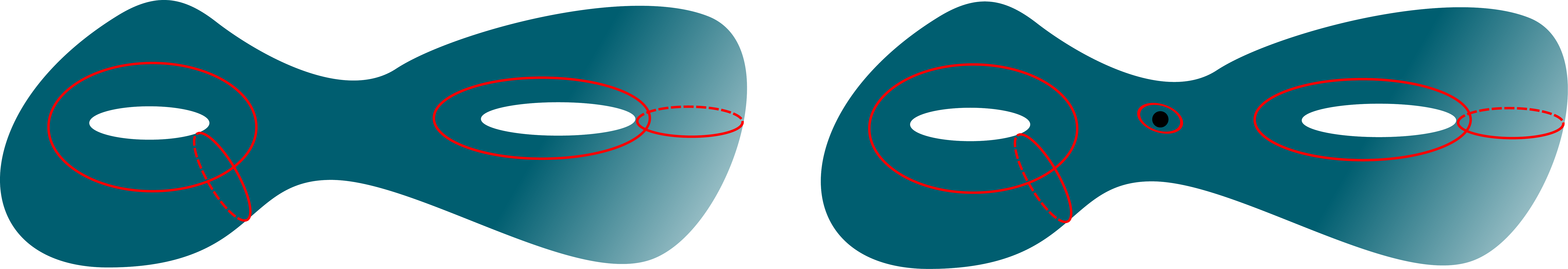}
\caption{On the left is a genus-two surface with four cycles shown in red. Each independent cycle gets an OSp$(1|2)$ holonomy group element, with a single relation between them: $(A_1B_1^{-1}A_1^{-1}B_1)(A_2B_2^{-1}A_2^{-1}B_2) = 1$. If each cycle is in the hyperbolic conjugacy class, then the surface is smooth. On the right is a genus-two surface with five cycles, one of which is a conical defect shown as a black dot. This surface is not smooth since one of the cycles has an elliptic holonomy matrix associated to it. It has to be excluded from the gravity path integral configurations, but is present in the OSp$(1|2)$ BF formulation.}
\label{ospcycle}
\end{figure}

A final suggestive argument in favor of the subsemigroup description comes from thinking about the BF formulation of the supergravity model on an arbitrarily complicated 2d super-Riemann surface $\Sigma$, possibly with geodesic boundaries. Performing the path integral over $\mathbf{B}$ reduces the amplitude to an integral over the moduli space of all flat connections $\mathbf{F}=0$ on $\Sigma$. Since a flat connection is specified by its holonomy around each nontrivial cycle, this reduces the integral to one over the moduli space of flat connections $\mathcal{M}(G,\Sigma) \equiv $ Hom$(\pi_1(\Sigma) \to \posp)/\posp$, where one simply specifies an OSp$(1|2)$ matrix for each cycle compatible with group multiplication for each three-holed sphere in the surface.\footnote{Very instructive examples of this construction can be found in \cite{Stanford:2019vob}.} The BF path integral hence boils down to the volume of $\mathcal{M}(G,\Sigma)$:
\begin{equation}
\int_{\mathcal{M}(G,\Sigma)} \left[\mathcal{D}(\text{moduli})\right] = \operatorname{Vol}\mathcal{M}(G,\Sigma), \qquad G = \text{OSp}(1|2,\mathbb{R}).
\end{equation}
Each such group element lies in one of the conjugacy classes of $\osp$, and hence encodes geometrical information (the geodesic length for a hyperbolic conjugacy class element, or the deficit angle for an elliptic element). However, only the hyperbolic conjugacy class elements correspond to smooth geometries, and are hence relevant for a gravitational description.\footnote{The elliptic class can appear, but only when we insert an operator that actively introduces a deficit angle in the surface. It should \emph{not} appear as an allowed ``intermediate'' configuration in the gravitational path integral.} An example of this is shown in Figure \ref{ospcycle}.

This restriction to smooth configurations corresponds to specializing to the so-called hyperbolic (or Hitchin) subset of $\mathcal{M}(G,\Sigma)$, the super-Teichm\"uller space $\mathcal{ST}(\Sigma)$.\footnote{Unlike bosonic Teichm\"uller space, the superanalogue $\mathcal{ST}(\Sigma)$ is not connected, but has multiple connected components labeling spin structures on $\Sigma$.  See, e.g., \cite{Penner_2019, Aghaei:2015bqi} for recent work.} We will argue next that this geometric restriction is naturally accommodated by restricting to the subsemigroup OSp$^+(1|2,\mathbb{R})$.

The holonomies of a generic OSp$(1|2,\mathbb{R})$ matrix $g$ can be classified by the value of the supertrace:
\begin{equation}
\operatorname{STr} g = a+d \pm (1+\beta\delta).
\end{equation}
In the NS sector, holonomies with $\left|\operatorname{STr} g\right| > 3$ are hyperbolic, those with $3 > \left|\operatorname{STr} g\right| > 0$ are elliptic, and those with $\left|\operatorname{STr} g\right| = 3$ are parabolic. In the R sector, the criteria are instead that holonomies with $\left|\operatorname{STr} g\right| > 1$ are hyperbolic, those with $1 > \left|\operatorname{STr} g\right| > 0$ are elliptic, and those with $\left|\operatorname{STr} g\right| = 1$ are parabolic.

Labeling the sector by $\epsilon=0$ for NS and $\epsilon=1$ for R, we compute that for a subsemigroup element $g$, where $a,b,c,d > 0$ and $ad-bc = 1+ \delta \beta$,
\begin{align}
\left|\operatorname{STr} g \right| &=  \left|a + d + (-)^\epsilon(1+\beta\delta) \right| = \left|a + \frac{1}{a} + \frac{bc}{a} + \frac{\delta\beta}{a} + (-)^\epsilon(1 + \beta\delta)\right| \nonumber \\
&> a + \frac{1}{a} + (-)^\epsilon \geq 2 + (-)^\epsilon,
\end{align}
where we again used that the absolute value of a supernumber is fully determined by its body. This result makes all holonomies automatically of hyperbolic class.

This result implies that a subsemigroup description is sufficient to exclude geometries that contain conical singularities (elliptic) or cusps (parabolic) from the very get-go, leaving only gravitational (smooth) configurations within the path integral. One furthermore needs to prove that such a description is also necessary, in the sense that \emph{all} hyperbolic super-Riemann surfaces can be accounted for by a flat $\ospp$ connection. In the $\slr$ case, evidence was provided for this converse statement in \cite{Blommaert:2018iqz} by looking at the three-holed sphere. We imagine that the same proof holds here, but postpone it for a deeper study.

In the next few subsections, we will show that once we commit to this structure, we indeed find the correct $\mathcal{N} = 1$ super-Schwarzian density of states.

\subsection{Gravitational Matrix Elements} \label{sect:whittaker}

It is well-known that the Plancherel measure of a Lie group can be extracted from the orthogonality relation obeyed by representation matrix elements of group elements with respect to the Haar measure.  In the gravitational scenario at hand, the key information is contained in representation matrices of group elements that lie in the maximal torus. The representation matrices themselves are special in that both indices are constrained to obey the Brown-Henneaux gravitational boundary conditions. This makes them mixed parabolic representation matrices, or \emph{Whittaker functions} \cite{Jacquet,Schiffmann,Hashizume1,Hashizume2}. We first compute these explicitly for OSp$^+(1|2,\mathbb{R})$, and then in the next few subsections explain their relation to JT supergravity.

The representation matrices themselves are taken within the only irreps of the subsemigroup: the principal series representations. To construct them, we proceed as follows.  We define the super half-line $\mathbb{R}^{+1|1}$ as the pair $(x|\vartheta)$ subject to the restriction $x > 0$:
\begin{equation}
\mathbb{R}^{+1|1} \equiv \{(x|\vartheta) \, | \, x > 0\}.
\end{equation}
Under the action of the semigroup on the super half-line, the bosonic coordinate $x$ maps to $\frac{ax+c+\beta \vartheta}{bx+d + \delta \vartheta}$. This new location is also positive since positivity is fully encoded within the body of a supernumber (see Appendix \ref{app:supernumbers}). Another way of appreciating this fact is to formally Taylor expand the Heaviside step function:
\begin{equation}
\Theta\left(\frac{ax+c}{bx+d} - \operatorname{sgn}(e)\frac{\alpha x + \gamma}{(bx+d)^2}\vartheta\right) = \Theta\left(\frac{ax+c}{bx+d} \right) - \delta\left(\frac{ax+c}{bx+d}\right)\operatorname{sgn}(e)\frac{\alpha x + \gamma}{(bx+d)^2}\vartheta.
\end{equation}
Hence it is indeed the case that, up to a delta function at the origin, only the bosonic parameters determine positivity (see, e.g., \cite{Witten:2012bg}).

We define the action of the semigroup OSp$^+(1|2,\mathbb{R})$ on functions $f(x, \vartheta)$ on $\mathbb{R}^{+1|1}$ as follows:\footnote{When working with the full group OSp$(1|2,\mathbb{R})$ rather than the subsemigroup, extra sign factors and absolute values must be included in this definition; see Appendix \ref{osprep}.}
\begin{equation}
\label{grouprepresentation}
(g\circ f)(x, \vartheta)\equiv (bx + d + \delta\vartheta)^{2j}f\left(\frac{ax + c + \beta\vartheta}{bx + d + \delta\vartheta}, -\frac{\alpha x + \gamma - e\vartheta}{bx + d + \delta\vartheta}\right).
\end{equation}
This definition corresponds to the supertranspose action of the group in the homogeneous $2|1$-dimensional space:
\begin{equation}
\left[\begin{array}{cc|c} x & z & -\vartheta \end{array}\right] \mapsto
\left[\begin{array}{cc|c} x & z & -\vartheta \end{array}\right]
\left[\begin{array}{cc|c}
a & b & \alpha \\
c & d & \gamma \\ \hline
\beta & \delta & e
\end{array}\right].
\label{supertransposeaction}
\end{equation}
It composes correctly under group multiplication and hence defines a representation of OSp$^+(1|2,\mathbb{R})$. The representation defined in this way is irreducible and unitary, just like the analogous principal series representation of the full group OSp$(1|2,\mathbb{R})$. These properties require independent proofs, and we present them in Appendix \ref{ospprep}. Infinitesimally, this action corresponds to the following representation of the generators in terms of first-order differential operators, which we call the Borel-Weil realization:
\begin{gather}
\hat{H} = x\partial_x + \frac{1}{2}\vartheta\partial_\vartheta - j, \qquad \hat{E}^- = \partial_x, \qquad \hat{E}^+ = -x^2\partial_x - x\vartheta\partial_\vartheta + 2jx, \\*
\hat{F}^- = \frac{1}{2}(\partial_\vartheta + \vartheta\partial_x), \qquad \hat{F}^+ = -\frac{1}{2}x\partial_\vartheta - \frac{1}{2}x\vartheta\partial_x + j\vartheta. \label{borelweil}
\end{gather}
These operators obey the commutation relations
\begin{alignat}{2}
[\hat{H}, \hat{E}^\pm] &= \pm \hat{E}^\pm, \quad & [\hat{E}^+, \hat{E}^-] &= 2\hat{H}, \nonumber \\
[\hat{H}, \hat{F}^\pm] &= \pm\frac{1}{2}\hat{F}^\pm, \quad & [\hat{E}^\pm, \hat{F}^\mp] &= -\hat{F}^\pm, \\
\{\hat{F}^+, \hat{F}^-\} &= -\frac{1}{2}\hat{H}, \quad & \{\hat{F}^\pm, \hat{F}^\pm\} &= \mp \frac{1}{2}\hat{E}^\pm, \nonumber
\end{alignat}
which differ in the anticommutators by a sign factor compared to the $\mathfrak{osp}(1|2)$ superalgebra \eqref{ospline} satisfied by the finite generators. Thus the infinitesimal group action leads to a representation of the \emph{opposite} superalgebra. This is consistent with the fact that the generators $\hat{F}^+$ and $\hat{F}^-$ have Grassmann statistics, unlike the bosonic matrices \eqref{defgen2} in finite-dimensional representations. More elaborate discussions of these issues are provided in Appendix \ref{osprep}. Demanding antihermiticity of the bosonic generators requires $\mathfrak{Re}(j) = -1/4$, which we write as $j = -1/4 + ik/2$ for $k\in \mathbb{R}$ (this should be contrasted with $\mathfrak{Re}(j) = -1/2$ for $\sltr$).  See Appendices \ref{app:unitarity} and \ref{app:infinitesimal} for details.

Finally, any OSp$^+(1|2,\mathbb{R})$ matrix can be written in the Gauss-Euler parametrization as
\begin{align}
g(\phi, \gamma_{\m}, \gamma_{\+}| \theta_{\m}, \theta_{\+}) &= e^{2\theta_{\m} F^-}e^{\gamma_{\m} E^-}e^{2\phi H}e^{\gamma_{\+}E^+}e^{2\theta_{\+}F^+} \label{gaussparam} \\
&= \left[\begin{array}{cc|c}
e^\phi & \gamma_{\+}e^\phi & e^\phi \theta_{\+} \\
\gamma_{\m} e^\phi & e^{-\phi} + \gamma_{\m}\gamma_{\+}e^\phi - \theta_{\m}\theta_{\+} & \gamma_{\m} e^\phi \theta_{\+} - \theta_{\m} \\ \hline
e^\phi\theta_{\m} & \gamma_{\+}e^\phi\theta_{\m} + \theta_{\+} & 1 + e^\phi\theta_{\m}\theta_{\+}
\end{array}\right], \label{gaussparamdefining}
\end{align}
where the second line is the formula in the defining representation. The condition \eqref{definingcond} can be explicitly verified to hold. Notice that in this form, the elements $a, b, c, d$ are positive when $\gamma_{\m}, \gamma_{\+} > 0$.

Our goal is to compute the gravitational matrix elements of OSp$^+(1|2,\mathbb{R})$. These are found by implementing the Brown-Henneaux supergravity boundary conditions \cite{Henneaux:1999ib} at the quantum level, which can be done by diagonalizing the parabolic generators in both the bra and ket states. This ``mixed parabolic'' matrix element diagonalizes the ``outer'' factors in the Gauss parametrization \eqref{gaussparam}. This is the supersymmetrization of the same statement in bosonic gravity, implemented in this language in \cite{Blommaert:2018oro, Blommaert:2018iqz}. At the infinitesimal level, we will end up diagonalizing the operators \eqref{borelweil} in the Borel-Weil realization of the \emph{opposite} superalgebra.\footnote{These exponentiate to a representation of the group, unlike those that furnish a Borel-Weil realization of $\mathfrak{osp}(1|2)$ itself (for the latter, see \eqref{badborelweil}).}

Therefore, working in the Gauss parametrization \eqref{gaussparam}, we wish to compute the mixed parabolic matrix element of a generic $\text{OSp}^+(1|2, \mathbb{R})$ element $g$ in the principal series representation defined by \eqref{grouprepresentation}.  In the bosonic case of $\sltr$, this involved diagonalizing $E^\pm$, but here, we must additionally diagonalize the fermionic generators $F^\pm$.  Consider the representation matrix element
\begin{align}
\label{swh}
\langle \psi_-|e^{2\theta_{\m} F^-} e^{\gamma_{\m} E^-} e^{2\phi H} e^{ \gamma_{\+} E^+} e^{2 \theta_{\+} F^+}|\psi_+\rangle,
\end{align}
which reads in coordinate space as
\begin{align}
\int dx\, d\vartheta\, \langle \psi_-|x, \vartheta\rangle e^{2\theta_{\m} \hat{F}^-} e^{\gamma_{\m} \hat{E}^-} e^{2\phi \hat{H}} e^{ \gamma_{\+} \hat{E}^+} e^{2 \theta_{\+} \hat{F}^+}\langle x, \vartheta|\psi_+\rangle.
\end{align}
We now choose the bra and ket states to be simultaneous eigenstates of the parabolic generators in the sense that:
\begin{alignat}{2}
\hat{E}^+ \langle x, \vartheta|\psi_+\rangle &= -\lambda \langle x, \vartheta|\psi_+\rangle, \qquad & \theta_+ \hat{F}^+ \langle x,\vartheta |\psi_+\rangle &= \langle x,\vartheta |\psi_+\rangle i\epsilon_{\+}\frac{\sqrt{\lambda}}{2} \theta_{\+}, \nonumber \\
\langle \psi_-| x,\vartheta \rangle \hat{E}^- &= \nu \langle \psi_-| x,\vartheta \rangle, \qquad & \langle \psi_-| x,\vartheta \rangle \theta_- \hat{F}^- &= i\epsilon_{\m}\frac{\sqrt{\nu}}{2}\theta_{\m} \langle \psi_-| x,\vartheta \rangle. \label{whvect}
\end{alignat}
Upon diagonalizing the bosonic parabolic generator $E^\pm$, by consistency with the algebra relation $\left\{F^\pm,F^\pm\right\} = \mp \frac{E^\pm}{2}$, ``diagonalization'' of the associated fermionic parabolic generator $F^\pm$ only allows for specifying a sign $\epsilon_{{\scalebox{.6}{$\pm$}}}\in \{+1, -1\}$.\footnote{We put the word ``diagonalization'' in quotes because it is only in the above sense that these operators are diagonalized, by including the Grassmann parameters $\theta_{{\scalebox{.6}{$\pm$}}}$ in the appropriate places. We will have more to say about this below.}

For bosonic Lie groups, states diagonalizing the parabolic generators are called \emph{Whittaker vectors}, and we will adhere to the same name for the supergroup case at hand. We can then write \eqref{swh} more explicitly as:
\begin{equation}
\label{whexp}
\left(1-i\epsilon_{\m}\sqrt{\nu}\theta_{\m} + i\epsilon_{\+}\sqrt{\lambda}\theta_{\+} + \epsilon_{\m}\epsilon_{\+}\sqrt{\lambda \nu}\theta_{\m}\theta_{\+} \right) e^{\gamma_{\m}\nu}e^{-\gamma_{\+}\lambda}\int dx\, d\vartheta\, \langle \psi_-| x,\vartheta \rangle e^{2\phi \hat{H}}\langle x,\vartheta |\psi_+\rangle.
\end{equation}
This results in what one would mathematically regard as the \emph{Whittaker function} \cite{Jacquet,Schiffmann,Hashizume1,Hashizume2}, in that the exponentiated Cartan element is the only factor in the Gauss decomposition that contributes nontrivially to the calculation of the matrix element.

We next explicitly construct the Whittaker vectors diagonalizing combinations of the parabolic generators $E^\pm$ and $F^\pm$, as in \eqref{whvect}. Note that taking the adjoint of a right eigenvector of $F^-$ does not yield a left eigenvector of $F^-$.  Therefore, to determine the right states with respect to which to compute the matrix element, we should first diagonalize $(F^-)^\dag = \frac{1}{2}(\partial_\vartheta - \vartheta\partial_x)$.\footnote{The formula for the adjoint of $F^-$ follows from the relation
\begin{equation}
\int dx\, d\vartheta\, f(x, \vartheta)^\ast(\partial_\vartheta + \vartheta\partial_x)g(x, \vartheta) = \int dx\, d\vartheta\, ((\partial_\vartheta - \vartheta\partial_x)f(x, \vartheta))^\ast g(x, \vartheta) + \text{(boundary terms)}.
\end{equation}} We can immediately write down the correct states:
\begin{align}
\langle x, \vartheta|\nu, \epsilon_{\m}\rangle &= \frac{1}{\sqrt{2\pi}}(e^{-\nu x}+ i\epsilon_{\m} \sqrt{\nu}\vartheta e^{-\nu x}), \label{whitplus1}
\end{align}
with properties
\begin{align}
&(E^-)^\dagger|\nu,\epsilon_{\m}\rangle = \nu|\nu,\epsilon_{\m}\rangle, \qquad (F^-)^\dagger|\nu, \epsilon_{\m}\rangle = \frac{i\epsilon_{\m}}{2}\sqrt{\nu}|\nu, -\epsilon_{\m}\rangle, 
\end{align}
and
\begin{align}
\langle x, \vartheta|\lambda, \epsilon_{\+}\rangle &= \frac{1}{\sqrt{2\pi}}(x^{2j}e^{-\lambda/x}- i\epsilon_{\+}\sqrt{\lambda}\vartheta x^{2j-1}e^{-\lambda/x}), \label{whitplus2}
\end{align}
satisfying
\begin{align}
&E^+|\lambda,\epsilon_{\+}\rangle = -\lambda|\lambda,\epsilon_{\+}\rangle, \qquad F^+|\lambda, \epsilon_{\+}\rangle = \frac{i\epsilon_{\+}}{2}\sqrt{\lambda}|\lambda, -\epsilon_{\+}\rangle, 
\end{align}
where $\nu, \lambda > 0$. Notice that these states do not \emph{literally} diagonalize the fermionic generators $F^\pm$, but instead map the states with different $\epsilon_{\m}$ or $\epsilon_{\+}$ into each other. One can check that this is equivalent to \eqref{whvect} and consistent with the opposite superalgebra relations $\left\{F^\pm,F^\pm\right\} = \mp \frac{E^\pm}{2}$. So the eigenvectors \eqref{whitplus1} and \eqref{whitplus2} are the Whittaker vectors of OSp$^+(1|2,\mathbb{R})$, found by diagonalizing parabolic generators. Note that the eigenfunctions, as written, do not have definite Grassmann parity.\footnote{The results for the full group $\osp$ correspond to taking $\nu\to -i\nu$ and $\lambda\to -i\lambda$ in the results for the semigroup, leading to imaginary rather than decaying exponentials.} All of these eigenfunctions are nor\-mal\-iz\-able on $\mathbb{R}^+$: this is clear for the $-$ generators, and it is true for the $+$ generators because the integral of $\smash{x^{4j - 1}e^{-2\lambda/x}}$ converges for $\mathfrak{Re}(j) = -1/4$ (in fact, for $\mathfrak{Re}(j) < 0$).  The eigenvalues are consistent with the relation $(F^\pm)^2 = \mp\frac{1}{4}E^\pm$.

The remaining Whittaker function is now readily computed:
\begin{align}
&\langle\nu, \epsilon_{\m}|e^{2\phi H}|\lambda, \epsilon_{\+}\rangle \equiv \int dx\, d\vartheta\, \langle \nu, \epsilon_{\m}| x,\vartheta \rangle e^{2\phi \hat{H}}\langle x,\vartheta |\lambda, \epsilon_{\+}\rangle \nonumber \\
&= \frac{1}{2\pi}\int_0^\infty dx\int d\vartheta\, (e^{-\nu x} - i\epsilon_{\m} \sqrt{\nu}\vartheta e^{-\nu x})\left(e^{2j\phi}x^{2j}e^{-\lambda e^{-2\phi}/x}- i\epsilon_{\+}\sqrt{\lambda}\vartheta e^{(2j - 1)\phi}x^{2j-1}e^{-\lambda e^{-2\phi}/x}\right) \nonumber \\
&= \frac{1}{\pi i} \frac{\lambda^{j + 1/2}}{\nu^j}e^{-\phi}\left(\epsilon_{\m} K_{2j + 1}(2e^{-\phi}\sqrt{\nu\lambda}) + \epsilon_{\+} K_{2j}(2e^{-\phi}\sqrt{\nu\lambda})\right). \label{whitcart}
\end{align}
We have used the fact that an element $e^{2\phi H}$ of the maximal torus acts via dilatations as in \eqref{grouprepresentation} (or specifically, \eqref{dilatation}), as well as the integral representation of the modified Bessel function of the second kind:
\begin{equation}
\label{besselint}
\int_0^\infty dx\, x^{2j - 1}e^{-\nu x - \lambda/x} = 2\left(\frac{\lambda}{\nu}\right)^j K_{2j}(2\sqrt{\nu\lambda})
\end{equation}
for $\mathfrak{Re}(\nu), \mathfrak{Re}(\lambda) > 0$. Inserting \eqref{whitcart} into \eqref{whexp}, we finally obtain
\begin{align}
\label{mixedparabospp} 
\langle\nu, \epsilon_{\m}|g|\lambda, \epsilon_{\+}\rangle &=\left(1-i\epsilon_{\m}\sqrt{\nu}\theta_{\m} + i\epsilon_{\+}\sqrt{\lambda}\theta_{\+} + \epsilon_{\m}\epsilon_{\+}\sqrt{\lambda \nu}\theta_{\m}\theta_{\+} \right) e^{\gamma_{\m}\nu}e^{-\gamma_{\+}\lambda} \nonumber \\*
&\phantom{==} \times \frac{1}{\pi i}\frac{\lambda^{\frac{1}{4} + \frac{ik}{2}}}{\nu^{-\frac{1}{4} + \frac{ik}{2}}}e^{-\phi}\left(\epsilon_{\m} K_{\frac{1}{2} + ik}(2e^{-\phi}\sqrt{\nu\lambda}) + \epsilon_{\+} K_{\frac{1}{2} - ik}(2e^{-\phi}\sqrt{\nu\lambda})\right),
\end{align}
where we have substituted $j = -1/4 + ik/2$ and used $K_\alpha(z) = K_{-\alpha}(z)$.  Upon stripping off the first line, these are indeed the known $\mathcal{N} = 1$ super-Liouville minisuperspace wavefunctions (involving both sign combinations, and ignoring the overall sign) \cite{Douglas:2003up}. Indeed, Whittaker functions have primarily appeared in the physics literature in an integrability context as solutions to Liouville and Toda equations of motion, e.g., in \cite{Sklyanin:1984sb,Gerasimov_1997,Kharchev:1999bh,Kharchev:2001rs}.

The distillation of the Virasoro algebra from the SL$(2,\mathbb{R})$ Kac-Moody algebra \cite{Bershadsky:1989mf,Bershadsky:1989tc,Verlinde:1989ua,Balog:1990mu} is the mechanism that extracts both 2d Liouville CFT and 3d gravity from the underlying SL$(2,\mathbb{R})$ WZW model. Dimensionally reducing this setup takes Liouville CFT to the Liouville minisuperspace eigenvalue problem, and takes 3d gravity to 2d JT gravity. It is hence no coincidence that JT (super)gravity is described by precisely the same objects (Whittaker functions) that govern (super-)Liouville minisuperspace models.

Moreover, the wavefunctions \eqref{mixedparabospp} are solutions to the Casimir eigenvalue equation for OSp$(1|2,\mathbb{R})$, and taking into account the sign choices, they span the entire eigenspace for fixed $j$. The treatment of the Casimir equation is given in Appendix \ref{app:regu}, with \eqref{Cassubsemi} being the particular solutions to compare to.

When specializing to gravity, we will set $\nu=\lambda=1$, corresponding to the entry ``$1$'' appearing in \eqref{bhs}.

\subsection{Gravitational Density of States}
\label{sect:plmeasu}

The Plancherel measure for the subsemigroup follows from the orthogonality relation of the mixed parabolic Whittaker function $\langle\nu_-, \epsilon_{\m}|e^{2\phi H}|\lambda_+, \epsilon_{\+}\rangle$.

As such, keeping $\nu, \lambda$ arbitrary and specializing to $j = -\frac{1}{4} + \frac{ik}{2}$, we write the mixed parabolic Whittaker function as a wavefunction:
\begin{align}
\psi^k_{\epsilon_{\+}\lambda, \epsilon_{\m}\nu}(\phi) &\equiv \langle\nu, \epsilon_{\m}|e^{2\phi H}|\lambda, \epsilon_{\+}\rangle \\
\label{endwhit}
&= \frac{1}{\pi i}\frac{\lambda^{\frac{1}{4} + \frac{ik}{2}}}{\nu^{-\frac{1}{4} + \frac{ik}{2}}}e^{-\phi}\left(\epsilon_{\m} K_{\frac{1}{2} + ik}(2e^{-\phi}\sqrt{\nu\lambda}) + \epsilon_{\+} K_{\frac{1}{2} - ik}(2e^{-\phi}\sqrt{\nu\lambda})\right).
\end{align}
The Haar measure of OSp$(1|2,\mathbb{R})$ is
\begin{equation}
d\mu(\phi) = \frac{1}{2}e^\phi\left[d\phi\, d\gamma_{\m}\, d\gamma_{\+} \,|\, d\theta_{\m}\, d\theta_{\+}\right],
\label{haarosp}
\end{equation}
which we prove in Appendix \ref{app:HM}.  The brackets denote an ``integration form'' on super\-space \cite{Witten:2012bg}.  Focusing on the $\phi$-dependent part, we get
\begin{align}
&\int_{-\infty}^\infty \left(\frac{1}{2}e^\phi\, d\phi\right)\psi^k_{\epsilon_{\+}\lambda, \epsilon_{\m}\nu}(\phi)^\ast \psi^{k'}_{\epsilon_{\+}\lambda, \epsilon_{\m}\nu}(\phi) \nonumber \\*
&=\frac{1}{2\pi^2}\frac{\lambda^{\frac{1}{2} - \frac{ik}{2} + \frac{ik'}{2}}}{\nu^{-\frac{1}{2} - \frac{ik}{2} + \frac{ik'}{2}}} \int_{-\infty}^\infty d\phi\, e^{-\phi}\left(\epsilon_{\m} K_{\frac{1}{2} - ik}(2e^{-\phi}\sqrt{\nu\lambda}) + \epsilon_{\+} K_{\frac{1}{2} + ik}(2e^{-\phi}\sqrt{\nu\lambda})\right) \nonumber \\*
&\hspace{5.5 cm} \times \left(\epsilon_{\m} K_{\frac{1}{2} + ik'}(2e^{-\phi}\sqrt{\nu\lambda}) + \epsilon_{\+} K_{\frac{1}{2} - ik'}(2e^{-\phi}\sqrt{\nu\lambda})\right).
\end{align}
To evaluate this integral, we use the identity\footnote{This identity follows from a regularized version of the $\alpha = 1/2$ case of \eqref{alphabessel}. One introduces a regulator $\epsilon$, as in Appendix B of \cite{Hikida:2007sz}, to evaluate
\begin{equation}
\int_0^\infty dx\, K_{\frac{1}{2} + is - \epsilon}(x)K_{\frac{1}{2} + is' - \epsilon}(x) = \frac{i}{4}\frac{\pi}{\sinh\pi(\frac{s + s'}{2} + i\epsilon)}\frac{\pi}{\cosh\pi(\frac{s - s'}{2})}.
\label{withepsilon}
\end{equation}
We have corrected a typo $\epsilon \to - \epsilon$ in \cite{Hikida:2007sz}. This is because the asymptotics as $x \to 0$ is of the form $K_\alpha(x) \sim x^{-\alpha}$ if $\mathfrak{Re}(\alpha)>0$. The $x\to 0$ region of the above integral is distributionally convergent $\sim \int_0 \frac{dx}{x}x^{i(s+s')}$. Regularizing it requires taking $1/2 \to 1/2 - \epsilon$, with $\epsilon>0$.}
\begin{equation}
\int_0^\infty dx \left(K_{\frac{1}{2} + ik}(x)\pm K_{\frac{1}{2} - ik}(x)\right)\left(K_{\frac{1}{2} - ik'}(x)\pm K_{\frac{1}{2} + ik'}(x)\right) = \frac{\pi^2\delta(k - k')}{\cosh(\pi k)},
\end{equation}
which holds for $k, k' > 0$, and conclude that
\begin{equation}
\int_{-\infty}^\infty \left(\frac{1}{2}e^\phi\, d\phi\right)\psi^k_{\epsilon_{\+}\lambda, \epsilon_{\m}\nu}(\phi)^\ast \psi^{k'}_{\epsilon_{\+}\lambda, \epsilon_{\m}\nu}(\phi) = \frac{\delta(k - k')}{4\cosh(\pi k)}.
\end{equation}
The resulting Plancherel measure is $\rho(k) = \cosh(\pi k)$, up to normalization. This is indeed the known result for $\mathcal{N}=1$ JT supergravity and the $\mathcal{N}=1$ super-Schwarzian model.\footnote{We stress that although considering the parabolic Whittaker function (matrix element of $e^{2\phi H}$) suffices to extract the Plancherel measure for $\ospp$, what would be the parabolic basis for the full group $\osp$ ceases to be a basis for the semigroup $\ospp$: that is, the eigenfunctions of the parabolic generators do not comprise a basis on $\mathbb{R}^+$. Strictly speaking, the above argument suffices only for regions connected to the boundary; otherwise, one needs a basis (completeness relation). To obtain an orthogonality relation for $\ospp$ matrix elements with respect to the full Haar measure, one can instead work in the hyperbolic basis (see Appendix \ref{app:whit}).}

\section{Gravitational Applications}
\label{applications}

In this section, we apply our previously acquired knowledge on the BF structure of $\mathcal{N}=1$ JT supergravity to find (or reproduce) gravitational amplitudes. Our treatment is rather concise, since it can be developed in complete parallel to the bosonic JT results. We will illustrate how the above calculated group-theoretic ingredients (the Whittaker functions, the Plancherel measure, and the characters) suffice to determine JT supergravity amplitudes.

\subsection{Application: Disk Amplitudes}

\begin{figure}[!htb]
\centering
\includegraphics[width=0.8\textwidth]{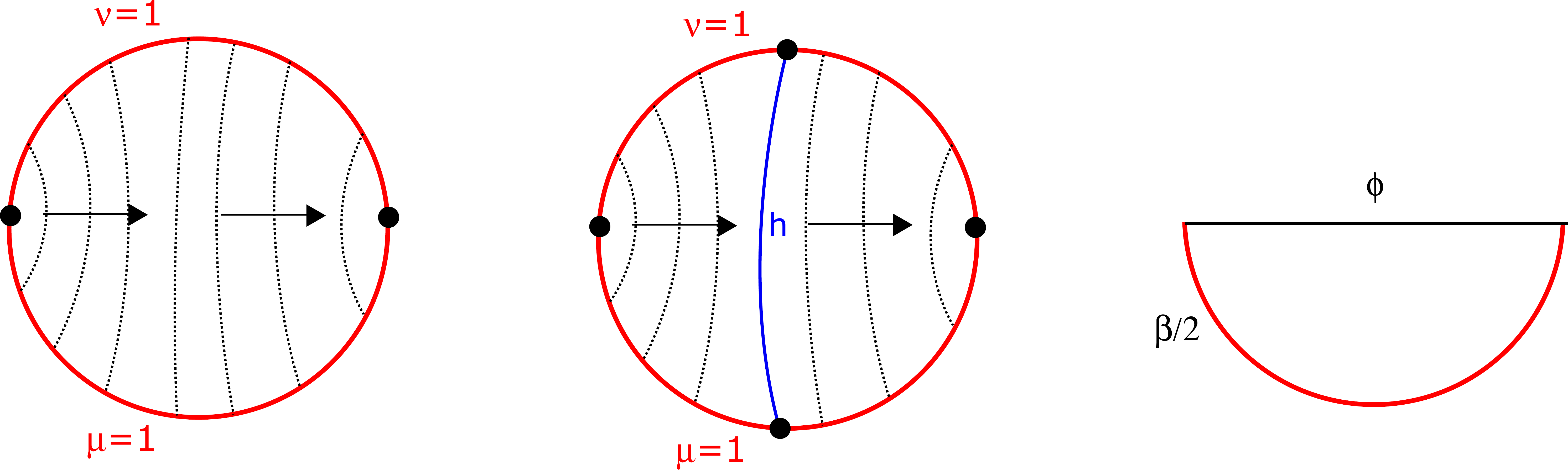}
\caption{BF evaluation of disk amplitudes. We use a Hamiltonian evaluation with the dashed lines being the fixed Euclidean time slices. Both endpoints of each time slice lie on the holographic boundary, where we impose the gravitational boundary conditions. Left: disk partition function. Middle: single boundary-anchored bilocal operator insertion on the disk. Right: Wheeler-DeWitt wavefunction $\Psi_{\beta/2}(\phi)$ as a function of geodesic distance $2\phi$, evolving from a half-circle with length $\beta/2$.}
\label{diskbilSJT}
\end{figure}

We first discuss the disk partition function, as well as the insertion of a single boundary bilocal operator on the disk (Figure \ref{diskbilSJT}).

Within the BF framework, the evaluation of such amplitudes was worked out in \cite{Blommaert:2018oro, Blommaert:2018iqz}. We refer the reader to those references for details. Here, we only summarize how knowledge of the above structural ingredients leads to a derivation of this class of boundary correlators.

The strategy is as follows. We time-slice the Euclidean disk as shown in Figure \ref{diskbilSJT}, where each time slice is an interval. The Hilbert space description of a BF model on an interval is known, where a complete set of wavefunctions consists of the representation matrix elements $R^j_{ab}(g)$ for each unitary irrep $j$ and for each pair of representation indices. This follows immediately from the Peter-Weyl theorem. These basis states are eigenstates of the Hamiltonian, which acts as the Casimir $\mathcal{C}_j$. In the case at hand, the boundary is the holographic boundary where gravitational constraints are imposed. Mathematically, this means the model is in fact a coset model, where the representation indices $a$ and $b$ are fixed to a specific choice. The constraints in our case are given in terms of two parabolic indices: $\nu=1$ and $\lambda=1$. This is precisely the Whittaker function we determined above. So
\begin{equation}
R^j_{ab}(g) \, \to \,  R^{-1/4+ik}_{\nu=1,\lambda=1}(\phi),
\end{equation}
and these form a complete set upon summing over the momentum index $k$.\footnote{To match the conventions in the Schwarzian literature, what we call $k$ in this section is what we call $k/2$ in the rest of the paper.  This should be kept in mind when comparing formulas between sections.}

These considerations immediately lead to the super-JT disk partition function
\begin{equation}
\label{sJTpf}
Z(\beta) = \int_{0}^\infty dk \cosh(2\pi k) e^{-\beta k^2} = \frac{1}{2} \left( \frac{\pi}{\beta} \right)^{1/2} e^{\frac{\pi^2}{\beta}}.
\end{equation}

When a boundary bilocal operator is present, we should insert a discrete representation matrix element in the calculation. The discrete representation Whittaker functions were determined up to normalization in \eqref{discrep} as solutions to the Casimir eigenvalue problem, and they take the form:
\begin{equation}
R(\phi) = e^{-\phi}J_{2j+1}\left(2\sqrt{-\nu\lambda}e^{-\phi}\right), \qquad e^{-\phi}J_{2j}\left(2\sqrt{-\nu\lambda}e^{-\phi}\right),
\end{equation}
where $j = -1/2, -1, \ldots$ to produce the lowest- or highest-weight discrete representations. The first entry above can be viewed as the bottom component, and the second as its superpartner.\footnote{This is not quite right: the actual superpartner is a suitable linear combination of both of these, as can be seen from the recursion relations in equation (E.8) of \cite{Mertens:2020pfe}.} Since this object plays the role of an operator insertion, we disregard the precise normalization, which is ultimately just a choice. It is convenient here to define $h = -j$. Discrete representations occur for $h$ a positive half-integer. Taking the limit $\nu, \lambda \to 0$ to obtain the lowest/highest-weight Whittaker vector, we obtain\footnote{We have used that for $\alpha\in \mathbb{Z}$, $J_\alpha(x)\sim x^{|\alpha|}$ as $x\to 0$.  We have also assumed that $h>1/2$, which is precisely the regime where the worldline description of Section \ref{s:gravwilson} is valid.}
\begin{equation}
\label{dislim}
R(\phi) \to e^{- 2h \phi}.
\end{equation}
This is to be identified with the bottom component of the bilocal operator \eqref{defbilocal}. Given the relation \eqref{aprogeod} between the bilocal operator and the geodesic distance $d$, we can identify
\begin{equation}
d_{\text{bot}} \simeq 2 \phi,
\end{equation}
providing a direct geometric interpretation of the group coordinate $\phi$. Notice in particular that $R(\phi) \to 1$ when we take the limit to the identity ($h=0$) insertion, as it should. The resulting vertex function (or 3$j$-symbol) is then a group (coset) integral of a product of two constrained principal series representation matrix elements \eqref{endwhit} and one discrete representation matrix element \eqref{dislim}. Setting $x=e^{-\phi}$, we can use the integral
\begin{align}
\int_0^{\infty} &dx \left( K_{1/2+2ik_1}(x) + \epsilon_{\m}\epsilon_{\+} K_{1/2-2ik_1}(x)\right)  \left( K_{1/2+2ik_2}(x) + \epsilon_{\m}\epsilon_{\+} K_{1/2-2ik_2}(x)\right) x^{2h} \nonumber \\
&= 4^{h-1}\frac{\left(\Gamma\bigl(\textstyle \frac{1}{2}+ h \pm i(k_1-k_2)\bigr)\, \Gamma\bigl(h \pm i(k_1+k_2)\bigr) + (k_2 \to -k_2)\right)}{\Gamma(2h)},
\end{align}
where a product over all four choices of $\pm$ is understood.  Notice that both choices of $\epsilon_{\m}\epsilon_{\+}$ give precisely the same result. These expressions are the known $3j$-symbols (or vertex functions) in $\mathcal{N}=1$ JT supergravity \cite{Mertens:2017mtv}. Inserting this quantity into the full answer for the correlation function then gives the bottom component of the boundary two-point function:
\begin{align}
\langle \mathcal{O}_{h}(\tau,0)\rangle_{\text{bot}} &= \frac{1}{Z(\beta)}\frac{1}{\pi^{2}}\int dk_1\, dk_2\, e^{-\tau k_1^2 - (\beta-\tau) k_2^2} \cosh(2\pi k_1) \cosh(2\pi k_2) \label{super2ptnoC} \\
&\hspace{2 cm} \times \frac{\Gamma\bigl(\textstyle \frac{1}{2}+h \pm i(k_1-k_2)\bigr)\, \Gamma\bigl(h \pm i(k_1+k_2)\bigr) + (k_2 \to -k_2)}{\Gamma(2h)}, \nonumber
\end{align}
in agreement with the known result obtained using super-Liouville techniques \cite{Mertens:2017mtv}.

Finally, just like in the bosonic case, it is interesting to note that one can write down a Wheeler-DeWitt wavefunction $\Psi_{\beta/2}(\phi)$ that creates a two-boundary state with geodesic separation $d_{\text{bot}} = 2 \phi$ between both boundaries, evolving from half of a Euclidean disk of boundary length $\beta/2$ (Figure \ref{diskbilSJT}), by writing:
\begin{equation}
\Psi_{\beta/2}(\phi) \equiv \int dk \cosh(2\pi k) e^{-\frac{\beta}{2}k^2} e^{-\phi}\left(K_{1/2+2ik}(2e^{-\phi}) + \epsilon_{\m}\epsilon_{\+} K_{1/2-2ik}(2e^{-\phi}) \right).
\end{equation}
Two copies of this ``half-disk'' wavefunction can be glued together to reproduce the disk partition function:
\begin{equation}
Z(\beta) = \int d\phi \left(\frac{e^{\phi}}{2}\right) \Psi_{\beta/2}^*(\phi) \Psi_{\beta/2}(\phi).
\end{equation}

\subsection{Application: Defect Insertions and Gluing}
\label{s:defglue}

As a further example, we discuss the insertion of hyperbolic defects in the disk and use them to glue surfaces together in the gauge-theoretic description. For a BF theory of a compact group $G$ described by an action of the type \eqref{jtac}, one can create a defect of holonomy $U$ in the disk by inserting a suitably normalized character in the region of the disk with representation $R$:
\begin{equation}
\frac{\chi_R(U)}{\dim R}.
\end{equation}
For instance, the disk amplitude with a single such insertion would be
\begin{equation}
Z_U(\beta) = \sum_R (\dim R)^2 \left(\frac{\chi_R(U)}{\dim R} \right) e^{-\beta \mathcal{C}_R}, 
\end{equation}
where one sums over all irreps of the group $G$, and where $\mathcal{C}_R$ is the quadratic Casimir of $R$. These equations are nearly identical to those of 2d Yang-Mills amplitudes \cite{Cordes:1994fc}. This analogy was studied more closely in several works \cite{Blommaert:2018oue, Blommaert:2018oro, Iliesiu:2019lfc, Kapec:2019ecr}.

Gravity differs from such a BF theory in two ways: firstly, as explained around Figure \ref{diskbilSJT}, it behaves as a coset model instead of a genuine group model. This coset restriction essentially strips off a factor of $\dim R$ from the above amplitude: see Section 2.3 of \cite{Blommaert:2018iqz} for an extensive discussion. Secondly, the relevant group is noncompact, where in our setup, the role of $\dim R$ is played by the Plancherel measure for the principal series representations of the positive semigroup.

Hence when applying the above procedure of inserting a defect to the gravitational case, we need only find the expression for the suitably normalized character and insert it into our super-JT disk partition function \eqref{sJTpf}. The relevant representation theory does not seem to be available in the mathematical literature, so we work it out from first principles. Within the NS sector, the character we need is computed in Appendix \ref{app:char} (in particular, see \eqref{chins}) and given by
\begin{equation}
\label{ncharns}
\chi_k(\phi) = \cos (2 \phi k).
\end{equation}
A few comments are in order. Here, as everywhere in this section, we have set $k\to2k$ to match the gravitational convention where the energy variable $E$ and the momentum parameter $k$ are related by $E = k^2 + \text{constant}$. We have stripped off the Weyl denominator of this character, and we likewise glue with a flat conjugacy class measure on the supergroup. This is merely a bookkeeping exercise, but the current normalization matches directly to the Schwarzian limit of the super-Virasoro modular S-matrices. Indeed, using the modular S-matrix between two nondegenerate super-Virasoro representations \cite{Ahn:2002ev}, we write:
\begin{equation}
\lim_{b \to 0} S_{s}{}^{P} = \lim_{b \to 0} \cos \left(4 \pi s P\right)  = \cos (2 \phi k) = \chi_k(\phi),
\end{equation}
where we fix $s = \phi/ 2\pi b$ and $P = b k$ in the limit as $b \to 0$ \cite{Mertens:2019tcm}. 

Inserting \eqref{ncharns} into the disk partition function gives the defect disk amplitude (or the single-trumpet amplitude):
\begin{equation}
Z_{\phi}(\beta) = \int_0^\infty dk \cos(2 \phi k) e^{-\beta k^2} = \frac{1}{2} \sqrt{\frac{\pi}{\beta}} e^{-\frac{\phi^2}{\beta}},
\end{equation}
geometrically interpretable as a single-trumpet geometry with a neck of length $2\phi$, as discussed in Section \ref{s:defint}. The length parameter $2\phi$ is related to the defect parameter $\Lambda$ by $\phi =\pi \Lambda$.

\begin{figure}[!htb]
\centering
\includegraphics[width=0.75\textwidth]{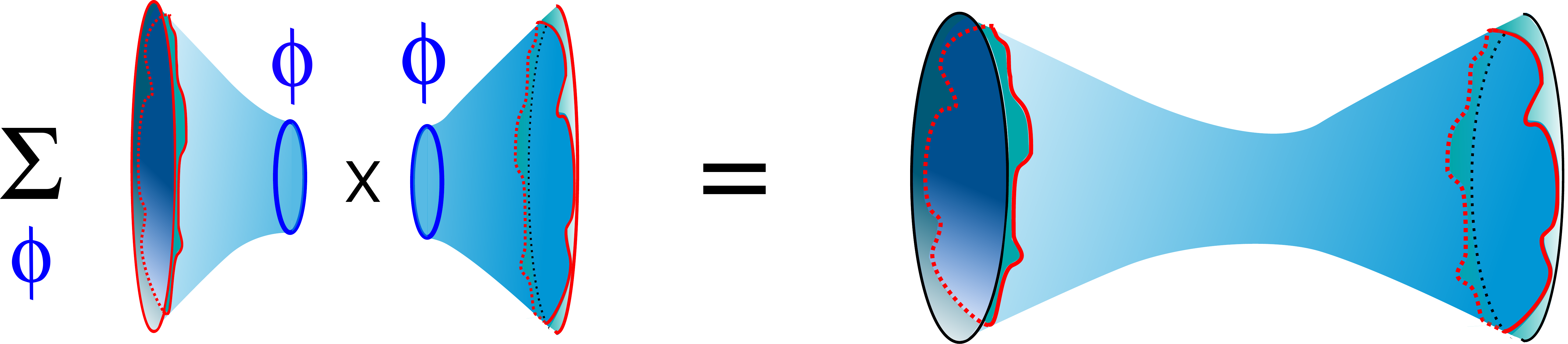}
\caption{Gluing together tubes by integrating over hyperbolic orbits with geodesic length $2\phi$ for a fixed spin structure.}
\label{SUSYhyperbolicorbit}
\end{figure}

Two such trumpets can be glued together in super-Teichm\"uller space by using character orthonormality \eqref{charotho}. This procedure is pictorially represented in Figure \ref{SUSYhyperbolicorbit}.  It gives the two-boundary amplitude:
\begin{equation}
Z(\beta_1,\beta_2) = \int_0^\infty dk\, e^{-(\beta_1+\beta_2)k^2} = \frac{1}{2}\sqrt{\frac{\pi}{\beta_1+\beta_2}}.
\end{equation}
Notice that this is \emph{not} the same double trumpet amplitude of \cite{Stanford:2019vob}. This discrepancy is due to our description in terms of super-Teichm\"uller space, compared to their description in terms of the moduli space of super-Riemann surfaces. The difference is a quotient by the mapping class group, which would lead to an additional factor of $\phi$ inserted in the gluing integral and matching to the computation of \cite{Stanford:2019vob}.\footnote{There are also extra OSp volume factors that are omitted here. See \cite{Blommaert:2018iqz} for details in the bosonic case.}

One can likewise find the amplitude with a single elliptic defect by analytically continuing $\phi \to i \phi$ to get
\begin{equation}
Z_{\phi}(\beta) = \int_0^\infty dk \cosh(2\phi k) e^{-\beta k^2} = \frac{1}{2} \sqrt{\frac{\pi}{\beta}} e^{\frac{\phi^2}{\beta}},
\end{equation}
interpretable as a disk with a conical defect of angular periodicity $2\phi$. We remark that, just as in bosonic JT gravity, this is a formal analytic continuation of the hyperbolic character insertion because the actual elliptic character vanishes, as shown in Appendix \ref{app:char}. This procedure can be phrased in many different ways. In particular, \cite{Mertens:2019tcm} used the language of coadjoint orbits of the Virasoro group and branes in Liouville CFT: there, analytically continued FZZT branes (dubbed $\mathfrak{i}$FZZT branes) were needed to access elliptic defects.

In the special case where $2\phi =n$ odd, we have the enhanced stabilizer OSp$(1|2,\mathbb{R})$, for which the super-Virasoro modular S-matrix element \cite{Ahn:2002ev} has the limit
\begin{equation}
\lim_{b \to 0} S_{(n,1)}{}^{P} = \lim_{b \to 0} 4 \cosh \left(2\pi n \frac{P}{b}\right) \cosh(2\pi P b) = 4\cosh(2\pi n k),
\end{equation}
which plays the role of the character insertion in the disk amplitude. For $n=1$, we get the vacuum orbit, leading to the ordinary thermal disk partition function.\footnote{For even $n$, we can instead work with the modular S-matrix
\begin{equation}
\lim_{b \to 0} S_{(n,2)}{}^{P} = \lim_{b \to 0} 4 \sinh \left(2\pi n \frac{P}{b}\right) \sinh(2\pi P b) \sim  k\sinh(2\pi n k),
\end{equation}
which is indeed the expected measure that gives three bosonic zero modes, corresponding to the stabilizer $\sltr$. We leave a more careful comparison for future work.}

Hence for all of the orbits discussed in Section \ref{sect:hill}, we can obtain suitable defect insertions to be inserted into the disk partition function. For the NS sector of interest, we summarize them below:
\begin{itemize}
\item Elliptic $H=U(1)_\Theta$:
\begin{equation}
F\circ_\Theta f = \tan \frac{\pi}{\beta} \Theta f, \qquad D_{\text{U(1)}_\Theta}(k) = \frac{\cosh( 2 \pi \Theta k)}{\cosh(2 \pi k)}.
\end{equation}
\item Special Elliptic $H=\text{OSp}^{n}(1|2,\mathbb{R})$, $n$ odd:
\begin{equation}
F\circ_n f = \tan \frac{\pi}{\beta} n f, \qquad D_{\text{OSp}^{n}(1|2,\mathbb{R})}(k) = \frac{\cosh( 2 \pi n k)}{\cosh(2 \pi k)}.
\end{equation}
\item Special Elliptic $H=\text{SL}^n(2,\mathbb{R})$, $n$ even:
\begin{equation}
F\circ_n f = \tan \frac{\pi}{\beta} n f, \qquad D_{\text{SL}^n(2,\mathbb{R})}(k) = \frac{k\sinh( 2 \pi n k)}{\cosh(2 \pi k)}.
\end{equation}
\item Hyperbolic $H=U(1)_\Lambda$:
\begin{equation}
F \circ_\Lambda f = \tanh \frac{\pi}{\beta}\Lambda f, \qquad D_{\text{U(1)}_\Lambda}(k) = \frac{\cos( 2 \pi \Lambda k)}{\cosh(2 \pi k)}.
\end{equation}
\item Parabolic $H=U(1)_0$:
\begin{equation}
\label{intro:par1}
F\circ_0 f =  f, \qquad D_{\text{U(1)}_0}(k) = \frac{1}{\cosh(2 \pi k)}.
\end{equation}
\end{itemize}

\subsection{Application: Wormhole Length and Complexity}
\label{sect:complex}

As a short application of our understanding of bilocal operators, and in particular the superspace geodesic formula \eqref{aprogeod}, we follow \cite{Yang:2018gdb} and utilize these correlators to compute the geometric boundary-to-boundary wormhole length in the two-sided black hole geometry. This geometric information is conjectured to encode the computational complexity of a putative boundary dual by way of the ``Complexity = Volume'' conjecture \cite{Susskind:2014rva}, which relates the complexity $\mathcal{C}(t)$ to the extremal wormhole volume $V(t)$ as $\mathcal{C}(t) = \frac{V(t)}{G L_{\text{AdS}}}$.

Since our goal in this subsection is to probe the strong-coupling regime of the super-Schwarzian theory, it is useful to make dimensionful parameters explicit:
\begin{equation}
S_\text{Sch}^{\mathcal{N} = 1} = -2C\oint_{\partial \mathcal{M}}d\tau\, T_{\B}(\tau),
\label{N1boundaryaction}
\end{equation}
where the coupling constant $C$ has dimensions of length.  We mostly work in units where $C=1/2$, and reinstate it here for clarity.  We see from \eqref{susyact} that the action \eqref{N1boundaryaction} is normalized such that for $\eta = 0$, it reproduces the bosonic result, with $C$ being the usual Schwarzian coupling.  In the standard second-order treatment of JT gravity, such a constant depends on the relative normalization of the boundary values of the dilaton and metric.  It would appear in our case as a constant factor in the boundary condition \eqref{mixedBC}, which we have suppressed.

Within bosonic JT gravity, the classical (renormalized) wormhole length was considered in \cite{Brown:2018bms}. The answer is essentially the logarithm of the semiclassical boundary two-point function, where the endpoints are separated by half a thermal circle in Euclidean time: 
\begin{equation}
\label{clworm}
d(t) = \ln \cosh \frac{2\pi}{\beta} t \approx \frac{2 \pi}{\beta} t.
\end{equation}
Famously, this quantity grows linearly in time for late times $t$ \cite{Hartman:2013qma}.

\begin{figure}[!htb]
\centering
\includegraphics[width=0.25\textwidth]{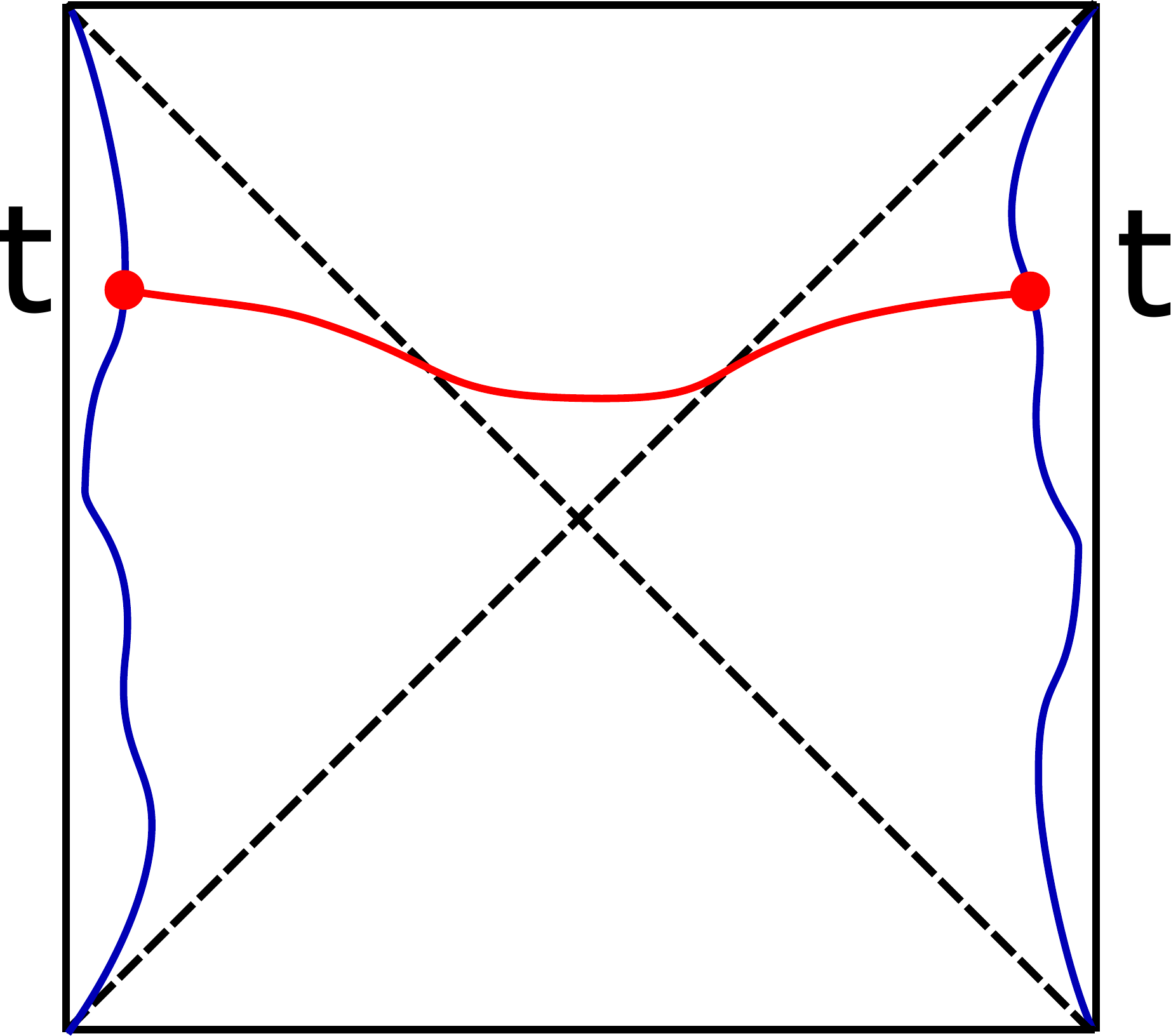}
\caption{Extremal wormhole in the bulk geometry of the thermal system, anchored at both boundaries at time $t$.}
\label{complvol}
\end{figure}

Within bosonic JT gravity, going beyond classical gravity can be done by realizing that in any off-shell gravitational background $F(t)$, the wormhole length is computed by the operator:
\begin{equation}
d(t) = \log \frac{(F(t)-F(t+i\beta/2))^2}{\partial_tF(t)\partial_tF(t+i\beta/2)} - \log \epsilon^2,
\end{equation}
with a term involving a UV regulator $\epsilon$ that is naturally measured with the proper boundary clock time $t$. Subtracting this quantity gives the renormalized wormhole length. The resulting time reparametrization $F(t)$ results in a wiggly boundary curve, but the bulk geometry is still a patch of AdS$_2$. The situation is sketched in Figure \ref{complvol}.

This operator is then inserted in the Schwarzian (or gravitational) path integral.  This calculation was done in \cite{Yang:2018gdb}, with the tantalizing result that the geodesic distance at late times $t \gg C$ still increases linearly with time:
\begin{equation}
d(t) \approx \frac{2\pi}{\beta} t.
\end{equation}
Hence this late-time growth persists even after classical gravity stops being valid. Here, we will show as an application that the same is true in JT supergravity.

Within classical JT supergravity, since the saddle solution of the geometry is not affected by the fermions, the wormhole length as a function of boundary time $t$ is given by precisely the same expression \eqref{clworm}. The difference between the bosonic and the supersymmetric theories is only visible when incorporating interactions with the boundary gravitino $\eta(t)$, and hence at the quantum level in the expansion in the gravitational coupling constant $G_N \sim 1/C$.

The tool that we use to go beyond classical gravity is the JT supergravity bilocal correlation function \eqref{super2ptnoC}, but with the coupling parameter $C$ made explicit:
\begin{align}
\langle \mathcal{O}_{h}(\tau,0)\rangle_{\text{bot}} &= \frac{1}{Z}\frac{1}{\pi^{2}(2C)^{2h}}\int dk_1\, dk_2\, e^{-\tau \frac{k_1^2}{2C} - (\beta-\tau) \frac{k_2^2}{2C}} \cosh(2\pi k_1) \cosh(2\pi k_2) \label{super2pt}  \\
&\hspace{2.6 cm} \times \frac{\Gamma\bigl(\textstyle \frac{1}{2}+h \pm i(k_1-k_2)\bigr)\, \Gamma\bigl(h \pm i(k_1+k_2)\bigr) + (k_2 \to -k_2)}{\Gamma(2h)}. \nonumber
\end{align}
Using the bottom component of the geodesic distance formula \eqref{aprogeod} in superspace, the (renormalized) wormhole length at real time $t$ is given by the following expression:
\begin{align}
\label{wlen}
d_{\text{bot}}(t) &= -\frac{1}{Z}\frac{1}{\pi^{2}}\int dk_1\, dk_2\, e^{-(\beta/2 + it) \frac{k_1^2}{2C} - (\beta/2-it) \frac{k_2^2}{2C}} \cosh(2\pi k_1) \cosh(2\pi k_2) \\*
&\hspace{1.7 cm} \times \left.\frac{\partial}{\partial h}\frac{\Gamma\bigl(\textstyle \frac{1}{2}+h \pm i(k_1-k_2)\bigr)\, \Gamma\bigl(h \pm i(k_1+k_2)\bigr) + (k_2 \to -k_2)}{(2C)^{2h}\Gamma(2h)}\right|_{h=0}. \nonumber 
\end{align}
Now we approximate this expression at late time $t\gg C$. We use the identity
\begin{equation}
\frac{\Gamma(h \pm i (k_1 - k_2))}{\Gamma(2h)} = 2\int_{-\infty}^{\infty} dy\, \frac{e^{2i(k_1-k_2)y}}{(2\cosh y)^{2h}},
\end{equation}
and differentiate with respect to $h$ to rewrite the second term of \eqref{wlen} as:
\begin{align}
d_{\text{bot}}(t) &= \frac{1}{Z}\frac{4}{\pi^{2}}\int dy \int dk_1\, dk_2\,  e^{-(\beta/2 + it) \frac{k_1^2}{2C} - (\beta/2-it) \frac{k_2^2}{2C}}e^{2i(k_1-k_2)y} \cosh(2\pi k_1) \cosh(2\pi k_2) \nonumber \\
&\hspace{2.5 cm} \times \ln (2\cosh y) \Gamma\bigl(\textstyle \frac{1}{2}\pm i(k_1+k_2)\bigr). \label{wsim}
\end{align}
Terms where the derivative $\partial/\partial h$ acts on any of the other factors are subdominant in the large-$t$ regime, or $t$-independent. The first term of \eqref{wlen} is subdominant as well. This can be seen since performing the same trick requires an extra factor of $2h$ in the numerator, which goes to 0 as we take $h\to 0$ in the end.

At late times $t \gg C$, the $k_i$-integrals in \eqref{wsim} are dominated by their nontrivial saddle point at $k_i^*/C = 2y/t$ where $y$ is positive (and large) as a consequence. One evaluates the two $k_i$-integrals then to be:
\begin{align}
d_{\text{bot}}(t) &\approx \frac{1}{Z}\frac{4}{\pi^{2}}\frac{2\pi C}{t}\int dy\, e^{-\beta \frac{(k^*)^2}{2C}} \cosh^2(2\pi k^*) y \Gamma\bigl(\textstyle \frac{1}{2} \pm 2ik^*\bigr) \nonumber \\
&= \frac{1}{Z} \int dk^*\, 2e^{-\beta \frac{(k^*)^2}{2C}} \cosh(2\pi k^*) \frac{k^*}{C}t,
\end{align}
where we have used $|\Gamma(1/2 + i k)|^2 = \pi/\cosh(\pi k)$.  This can be viewed as the thermal ensemble version of the microcanonical $\frac{k^*}{C} t$. For a macroscopic black hole (the thermodynamic limit) where $\beta \ll C$, this integral gets further evaluated on its saddle $k^* \approx \frac{2\pi C}{\beta}$, leading indeed to 
\begin{equation}
d_{\text{bot}}(t) \approx \frac{2\pi}{\beta} t.
\end{equation}
This is precisely equal to the semiclassical wormhole length, but now valid at late times $t \gg C$ where quantum gravity is strongly coupled.

These calculations, however, only use the lowest disk topology for the boundary bilocal correlator. It is known that higher-genus corrections to boundary correlators exist and that they are important at late times \cite{Saad:2019lba, Blommaert:2019hjr, Saad:2019pqd}. It is natural to suspect that these will lead to a saturation of the complexity $\mathcal{C}(t)$ at very late times and cause the complexity plateau to appear \cite{Brown:2017jil}. It would be interesting to understand this effect in more detail.

\section{Discussion and Open Problems}
\label{discussion}

In this work, we have advocated for a group-theoretic perspective on $\mathcal{N}=1$ JT supergravity. This required a great deal of the relevant OSp$(1|2)$ supergroup theory, which we developed independently mostly in the appendices.

Our results were obtained in the framework of gauge theory.  At a basic level, one can ask: how much of the gravitational theory does a gauge theory description even capture?  One point of view is that the gauge theory only describes the limit $S_0\to\infty$, where $S_0$ is the coefficient of the purely geometric term in the Euclidean JT action that weights topologies by $(e^{S_0})^{\chi(\mathcal{M})}$.  This limit simplifies the statement of holographic duality to an equivalence between a BF theory on a fixed topology and edge modes on the boundary.  For instance, only when restricting to the disk is JT gravity equivalent to the Schwarzian theory at finite temperature.  When the path integral implements a sum over topologies, JT gravity is dual to an ensemble of random Hamiltonians with Schwarzian density of states \cite{Saad:2019lba}.

Therefore, it may seem that the disk probes a very limited sector of the full gravitational theory.  However, the disk observables (e.g., vertex functions), when supplied as input to the dual matrix model, seem to provide all the data needed to compute multi-boundary and higher-genus amplitudes, including amplitudes with bilocal lines \cite{Saad:2019pqd}.  Indeed, random matrix theory gives all multi-level spectral densities in terms of the single-level spectral density $\rho_0(E)$, which is computed as the inverse Laplace transform of the disk partition function.  This fact, combined with suitable gluing rules \cite{Saad:2019lba} and known results on the disk \cite{Blommaert:2018iqz, Mertens:2019tcm}, leads to a recipe for JT gravity amplitudes on surfaces with handles, boundaries, Wilson lines, and defects.  Our results provide the ingredients in a similar recipe for JT supergravity.

Of course, the major caveat is that the gluing rules relevant to gravity transcend gauge theory.  To appreciate this caveat, we reprise the points in the introduction and summarize their resolution in JT gravity.  We phrase the following for bosonic JT gravity, but similar statements hold if one replaces $\text{PSL}(2, \mathbb{R})$ with $\posp$ and $\mathcal{T}$ with $\mathcal{ST}$ and inserts the word ``super'' as appropriate.

Low-dimensional gravity is a gauge theory at the level of the classical action and in perturbation theory around classical solutions, but it differs substantially from gauge theory at the level of the nonperturbative path integral \cite{Witten:2007kt, Blommaert:2018iqz}.  Perhaps most obviously, gauge theory is formulated on a fixed background, so any sum over topologies that is needed to match with the gravitational path integral must be implemented by hand.  On top of this discrepancy, the gauge and gravity path integrals also differ in the integration space within a given topological class:
\begin{itemize}
\item First, classical solutions of the gauge theory do not necessarily map onto classical solutions of the gravitational theory (nonsingular metrics).  In JT gravity, we solve this problem by restricting the BF path integral to the \emph{hyperbolic component} of the moduli space of flat $\text{(P)SL}(2, \mathbb{R})$ connections, namely the component in which all holonomies are conjugate to hyperbolic elements.  This is precisely Teichm\"uller space $\mathcal{T}(\Sigma)$: the space of (smooth) hyperbolic metrics on $\Sigma$, modulo diffeomorphisms that are connected to the identity \cite{Dijkgraaf:2018vnm}.
\item Second, gauge transformations can be identified with diffeomorphisms that are connected to the identity, but they do not account for large diffeomorphisms.  In JT gravity, we must perform a further quotient on the path integration space over geometries, thereby restricting it to the moduli space of Riemann surfaces $\mathcal{M}(\Sigma)$.
\end{itemize}

We have argued that the first point can be addressed purely within gauge theory by carefully identifying the global form of the gauge group in the gauge theory description of JT gravity.  Namely, the moduli space of flat $\slr$ connections is contained in Teichm\"uller space, and is conjecturally equal to it \cite{Blommaert:2018iqz}.  Settling both this conjecture and the corresponding one for $\ospp$ are outstanding problems.

Addressing the second point seems to require going beyond gauge theory.  Namely, from the gauge theory perspective, it is more natural that higher-genus amplitudes should be computed by integrating over Teichm\"uller space $\mathcal{T}(\Sigma)$, but in practice, the results converge only for very low genus ($\chi\geq 0$) \cite{Blommaert:2018iqz}.  Matching with the true JT gravity amplitudes requires a different gluing measure appropriate for the moduli space of Riemann surfaces $\mathcal{M}(\Sigma)$ \cite{Saad:2019lba}.  In the latter approach, one treats the Weil-Petersson volumes as external data and glues them to gauge theory disk amplitudes (decorated with hyperbolic defects) by restricting the integration range of the length parameter using input that goes beyond gauge theory.

Finally, at least two different gauge theory descriptions of bosonic JT gravity have been proposed in the literature: one involves restricting to a subsemigroup \cite{Blommaert:2018oro, Blommaert:2018iqz}, and another involves passing to the universal cover of $\sltr$ \cite{Iliesiu:2019xuh}.  They are different quantizations of the same classical theory that agree on the disk (the ``gauge sector'' of JT gravity), but may disagree on multi-boundary or higher-genus surfaces. It remains an open question to understand the relation between these proposals, and to test them against each other in different situations.  It also begs the question: how unique is the gauge theory description of JT gravity?

We next survey some open problems that are worthy of further investigation, some of which will be addressed in upcoming work \cite{Fan:2021bwt}.

\subsubsection*{Local Operators and Interpretation of Representation Carrier Space}

The carrier space of the representations discussed in this work is built on the superline $(x|\vartheta)$, and is introduced as an auxiliary object in order to construct the representation. To distill a physical interpretation, it is useful to observe that there exist $\sltr$-covariant local operators in the Schwarzian models that take the form
\begin{equation}
\label{bosop}
\phi_j(x,\tau) \equiv g^{-1}(\tau) \cdot x^{2j} = \left[\frac{f'(\tau)}{(x - f(\tau))^2}\right]^j. 
\end{equation}
The first equality shows that this operator is found by applying ``half'' of the Wilson line in \eqref{tocom} and surrounding expressions. This operator depends on a single coordinate $x$, and under 
\begin{equation}
f(\tau) \to \frac{a f(\tau) + c}{b f(\tau) + d},
\end{equation}
it transforms into a local operator within the same representation:
\begin{equation}
\phi_j(x,\tau) \to (bx'+d)^{2j} \phi_j (x',\tau), \qquad x= \frac{ax'+c}{bx'+d},
\end{equation}
namely the spin-$j$ representation of $\sltr$.

We can play the same game in the supersymmetric case to define OSp$(1|2,\mathbb{R})$-covariant local operators. The local operator
\begin{equation}
\label{susyop}
\phi_j(x,\vartheta,\tau) \equiv g^{-1}(\tau) \cdot x^{2j} = \left(\frac{D \theta'}{x - \tau' + \theta' \vartheta}\right)^{2j}
\end{equation}
transforms in the spin-$j$ representation of OSp$(1|2,\mathbb{R})$. Indeed, under\footnote{The minus signs may look a bit odd here.  They are present because it is $(-\tau', \theta')$ and not $(\tau', \theta')$ that transforms naturally under super-M\"obius transformations as in, e.g., \eqref{grouprepresentation}.  That action, however, preserves $\tilde{D}\tau' = -\theta'\tilde{D}\theta'$ with $\tilde{D}\equiv \partial_\theta - \theta\partial_\tau$.  While this action is consistent, the current choice makes contact with the conventions in gravity (Section \ref{supersch}).  It is easy to go between these conventions by simply letting $\tau' \to - \tau'$.  We make similar adjustments in the super-M\"obius transformations of the discrete representation carrier space coordinates $(x, \vartheta)$.

In the literature, one sometimes encounters the alternative super-M\"obius transformations
\begin{equation}
\tau'\to \tau''\equiv \frac{a\tau' + b - \alpha\theta'}{c\tau' + d - \gamma\theta'}, \qquad \theta'\to \theta''\equiv \frac{\beta\tau' + \delta + \theta'}{c\tau' + d + \gamma\theta'}. \label{conventionalmobius}
\end{equation}
These likewise respect the condition $D\tau' = \theta' D\theta'$ (i.e., $D\tau'' = \theta'' D\theta''$) on account of the property $D'\tau'' = \theta'' D'\theta''$ (and hence $D' = (D'\theta'')D''$ and $(D'\theta'')(D''\theta') = 1$), where in addition to $D\equiv \partial_\theta + \theta\partial_\tau$, we have defined the derivatives $D'\equiv \partial_{\theta'} + \theta'\partial_{\tau'}$ and $D''\equiv \partial_{\theta''} + \theta''\partial_{\tau''}$ with respect to the superfields $(\tau', \theta')$ and the transformed superfields $(\tau'', \theta'')$.  Moreover, it is common to rescale the bosonic parameters so that the bosonic part of \eqref{conventionalmobius} coincides with an ordinary M\"obius transformation \cite{Arvis:1982tq, Gieres:1992sc}:
\begin{equation}
(\tilde{a}, \tilde{b}, \tilde{c}, \tilde{d})\equiv e^{1/2}(a, b, c, d) = (1 + \beta\delta/2)(a, b, c, d), \qquad \tilde{a}\tilde{d} - \tilde{b}\tilde{c} = 1.
\end{equation}
Rescaling the fermionic parameters $\alpha, \beta, \gamma, \delta$ by powers of $e$ makes no difference.  Unlike those used throughout the text, the transformations \eqref{conventionalmobius} do not manifestly compose as a group.}
\begin{equation}
\tau' \to \frac{a\tau'-c-\beta \theta'}{-b\tau'+d+\delta \theta'}, \qquad \theta' \to \frac{\alpha\tau'-\gamma+e\theta'}{-b\tau'+d+\delta \theta'},
\end{equation}
which is a superconformal transformation that preserves the condition $D\tau' = \theta' D\theta'$, the operator \eqref{susyop} transforms as follows:
\begin{equation}
\phi_j(x,\vartheta,\tau) \to (-bx'+d+\delta \vartheta')^{2j} \phi_j(x',\vartheta',\tau), \quad x = \frac{ax'-c-\beta \vartheta'}{-bx'+d+\delta \vartheta'}, \quad \vartheta = \frac{\alpha x'-\gamma+e\vartheta'}{-bx'+d+\delta \vartheta'},
\end{equation}
as appropriate for the spin-$j$ representation.

\begin{figure}[!htb]
\centering
\includegraphics[width=0.2\textwidth]{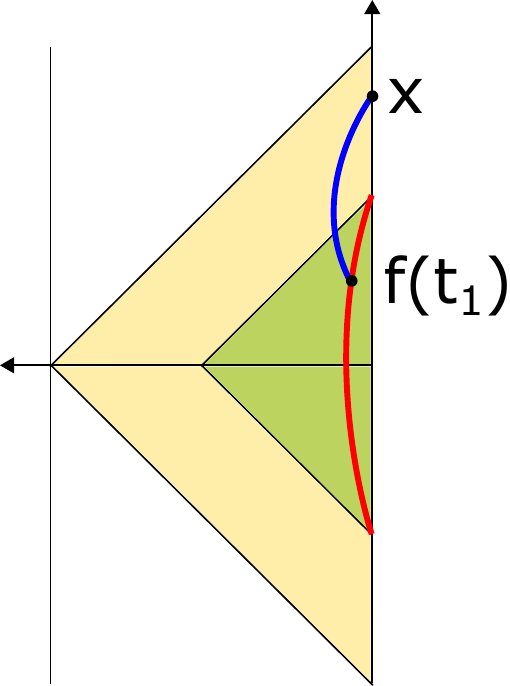}
\caption{Penrose diagram of the AdS$_2$ Poincar\'e patch (large triangle), with the black hole Penrose diagram embedded within (smaller green triangle). A bilocal operator \eqref{bosop} (blue) has one endpoint on the wiggly boundary curve (red) at proper time $t_1$ and another endpoint at Poincar\'e time $x$, which might be beyond the black hole horizon as it is here. The vertical axis measures global time, and the horizonal axis the global radial coordinate.}
\label{patchesLocalOp}
\end{figure}

From the above, it is clear that the carrier space labels $x$ and $(x|\vartheta)$ play the role of (super-)Poincar\'e coordinates of the second boundary point that is not reparametrized in the Schwarzian path integration. This has an interesting interpretation in Lorentzian time: whereas the first location parametrized by, e.g., $F(t_1) = \tanh \frac{\pi}{\beta} f(t_1)$ is always contained in the exterior of the black hole $-\infty < f < +\infty$, this is not so for the boundary time $x$. Hence these local operators seem to be able to probe behind-the-horizon physics. Preliminary studies of correlators of these objects appeared in Appendix D of \cite{Mertens:2017mtv}, but a more in-depth study would be worthwhile.\footnote{Such operators may also play a role in bulk reconstruction, namely in computing bulk-boundary correlators rather than just boundary correlators.} We depict the situation in Figure \ref{patchesLocalOp}.

\subsubsection*{$\mathcal{N}=2$ JT Supergravity and Beyond}

Since the techniques pursued in this work are very specific to the model of interest, it is important to learn whether the structural lessons that we draw here are broadly applicable. We make several comments along these lines.

A first hope is that the techniques employed here can be generalized to the $\mathcal{N}=2$ case. The corresponding boundary correlation functions are not known, and it would be very useful to make progress using group-theoretic techniques.

In this setting, the holographic boundary line and carrier space of the representations are $1|2$ dimensional, parametrized by coordinates $(x|\vartheta,\bar{\vartheta})$ where the Grassmann coordinates are related by conjugation. The super-M\"obius transformations in this case can be written in a suggestive way using the combinations $x_{\pm} = x \pm \vartheta \bar{\vartheta}$, as in \cite{Fu:2016vas}:
\begin{alignat}{2}
\label{n2tf}
x_+' &= \frac{ax_+ + c + \beta\vartheta}{bx_+ + d + \delta\vartheta}, \qquad & x_-' &= \frac{\bar{a}x_- + \bar{c} + \bar{\beta}\bar{\vartheta}}{\bar{b}x_- + \bar{d} + \bar{\delta}\bar{\vartheta}}, \\
\label{n3tf}
\vartheta' &= \frac{-\alpha x_+ - \gamma + e\vartheta}{bx_+ + d + \delta\vartheta}, \qquad & \bar{\vartheta}' &= \frac{-\bar{\alpha} x_- - \bar{\gamma} + \bar{e}\bar{\vartheta}}{\bar{b}x_- + \bar{d} + \bar{\delta}\bar{\vartheta}}.
\end{alignat}
The $5|4$ complex supernumbers $a,b,c,d,e$ and $\alpha,\beta,\gamma,\delta$ that parametrize the group element satisfy $6|4$ real relations between them, leading to the real $4|4$-dimensional supergroup
\begin{equation}
\text{SU}(1,1|1) \simeq \text{SL}(2|1) \simeq \text{OSp}(2|2)
\end{equation}
relevant for $\mathcal{N}=2$ JT supergravity. The above parametrization is that of SU$(1,1|1)$.

Given this transformation, it is not hard to propose a formula that constructs the principal series representations:
\begin{equation}
(g\circ f)(x_+,x_-, \vartheta,\bar{\vartheta}) \equiv (bx_+ + d + \delta\vartheta)^{j+q}(\bar{b}x_- + \bar{d} + \bar{\delta}\bar{\vartheta})^{j-q} f(x_+',x_-',\vartheta',\bar{\vartheta}'),
\end{equation}
where the transformed supercoordinates \eqref{n2tf} and \eqref{n3tf} appear on the right-hand side. We have introduced two representation labels $j=ik$ and $q$, the first interpretable in terms of an energy label $k^2$ and the latter interpretable in terms of an electric charge. This would be the starting point of a similar endeavor as the one we pursued for the simpler $\mathcal{N}=1$ case in this work. We leave it for future study.

Physical considerations (e.g., the dynamics of a particle in the hyperbolic superplane \cite{Uehara:1987xx, Zirnbauer1991}) may allow for an alternative derivation of the Plancherel measure for OSp$(2|2)$.  In particular, in \cite{Zirnbauer1991}, the following expression is found:
\begin{equation}
\label{N2P}
d\mu(k) = \frac{k\tanh(\pi k)}{k^2 + 1/4}\, dk,
\end{equation}
which contracts to the scale-invariant form $k^{-1}\, dk$ as $k\to\infty$ and which is reminiscent of the formula in Appendix C.2 of \cite{Mertens:2017mtv} for the density of states in the $\mathcal{N} = 2$ super-Schwarzian theory. The only replacement necessary is $\tanh \pi k \to \sinh 2 \pi k$ (see below).

Going beyond $\mathcal{N}=2$ is daunting, but recent progress for the $\mathcal{N}=4$ case is promising \cite{Heydeman:2020hhw} and highlights applications to higher-dimensional near-horizon black hole physics.

\subsubsection*{Plancherel Measure: General Real Semisimple Groups}

Our story mimics the bosonic story in which the gravitational density of states $\sinh 2 \pi \sqrt{E}$ follows from a Plancherel measure on the subsemigroup, whereas the measure on the full group would be $\tanh \pi \sqrt{E}$. One can then ask how generic this phenomenon is.

In the mathematics literature, the Plancherel measure for \emph{any} real semisimple group has been determined \cite{Helgason,Beerends}:
\begin{equation}
d\mu(\mathbf{k}) = d\mathbf{k} \, \dim(\mathbf{j}) \prod_{\alpha \in \Delta^+} \tanh \frac{\pi (\mathbf{k},\alpha)}{(\alpha,\alpha)},
\end{equation}
where the dimension of a finite-dimensional irrep with highest weight $\mathbf{j}$ is given by Weyl's dimension formula:
\begin{equation}
\dim(\mathbf{j}) = \frac{\prod_{\alpha \in \Delta^+}(\mathbf{j} + \boldsymbol{\rho},\alpha)}{\prod_{\alpha \in \Delta^+}(\boldsymbol{\rho},\alpha)}.
\end{equation}
Using that the highest weight vector for the principal series representations can generically be written as $\mathbf{j} = -\boldsymbol{\rho} + i \mathbf{k}$, we plug into the dimension formula and simplify:
\begin{equation}
\label{genPbos}
d\mu(\mathbf{k}) \sim d\mathbf{k} \prod_{\alpha \in \Delta^+} (\mathbf{k},\alpha) \tanh \frac{\pi (\mathbf{k},\alpha)}{(\alpha,\alpha)}.
\end{equation}
In particular, this formula holds for SL$(N,\mathbb{R})$, where $(\alpha,\alpha)=2$. However, the equation that one finds by taking the Schwarzian/JT limit of the $\mathcal{W}_N$ character yields (see, e.g., \cite{Datta:2021efl}):
\begin{equation}
d\mu(\mathbf{k}) \sim d\mathbf{k} \prod_{\alpha \in \Delta^+}(\mathbf{k},\alpha) \sinh \pi(\mathbf{k},\alpha),
\end{equation}
showing that just like for $N=2$ (and up to overall prefactors), the same question arises as to how one would effectively make the replacement $\tanh x/2 \to \sinh x$.

This observation is significant for black hole physics, where the large-$k$ regime probes semiclassical (large) black holes. Taking inspiration from the different Plancherel measure results \eqref{bosP}, \eqref{genPbos}, \eqref{N1P}, and \eqref{N2P}, we are led to propose a formula for the large-$k$ asymptotics of the Plancherel measure on the principal series representations for any semisimple Lie (super)group:\footnote{We assume the weight vector scales as $\mathbf{k} \sim k \mathbf{\lambda}$, with $\mathbf{\lambda}$ an order-one weight vector, as $k\to +\infty$. This is hence a ``generic'' or worst-case result that does not necessarily hold in every direction in weight space.}
\begin{equation}
\rho(k) \sim k^{\left|\Delta_B^+\right| - \left|\Delta_F^+\right|},
\end{equation}
in terms of the number of positive bosonic roots minus the number of positive fermionic roots. Since in these models, we identify the spacetime energy $E$ with the quadratic Casimir, we have the scaling $E \sim k^2$ at large $k$. Hence such a large-$k$ polynomial density of states can never account for the exponentially large number of microstates of a large black hole.

\subsubsection*{Dilaton (Super)gravities}

It is a well-known fact that \emph{any} 2d dilaton gravity theory of the type
\begin{equation}
S = - \frac{1}{2} \int d^2 x\sqrt{g}\, (\Phi R + V(\Phi)),
\end{equation}
including the recently considered deformations of JT gravity \cite{Witten:2020ert, Maxfield:2020ale, Witten:2020wvy}, has a first-order gauge theory description with nonlinear gauge algebra, i.e., as a Poisson sigma model with Poisson-Lie symmetry \cite{Schaller:1994es}.  The action is
\begin{equation}
S = \int_M A_i \wedge dX^i + P(X)^{ij} A_i \wedge A_j,
\end{equation}
with Poisson tensor that encodes the dilaton potential $V(\Phi)$:
\begin{equation}
P(X)^{12} = V(X^3), \qquad P(X)^{13} = -X^2, \qquad P(X)^{23} = X^1.
\end{equation}
The underlying nonlinear gauge algebra is \cite{Ikeda:1993aj, Ikeda:1993fh}:
\begin{align}
[J,P_a] = \epsilon_{ab}\eta^{bc} P_c, \qquad [P_a,P_b] = -\frac{1}{2}\epsilon_{ab}V(J).
\end{align}
Some explicit examples for which the resulting gauge algebra reduces to a known structure are:
\begin{alignat}{2}
V(\Phi) &= 2\Lambda \Phi \quad &&\rightarrow \quad \mathfrak{sl}(2,\mathbb{R}) \text{ Lie algebra (JT)}, \\
V(\Phi) &= \sinh b^2 \Phi \quad &&\rightarrow \quad \text{U}_q(\mathfrak{sl}(2,\mathbb{R})) \text{ quantum algebra with } q=e^{\pi i b^2}. \label{sinhpot}
\end{alignat}
Group-theoretic techniques have been successfully applied to both of these cases. It would be very interesting to learn whether generic dilaton (super)gravity models can be formulated and solved using techniques akin to these group-theoretic techniques (and their $q$-deformed cousins). For some relevant classical results on more generic dilaton gravity models, see \cite{Kyono:2017jtc, Kyono:2017pxs, Ecker:2021guy}. For the concrete result on the disk partition function, see \cite{Hirshfeld:1999xm}. See also \cite{Verlinde:2021kgt} for recent ideas in this direction.

This entire discussion pertained to the bosonic case. It would be interesting to write down the analogous class of dilaton supergravity models following \cite{Ikeda:1993fh}, in particular having in mind the model associated to U$_q(\mathfrak{osp}(1|2,\mathbb{R}))$, which would be interpretable in terms of Liouville supergravity and the minimal superstring, as we discuss next.

\subsubsection*{Liouville Supergravity}

As an immediate example of the previous goal, Liouville gravity and supergravity provide interesting setups. These theories can be formulated as ($\mathcal{N} = 1$ supersymmetric) Liouville CFT coupled to a matter CFT and ghosts, and it was further argued in \cite{Mertens:2020hbs} that Liouville gravity is equivalent to a 2d dilaton gravity with potential \eqref{sinhpot}. In recent works, it has become clear that one can formulate disk amplitudes with boundary tachyon vertex operators in a very similar language to JT (super)gravity \cite{Mertens:2020hbs, Mertens:2020pfe}.

In fact, for bosonic Liouville gravity, it was shown in \cite{Mertens:2020hbs} that the amplitudes themselves are quantum ($q$) deformations of those in JT gravity. In particular, the structures highlighted in this work (the Plancherel measure and the Whittaker function) are still present in the $q$-deformed case. The Whittaker function itself was first derived in the context of the $q$-deformed Toda chain \cite{Kharchev:2001rs}, and was shown to lead to the correct Liouville gravity vertex function in \cite{Mertens:2020hbs}.

We can follow similar logic for Liouville supergravity, for which the required Whittaker function has not yet been derived in the mathematics literature. We propose the following expression for the $q$-deformed Whittaker function of U$_q(\mathfrak{osp}(1|2))$:
\begin{align}
\label{qmelb}
\psi^{\epsilon, \pm}_s(x) = e^{\pi i s x}\int_{-\infty}^{+\infty} &\frac{d\zeta}{(4\pi b)^{-i\zeta/b-is/b}}e^{-\pi i \frac{\epsilon}{2} (\zeta^2 + 2s \zeta)} e^{\pi i \zeta x} \\
&\times \left[S_{\NS}(-i\zeta) S_{\R}(-2i s -i \zeta ) \pm S_{\R}(-i\zeta) S_{\NS}(-2i s -i \zeta )\right], \nonumber
\end{align}
where $\epsilon=-1,0,1$ is an additional parameter allowed by the deformation.\footnote{The supersymmetric double sine functions that appear in this expression are defined in terms of the ordinary double sine function $S_b$ as follows:
\begin{align}
\label{Sindef}
S_{\NS}(x) &= S_b\left(\frac{x\vphantom{1}}{2}\right) S_b\left(\frac{x}{2} + \frac{Q}{2}\right), \qquad S_{\R}(x) = S_b\left(\frac{x}{2} + \frac{b}{2}\right) S_b\left(\frac{x}{2} + \frac{1}{2b}\right).
\end{align}
They have the following $b\to 0$ limits:
\begin{equation}
\label{bnullim}
S_{\NS}(bx) \, \to \, \frac{1}{\sqrt{2\pi}} 2^{\frac{x}{2}} (2\pi b^2)^{\frac{x}{2}-\frac{1}{2}} \Gamma\left(\frac{x\vphantom{1}}{2} \right), \qquad
S_{\R}(bx) \, \to \, \frac{1}{\sqrt{2\pi}} 2^{\frac{x}{2} - \frac{1}{2}} (2\pi b^2)^{\frac{x}{2}} \Gamma\left(\frac{x}{2} + \frac{1}{2} \right).
\end{equation}
For more details on these definitions, we refer to the above references.} The deformation parameter is $q=e^{\pi i b^2}$, with the classical limit corresponding to $q\to 1$ or $b\to 0$.\footnote{It would be interesting to clarify the relation between this deformation and the one used in \cite{Berkooz:2020xne} in the context of double-scaled supersymmetric SYK models.  The two deformations seem to go in different directions away from $q = 1$ in the complex $q$-plane.} Here, we will illustrate that the proposal \eqref{qmelb} has the correct classical limit \eqref{whitcart} determined in this work. Further implications and a derivation will be treated in upcoming work \cite{Fan:2021bwt}. To arrive at the classical limit, we scale the variables as
\begin{equation}
\zeta = 2ibt, \qquad e^{x} = be^{-\phi}, \qquad s = bk,
\end{equation}
in the limit $b\to 0$. Using suitable $b\to 0$ limits of the double sine functions \eqref{bnullim}, we need to evaluate the integrals
\begin{align}
e^{-ik\phi} \int_{i\mathbb{R}} dt\, \Gamma(t) \Gamma(t-ik+1/2) e^{2\phi t} &= 4\pi ie^{-\phi/2} K_{ik-1/2}(2 e^{-\phi}), \\
e^{-ik\phi} \int_{i\mathbb{R}} dt\, \Gamma(t+1/2) \Gamma(t-ik) e^{2\phi t} &= 4\pi ie^{-\phi/2} K_{ik+1/2}(2 e^{-\phi}),
\end{align}
leading to the limit:
\begin{equation}
\psi^{\epsilon, \pm}_s\left(\frac{x}{\pi b}\right) \to e^{-\phi/2} \left(K_{ik-1/2}(2 e^{-\phi}) \pm K_{ik+1/2}(2 e^{-\phi}) \right).
\end{equation}
We have suppressed some constant factors and details about the integration contour.  This indeed yields the expression \eqref{whitcart} upon setting $\nu=\lambda=1$ and absorbing the square root of the Haar measure $e^{\phi/2}$ into the Whittaker functions themselves, such that the remaining $\phi$-integral has a flat measure. With this Whittaker function, one can indeed reproduce the $\mathcal{N}=1$ Liouville supergravity vertex functions present in the boundary tachyon two-point function \cite{Mertens:2020pfe}, as will be shown elsewhere \cite{Fan:2021bwt}.

\section*{Acknowledgements}

We thank A.\ Blommaert and G.\ J.\ Turiaci for discussions, and G.\ J.\ Turiaci for comments on a preliminary draft.  YF thanks the participants of the Geometry and String Theory Seminar at UT Austin for some excursions into the mathematical aspects of JT gravity.  The work of YF was supported by the National Science Foundation under Grant No.\ PHY-1914679.  TM gratefully acknowledges financial support from Research Foundation Flanders (FWO Vlaanderen).

\appendix

\section{Supernumbers}
\label{app:supernumbers}

In this appendix, we define and collect our conventions for supernumbers as elements of a Grassmann algebra.

Within a Grassmann algebra $\Lambda_n$ over $\mathbb{C}$, spanned by $n$ Grassmann variables $\vartheta_1, \ldots, \vartheta_n$ satisfying
\begin{equation}
\vartheta_i \vartheta_j = - \vartheta_j \vartheta_i, \qquad \vartheta_i^2 = 0,
\end{equation}
we can expand an arbitrary supernumber as
\begin{equation}
\mathbf{z} = z_0 + \sum_{\alpha} z_{\alpha}e_\alpha,
\end{equation}
where the prefactors $z_0$ and $z_\alpha$ are complex numbers and the basis elements $e_\alpha$ span the set of all elementary Grassmann elements $\vartheta_{i_1} \cdots \vartheta_{i_t}$ with $t = 1, \ldots, n$. To avoid spurious cancellations, we take $n\to \infty$ throughout this work.

Following common convention \cite{DeWitt:1992cy}, we refer to the purely numerical piece $z_0$ as the \emph{body}, and the remainder as the \emph{soul} of the supernumber.  While the body is Grassmann-even, the soul may have both even and odd parts. In \cite{Alpay_20192}, a definition of \emph{positive} supernumber was formulated. One first defines the conjugate supernumber as\footnote{Our definition is modified from that of \cite{Alpay_20192} because we use a different definition for complex conjugation of fermionic variables. Many different definitions are possible, and arguments can be made for each one \cite{Alpay_2019}.}
\begin{equation}
\mathbf{z}^* = z_0^* + \sum_{\alpha} z_{\alpha}^* e_\alpha.
\end{equation}
This definition corresponds to taking the complex conjugate of all numerical factors, and defining complex conjugation on Grassmann numbers as preserving the order:
\begin{equation}
(\vartheta_1 \vartheta_2 \cdots \vartheta_i)^* = \vartheta_1 \vartheta_2 \cdots \vartheta_i.
\end{equation}
A \emph{real} supernumber is defined as one satisfying $\mathbf{z}^* = \mathbf{z}$. Our choice of conjugation ensures that combinations of the following form are real:\footnote{Another possible choice consistent with this reality condition is to take $\vartheta^* = i \vartheta$ and to reverse the order of a product of Grassmann numbers: $(\vartheta_i \vartheta_j)^* = \vartheta_j \vartheta_i$. This choice was adopted in \cite{Matsumoto:1989hc}.}
\begin{equation}
(\tau_1-\tau_2-\vartheta_1\vartheta_2)^* = \tau_1-\tau_2-\vartheta_1\vartheta_2,
\end{equation}
where $\tau_1, \tau_2\in \mathbb{R}$.  A nonnegative supernumber is then defined as one for which there exists another supernumber $\mathbf{w}$ such that
\begin{equation}
\mathbf{z} = \mathbf{w} \mathbf{w}^*.
\end{equation}
It was shown in \cite{Alpay_20192} that a positive supernumber is automatically real ($\mathbf{z}^* = \mathbf{z}$), and more\-over that its positivity is equivalent to the positivity of its body:
\begin{equation}
\mathbf{z} > 0 \Longleftrightarrow z_0 > 0.
\end{equation}
This means that only the purely numerical piece $z_0$ of a supernumber determines whether it is positive or negative: the soul is ``infinitesimal'' and therefore irrelevant to positivity. An ordering is then naturally implemented for supernumbers that have different bodies $z_0$, where $\mathbf{z}_1 > \mathbf{z}_2$ iff $ \mathbf{z}_1-\mathbf{z}_2 > 0$. Finally, the absolute value of a supernumber can be defined as
\begin{equation}
\mathbf{z} \equiv \operatorname{sgn}(\mathbf{z})|\mathbf{z}|,
\end{equation}
where $\operatorname{sgn}(\mathbf{z})=1$ if $\mathbf{z}>0$ and $\operatorname{sgn}(\mathbf{z})=-1$ if $\mathbf{z}<0$.

\section{Bosonic JT Gravity} \label{app:bosonic}

To orient ourselves with respect to JT supergravity, it is helpful to recall some of the cor\-res\-pon\-ding results in the bosonic case.  For all groups and semigroups considered, we focus on the matrix elements and Plancherel measure for the continuous series irreps.  Up to numerical factors, the Plancherel measure $\rho(k)$ is $k\tanh(\pi k)$ for $\sltr$ and $k\sinh(2\pi k)$ for $\slr$.  We can derive $\rho(k)$ from the Haar measure $dg$ and the orthogonality relation of group matrix elements in an appropriate basis (e.g., parabolic or hyperbolic).

\subsection{\texorpdfstring{$\mathfrak{sl}(2, \mathbb{R})$}{sl(2, R)} BF Theory}

To set our conventions for 2d geometry, we first recall the formulation of bosonic JT gravity as an $\mathfrak{sl}(2, \mathbb{R})$ BF theory (summarized in, e.g., \cite{Saad:2019lba}).  We work in Euclidean signature and set the cosmological constant to $\Lambda = 2$.  The Euclidean JT gravity action is
\begin{equation}
S_\text{JT} = -\frac{1}{16\pi G}\left[\int_{\mathcal{M}} d^2 x\sqrt{g}\, \phi(R + 2) + 2\int_{\partial\mathcal{M}} dt\sqrt{\gamma}\, \phi_b(K - 1)\right],
\label{JTaction}
\end{equation}
where $\phi_b\equiv \phi|_{\partial\mathcal{M}}$.  We define an orthonormal frame by
\begin{equation}
g_{\mu\nu} = e_\mu^a e_\nu^b\delta_{ab},
\end{equation}
where $a, b\in \{0, 1\}$.  Writing the zweibein as a one-form $e^a = e_\mu^a\, dx^\mu$, the torsion-free spin connection $\omega^{ab} = \smash{\omega_\mu^{[ab]}}\, dx^\mu$ is determined by
\begin{equation}
de^a + \omega^a{}_b\wedge e^b = 0.
\end{equation}
In 2d, we have $\omega^{ab} = \epsilon^{ab}\omega$ as well as
\begin{equation}
d^2 x\, \sqrt{g} = e^0\wedge e^1, \qquad d^2 x\, \sqrt{g}R = 2\, d\omega.
\end{equation}
Therefore, in the first-order formulation,
\begin{equation}
\frac{1}{4}\int_{\mathcal{M}} d^2 x\, \sqrt{g}\phi(R + 2)\sim \frac{1}{2}\int_{\mathcal{M}} [\phi(d\omega + e^0\wedge e^1) + \phi_a(de^a + \epsilon^a{}_b\omega\wedge e^b)] = \int_{\mathcal{M}} \Tr(\mathbf{B}\mathbf{F}),
\end{equation}
where we introduced the Lagrange multipliers $\phi^a$ to enforce the torsion constraint as well as the $\mathfrak{sl}(2, \mathbb{R})$-valued fields $\mathbf{B}, \mathbf{A}$ (the latter with field strength $\mathbf{F} = d\mathbf{A} + \mathbf{A}\wedge \mathbf{A}$) given by
\begin{equation}
B_I = (\phi_a, \phi), \quad A^I = (e^a, \omega), \quad \mathbf{B} = B^I J_I, \quad \mathbf{A} = A^I J_I.
\end{equation}
The generators $J_0, J_1, J_2$ satisfy
\begin{equation}
[J_I, J_J] = \epsilon_{IJK}J^K, \qquad \Tr(J_I J_J) = \frac{1}{2}\eta_{IJ},
\end{equation}
where $\epsilon_{012} = -1$ and $\mathfrak{sl}(2, \mathbb{R})$ indices $I, J, K\in \{0, 1, 2\}$ are raised and lowered by $\eta_{IJ} = \operatorname{diag}(1, 1, -1)$.  This basis is related to the Cartan-Weyl basis
\begin{equation}
H = \frac{1}{2}\left[\begin{array}{cc} 1 & 0 \\ 0 & -1 \end{array}\right], \qquad E^- = \left[\begin{array}{cc} 0 & 0 \\ 1 & 0 \end{array}\right], \qquad E^+ = \left[\begin{array}{cc} 0 & 1 \\ 0 & 0 \end{array}\right]
\label{sltrbasis}
\end{equation}
for the $\mathfrak{sl}(2, \mathbb{R})$ algebra as follows:
\begin{equation}
J_0 = -H, \qquad J_1 = \frac{1}{2}(E^- + E^+), \qquad J_2 = \frac{1}{2}(E^- - E^+).
\label{sltridentification}
\end{equation}
The contour of integration for $\mathbf{B}$ is understood to be imaginary in Euclidean signature.  Integrating out $\mathbf{B}$ implements the constraint $\mathbf{F} = 0$, which reduces the path integral to an integral over flat $\mathfrak{sl}(2, \mathbb{R})$ connections $\mathbf{A}$.  Infinitesimal gauge transformations take the form
\begin{equation}
\delta_\epsilon\mathbf{A} = d\epsilon + [\mathbf{A}, \epsilon], \qquad \delta_\epsilon\mathbf{B} = [\mathbf{B}, \epsilon],
\label{infinitesimalgauge}
\end{equation}
where $\epsilon$ is an $\mathfrak{sl}(2, \mathbb{R})$-valued parameter.  For flat $\mathbf{A}$, these transformations are interpretable as infinitesimal diffeomorphisms and local Lorentz transformations.

\subsection{\texorpdfstring{$\sltr$}{SL(2, R)} Group Theory}
\label{app:ha}
In the bosonic case $\sltr$, we choose the carrier space of the spin-$j$ representation (on which the Casimir evaluates to $j(j + 1)$) to be $L^2(\mathbb{R})$, with group action defined by
\begin{equation}
(g\circ f)(x) = |bx + d|^{2j}f\left(\frac{ax + c}{bx + d}\right).
\end{equation}
We let $j\in \mathbb{C}$, with an eye toward the principal series representations.  Using \eqref{sltrbasis}, we get
\begin{align}
(e^{2\phi H}\circ f)(x) &= e^{-2j\phi}f(e^{2\phi}x), \label{sltrdilatation} \\
(e^{\gamma_{\m} E^-}\circ f)(x) &= f(x + \gamma_{\m}), \\
(e^{\gamma_{\+} E^+}\circ f)(x) &= \textstyle |\gamma_{\+} x + 1|^{2j}f\left(\frac{x}{\gamma_{\+} x + 1}\right), \label{sltrlastaction}
\end{align}
and hence the Borel-Weil realization
\begin{equation}
\hat{H} = x\partial_x - j, \qquad \hat{E}^- = \partial_x, \qquad \hat{E}^+ = -x^2\partial_x + 2jx
\label{sltrborelweil}
\end{equation}
satisfying the $\mathfrak{sl}(2, \mathbb{R})$ algebra
\begin{equation}
[H, E^\pm] = \pm E^\pm, \qquad [E^+, E^-] = 2H.
\end{equation}
Note that our conventions differ from those in Appendix G of \cite{Blommaert:2018iqz}.

Unitarity of the representation constrains the value of $j$.  We compute that antihermiticity of the generators \eqref{sltrborelweil} with respect to the measure $dx$ requires that $j = -1/2 + ik$ for $k\in \mathbb{R}$. (By assumption, the relevant functions on $\mathbb{R}$ decay sufficiently fast that integrating $\partial_x$, $x\partial_x$, $x^2\partial_x$ by parts produces no boundary terms.) Moreover, a change of variables $x = \frac{dx' - c}{-bx' + a}$ in the inner product $\int dx\, F(x)^\ast G(x)$ shows that if $j = -1/2 + ik$, then
\begin{equation}
\int dx\, F(x)^\ast|bx + d|^{2j}G\left(\frac{ax + c}{bx + d}\right) = \int dx'\left(|{-bx' + a}|^{2j}F\left(\frac{dx' - c}{-bx' + a}\right)\right)^\ast G(x'),
\end{equation}
so that the adjoint action is precisely by $g^{-1}$.

The Plancherel measure can be computed as follows \cite{Blommaert:2018iqz}.  We first compute, using the Gauss decomposition
\begin{equation}
g = e^{\gamma_{\m}E^-}e^{2\phi H}e^{\gamma_{\+} E^+} = \left[\begin{array}{cc}
1 & 0 \\
\gamma_{\m} & 1
\end{array}\right]\left[\begin{array}{cc}
e^\phi & 0 \\
0 & e^{-\phi}
\end{array}\right]\left[\begin{array}{cc}
1 & \gamma_{\+} \\
0 & 1
\end{array}\right] = \left[\begin{array}{cc} e^\phi & \gamma_{\+}e^\phi \\ \gamma_{\m} e^\phi & e^{-\phi} + \gamma_{\m}\gamma_{\+} e^\phi \end{array}\right],
\label{gausssltr}
\end{equation}
the bi-invariant metric for the Poincar\'e patch of $\sltr$:
\begin{equation}
ds^2 = \frac{1}{2}\Tr((g^{-1}dg)^{\otimes 2}) = d\phi^2 + e^{2\phi}\, d\gamma_{\m}\, d\gamma_{\+},
\end{equation}
and hence the Haar measure $dg = \frac{1}{2}e^{2\phi}\, d\phi\, d\gamma_{\m}\, d\gamma_{\+}$.  From the normalized wavefunctions
\begin{equation}
\langle x|\nu_-\rangle = \frac{1}{\sqrt{2\pi}}e^{i\nu x}, \quad \langle x|\lambda_+\rangle = \frac{1}{\sqrt{2\pi}}|x|^{2j}e^{i\lambda/x},
\end{equation}
we deduce the mixed parabolic matrix elements
\begin{align}
\label{repmatbos}
R_{k, \nu\lambda}(g) &\equiv \langle\nu_-|g|\lambda_+\rangle = e^{i\gamma_{\m}\nu}e^{i\gamma_{\+}\lambda}\langle\nu_-|e^{2\phi H}|\lambda_+\rangle \\
&= \frac{2}{\pi}e^{i\gamma_{\m}\nu}e^{i\gamma_{\+}\lambda}e^{-\phi}\cosh(\pi k)\left(\frac{\lambda}{\nu}\right)^{ik}K_{2ik}(2e^{-\phi}\sqrt{\nu\lambda}),
\end{align}
where we have used \eqref{sltrdilatation} and substituted $j = -1/2 + ik$.  In the second line, we assumed $\nu\lambda >0$ for simplicity. Via the orthogonality relation\footnote{Assuming $\mu, \nu > 0$, the relation \eqref{orthobessel} follows from the identity
\begin{equation}
\lim_{\alpha\to 0}\frac{\Gamma(\alpha + ix)\Gamma(\alpha - ix)}{\Gamma(2\alpha)} = 2\pi\delta(x)
\end{equation}
(see Appendix A of \cite{Hogervorst:2017sfd}) applied to the $\alpha\to 0$ limit of
\begin{equation}
\int_0^\infty dx\, x^{2\alpha - 1}K_{2i\mu}(x)K_{2i\nu}(x) = \frac{2^{2\alpha - 3}}{\Gamma(2\alpha)}\Gamma(\alpha\pm i\mu\pm i\nu).
\label{alphabessel}
\end{equation}
This relation is implicit in the Kontorovich-Lebedev integral transform.}
\begin{equation}
\int_0^\infty \frac{dx}{x}\, K_{2i\mu}(x)K_{2i\nu}(x) = \frac{\pi^2}{8\mu\sinh(2\pi\mu)}\delta(\mu - \nu),
\label{orthobessel}
\end{equation}
we conclude that
\begin{equation}
\int dg\, R_{k, \nu\lambda}(g)^\ast R_{k', \nu'\lambda'}(g) = \frac{\pi^2}{2k\tanh(\pi k)}\delta(k - k')\delta(\nu - \nu')\delta(\lambda - \lambda').
\end{equation}
This is not the whole story because the $\sltr$ group manifold is actually covered by four patches \cite{Forgacs:1989ac}. It can be shown that summing the contributions of all patches just gives a factor of four, so we obtain finally
\begin{equation}
\label{bosP}
\langle R_{k, \nu\lambda}, R_{k', \nu'\lambda'}\rangle = \frac{\delta(k - k')\delta(\nu - \nu')\delta(\lambda - \lambda')}{\rho(k)}, \quad \rho(k) = \frac{2k\tanh(\pi k)}{(2\pi)^2}.
\end{equation}
This is the desired Plancherel measure.

Harmonic analysis on $\sltr$ is a well-studied subject. Let us quickly review the salient points; we will elaborate more on the supergroup case in Appendices \ref{app:regu} and \ref{app:hasu}.

Every Lie group has a left regular representation in which the group acts on itself by left multiplication. Infinitesimally, this action corresponds to a realization in terms of differential operators on the group manifold:
\begin{align}
\hat{L}_{E^-} = -\partial_{\gamma_{\m}}, \qquad \hat{L}_H = -\frac{1}{2} \partial_\phi + \gamma_{\m} \partial_{\gamma_{\m}}, \qquad \hat{L}_{E^{+}} = -\gamma_{\m} \partial_\phi +\gamma_{\m}^2 \partial_{\gamma_{\m}} - e^{-2\phi}\partial_{\gamma_{\+}},
\end{align}
from which one computes the Casimir operator to be
\begin{align}
\mathcal{C} = \hat{L}_H ^2 + \frac{1}{2} \left(\hat{L}_{E^{+}}\hat{L}_{E^-} + \hat{L}_{E^-}\hat{L}_{E^{+}} \right) = \frac{1}{4}\partial_\phi^2 + \frac{1}{2}\partial_\phi + e^{-2\phi}\partial_{\gamma_{\m}}\partial_{\gamma_{\+}}.
\end{align}
To find the eigenfunctions $f$ with eigenvalues $j(j + 1)$, we choose to diagonalize $\partial_{\gamma_{\m}} = i \nu$ and $\partial_{\gamma_{\+}} = i\lambda$ (this corresponds to working in the mixed parabolic basis). Upon setting $f(\phi) = e^{-\phi}g(\phi)$, we get the Liouville minisuperspace eigenvalue problem:
\begin{equation}
\left( - \frac{1}{4} \partial_\phi^2 + \nu\lambda e^{-2\phi} \right) g(\phi) = k^2 g(\phi)
\end{equation}
with potential $V(\phi) = \nu\lambda e^{-2\phi}$, where we have written $j=-1/2+ik$. From this equation, we obtain the full (delta-function normalizable) Casimir eigenfunctions:
\begin{align}
\nu\lambda >0: \qquad &e^{i\nu \gamma_{\m}} e^{i\lambda \gamma_{\+}} e^{-\phi} K_{2ik}(2\sqrt{\nu\lambda}e^{-\phi}), \\
\nu\lambda <0: \qquad &e^{i\nu \gamma_{\m}} e^{i\lambda \gamma_{\+}} e^{-\phi} J_{2ik}(2\sqrt{-\nu\lambda}e^{-\phi}),
\end{align}
all of which have positive energy $k^2$ (and negative Casimir eigenvalue $-1/4-k^2$). These solutions can be interpreted as the continuous series representation matrices; the first one, for instance, matches (up to normalization) with \eqref{repmatbos}.

For the second case where $\nu\lambda <0$, negative-energy solutions exist as well, leading to the Casimir eigenfunctions:
\begin{align}
\nu\lambda <0: \qquad &e^{i\nu \gamma_{\m}} e^{i\lambda \gamma_{\+}} e^{-\phi} J_{2j+1}(2\sqrt{-\nu\lambda}e^{-\phi}).
\end{align}
These can be interpreted as the discrete series representation matrices, with positive Casimir eigenvalue $j(j+1) > 0$.

\begin{figure}[!htb]
\centering
\includegraphics[width=0.9\textwidth]{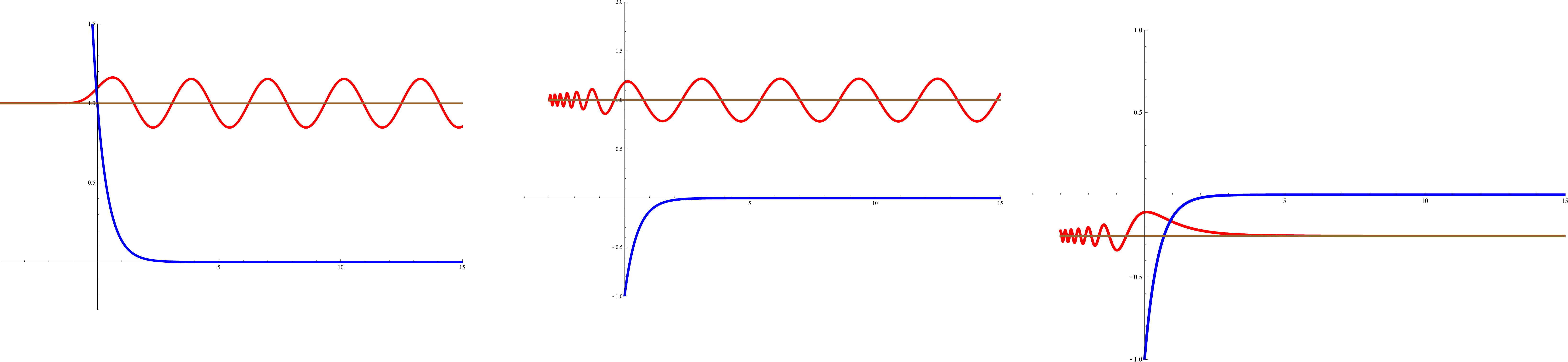}
\caption{Blue: Liouville potential $V(\phi)$. Red: wavefunction solution with energy eigenvalue marked by the brown line. Left: $\nu\lambda >0$ for $k=1$ with energy value $k^2=1$. Middle: $\nu\lambda <0$ for $k=1$ with energy value $k^2=1$. Right: $\nu\lambda <0$ for $j=0$ with energy value $k^2=-1/4$.}
\label{LiouPanel}
\end{figure}

The Liouville eigenfunctions $g(\phi)$ are illustrated in Figure \ref{LiouPanel}.

The spin label $j$ of the discrete representations is not discretized within this setup. The reason can be traced back to the fact that we are actually finding all representation matrices for the universal cover $\widetilde{\text{SL}}(2,\mathbb{R})$ of $\sltr$. For the former, it is indeed known that the discrete representations are not truly discrete. However, because we do know from direct computation that the discrete representations of $\sltr$ are restricted to $2j \in -\mathbb{N}$, we can formulate a rule of thumb to immediately find the correct values of $j$ within the above analysis. For the discrete representations where $\nu\lambda <0$, we demand single-valuedness of the representation matrix element when $e^{-\phi} \to e^{2\pi i}e^{-\phi}$. The BesselJ function is generically a multi-valued function except when its index $2j+1$ is an integer. This effectively causes a restriction to $2j \in - \mathbb{N}$.\footnote{Representations where $j \to -1-j$ are equivalent, allowing us to choose this range of $j$.} We will use this trick also for the supergroup case in Appendix \ref{app:hasu}. It would be instructive to understand this rule a bit better, for instance by comparing with computations done in the elliptic basis \cite{Kitaev:2017hnr}, but for our purposes, it is sufficient.

\subsection{\texorpdfstring{$\slr$}{SL+(2, R)} Semigroup Theory}
\label{app:slr}

For either $\sltr$ or $\slr$, a basis for the spin-$j$ representation is obtained by diagonalizing a chosen generator.  For $\sltr$, it is convenient to work in the mixed parabolic basis, where matrix elements of group elements are evaluated in the basis of eigenstates of $E^+$ on the right and $E^-$ on the left.  On the other hand, the natural basis for $\slr$ is the hyperbolic basis, which corresponds to diagonalizing the hyperbolic generator $H$.  See Appendix H of \cite{Blommaert:2018iqz}, as well as \cite{Blommaert:2018oro, Dijkgraaf:1991ba} for earlier discussions.

The carrier space of the spin-$j$ representation of $\slr$ is $L^2(\mathbb{R}^+)$, with inner product $\int_0^\infty dx\, f(x)^\ast g(x)$.  Consider taking the adjoints of the generators \eqref{sltrborelweil}.  We have
\begin{equation}
\int_0^\infty dx\, f(x)^\ast(x^n\partial_x g(x)) = [x^n f(x)^\ast g(x)]_0^\infty - \int_0^\infty dx\, (x^{n-1}(x\partial_x + n)f(x))^\ast g(x).
\end{equation}
If $f\in L^2(\mathbb{R}^+)$, then $|f(x)|^2$ must decay faster than $1/x$ as $x\to\infty$ and grow more slowly than $1/x$ as $x\to 0$.  Therefore, to ensure that the boundary terms vanish for all $f, g\in L^2(\mathbb{R}^+)$, we must have $n = 1$, which implies that $H^\dag = -H$ for $j = -1/2 + ik$.  Hence the only $\mathfrak{sl}(2, \mathbb{R})$ generator that is antihermitian on $\mathbb{R}^+$ is the hyperbolic generator $H$: the parabolic generators $E^\pm$ are not.  Correspondingly, the eigenfunctions of $E^\pm$ are not delta-function normalizable on $\mathbb{R}^+$.  Indeed, these are (ignoring the overall prefactor)
\begin{equation}
\langle x|\nu_-\rangle = e^{-\nu x}, \quad \langle x|\nu_+\rangle = x^{2is - 1}e^{-\nu/x}
\end{equation}
for $E^-$ and $E^+$, respectively, written with eigenvalues $-\nu$.  These two sets of eigenfunctions do not satisfy orthogonality relations on $\mathbb{R}^+$ (by contrast, we would have $\int_{\mathbb{R}} \frac{dx}{x^2}\, e^{i(\nu - \nu')/x} = \int_{\mathbb{R}} dx\, e^{i(\nu - \nu')x} = 2\pi\delta(\nu - \nu')$ for real $\nu, \nu'$).  So for $\slr$, only the eigenfunctions of $H$ furnish a basis for the carrier space.

Consider the hyperbolic basis for $\sltr$.  The properly normalized eigenfunctions of $H$ on $\mathbb{R}^+$ or $\mathbb{R}^-$ are
\begin{equation}
\langle x|s, \pm\rangle = \frac{1}{\sqrt{2\pi}}(\pm x)^{is - 1/2} \quad (\pm x > 0),
\label{hypereigenfunctions}
\end{equation}
with eigenvalue $i(s - k)$ where $s\in \mathbb{R}$.  These form a basis on either $\mathbb{R}^+$ or $\mathbb{R}^-$:
\begin{align}
\delta(s_1 - s_2) &= \pm\int_0^{\pm\infty} dx\, \langle s_1, \pm|x\rangle\langle x|s_2, \pm\rangle = \frac{1}{2\pi}\int_0^{\pm\infty} \frac{dx}{x}\, (\pm x)^{-i(s_1 - s_2)}, \label{delta1} \\
\delta(x - x') &= \int_{-\infty}^\infty ds\, \langle x|s, \pm\rangle\langle s, \pm|x'\rangle = \frac{1}{2\pi}\int_{-\infty}^\infty d\alpha\, (\pm x)^{is - 1/2}(\pm x')^{-is - 1/2}. \label{delta2}
\end{align}
We focus on $\mathbb{R}^+$.  The representation matrix elements of $\slr$ on $L^2(\mathbb{R}^+)$ in the hyperbolic basis are denoted by
\begin{equation}
K_{s_1 s_2}^{++}(g)\equiv \langle s_1, +|g|s_2, +\rangle.
\end{equation}
Their composition law is
\begin{equation}
K_{s_1 s_2}^{++}(g_1 g_2) = \int_{-\infty}^\infty ds\, K_{s_1 s}^{++}(g_1)K_{s s_2}^{++}(g_2).
\end{equation}
The Gauss decomposition and metric are the same as for the Poincar\'e patch of $\sltr$, but now with the additional restriction that $\gamma_{\m}, \gamma_{\+} > 0$.  By inserting a resolution of the identity $\int_0^\infty |x\rangle\langle x|$, applying \eqref{sltrdilatation}--\eqref{sltrlastaction} to \eqref{hypereigenfunctions}, and using the beta function integral
\begin{equation}
B(x, y) = \int_0^\infty \frac{t^{x-1}}{(1 + t)^{x+y}}\, dt = \frac{\Gamma(x)\Gamma(y)}{\Gamma(x + y)},
\label{betafunction}
\end{equation}
we compute that\footnote{These expressions all have the correct $\phi, \gamma_{\m}, \gamma_{\+}\to 0$ limits $K_{s_1 s_2}^{++}(\mathbf{1}) = \delta(s_1 - s_2)$ as a consequence of the distributional identity $\lim_{y\to 0} \Gamma(ix)y^{-ix} = 2\pi\delta(x)$, which in turn follows from \eqref{cahenmellin}.}
\begin{align}
K_{s_1 s_2}^{++}(e^{2\phi H}) &= e^{2i(s_1 - k)\phi}\delta(s_1 - s_2), \label{diagonalKpp} \\
K_{s_1 s_2}^{++}(e^{\gamma_{\m} E^-}) &= \frac{1}{2\pi}\frac{\Gamma(1/2 - is_1)\Gamma(is_1 - is_2)}{\Gamma(1/2 - is_2)}\gamma_{\m}^{is_2 - is_1}, \label{Kppminus} \\
K_{s_1 s_2}^{++}(e^{\gamma_{\+}E^+}) &= \frac{1}{2\pi}\frac{\Gamma(1/2 - 2ik + is_1)\Gamma(is_2 - is_1)}{\Gamma(1/2 - 2ik + is_2)}\gamma_{\+}^{is_1 - is_2}. \label{Kppplus}
\end{align}
We then have the generic matrix element
\begin{align}
\label{genericKpp}
K_{s_1 s_2}^{++}(g) &= \int_{-\infty}^\infty ds\, ds'\, K_{s_1 s}^{++}(e^{\gamma_{\m} E^-})K_{ss'}^{++}(e^{2\phi H})K_{s' s_2}^{++}(e^{\gamma_{\+}E^+}) \\
 &=\frac{1}{2\pi} \gamma_+^{n}\gamma_-^{m}\sinh^{2j}\zeta \frac{\Gamma(-j-m)\Gamma(-j+m)}{\Gamma(-2j)}{}_2F_1\left(-j-m,-j-n;-2j;-\frac{1}{\sinh^2\zeta}\right), \nonumber
\end{align}
evaluated using the Barnes integral representation of the hypergeometric function. We have denoted $m=ik-is_1$ and $n=ik-is_2$, and we have introduced the coordinate $\zeta$ through $\sinh^2\zeta \equiv \gamma_{\+}\gamma_{\m} e^{2\phi}$. The quantum numbers $m$ and $n$ are the eigenvalues of the hyperbolic generator $H$ \cite{V}.  The above result can also be found as a classical limit of the $q$-deformed hyperbolic representation matrix element constructed in \cite{ip2012representation}. One can show very explicitly that this expression is indeed a solution to the Casimir eigenvalue equation. Intriguingly, it is also precisely equal to a global conformal block. Besides both expressions being solutions to the Casimir eigenvalue equation, the deeper meaning of this observation eludes us, but we will show in Appendix \ref{app:unit} that a similar observation is true of the hyperbolic representation matrices for OSp$^{+}(1|2,\mathbb{R})$.

Unitarity of this representation, namely
\begin{equation}
\int_{-\infty}^\infty ds\, K_{s_1 s}^{++}(g)K_{s_2 s}^{++}(g)^\ast = \delta(s_1 - s_2),
\label{unitarity}
\end{equation}
can be readily established using the integral representations \eqref{betafunction} of \eqref{Kppminus}--\eqref{Kppplus}, as was done in \cite{Blommaert:2018iqz}. Irreducibility of this representation requires a separate proof. We provide it in Appendix \ref{app:irre} as a warmup for the analogous supergroup proof.

The mixed parabolic matrix elements are 
\begin{equation}
R_{k, \nu\lambda}(g)\equiv \langle\nu_-|g|\lambda_+\rangle = \langle\nu_-|e^{-\gamma_{\m}E^-}e^{2\phi H}e^{\gamma_{\+}E^+}|\lambda_+\rangle = e^{\gamma_{\m}\nu}e^{-\gamma_{\+}\lambda}\psi^{k}_{\lambda,\nu}(\phi),
\end{equation}
where we have defined the Whittaker functions
\begin{equation}
\psi^{k}_{\lambda,\nu}(\phi) \equiv \langle\nu_-|e^{2\phi H}|\lambda_+\rangle = 2e^{-\phi}\left(\frac{\lambda}{\nu}\right)^{ik}K_{2ik}(2e^{-\phi}\sqrt{\nu\lambda})
\label{whitslr}
\end{equation}
(see \eqref{besselint}).  By virtue of \eqref{orthobessel}, the latter satisfy the orthogonality relation
\begin{equation}
\int \left(\frac{1}{2}e^{2\phi}\, d\phi\right)\psi^{k}_{\lambda,\nu}(\phi)^\ast \psi^{k'}_{\lambda,\nu}(\phi) = \frac{\pi^2}{4k\sinh(2\pi k)}\delta(k - k'),
\end{equation}
from which we read off the Plancherel measure.

Note that we can always expand a parabolic state in the hyperbolic basis (but not vice versa).  Using the Cahen-Mellin integral
\begin{equation}
e^{-y} = \frac{1}{2\pi i}\int_{c - i\infty}^{c + i\infty} ds\, \Gamma(s)y^{-s} = \frac{1}{2\pi}\int_{-\infty}^{\infty} dt\, \Gamma(it)y^{-it} \quad (y > 0)
\label{cahenmellin}
\end{equation}
and the delta function identity \eqref{delta1}, we compute the overlaps
\begin{align}
\langle s, +|\nu_-\rangle &= \frac{1}{\sqrt{2\pi}}\Gamma(1/2 - is)\nu^{is - 1/2}, \\
\langle s, +|\lambda_+\rangle &= \frac{1}{\sqrt{2\pi}}\Gamma(1/2 - 2ik + is)\lambda^{2ik - is - 1/2}.
\end{align}
Now we can insert complete sets of hyperbolic states:
\begin{equation}
\langle\nu_-|g|\lambda_+\rangle = \int_{-\infty}^\infty ds_1\, ds_2\, \langle\nu_-|s_1, +\rangle\langle s_1, +|g|s_2, +\rangle\langle s_2, +|\lambda_+\rangle.
\end{equation}
In particular, we have
\begin{equation}
\psi^{k}_{\lambda,\nu}(\phi) = \frac{\lambda^{2ik}}{2\pi\sqrt{\nu\lambda}}\int_{-\infty}^\infty ds_1\, ds_2\, \nu^{-is_1}\lambda^{-is_2}\Gamma(1/2 + is_1)\Gamma(1/2 - 2ik + is_2)K_{s_1 s_2}^{++}(e^{2\phi H}).
\end{equation}
Indeed, inserting the explicit expression \eqref{diagonalKpp} and using the identity
\begin{equation}
\frac{1}{2\pi}\int_{-\infty}^\infty ds\, \Gamma(1/2 + is + ik)\Gamma(1/2 + is - ik)x^{-2is} = 2xK_{2ik}(2x)
\end{equation}
reproduces the previous expression \eqref{whitslr}.

Whereas the orthogonality relation for the mixed parabolic matrix elements (more precisely, Whittaker functions) $\psi^{k}_{\lambda,\nu}(\phi)$ follows from an integral over the single group parameter $\phi$ of the Cartan generator $H$, performing the full semigroup integral would be necessary to derive the orthogonality relation for the hyperbolic matrix elements $K^{++}$.

The harmonic analysis presented in Appendix \ref{app:ha} can be restricted to the subsemigroup $\slr$. This merely requires setting $\lambda \to i\lambda$ and $\nu \to -i \nu$ with $\nu, \lambda > 0$. This means we only have the case where $\nu\lambda >0$, and the Casimir eigenfunctions have to be proportional to:
\begin{align}
e^{\nu \gamma_{\m}} e^{-\lambda \gamma_{\+}} e^{-\phi} K_{2ik}(2\sqrt{\nu\lambda}e^{-\phi}),
\end{align}
which is indeed our \eqref{whitslr}. It is important to notice that the discrete representation matrices cannot be found in the regime $\nu\lambda >0$ relevant for the subsemigroup. This is one way of appreciating \eqref{pldecom}, where only the principal series representations appear in the decomposition.

\subsection{Schwarzian Correlation Functions}
\label{app:boscor}

Finally, we discuss the gauge theory description of the boundary dynamics in JT gravity and the corresponding observables.  The setup is completely analogous to that in Section \ref{gravboundact} for the $\mathcal{N} = 1$ case, so we will be brief.  Including the boundary term, the BF form of the JT gravity action \eqref{JTaction} is
\begin{equation}
S_\text{JT} = -\frac{1}{4\pi G}\left[\int_\mathcal{M} \Tr(\mathbf{B}\mathbf{F}) - \frac{1}{2} \oint_{\partial \mathcal{M}} d\tau \Tr(\mathbf{B}\mathbf{A}_\tau)\right],
\label{jtac}
\end{equation}
where $\mathcal{M}$ is assumed to be a disk.  If \eqref{jtac} were merely a $G$-BF theory, then imposing the boundary condition $\mathbf{B}|_{\partial\mathcal{M}} = \gamma\mathbf{A}_\tau|_{\partial\mathcal{M}}$ and integrating out $\mathbf{B}$ would reduce the dynamics to that of a boundary action for a particle on the group manifold $G$:
\begin{equation}
Z = \int_{LG/G} [\mathcal{D}g]\, e^{-S[g]}, \quad S[g] = \frac{\gamma}{8\pi G}\oint_{\partial \mathcal{M}} d\tau \Tr((g^{-1}\partial_\tau g)^2),
\end{equation}
where $\gamma$ is a constant with dimensions of length and we have substituted $\mathbf{A}_\tau = -\partial_\tau gg^{-1}$ with $g(\tau + \beta) = g(\tau)$.  Instead, JT gravity is a constrained BF theory.  The gravitational degrees of freedom $g$ take values in $\sltr$, subject to the constraint
\begin{equation}
\mathbf{A}_\tau|_{\partial\mathcal{M}} = \left[\begin{array}{cc} 0 & -T(\tau)/2 \\ 1 & 0 \end{array}\right]
\label{gravBCs}
\end{equation}
where $T(\tau + \beta) = T(\tau)$ is the boundary stress tensor.  Upon writing $T(\tau) = \{F(\tau), \tau\}$ (up to an $\sltr$ redundancy), the dynamics of this constrained group element becomes that of the Schwarzian theory:\footnote{This result coincides exactly with that derived in the second-order formalism \cite{Maldacena:2016upp, Sarosi:2017ykf}, starting from Euclidean AdS$_2$ in Poincar\'e coordinates: $ds^2 = Z^{-2}(dF^2 + dZ^2)$.  Namely, one fixes a boundary curve $(F(\tau), Z(\tau))$ such that $g_{\tau\tau} = 1/\epsilon^2$ and $\phi_b = \gamma/\epsilon$, where $\epsilon$ is a UV cutoff.  Then $F(\tau)$ is the only dynamical variable since $Z = \epsilon F' + \mathcal{O}(\epsilon^3)$, and its action $S_\text{Sch}[F]$ comes solely from the boundary term in \eqref{JTaction}.  The ``$-1$'' in \eqref{JTaction} subtracts a $1/\epsilon^2$ divergence.  The reparametrization mode $F(\tau)$, or the Poincar\'e time as a function of proper time, describes fluctuations in the shape of the boundary curve.  The isometry group $\sltr$ of the hyperbolic disk preserves the boundary curve and is regarded as a gauge symmetry.}
\begin{equation}
Z = \int_{\operatorname{Diff}(S^1)/\sltr} [\mathcal{D}F]\, e^{-S_\text{Sch}[F]}, \quad S_\text{Sch}[F] = -C\oint_{\partial \mathcal{M}} d\tau\, \{F(\tau), \tau\}, \quad C\equiv \frac{\gamma}{8\pi G}.
\end{equation}
Moving beyond the disk, the BF boundary condition $\mathbf{B}|_{\partial\mathcal{M}} = \gamma\mathbf{A}_\tau|_{\partial\mathcal{M}}$ and its constrained version \eqref{gravBCs} define gluing boundaries and holographic boundaries, respectively \cite{Blommaert:2018iqz}.

Further discussion of boundary conditions for JT gravity, and their relation to its first-order formulation and boundary Schwarzian description, can be found in \cite{Gonzalez:2018enk, Gaikwad:2018dfc, Grumiller:2020elf, Goel:2020yxl, Ferrari:2020yon}.

Coupling the bulk theory to a massive scalar field whose boundary value sources an operator of dimension $h$, we see that the natural bilocal operators to consider in the Schwarzian quantum mechanics are
\begin{equation}
\left[\frac{F'(\tau)F'(\tau')}{(F(\tau) - F(\tau'))^2}\right]^h.
\end{equation}
These Schwarzian bilocal operators are precisely equivalent to boundary-anchored Wilson lines in the constrained BF theory, in discrete representations of lowest weight $j = -h$.  We now consider the group-theoretic representation of these bilocal operators.  Specifically, we consider the most general operators that can be obtained as matrix elements in arbitrary states of a given $\sltr$ representation, rather than just those corresponding to mixed matrix elements between lowest- and highest-weight states.  These general operators can be packaged into a Gram matrix of inner products, extending the results in Appendix D of \cite{Mertens:2019tcm}.

We use the following parametrization for the boundary $\sltr$ group element and the gravitational boundary conditions:
\begin{equation}
g^{-1} = \left[\begin{array}{cc} A & B \\ C & D \end{array}\right], \qquad \left[\begin{array}{cc} A & B \\ C & D \end{array}\right]\left[\begin{array}{cc} 0 & -T(\tau)/2 \\ 1 & 0 \end{array}\right] = \left[\begin{array}{cc} A' & B' \\ C' & D' \end{array}\right].
\end{equation}
Hence $A$ and $C$ are seen to be the two linearly independent solutions to Hill's equation,
\begin{equation}
A'' + \frac{1}{2}T(\tau)A = 0, \qquad C'' + \frac{1}{2}T(\tau)C = 0,
\end{equation}
satisfying the Wronskian condition $AC' - A'C = 1$.  Up to M\"obius transformations, we have
\begin{equation}
A = \frac{1}{\sqrt{F'}}, \qquad C = \frac{F}{\sqrt{F'}},
\end{equation}
with $\{F, \tau\} = T(\tau)$.  Using the Gauss decomposition $g^{-1} = e^{\gamma_{\m} E^-}e^{2\phi H}e^{\gamma_{\+}E^+}$ (written here for $g^{-1}$ rather than $g$), we identify\footnote{Since the defining representation of $\sltr$ is faithful, this identification of Gauss parameters is independent of representation. We can also show this explicitly as follows. We write Hill's equation as
\begin{align}
\textstyle E^- - \frac{1}{2}T(\tau)E^+ &= e^{-\gamma_{\+}(\tau)E^+}e^{-2\phi(\tau)H}e^{-\gamma_{\m}(t)E^-}\partial_\tau(e^{\gamma_{\m}(\tau)E^-}e^{2\phi(\tau)H}e^{\gamma_{\+}(\tau)E^+}) \\
&= e^{-\gamma_{\+}(\tau)E^+}e^{-2\phi(\tau)H}E^- e^{2\phi(\tau)H}e^{\gamma_{\+}(\tau)E^+}\cdot \gamma_{\m}'(\tau) + e^{-\bar{\gamma}(\tau)E^+}He^{\gamma_{\+}(\tau)E^+}\cdot 2\phi'(\tau) + E^+\cdot \gamma_{\+}'(\tau) \nonumber \\
&= (E^- - 2\gamma_{\+}(\tau)H - \gamma_{\+}(\tau)^2 E^+)\cdot e^{2\phi(\tau)}\gamma_{\m}'(\tau) + (H + \gamma_{\+}(t)E^+)\cdot 2\phi'(t) + E^+\cdot \gamma_{\+}'(\tau), \nonumber
\end{align}
where we have used the Baker-Campbell-Hausdorff formula
\begin{equation}
e^Y Xe^{-Y} = e^{\operatorname{ad}_Y}X = X + [Y, X] + \frac{1}{2!}[Y, [Y, X]] + \cdots
\end{equation}
in the last step.  Matching the coefficients of the generators gives three equations, which imply that both $e^{\phi(\tau)}$ and $\gamma_{\+}(\tau)$ are determined in terms of $\gamma(\tau)$ as
\begin{equation}
e^{\phi(t)} = \frac{1}{\sqrt{\gamma_{\m}'(\tau)}}, \qquad \gamma_{\+}(\tau) = -\frac{1}{2}\frac{\gamma_{\m}''(\tau)}{\gamma_{\m}'(\tau)},
\end{equation}
while $\gamma_{\m}(\tau)$ satisfies $\{\gamma_{\m}(\tau), \tau\} = T(\tau)$, just as in \eqref{gaussparamssltr}.}
\begin{equation}
e^\phi = \frac{1}{\sqrt{F'}}, \qquad \gamma_{\m} = F, \qquad \gamma_{\+} = -\frac{1}{2}\frac{F''}{F'},
\label{gaussparamssltr}
\end{equation}
and therefore
\begin{equation}
g^{-1}(\tau) = \frac{1}{\sqrt{F'(\tau)}}\left[\begin{array}{cc} 1 & -F''(\tau)/2F'(\tau) \\ F(\tau) & F'(\tau) - F(\tau)F''(\tau)/2F'(\tau) \end{array}\right].
\end{equation}
So in the spin-$1/2$ representation, with $|\text{l.w.}\rangle = \left[\begin{smallmatrix} 0 \\ 1 \end{smallmatrix}\right]$ and $|\text{h.w.}\rangle = \left[\begin{smallmatrix} 1 \\ 0 \end{smallmatrix}\right]$, we compute that\footnote{For simplicity, we consider a single holographic boundary: Wilson line endpoints on different holographic boundaries could give rise to different functions $F$ associated to the group elements at $\tau_1$ and $\tau_2$.}
\begin{align}
R_{1/2}(g(t_2)g^{-1}(t_1)) &= \left[\begin{array}{cc} \langle\text{h.w.}|g(\tau_2)g^{-1}(\tau_1)|\text{h.w.}\rangle & \langle\text{h.w.}|g(\tau_2)g^{-1}(\tau_1)|\text{l.w.}\rangle \\ \langle\text{l.w.}|g(\tau_2)g^{-1}(\tau_1)|\text{h.w.}\rangle & \langle\text{l.w.}|g(\tau_2)g^{-1}(\tau_1)|\text{l.w.}\rangle \end{array}\right] \\
&= \left[\begin{array}{cc} -\partial_{\tau_2} & -\partial_{\tau_2}\partial_{\tau_1} \\ 1 & \partial_{\tau_1} \end{array}\right]\frac{F(\tau_1) - F(\tau_2)}{\sqrt{F'(\tau_1)F'(\tau_2)}}.
\end{align}
To see why this is true, note that Hill's equation can be written in the equivalent forms
\begin{equation}
g^{-1}(\tau)\left(E^- - \frac{T(\tau)}{2}E^+\right) = \partial_\tau g^{-1}(\tau) \Longleftrightarrow \left(E^- - \frac{T(\tau)}{2}E^+\right)g(\tau) = -\partial_\tau g(\tau),
\end{equation}
which give
\begin{align}
g^{-1}(\tau)|\text{l.w.}\rangle &= g^{-1}(\tau)E^-|\text{h.w.}\rangle = \partial_\tau g^{-1}(\tau)|\text{h.w.}\rangle, \\
\langle\text{h.w.}|g(\tau) &= \langle\text{l.w.}|E^- g(\tau) = -\partial_\tau\langle\text{l.w.}|g(\tau),
\end{align}
respectively.  Hence all matrix elements of $R_{1/2}(g(\tau_2)g^{-1}(\tau_1))$ can be obtained as derivatives of $\langle\text{l.w.}|g(\tau_2)g^{-1}(\tau_1)|\text{h.w.}\rangle$.

In the finite-dimensional spin-$j$ representation, we have
\begin{equation}
H = \operatorname{diag}(j, j - 1, \ldots, -j), \quad E^- = \left[\begin{smallmatrix}
\mathbf{0}_{1\times 2j} & 0 \\
\operatorname{diag}(1, 2, \ldots, 2j) & \mathbf{0}_{2j\times 1}
\end{smallmatrix}\right], \quad E^+ = \left[\begin{smallmatrix}
\mathbf{0}_{2j\times 1} & \operatorname{diag}(2j, 2j - 1, \ldots, 1) \\
0 & \mathbf{0}_{1\times 2j}
\end{smallmatrix}\right].
\label{spinjgens}
\end{equation}
Identifying the states $|j\rangle, |j - 1\rangle, \ldots, |{-j}\rangle$ with the standard basis in $\mathbb{R}^{2j + 1}$, we have
\begin{equation}
H|m\rangle = m|m\rangle, \qquad E^\pm|m\rangle = (j\pm m + 1)|m\pm 1\rangle.
\label{spinjaction}
\end{equation}
It is convenient to note that
\begin{equation}
(E^\pm)^n|m\rangle = \frac{(j\pm m + n)!}{(j\pm m)!}|m\pm n\rangle, \qquad \langle m|(E^\pm)^n = \frac{(j\pm m)!}{(j\pm m - n)!}\langle m\mp n|. \label{Efactorials}
\end{equation}
One can compute the matrix element $\langle-j|g(\tau_2)g^{-1}(\tau_1)|j\rangle$ directly.  Using \eqref{Efactorials}, we have
\begin{align}
g^{-1}(\tau_1)|j\rangle &= e^{\gamma_{\m}(\tau_1)E^-}e^{2\phi(\tau_1)H}e^{\gamma_{\+}(\tau_1)E^+}|j\rangle = e^{2j\phi(\tau_1)}\sum_{n=0}^{2j} \gamma_{\m}(\tau_1)^n|j - n\rangle, \\
\langle-j|g(\tau_2) &= \langle-j|e^{-\gamma_{\+}(\tau_2)E^+}e^{-2\phi(\tau_2)H}e^{-\gamma_{\m}(\tau_2)E^-} = e^{2j\phi(\tau_2)}\sum_{n=0}^{2j} \binom{2j}{n}(-\gamma_{\m}(\tau_2))^n\langle-j + n|. \nonumber
\end{align}
We then find that
\begin{equation}
\langle-j|g(\tau_2)g^{-1}(\tau_1)|j\rangle = [e^{\phi(\tau_1) + \phi(\tau_2)}(\gamma_{\m}(\tau_1) - \gamma_{\m}(\tau_2))]^{2j} = \left[\frac{F(\tau_1) - F(\tau_2)}{\sqrt{F'(\tau_1)F'(\tau_2)}}\right]^{2j}. \label{fiducialsltr}
\end{equation}
Given this result, one can determine the other matrix elements recursively.  Letting $|m; \tau\rangle\equiv g^{-1}(\tau)|m\rangle$, Hill's equation gives
\begin{equation}
(j - m)|m; \tau\rangle = \partial_\tau|m + 1; \tau\rangle + \frac{T(\tau)}{2}(j + m + 2)|m + 2; \tau\rangle.
\end{equation}
Therefore, for arbitrary $j$, a general matrix element does not only involve derivatives of the fiducial matrix element \eqref{fiducialsltr}.  Alternatively, one can again compute $\langle m'|g(\tau_2)g^{-1}(\tau_1)|m\rangle$ directly from \eqref{Efactorials}.

Before moving on to the discrete lowest/highest-weight representations, we present an alternative point of view on the finite-dimensional representations.  In the basis of monomials $1, x, \ldots, x^{2j}$, the action of the differential operators in the Borel-Weil representation is equivalent to that of the matrices \eqref{spinjgens} \cite{Hikida:2017ehf}.  Starting with
\begin{equation}
\langle m|m'\rangle = \int dx\, \langle m|x\rangle\langle x|m'\rangle = \delta_{mm'},
\label{spinjorthonormality}
\end{equation}
we fix $\langle x|j\rangle = x^{2j}$ by demanding that $E^+|j\rangle = 0$ (the overall normalization is arbitrary).  Stipulating that the action of $E^\pm$ is as in \eqref{spinjaction} then gives
\begin{equation}
\langle x|m\rangle = \frac{(2j)!}{(j - m)!(j + m)!}x^{j+m}.
\end{equation}
This fixes the wavefunctions of the dual states to be
\begin{equation}
\langle m|x\rangle = \frac{(j - m)!}{(2j)!}(-\partial_x)^{j+m}\delta(x),
\end{equation}
and in particular, $\langle -j|x\rangle = \delta(x)$.  It is important that $j > 0$, so that $x^{2j}$ is annihilated by sufficiently high powers of $E^- = \partial_x$ and this is a finite-dimensional representation.  Note that while $\delta(x)$ is never annihilated by powers of $\partial_x$, putative bra states $\langle m > j|$ have zero overlap with all ket states.  We thus have
\begin{equation}
\langle x|\text{h.w.}\rangle_j = x^{2j}, \quad {}_j\langle\text{l.w.}|x\rangle = \delta(x), \quad \langle\text{l.w.}|g(\tau_2)g^{-1}(\tau_1)|\text{h.w.}\rangle = \int dx\, \delta(x)g(\tau_2)g^{-1}(\tau_1)x^{2j}
\end{equation}
for an ordinary Wilson line, which we can evaluate as follows:
\begin{align}
x^{2j} &\xrightarrow{g^{-1}(\tau_1)} \frac{(x + F(\tau_1))^{2j}}{F'(\tau_1)^j} \\*
&\xrightarrow{g(t_2)} \frac{(F'(\tau_2)x + (F(\tau_1) - F(\tau_2))(\frac{F''(\tau_2)}{2F'(\tau_2)}x + 1))^{2j}}{(F'(\tau_1)F'(\tau_2))^j} \\*
&\xrightarrow{x=0} \frac{(F(\tau_1) - F(\tau_2))^{2j}}{(F'(\tau_1)F'(\tau_2))^j}.
\end{align}
It is no more difficult to compute a general matrix element:
\begin{align}
&\langle m|g(\tau_2)g^{-1}(\tau_1)|m'\rangle = \frac{(j - m)!}{(j - m')!(j + m')!}\int dx\, \delta(x)\partial_x^{j+m}(g(\tau_2)g^{-1}(\tau_1)x^{j+m'}) \\
&= \frac{(j - m)!}{(j - m')!(j + m')!}e^{2j(\phi(\tau_1) + \phi(\tau_2))}\partial_x^{j + m}\Big[\left(e^{-2\phi(\tau_2)}x + (\gamma_{\m}(\tau_1) - \gamma_{\m}(\tau_2))(1 - \gamma_{\+}(\tau_2)x)\right)^{j + m'} \nonumber \\
&\phantom{==} \times \left(e^{-2\phi(\tau_1)}(1 - \gamma_{\+}(\tau_2)x) + \gamma_{\+}(\tau_1)(e^{-2\phi(\tau_2)}x + (\gamma_{\m}(\tau_1) - \gamma_{\m}(\tau_2))(1 - \gamma_{\+}(\tau_2)x))\right)^{j - m'}\Big]\Big|_{x=0}. \nonumber
\end{align}
This approach has the virtue of allowing one to derive simple closed-form expressions.

We now generalize to the infinite-dimensional lowest/highest-weight representations by setting $j= -h$ where $h>0$.  The state $|{j}\rangle_{j} \equiv |\text{h.w.}\rangle_{j} $ is still annihilated by the corresponding $E^+$, but is never annihilated by powers of $E^-$.  In terms of the new variable $h=-j>0$, the Wilson line is
\begin{equation}
{}_j\langle\text{l.w.}|g(\tau_2)g^{-1}(\tau_1)|\text{h.w.}\rangle_{j} = \int dx\, \delta(x)g(\tau_2)g^{-1}(\tau_1)\frac{1}{x^{2h}} = \frac{(F'(\tau_1)F'(\tau_2))^h}{(F(\tau_1) - F(\tau_2))^{2h}}.
\end{equation}
This Schwarzian bilocal has a pole of the form $(\tau_1 - \tau_2)^{-2h}$ as $\tau_1\to \tau_2$, as required for a 1d CFT correlator \cite{Fitzpatrick:2016mtp}.

\section{First-Order Formalism} \label{firstorderformalism}

We begin by reviewing the first-order formulation of $\mathcal{N} = 1$ (or $\mathcal{N} = (1, 1)$) JT supergravity as an $\osp$ gauge theory described in \cite{Cardenas:2018krd}, building on \cite{Montano:1990ru}.  Such an action can also be written in superspace \cite{Chamseddine:1991fg, DHoker:1988pdl}, where the connection to the bosonic second-order formulation is clearer.  For a derivation of the boundary super-Schwarzian theory from JT supergravity directly in superspace, see \cite{Forste:2017kwy}.

\subsection{\texorpdfstring{$\mathcal{N} = 1$}{N = 1} Supergravity}

Our goal is to arrive at the BF theory description of JT supergravity, starting from the superspace action.  Due to the presence of spinors, it is convenient to work from the start with frame fields rather than the metric.

To begin, we review a few standard facts about 2d supergeometry, as developed in \cite{Howe:1978ia} and recounted in \cite{DHoker:1988pdl}.  We parametrize $\mathcal{N} = 1$ superspace by local coordinates
\begin{equation}
Z^M = (z^m, \theta^\mu) = (z, \bar{z}, \theta, \bar{\theta}),
\end{equation}
and we additionally introduce tangent space coordinates carrying local $U(1)$ frame indices $A = (a, \alpha)$.  Bosonic frame indices $a, b\in \{0, 1\}$ are raised and lowered by $\delta_{ab}$ (with $\epsilon_{01} = 1$), while fermionic frame indices $\alpha, \beta\in \{+, -\}$ are lowered from the right by $\epsilon_{\alpha\beta}$ (with $\epsilon_{+-} = -1$).  Following the notational conventions of \cite{Howe:1978ia}, spacetime and frame indices are denoted by letters from the middle and the beginning of the alphabet, respectively.  Lowercase Latin (Greek) letters indicate bosonic (fermionic) components, while uppercase letters span both types.  Our conventions for differential forms are that
\begin{align}
Z^M Z^N &= (-1)^{|M||N|}Z^N Z^M, \nonumber \\
dZ^M\wedge dZ^N &= -(-1)^{|M||N|}dZ^N\wedge dZ^M, \label{formconventions} \\
dZ^M Z^N &= (-1)^{|M||N|}Z^N dZ^M, \nonumber
\end{align}
where $|M|, |N|$ are $\mathbb{Z}_2$-valued and indicate whether the coordinates are even or odd.

The supergeometry is characterized by a superzweibein, which we can write as a one-form:
\begin{equation}
E^A = E^A{}_M\, dZ^M.
\end{equation}
The superzweibein and its inverse are related by
\begin{align}
E^A{}_M E^M{}_B &= \delta^A_B, \\
E^M{}_A E^A{}_N &= \delta^M_N.
\end{align}
Under local Lorentz transformations, Lorentz vectors and covectors (which can be differential forms of arbitrary degree) transform as
\begin{align}
\delta V^A &= L^A{}_B V^B, \\
\delta V_A &= -V_B L^B{}_A.
\end{align}
By introducing a superconnection that transforms inhomogeneously as
\begin{equation}
\delta\Omega^A{}_B = L^A{}_C\Omega^C{}_B - \Omega^A{}_C L^C{}_B - dL^A{}_B,
\label{inhomogeneous}
\end{equation}
we define a Lorentz-covariant superderivative that acts on Lorentz tensors as follows:
\begin{align}
DV^A &= dV^A + \Omega^A{}_B\wedge V^B, \\
DV_A &= dV_A - \Omega^B{}_A\wedge V_B.
\end{align}
We write $D = dZ^M D_M$, with the usual exterior derivative being $d = dZ^M\partial_M$.  Our derivatives act from the left, unlike in \cite{Wess:1992cp} and much of the supergravity literature.

In light of the fact that the 2d Lorentz group is $U(1)$, local Lorentz transformations and the superconnection simplify to
\begin{equation}
L^A{}_B = LE^A{}_B, \qquad \Omega^A{}_B = \Omega E^A{}_B,
\label{connection2d}
\end{equation}
where $\Omega = \Omega_M\, dZ^M$ and $E^A{}_B$ is defined as follows:
\begin{equation}
E^a{}_b = \epsilon^a{}_b, \qquad E^\alpha{}_b = E^a{}_\beta = 0, \qquad E^\alpha{}_\beta = -\frac{1}{2}(\gamma_5)^\alpha{}_\beta.
\label{lorentzdef}
\end{equation}
Thus we have, for instance, $D_M V^A = \partial_M V^A + \Omega_M E^A{}_B V^B$ and $D_M V_A = \partial_M V_A - \Omega_M V_B E^B{}_A$.  The transformation rule \eqref{inhomogeneous} then becomes simply $\delta\Omega = -dL$.  Our 2d gamma matrices satisfy
\begin{equation}
\{\gamma_a, \gamma_b\} = 2\delta_{ab}, \quad \gamma_5 = -\gamma_0\gamma_1,
\label{gammadef}
\end{equation}
from which it follows that $\{\gamma_a, \gamma_5\} = 0$ and $\gamma_5^2 = -1$.  We also have
\begin{equation}
\gamma_a\gamma_b = \delta_{ab} - \epsilon_{ab}\gamma_5, \quad \gamma_a\gamma_5 = -\epsilon_a{}^b\gamma_b,
\label{gammaprops}
\end{equation}
and in particular $[\gamma_a, \gamma_b] = -2\epsilon_{ab}\gamma_5$ and $[\gamma_a, \gamma_5] = -2\epsilon_a{}^b\gamma_b$.  The sign in the definition of $E^\alpha{}_\beta$ \eqref{lorentzdef} is correlated with the sign in the definition of $\gamma_5$ \eqref{gammadef}.

We define the supertorsion and the supercurvature as follows:
\begin{align}
T^A &= DE^A = \frac{1}{2}T^A{}_{BC}E^B\wedge E^C, \\
R^A{}_B &= d\Omega^A{}_B + \Omega^A{}_C\wedge \Omega^C{}_B = \frac{1}{2}R^A{}_{BCD}E^C\wedge E^D.
\end{align}
These two-forms satisfy two Bianchi identities, namely
\begin{align}
DT^A &= R^A{}_B\wedge E^B, \label{firstbianchi} \\
DR^A{}_B &= 0. \label{secondbianchi}
\end{align}
In components, the first Bianchi identity \eqref{firstbianchi} reads
\begin{equation}
R^A{}_{[BCD]} = D_{[B}T^A{}_{CD]} + T^A{}_{E[B}T^E{}_{CD]},
\label{bianchicomponents}
\end{equation}
where generalized (graded) antisymmetrization is understood.  This entails, e.g.,
\begin{equation}
T_{[AB]} = \frac{1}{2}(T_{AB} - (-1)^{|A||B|}T_{BA}),
\end{equation}
in accordance with the conventions \eqref{formconventions}.  Using \eqref{connection2d} gives
\begin{equation}
R^A{}_B = FE^A{}_B, \quad F = d\Omega
\end{equation}
since $\Omega\wedge \Omega = 0$, so that the second Bianchi identity \eqref{secondbianchi} becomes $dF = 0$.

By imposing appropriate constraints on the supertorsion \cite{Howe:1978ia}, all components of the supertorsion and supercurvature can be expressed in terms of a single scalar superfield $R_{+-}$.  As a consequence, one can also write the superconnection entirely in terms of the superzweibein.

Namely, we define the constrained supergeometry by imposing
\begin{equation}
T^a{}_{bc} = T^\alpha{}_{\beta\gamma} = 0, \quad T^a{}_{\beta\gamma} = 2i(\gamma^a)_{\beta\gamma}.
\label{torsionconstraints}
\end{equation}
These ``kinematics'' are motivated by the global frame for the supervielbein in flat superspace, in which $T^a{}_{\beta\gamma}$ is nonzero and all other components vanish \cite{Wess:1992cp}.  The algebraic Bianchi identities for the Riemann tensor, in conjunction with \eqref{bianchicomponents}, fix the remaining components of the supertorsion in addition to those of the supercurvature.  For example, \eqref{gammaprops} implies that
\begin{equation}
D_A(\gamma^a)_{\alpha\beta} = 0.
\end{equation}
Hence, given \eqref{torsionconstraints}, the identity $R^a{}_{[\beta\gamma\delta]} = 0$ reduces to
\begin{equation}
T^a{}_{b[\beta}(\gamma^b){}_{\gamma\delta]} = 0 \,\, \Longleftrightarrow \,\, T^a{}_{b\beta} = 0.
\end{equation}
Similarly, the identities $R^a{}_{[b\gamma\delta]} = 0$ and $R^\alpha{}_{[\beta\gamma d]} = 0$ show that $T^\alpha{}_{a\beta}$ and $T^\alpha{}_{ab}$ are proportional to $(\gamma_a)^\alpha{}_\beta R_{+-}$ and $\epsilon_{ab}(\gamma_5)^{\alpha\beta}D_\beta R_{+-}$, respectively.  This determines all of the $T^A{}_{BC}$.

Transformations of the superzweibein that preserve the supertorsion constraints \eqref{torsionconstraints} are symmetries of the supergeometry.  They consist of super diffeomorphisms and local Lorentz transformations, as well as the super Weyl transformations developed in \cite{Howe:1978ia}.  Just as a Riemann surface depends only on a conformal class of Riemannian metrics, a super Riemann surface is a supersurface endowed with a supercomplex structure, where the latter depends only on the superconformal class of the superzweibein (i.e., is invariant under super Weyl transformations) \cite{DHoker:1988pdl}.  Moreover, just as all two-dimensional Riemannian manifolds are locally conformally flat, all two-di\-men\-sion\-al supergeometries are locally superconformally flat (i.e., related by super Weyl and Lorentz transformations to a flat superspace with $R_{+-} = 0$) \cite{Howe:1978ia}.

The passage from superspace to the component formalism is aided substantially by bringing the su\-per\-zwei\-bein and superconnection to Wess-Zumino gauge via super diffeomorphism and super Lorentz transformations, respectively.  This gauge, along with the supertorsion constraints, suffices to express all geometrical quantities in $\mathcal{N} = 1$ superspace (and in particular, all components of the superzweibein) in terms of three fields:
\begin{itemize}
\item a zweibein $e^a_m$ (the bottom component of $E^a{}_m$),
\item a gravitino/Rarita-Schwinger field $\psi^\alpha_m$ (the bottom component of $E^\alpha{}_m$),
\item and an auxiliary field $A$ (the bottom component of $R_{+-}$).
\end{itemize}
These three fields comprise the $\mathcal{N} = 1$ supergravity multiplet.  We will not need the full component expansion of the superzweibein in this gauge, but instead record here only those of the superdeterminant $E = \operatorname{sdet}(E^A{}_M)$ and the supercurvature:
\begin{align}
E &= e\left[1 + \frac{1}{2}\bar{\theta}\gamma^m\psi_m' - \frac{1}{4}\bar{\theta}\theta\left(A + \frac{1}{2}\epsilon^{mn}\bar{\psi}_m'\gamma_5\psi_n'\right)\right], \\
R_{+-} &= A + \bar{\theta}\chi + \frac{1}{2}\bar{\theta}\theta\left(R - \frac{1}{2}\bar{\psi}_m'\gamma^m\chi + \frac{1}{4}\epsilon^{mn}\bar{\psi}_m'\gamma_5\psi_n' A + \frac{1}{2}A^2\right),
\end{align}
where $R = 2\epsilon^{mn}\partial_m\omega_n$ is the curvature of the spin connection
\begin{equation}
\omega_m = -\epsilon^{n\ell}e_{am}\partial_n e^a_\ell + \frac{1}{2}\bar{\psi}_m'\gamma_5\gamma^n\psi_n'
\label{omegafromtorsion}
\end{equation}
(the bottom component of $\Omega_m$) and the middle component of $R_{+-}$ is given by
\begin{equation}
\chi = -2\gamma_5\epsilon^{mn}D_m\psi_n' - \frac{1}{2}\gamma^m\psi_m' A
\label{chifromtorsion}
\end{equation}
in terms of the bosonic Lorentz-covariant derivative
\begin{equation}
D_m\psi_n' = \partial_m\psi_n' - \frac{1}{2}\omega_m\gamma_5\psi_n'.
\end{equation}
The constraint \eqref{omegafromtorsion} follows from $T^a{}_{bc} = 0$, while the constraint \eqref{chifromtorsion} follows from that on $T^\alpha{}_{bc}$.  We have written the gravitino as $\psi'$ in anticipation of a later change of variables.  Note that \eqref{omegafromtorsion} implies nonvanishing torsion in bosonic spacetime: $D_m e^a_n - D_n e^a_m = \frac{1}{2}\bar{\psi}_m'\gamma^a\psi_n'$.

Above, spinor contractions are defined with respect to the Majorana conjugate
\begin{equation}
\bar{\psi}_\alpha = \psi^\beta C_{\beta\alpha}, \quad C_{\alpha\beta} = \epsilon_{\alpha\beta}, \quad \epsilon_{+-} = -1.
\label{chargeconjugation}
\end{equation}
Equivalently, if we regard spinors with upper indices as column vectors, then $\bar{\psi} = \psi^T C$.  We have the following exchange properties of spinor bilinears:
\begin{equation}
\bar{\psi}\chi = \bar{\chi}\psi, \quad \bar{\psi}\gamma_a\chi = -\bar{\chi}\gamma_a\psi, \quad \bar{\psi}\gamma_5\chi = -\bar{\chi}\gamma_5\psi.
\label{exchangeprops}
\end{equation}
These follow from the antisymmetry of $C$ and the symmetry of $C\gamma_a$ and $C\gamma_5$, using that matrix transpose reverses the order of Grassmann variables.

Finally, we can present the $\mathcal{N} = 1$ JT supergravity action in superspace (adapted from \cite{Chamseddine:1991fg, Forste:2017kwy}):
\begin{equation}
S_\text{JT}^{\mathcal{N} = 1} = \frac{1}{4}\left[\int_\Sigma d^2 z\, d^2\theta\, E\Phi(R_{+-} + 2) + 2\int_{\partial\Sigma} d\tau\, d\theta\, \Phi K\right],
\label{sugraaction}
\end{equation}
where we have chosen a convenient overall normalization and omitted a factor of $-1/4\pi G$.  For now, we restrict our attention to the bulk term.  The dilaton superfield contains the dilaton $\phi$, the dilatino $\lambda$, and an auxiliary field $F$:
\begin{equation}
\Phi = \phi + \bar{\theta}\lambda + \bar{\theta}\theta F.
\end{equation}
Assuming the order-\emph{preserving} convention for complex conjugation of Grassmann variables, this is a real superfield.  The path integral is taken over $\Phi$ and over all supergeometries $E^A{}_M$ satisfying the supertorsion constraints.  In gravity, we would also sum over genera, but to compare to gauge theory, we restrict to the disk.

To unpack the action \eqref{sugraaction}, we use
\begin{equation}
\bar{\theta}\theta = \bar{\theta}_\alpha\theta^\alpha = 2\theta^-\theta^+, \qquad (\bar{\theta}\psi)(\bar{\theta}\chi) = -\frac{1}{2}(\bar{\theta}\theta)(\bar{\psi}\chi), \qquad \int d^2\theta\, \bar{\theta}\theta = 2,
\end{equation}
where the last equation is a convenient definition.  We see that
\begin{equation}
\frac{1}{4\pi}\int_\Sigma d^2 z\, d^2\theta\, ER_{+-} = \frac{1}{4\pi}\int_\Sigma d^2 z\, eR = \chi(\Sigma).
\end{equation}
Integration over the dilaton superfield in \eqref{sugraaction} localizes the path integral to surfaces of constant negative supercurvature: $R_{+-} + 2 = 0$.  In components, this constraint reads:
\begin{equation}
A = -2, \qquad \chi = 0, \qquad R = \frac{1}{2}\epsilon^{mn}\bar{\psi}_m'\gamma_5\psi_n' - 2.
\label{dilatonconstraints}
\end{equation}
Substituting \eqref{chifromtorsion}, we compute that the top component of $E\Phi(R_{+-} + 2)$ is
\begin{equation}
e\bar{\theta}\theta\left[F(A + 2) + \frac{1}{2}\phi\left(R - A - \frac{1}{2}\epsilon^{mn}\bar{\psi}_m'\gamma_5\psi_n'\right) + \bar{\lambda}\gamma_5\epsilon^{mn}D_m\psi_n' - \frac{1}{2}\bar{\lambda}\gamma^m\psi_m'\right].
\label{topcomponent}
\end{equation}
To match to the gauge theory formulation, it is convenient to define $\psi_m = \gamma_5\psi_m'$ so that
\begin{equation}
\epsilon^{mn}\bar{\psi}_m\gamma_5\psi_n = \epsilon^{mn}\bar{\psi}_m'\gamma_5\psi_n', \quad \bar{\lambda}\epsilon^{mn}D_m\psi_n = \bar{\lambda}\gamma_5\epsilon^{mn}D_m\psi_n', \quad \bar{\lambda}\epsilon^{mn}\gamma_m\psi_n = -\bar{\lambda}\gamma^m\psi_m',
\end{equation}
where we have used \eqref{gammaprops} and that $D_m$ commutes with $\gamma_5$.  Integrating \eqref{topcomponent} over superspace and also integrating out $F$, we thus obtain the action in terms of component fields:
\begin{equation}
S_\text{JT}^{\mathcal{N} = 1} = \frac{1}{2}\int_\Sigma d^2 z\, e\bigg[\frac{1}{2}\phi\left(R + 2 - \frac{1}{2}\epsilon^{mn}\bar{\psi}_m\gamma_5\psi_n\right) + \bar{\lambda}\epsilon^{mn}D_m\psi_n + \frac{1}{2}\bar{\lambda}\epsilon^{mn}\gamma_m\psi_n\bigg],
\end{equation}
with the constraint \eqref{omegafromtorsion} implicit (the component form of the action already assumes the other constraints).  More usefully, we may regard $\omega_m$ as an independent field, which requires making the constraint \eqref{omegafromtorsion} explicit.  The first-order action can then be written as
\begin{align}
S_\text{JT}^{\mathcal{N} = 1} &= \frac{1}{2}\int_\Sigma d^2 z\, e\bigg[\frac{1}{2}\phi\left(R + 2 - \frac{1}{2}\epsilon^{mn}\bar{\psi}_m\gamma_5\psi_n\right) \label{componentswithconstraints} \\
&\phantom{==} + \phi_a\left(\epsilon^{mn}\partial_m e^a_n + \epsilon^a{}_b\epsilon^{mn}\omega_m e^b_n - \frac{1}{4}\epsilon^{mn}\bar{\psi}_m\gamma^a\psi_n\right) + \bar{\lambda}\epsilon^{mn}D_m\psi_n + \frac{1}{2}\bar{\lambda}\epsilon^{mn}\gamma_m\psi_n\bigg]. \nonumber
\end{align}
To see that the Lagrange multipliers in \eqref{componentswithconstraints} indeed enforce \eqref{omegafromtorsion}, note that integrating out $\phi_a$ results in the component-wise torsion constraints
\begin{equation}
T^a{}_{mn}|\equiv \partial_{[m}e^a_{n]} + \epsilon^a{}_b\omega_{[m}e^b_{n]} - \frac{1}{4}\bar{\psi}_{[m}\gamma^a\psi_{n]} = 0,
\end{equation}
where the bar indicates the bottom component.  We solve for the spin connection in the standard way \cite{Freedman:2012zz} by writing $T_{\ell[mn]}| - T_{m[n\ell]}| + T_{n[\ell m]}| = 0$ with $T_{\ell[mn]}|\equiv e_{a\ell}T^a{}_{mn}|$, giving:
\begin{align}
\omega_m &= \frac{1}{2}\epsilon^{cd}e^n_c(\partial_m e_{dn} - \partial_n e_{dm}) - \frac{1}{2}\epsilon^{n\ell}e_{am}\partial_n e^a_\ell + \frac{1}{8}\epsilon^{cd}(\bar{\psi}_m\gamma_c\psi_d + \bar{\psi}_c\gamma_m\psi_d + \bar{\psi}_c\gamma_d\psi_m) \label{standardomega} \\
&= -\epsilon^{n\ell}e_{am}\partial_n e^a_\ell - \frac{1}{2}\bar{\psi}_a\gamma_5\gamma^b\psi_b. \label{simpomega}
\end{align}
In 2d, the first term of \eqref{standardomega} equals the second, while the last term of \eqref{standardomega} simplifies via \eqref{exchangeprops} and \eqref{gammaprops}.  Hence we obtain the simplified form \eqref{simpomega}, or equivalently, \eqref{omegafromtorsion}.

So far, we have written the action in terms of zero-form fields.  To pass to the BF description, it is more convenient to work in terms of the one-form fields $\omega = \omega_m\, dz^m$, $e^a = e^a_m\, dz^m$, and $\psi = \psi_m\, dz^m$.  Note that for arbitrary one-forms $\xi$ and $\xi'$, we have
\begin{align}
\xi\wedge \xi' &= d^2 z\, \tilde{\epsilon}^{mn}\xi_m\xi'_n = d^2 z\, e\epsilon^{mn}\xi_m\xi'_n, \\
d\xi &= d^2 z\, \tilde{\epsilon}^{mn}\partial_m\xi_n = d^2 z\, e\epsilon^{mn}\partial_m\xi_n,
\end{align}
where the Levi-Civita symbol and tensor are related by
\begin{equation}
\epsilon_{mn} = e\tilde{\epsilon}_{mn}\,\, \Longleftrightarrow \,\, \tilde{\epsilon}^{mn} = e\epsilon^{mn}, \quad e\equiv \det e^a_m.
\end{equation}
In particular, we have
\begin{equation}
e^0\wedge e^1 = d^2 z\, e, \quad d\omega = \frac{1}{2}d^2 z\, eR.
\end{equation}
Hence we have, in shorthand notation (omitting wedge products),
\begin{align}
S_\text{JT}^{\mathcal{N} = 1} = \frac{1}{2}\int_\Sigma \bigg[&\phi\left(d\omega + e^0 e^1 - \frac{1}{4}\bar{\psi}\gamma_5\psi\right) \nonumber \\
&+ \phi_a\left(de^a + \epsilon^a{}_b\omega e^b - \frac{1}{4}\bar{\psi}\gamma^a\psi\right) + \bar{\lambda}D\psi + \frac{1}{2}\bar{\lambda}e^a\gamma_a\psi\bigg]. \label{firstorderaction}
\end{align}
The second term enforces the constraints on the bosonic components of the supertorsion.

\subsection{\texorpdfstring{$\mathfrak{osp}(1|2)$}{osp(1|2)} BF Theory}

We consider the first-order action \eqref{firstorderaction} where $\Sigma$ is a disk with compact time coordinate $\tau\sim \tau + \beta$ and radial coordinate $r > 0$.  To rewrite it as a BF action, we again introduce $\mathfrak{sl}(2, \mathbb{R})$ (or $\mathfrak{so}(2, 1)$) generators via the identification \eqref{sltridentification}, where the objects on the right now belong to $\mathfrak{osp}(1|2)$.  Hence
\begin{equation}
J_I = \left[\begin{array}{c|c}
\frac{1}{2}\Gamma_I & \mathbf{0}_{2\times 1} \\ \hline
\mathbf{0}_{1\times 2} & 0
\end{array}\right]
\end{equation}
where
\begin{equation}
\Gamma_0 = \left[\begin{array}{cc}
-1 & 0 \\
0 & 1
\end{array}\right], \qquad
\Gamma_1 = \left[\begin{array}{cc}
0 & 1 \\
1 & 0
\end{array}\right], \qquad
\Gamma_2 = \left[\begin{array}{cc}
0 & -1 \\
1 & 0
\end{array}\right],
\end{equation}
which satisfy $\{\Gamma_I, \Gamma_J\} = 2\eta_{IJ}$ where $\eta_{IJ} = \operatorname{diag}(1, 1, -1)$ (our conventions differ slightly from those of \cite{Cardenas:2018krd}, so as to mimic those of bosonic JT gravity).  We can write the $\mathfrak{osp}(1|2)$ algebra as
\begin{equation}
[J_I, J_J] = \epsilon_{IJK}J^K, \qquad [J_I, Q_\alpha] = \frac{1}{2}(\Gamma_I)^\beta{}_\alpha Q_\beta, \qquad \{Q_\alpha, Q_\beta\} = -\frac{1}{2}(C\Gamma^I)_{\alpha\beta}J_I
\label{postulatedosp}
\end{equation}
with $\epsilon_{012} = -1$ and $C$ as in \eqref{chargeconjugation}, so that
\begin{equation}
C = \left[\begin{array}{cc}
0 & -1 \\
1 & 0
\end{array}\right], \quad C\Gamma_0 = \left[\begin{array}{cc}
0 & -1 \\
-1 & 0
\end{array}\right], \quad C\Gamma_1 = \left[\begin{array}{cc}
-1 & 0 \\
0 & 1
\end{array}\right], \quad C\Gamma_2 = \left[\begin{array}{cc}
-1 & 0 \\
0 & -1
\end{array}\right].
\end{equation}
Given \eqref{sltridentification}, the algebra \eqref{postulatedosp} fixes
\begin{equation}
Q_- = -F^-, \qquad Q_+ = F^+,
\end{equation}
up to an overall sign ambiguity $Q_\pm\leftrightarrow -Q_\pm$.  In this basis, the generators satisfy
\begin{equation}
\operatorname{STr}(J_I J_J) = \frac{1}{2}\eta_{IJ}, \qquad \operatorname{STr}(Q_\alpha Q_\beta) = \frac{1}{2}\epsilon_{\alpha\beta}.
\end{equation}
We now write the zero-form dilaton supermultiplet and the one-form superconnection as $\mathfrak{osp}(1|2)$-valued fields:
\begin{equation}
B_I = (\phi_a, \phi), \quad A^I = (e^a, \omega), \quad \mathbf{B} = B^I J_I + \lambda^\alpha Q_\alpha, \quad \mathbf{A} = A^I J_I + \psi^\alpha Q_\alpha.
\end{equation}
The latter has field strength
\begin{align}
\mathbf{F} &= \left(F^I - \frac{1}{4}\bar{\psi}_\alpha\wedge (\Gamma^I\psi)^\alpha\right)J_I + \mathcal{D}\psi^\alpha Q_\alpha, \\
F^I &\equiv dA^I + \frac{1}{2}\epsilon^{IJK}A_J\wedge A_K,
\end{align}
in contrast to the bosonic case where $\mathbf{F} = F^I J_I$.  Here, we have defined the $\mathfrak{so}(2, 1)$-covariant derivative
\begin{equation}
\mathcal{D}\psi^\alpha\equiv d\psi^\alpha + \frac{1}{2}A^I\wedge (\Gamma_I)^\alpha{}_\beta\psi^\beta,
\end{equation}
in contrast to the Lorentz- or dif\-feo\-mor\-phism-covariant exterior derivative
\begin{equation}
D\psi^\alpha = d\psi^\alpha - \frac{1}{2}\omega\wedge (\gamma_5\psi)^\alpha.
\end{equation}
Lastly, we set
\begin{equation}
\Gamma_I = (\gamma_a, -\gamma_5).
\end{equation}
In this basis, the $\gamma_a$ are symmetric and $\gamma_5$ is antisymmetric.  We then obtain
\begin{align}
\operatorname{STr}(\mathbf{B}\mathbf{F}) &= \frac{1}{2}B^I F_I - \frac{1}{8}B^I\bar{\psi}_\alpha\wedge (\Gamma_I\psi)^\alpha + \frac{1}{2}\bar{\lambda}_\alpha\mathcal{D}\psi^\alpha \\
&= \frac{1}{2}\phi\left(d\omega + e^0\wedge e^1 - \frac{1}{4}\bar{\psi}_\alpha\wedge (\gamma_5\psi)^\alpha\right) + \frac{1}{2}\phi_a\left(de^a + \epsilon^a{}_b\omega\wedge e^b - \frac{1}{4}\bar{\psi}_\alpha\wedge (\gamma^a\psi)^\alpha\right) \nonumber \\
&\phantom{==} + \frac{1}{2}\bar{\lambda}_\alpha D\psi^\alpha + \frac{1}{4}\bar{\lambda}_\alpha e^a\wedge (\gamma_a\psi)^\alpha. \label{strBF}
\end{align}
Comparing \eqref{strBF} to \eqref{firstorderaction}, we see that the first-order action of $\mathcal{N} = 1$ JT supergravity is precisely an $\mathfrak{osp}(1|2)$ BF theory:
\begin{equation}
S_\text{JT}^{\mathcal{N} = 1} = \int_\Sigma \operatorname{STr}(\mathbf{B}\mathbf{F}).
\label{BFaction}
\end{equation}
With our definition of order-preserving complex conjugation for Grassmann variables, the spinor bilinears appearing in Lagrangians are real with real coefficients. (Reality of spinor bilinears under the order-reversing convention can be achieved by adjusting the phase of the charge conjugation matrix, $C\to iC$.  This amounts to pulling out a factor of $-i$ from the anticommutators of fermionic generators in the $\mathfrak{osp}(1|2)$ algebra.)

The action \eqref{BFaction} is manifestly invariant under gauge transformations whose infinitesimal form is \eqref{infinitesimalgauge}, with $\epsilon$ now valued in $\mathfrak{osp}(1|2)$.  Such transformations with $\epsilon = \xi^\alpha Q_\alpha$ give, via the algebra \eqref{postulatedosp},
\begin{alignat}{2}
\delta_\xi A^I &= \frac{1}{2}\bar{\xi}\Gamma^I\psi, \qquad & \delta_\xi B^I &= \frac{1}{2}\bar{\xi}\Gamma^I\lambda, \\
\delta_\xi\psi^\alpha &= \mathcal{D}\xi^\alpha, \qquad & \delta_\xi\lambda^\alpha &= \frac{1}{2}B^I(\Gamma_I)^\alpha{}_\beta\xi^\beta.
\end{alignat}
These are equivalent to the local supersymmetry transformations
\begin{alignat}{3}
\delta_\xi e^a &= \frac{1}{2}\bar{\xi}\gamma^a\psi, \qquad & \delta_\xi\omega &= \frac{1}{2}\bar{\xi}\gamma_5\psi, \qquad & \delta_\xi\psi^\alpha &= D\xi^\alpha + \frac{1}{2}e^a(\gamma_a)^\alpha{}_\beta\xi^\beta, \label{localSUSY1} \\
\delta_\xi\phi^a &= \frac{1}{2}\bar{\xi}\gamma^a\lambda, \qquad & \delta_\xi\phi &= -\frac{1}{2}\bar{\xi}\gamma_5\lambda, \qquad & \delta_\xi\lambda^\alpha &= \frac{1}{2}\phi^a(\gamma_a)^\alpha{}_\beta\xi^\beta + \frac{1}{2}\phi(\gamma_5)^\alpha{}_\beta\xi^\beta. \label{localSUSY2}
\end{alignat}
By an appropriate Fierz identity, the variation of the action \eqref{firstorderaction} under \eqref{localSUSY1}--\eqref{localSUSY2} vanishes exactly, incurring no boundary terms: $\delta_\xi S_\text{JT}^{\mathcal{N} = 1} = 0$.  The transformations \eqref{localSUSY1} of the supergravity multiplet, written in components as
\begin{equation}
\delta_\xi e^a_m = \frac{1}{2}\bar{\xi}\gamma^a\psi_m, \qquad \delta_\xi\omega_m = \frac{1}{2}\bar{\xi}\gamma_5\psi_m, \qquad \delta_\xi\psi_m = D_m\xi + \frac{1}{2}\gamma_m\xi,
\end{equation}
are equivalent to the local supersymmetry transformations of \cite{Howe:1978ia} after imposing the dilaton constraints \eqref{dilatonconstraints} (eliminating the auxiliary field $A$).\footnote{The subgroup of super diffeomorphisms and local Lorentz transformations that preserves Wess-Zumino gauge consists of ordinary diffeomorphisms, local $\mathcal{N} = 1$ supersymmetry, and ordinary local Lorentz transformations.  By tracking the fermionic part of the allowed super diffeomorphisms, one derives the $\mathcal{N} = 1$ supersymmetry transformations of the supergravity multiplet \cite{Howe:1978ia}.}

\subsection{Worldline Action for Wilson Lines} \label{app:orbits}

Whereas the BF description involves integrating over the fermionic half of superspace, it is most natural to interpret these Wilson lines directly in the full superspace.  We do so by expanding the gauge field in terms of the superzweibein and superconnection as \cite{Montano:1990ru}
\begin{equation}
A_M\equiv E^A{}_M J_A + \Omega_M J_2 \quad (A = 0, 1, +, -), \qquad J_\alpha\equiv Q_\alpha,
\end{equation}
and then rewriting the Wilson line as a path integral for the worldline action of a massive probe particle in superspace, along the lines of \cite{Iliesiu:2019xuh}.  For earlier discussions in the context of 3d gravity, see \cite{Ammon:2013hba, Castro:2018srf} and references therein.

We start by writing a Wilson line along a fixed contour $\mathcal{C}$ and in a given representation $h$ in a standard way as
\begin{equation}
\int D_{\boldsymbol{\Lambda}}g\, e^{-S_{\boldsymbol{\Lambda}}[g, A]},
\end{equation}
where $\boldsymbol{\Lambda}$ is the (dual of the) highest weight of the representation $h$, $g(s)$ is a map $\mathcal{C}\to \osp$ that (redundantly) parametrizes the orbit of $\boldsymbol{\Lambda}$ in $\mathfrak{osp}(1|2)$ under the adjoint action of $\osp$, $s$ is the worldline coordinate along $\mathcal{C}$, and $S_{\boldsymbol{\Lambda}}[g, A]$ is the first-order action\footnote{This is the worldline action for a bosonic particle on a coadjoint orbit of a supergroup \cite{Mikhaylov:2014aoa}.  The supersymmetrization of a standard coadjoint orbit would instead lead to the worldline action for a \emph{super}particle on a coadjoint orbit of an ordinary group \cite{Fan:2018wya}.}
\begin{equation}
S_{\boldsymbol{\Lambda}}[g, A] = \int ds \operatorname{STr}(\boldsymbol{\Lambda}g^{-1}D_A g), \quad D_A\equiv \partial_s + A_s, \quad A_s(s)\equiv A_M(Z(s))\dot{Z}^M(s).
\end{equation}
The gauge redundancy in $g$, which amounts to right multiplication by the stabilizer of $\boldsymbol{\Lambda}$ in $\osp$, is implicit in the domain of path integration.  From the perspective of the worldline path integral, $A_s$ (the restriction of the bulk gauge field) is a background gauge field, and it transforms in such a way that the action $S_{\boldsymbol{\Lambda}}[g, A]$ is gauge-invariant under left multiplication of $g$ by elements of $\osp$.

The length of
\begin{equation}
\boldsymbol{\Lambda} = \Lambda^I J_I + \Xi^\alpha Q_\alpha\in \mathfrak{osp}(1|2)
\end{equation}
is determined by the quadratic Casimir of the representation:
\begin{equation}
\operatorname{STr}(\boldsymbol{\Lambda}^2) = \frac{1}{2}(\eta_{IJ}\Lambda^I\Lambda^J + \epsilon_{\alpha\beta}\Xi^\alpha\Xi^\beta) = \frac{1}{2}(\Lambda^I\Lambda_I + \overline{\Xi}\Xi) = 2h(h - 1/2)\equiv 2m^2.
\label{lengthconstraint}
\end{equation}
(We assume $h > 1/2$ so that $m$ is interpretable as the mass of a genuine probe particle.) Since the adjoint action of $\osp$ is transitive on all Lie algebra elements of a given length, we may further introduce a functional integral over all elements of the form
\begin{equation}
\boldsymbol{\Lambda}(s) = \Lambda^a(s)J_a + \Xi^\alpha(s)Q_\alpha(s)\equiv \Lambda^A(s) J_A,
\label{integralelement}
\end{equation}
with length constrained by \eqref{lengthconstraint}, to write the worldline path integral equivalently as
\begin{equation}
\int D\boldsymbol{\Lambda}\, D_{\boldsymbol{\Lambda}}g\, D\Theta\, e^{-S_{\boldsymbol{\Lambda}}[g, A, \Theta]}
\end{equation}
where $\Theta$ is a Lagrange multiplier:
\begin{equation}
S_{\boldsymbol{\Lambda}}[g, A, \Theta]\equiv \int ds\left[\operatorname{STr}(\boldsymbol{\Lambda}g^{-1}D_A g) + \frac{i}{2}\Theta(\Lambda^a\Lambda_a + \overline{\Xi}\Xi - 4m^2)\right].
\end{equation}
Note that we have omitted the $J_2$-component of $\boldsymbol{\Lambda}$ in \eqref{integralelement}, the motivation being that the bulk path integral imposes flatness of the super gauge field, which implements the supertorsion constraint(s) and expresses $\Omega$ in terms of $E$, so that the $J_2$-component is not independent of the $J_{0, 1}$-components in our setting.

We now fix a gauge on the disk in which $g = 1$ along the curve $\mathcal{C}$, so that $g^{-1}D_A g$ reduces to $A_s = (E^A{}_M J_A + \Omega_M J_2)\dot{Z}^M$ and the action becomes
\begin{equation}
\frac{1}{2}\int ds\left[\Lambda_A E^A{}_M\dot{Z}^M + i\Theta(\Lambda^a\Lambda_a + \overline{\Xi}\Xi - 4m^2)\right].
\end{equation}
Integrating out $\Lambda^a$ and $\Xi$ reduces this to
\begin{equation}
\frac{i}{2}\int ds\left(\frac{1}{4\Theta}g_{MN}\dot{Z}^M\dot{Z}^N - 4m^2\Theta\right),
\end{equation}
where we have written the supermetric in terms of the superzweibein:
\begin{equation}
g_{MN} = g_{(MN)} = \mathcal{E}_{AB}E^A{}_M E^B{}_N\equiv \delta_{ab}E^a{}_M E^b{}_N + \epsilon_{\alpha\beta}E^\alpha{}_M E^\beta{}_N.
\end{equation}
This is precisely the first-order form of the standard point-particle action in superspace, with $\Theta$ playing the role of the einbein.  Further integrating out $\Theta$ gives the second-order form of the point-particle action:
\begin{equation}
m\int ds\, (g_{MN}\dot{Z}^M\dot{Z}^N)^{1/2}.
\label{secondorderpp}
\end{equation}
(Note that we have implicitly shifted the \emph{a priori} real integration contour for $\Theta$ to account for the imaginary saddle points.)

Finally, the identification of superdiffeomorphisms with gauge transformations in the BF description for flat $A$ shows that the Wilson line, as an operator insertion inside the path integral of JT supergravity, is equivalent to a path integral for the action \eqref{secondorderpp} taken over all paths superdiffeomorphic to $\mathcal{C}$.

\section{Details on Orbits and Operators}
 \label{app:details}

This appendix collects more details on super-Virasoro coadjoint orbits and on bilocal operator calculations.

\subsection{Classification of Super-Virasoro Coadjoint Orbits} \label{app:coadj}

To find and characterize the different coadjoint orbits, we need to find all super-Virasoro transformations that leave $(T_{\B}(\tau), T_{\F}(\tau))$ invariant. If one is interested only in orbits that contain a constant representative (as we are here), then there is a shortcut to the analysis. Due to the quotient structure of the orbit itself, it suffices to look at this constant representative to deduce the stabilizer for the orbit of interest. This constant representative is a solution to the super-Schwarzian equations of motion. We hence look only for the most general classical solution with the prescribed periodicity constraints \eqref{perorbit}, which we repeat here for convenience: 
\begin{equation}
F(\tau + \beta) = M \cdot F(\tau), \qquad \eta(\tau+ \beta) = \pm \eta(\tau), \qquad M \in \text{SL}(2,\mathbb{R}).
\end{equation}
The solution to these equations of motion will generate all possible constant values of the super-Schwarzian derivative, and we merely have to compare this constant value between solutions to find the stabilizer of each orbit. The remaining undetermined coefficients in the classical solution that retain this value then parametrize the stabilizer. The super-Schwarzian equations of motion, associated with the Lagrangian \eqref{susyact}, are given by:
\begin{align}
&\delta \eta \quad \rightarrow \quad 2\eta''' + \left\{F,\tau\right\}\eta'=0, \\
&\delta f \quad \rightarrow \quad \left\{F,\tau\right\}'\left(1-\eta\eta'\right)=0.
\end{align}
The general classical solution is given by:
\begin{align}
\label{clsol}
\left\{F,\tau\right\} &= \frac{2\pi^2}{\beta^2} \Theta^2 = \text{constant}, \\
F(\tau) &= \frac{a \tan \frac{\pi}{\beta} \Theta \tau + b}{c \tan \frac{\pi}{\beta} \Theta \tau +d }, \qquad ad-bc=1, \\
\eta(\tau) &= \Gamma_1 \cos \frac{\pi}{\beta} \Theta \tau + \Gamma_2 \sin \frac{\pi}{\beta} \Theta \tau + \Gamma_3,
\end{align}
in terms of \emph{a priori} six parameters $a$, $b$, $c$, $\Gamma_1$, $\Gamma_2$, and $\Gamma_3$. We choose to parametrize the constant value of $\left\{F,\tau\right\}$ as $\frac{2\pi^2}{\beta^2} \Theta^2$ in terms of the parameter $\Theta$ (which may be imaginary). Plugging this solution into \eqref{susyact} and \eqref{susyactother}, we get
\begin{align}
T_{\B}(\tau) &= \frac{\pi^2}{\beta^2} \Theta^2, \\
T_{\F}(\tau) &= \frac{\pi^2}{\beta^2} \Theta^2 \Gamma_3 + \frac{\pi^3}{2\beta^3} \Theta^3 \Gamma_1\Gamma_2\Gamma_3.
\end{align}
These indeed satisfy $\partial_\tau \operatorname{Sch} = 0$. The Grassmann parameter $\Gamma_3 = \eta_0$ changes the on-shell value of $T_{\F}(\tau)$, and hence is not in the stabilizer generically; instead, it parametrizes different orbits. This means that the fermionic parts of the stabilizer are parametrized by $\Gamma_1$ and $\Gamma_2$ if they are allowed by the periodicity constraints when $\Theta \in \mathbb{N}$. For the bosonic pieces, for $\Theta \in \mathbb{N}$, the full $\sltr$ group is compatible with the periodicity constraints.  Otherwise, only $F \to F + b$ remains. We hence deduce the orbit stabilizers and on-shell stress tensors in Table \ref{orbits}.  These different orbits match with the constant-representative orbits in \cite{Delius:1990pt}.\footnote{Specifically, these are the orbits denoted by (a), (b), (c), and (e) in \cite{Delius:1990pt}, where we have corrected what we believe is an error in orbit (c).}

\begin{table}[!htb]
\centering
\begin{equation*}
\begin{array}{c|c||c|c|c}
\text{Sector}& \text{Orbit} & \text{Stabilizer} & T_{\B} & T_{\F} \\
\hline \hline
\text{NS} & \Theta = \text{$n$ odd} & H = \text{OSp}(1|2,\mathbb{R}) &   \frac{\pi^2}{\beta^2} n^2 & 0 \\
\text{R} &\Theta = \text{$n$ even} & H = \text{OSp}(1|2,\mathbb{R}) &   \frac{\pi^2}{\beta^2} n^2 & 0 \\
\hline
\text{NS} & \Theta = \text{$n$ even} & H = \text{SL}(2,\mathbb{R}) &   \frac{\pi^2}{\beta^2} n^2 & 0 \\
\text{R} & \Theta = \text{$n$ even, with } \eta_0 \neq 0 & H = \text{SL}(2,\mathbb{R}) &   \frac{\pi^2}{\beta^2} n^2 &  \frac{\pi^2}{\beta^2} n^2  \eta_0 \\
\text{R} & \Theta = \text{$n$ odd} & H = \text{SL}(2,\mathbb{R}) &   \frac{\pi^2}{\beta^2} n^2 & \frac{\pi^2}{\beta^2} n^2  \eta_0 \\
\hline
\text{R/NS} & \Theta \not\in \mathbb{Z} &  H = U(1) &  \frac{\pi^2}{\beta^2} \Theta^2 &  \frac{\pi^2}{\beta^2} \Theta^2  \eta_0 \\
\text{NS} & \Theta = 0 & H = \mathbb{R} & 0 & 0 \\
\text{R} & \Theta = 0 &  H =  \mathbb{R}^{1|1} & 0 & 0
\end{array}
\end{equation*}
\caption{Constant-representative sVirasoro orbits.}
\label{orbits}
\end{table}

By looking at the infinitesimal version of this solution space around a saddle, we can see that the resulting zero modes match directly to specific subalgebras of the $\mathcal{N}=1$ super-Virasoro algebra, as written in the last column of Table \ref{monodromies}. One of the cases is worked out in the next subsection.

The total energy of the constant representative (or the saddle solution) within each orbit is given by the expression:
\begin{equation}
T_{\B}(\tau) = \frac{\pi^2}{\beta^2} \Theta^2, \qquad T_{\F}(\tau) = \frac{\pi^2}{\beta^2} \Theta^2 \eta_0.
\end{equation}
As such, all orbits discussed here have on-shell action
\begin{equation}
S_{\text{on-shell}} = \oint d\tau\, T_{\B}(\tau) = \frac{\pi^2}{\beta} \Theta^2.
\end{equation}
The zero mode $\eta_0$ is an additional label of the orbit, only present in the Ramond sector. Moreover, only for the R parabolic orbit where $\Theta=0$ is $\eta_0$ a gauge mode and therefore dropped. For all other R cases, $\eta_0$ is not a gauge mode, and its fermionic integral causes the path integral to vanish \cite{Stanford:2019vob}.

\subsection{Special Elliptic Orbits in the NS Sector}
\label{app:detNS}

We now assume odd $n\in \mathbb{N}$, and look at the gauge zero modes present in the special elliptic coadjoint orbit. The generic solution was written above in \eqref{clsol}, but here we work it out at the infinitesimal level. In terms of the variable $f(\tau)$ defined through $F(\tau) = \tan \frac{\pi n}{\beta}f(\tau)$, the saddle solution is $f(\tau)=\tau$ and the gauge zero modes in the bosonic sector are parametrized by $\epsilon(\tau)$:
\begin{equation}
\epsilon(\tau) = \epsilon_1 + \epsilon_2 \cos \frac{2\pi n}{\beta} \tau + \epsilon_3 \sin \frac{2\pi n}{\beta} \tau,
\end{equation}
where $f(\tau) = \tau + \epsilon(\tau)$. In terms of the $F$ variable, this is equivalent to
\begin{equation}
F(\tau) = \tan \frac{\pi n}{\beta}\tau + \frac{\pi n}{\beta}\left[\epsilon_1 \left(1+\tan^2 \frac{n \pi}{\beta}\tau\right) + \epsilon_2 \left(1-\tan^2 \frac{n \pi}{\beta}\tau\right) + 2\epsilon_3 \tan \frac{n \pi}{\beta}\tau\right] + \mathcal{O}(\epsilon^2).
\end{equation}
This can be interpreted as the infinitesimal expansion
\begin{equation}
F \to F + b + (a-1) F - c F^2
\end{equation}
of the SL$(2,\mathbb{R})$ subgroup acting through M\"obius transformations on $F$ as in \eqref{clsol}. From the expansion of the (bosonic parts of the) Virasoro generators of Diff$(S^{1|1})$ as $L_n = \frac{1}{i}e^{i n \tau} \partial_\tau$, we can immediately identify these zero modes as generated by the SL$^n(2,\mathbb{R})$ subalgebra generated by $L_0,L_n,L_{-n}$.\footnote{The super Witt algebra on $\operatorname{Diff}(S^{1|1})$, with supercircle coordinates $\tau|\vartheta$ and periodicity $\tau \sim \tau +2 \pi$, can be represented by the superspace differential operators
\begin{equation}
L_n = \frac{1}{i}e^{in\tau} \left( \partial_\tau + i \frac{n}{2} \vartheta \partial_\vartheta\right), \qquad G_n = \frac{1}{\sqrt{i}} e^{in\tau} ( \partial_\vartheta + \vartheta \partial_\tau).
\end{equation}}

The fermionic variable $\eta$ has the gauge zero modes
\begin{equation}
\eta(\tau) = \epsilon_4 \cos \frac{n \pi}{\beta} \tau + \epsilon_5\sin \frac{n \pi}{\beta} \tau,
\end{equation}
which are antiperiodic for odd $n$. This corresponds to the infinitesimal action by the sVirasoro generators $G_{\frac{n}{2}}, G_{-\frac{n}{2}}$. This leads to the relation
\begin{equation}
\label{signproblem}
\alpha_{\F}(\tau) \equiv \sqrt{\partial_\tau F} \eta = \epsilon_4\operatorname{sgn} \left(\cos \frac{n \pi}{\beta} \tau\right) + \epsilon_5 \operatorname{sgn}\left(\cos \frac{n \pi}{\beta} \tau\right) \tan \frac{n \pi}{\beta} \tau + \mathcal{O}(\epsilon^2)
\end{equation}
for the bottom (fermionic) part of \eqref{alphareparam}. Notice in particular the sign functions. If they were absent, then all of these gauge zero modes $\epsilon_1, \ldots, \epsilon_5$ would combine into the infinitesimal expansion of OSp$(1|2)$ acting through super-M\"obius transformations on $(\tau',\theta')$ as in \eqref{ospredun}. With these sign functions, the fermionic parameters in this OSp$(1|2)$ group flip sign when $\tau \to \tau+\beta$. This is further implemented at the level of the group variables in the main text. 

\subsection{\texorpdfstring{$\osp$}{OSp(1|2)}-Invariant Bilocal Operators} \label{app:bilocal}

As mentioned in Section \ref{finitereps}, the supersymmetric Hill's equation yields recursion relations that allow for the computation of general matrix elements of bilocal operators: letting $|m; \tau\rangle\equiv g^{-1}(\tau)|m\rangle$, we have
\begin{equation}
(j - m)|m; \tau\rangle = \partial_\tau|m + 1; \tau\rangle - T_{\F}(\tau)|{\textstyle m + \frac{3}{2}; \tau}\rangle - (j + m + 2)T_{\B}(\tau)|m + 2; \tau\rangle
\end{equation}
if $j - m\in \mathbb{Z}$ and
\begin{equation}
\textstyle \left(j - m - \frac{1}{2}\right)|m; \tau\rangle = \partial_\tau|m + 1; \tau\rangle - \left(j + m + \frac{3}{2}\right)(T_{\F}(\tau)|{\textstyle m + \frac{3}{2}; \tau}\rangle + T_{\B}(\tau)|m + 2; \tau\rangle)
\end{equation}
if $j - m\in \mathbb{Z} + \frac{1}{2}$.  Note that unlike in the bosonic case, we generally need both the highest-weight state $j$ and the next-highest-weight state $j - \frac{1}{2}$ as base cases.

For example, to derive the form of the bilocal operator in arbitrary states in the spin-$1/2$ representation, we write the supersymmetric Hill's equation independently of representation as
\begin{equation}
g^{-1}(\tau)(E^- + T_{\B}(\tau)E^+ + 2T_{\F}(\tau)F^+) = \partial_\tau g^{-1}(\tau),
\end{equation}
or equivalently as
\begin{equation}
(E^- + T_{\B}(\tau)E^+ + 2T_{\F}(\tau)F^+)g(\tau) = -\partial_\tau g(\tau).
\end{equation}
Applying Hill's equation to $|{\textstyle \pm\frac{1}{2}}\rangle$ gives
\begin{equation}
g^{-1}(\tau)|{\textstyle -\frac{1}{2}}\rangle = \partial_\tau g^{-1}(\tau)|{\textstyle \frac{1}{2}}\rangle, \qquad g^{-1}(\tau)|0\rangle = \frac{\partial_\tau^2 g^{-1}(\tau)|{\textstyle \frac{1}{2}}\rangle - T_{\B}(\tau)g^{-1}(\tau)|{\textstyle \frac{1}{2}}\rangle}{T_{\F}(\tau)}.
\end{equation}
Similarly, applying Hill's equation to $\langle{\textstyle \pm\frac{1}{2}}|$ gives
\begin{equation}
\langle{\textstyle \frac{1}{2}}|g(\tau) = -\partial_\tau\langle{\textstyle -\frac{1}{2}}|g(\tau), \qquad \langle 0|g(\tau) = \frac{\partial_\tau^2\langle{\textstyle -\frac{1}{2}}|g(\tau) - T_{\B}(\tau)\langle{\textstyle -\frac{1}{2}}|g(\tau)}{T_{\F}(\tau)}.
\end{equation}
From these relations, we deduce that
\begin{equation}
R_{1/2}(g(\tau_2)g^{-1}(\tau_1)) = \left[\begin{array}{ccc}
-\partial_{\tau_2} & -\partial_{\tau_2}\Delta_{\tau_1} & -\partial_{\tau_2}\partial_{\tau_1} \\
\Delta_{\tau_2} & \Delta_{\tau_2}\Delta_{\tau_1} & \Delta_{\tau_2}\partial_{\tau_1} \\
1 & \Delta_{\tau_1} & \partial_{\tau_1}
\end{array}\right]\langle{\textstyle -\frac{1}{2}}|g(\tau_2)g^{-1}(\tau_1)|{\textstyle \frac{1}{2}}\rangle
\end{equation}
where $\Delta_\tau\equiv T_{\F}^{-1}(\tau)(\partial_\tau^2 - T_{\B}(\tau))$, as stated in the main text.

We now turn our attention to higher-dimensional finite representations.  The relevant representation theory is summarized in Appendix \ref{app:finiterep}.  For such representations, we can again compute the matrix element $\langle-j|g(\tau_2)g^{-1}(\tau_1)|j\rangle$ directly.  We have
\begin{align}
g^{-1}(\tau_1)|j\rangle &= e^{2\theta_{\m}(\tau_1)F^-}e^{\gamma_{\m}(\tau_1)E^-}e^{2\phi(\tau_1)H}e^{\gamma_{\+}(\tau_1)E^+}e^{2\theta_{\+}(\tau_1)F^+}|j\rangle \\
&= e^{2j\phi(\tau_1)}\sum_{n=0}^{2j} \gamma_{\m}(\tau_1)^n(|j - n\rangle + \theta_{\m}(\tau_1)|{\textstyle j - n - \frac{1}{2}}\rangle),
\end{align}
in addition to
\begin{align}
\langle-j|g(\tau_2) &= \langle-j|e^{-2\theta_{\+}(\tau_2)F^+}e^{-\gamma_{\+}(\tau_2)E^+}e^{-2\phi(\tau_2)H}e^{-\gamma_{\m}(\tau_2)E^-}e^{-2\theta_{\m}(\tau_2)F^-} \\
&= e^{2j\phi(\tau_2)}\sum_{n=0}^{2j} \binom{2j}{n}(-\gamma_{\m}(\tau_2))^n(\langle-j + n| + \theta_{\m}(\tau_2)(2j - n)\langle{\textstyle -j + n + \frac{1}{2}}|).
\end{align}
Taking the inner product gives
\begin{align}
&\hspace{-0.5 cm} \langle-j|g(\tau_2)g^{-1}(\tau_1)|j\rangle \nonumber \\
&= e^{2j\phi(\tau_1)}e^{2j\phi(\tau_2)}\sum_{n=0}^{2j} \binom{2j}{n}\gamma_{\m}(\tau_1)^n(-\gamma_{\m}(\tau_2))^{2j - n} \nonumber \\
&\phantom{==} + 2je^{2j\phi(\tau_1)}e^{2j\phi(\tau_2)}\sum_{n=0}^{2j - 1} \binom{2j - 1}{n}\gamma_{\m}(\tau_1)^n(-\gamma_{\m}(\tau_2))^{2j - n - 1}\theta_{\m}(\tau_2)\theta_{\m}(\tau_1) \\
&= e^{2j(\phi(\tau_1) + \phi(\tau_2))}[(\gamma_{\m}(\tau_1) - \gamma_{\m}(\tau_2))^{2j} + 2j(\gamma_{\m}(\tau_1) - \gamma_{\m}(\tau_2))^{2j - 1}\theta_{\m}(\tau_2)\theta_{\m}(\tau_1)] \\
&= [e^{\phi(\tau_1) + \phi(\tau_2)}(\gamma_{\m}(\tau_1) - \gamma_{\m}(\tau_2) + \theta_{\m}(\tau_2)\theta_{\m}(\tau_1))]^{2j} \\
&= [\psi_{1, \text{bot}}(\tau_2) \psi_{2, \text{bot}}(\tau_1) - \psi_{2, \text{bot}}(\tau_2)\psi_{1, \text{bot}}(\tau_1) + \psi_{3, \text{bot}}(\tau_2)\psi_{3, \text{bot}}(\tau_1)]^{2j}, \label{fiducialspinjosp}
\end{align}
which is simply the $j = 1/2$ result \eqref{fiducialspin12osp} to the power of $2j$.  More general matrix elements can be computed with the aid of the supersymmetric Hill's equation as above.

\section{\texorpdfstring{$\text{OSp}(1|2, \mathbb{R})$}{OSp(1|2, R)} Representation Theory} \label{osprep}

In this lengthy appendix, we give an overview of OSp$(1|2,\mathbb{R})$ representation theory. A particular emphasis is placed on the principal series representations. Some of the results presented here are known in the literature, but we are not aware of a comprehensive treatment. We base our methods largely on those for SL$(2,\mathbb{R})$, as written, for instance, in the textbooks \cite{V, VK}.\footnote{The fact that the representation theory of OSp$(1|2,\mathbb{R})$ is so closely related to bosonic representation theory is specific to the $B(0,n) \equiv$ OSp$(1|2n)$ Lie supergroups: see \cite{Mikhaylov:2014aoa} for tangentially related comments.}

\subsection{\texorpdfstring{$\text{OSp}(1|2, \mathbb{R})$}{OSp(1|2, R)} Supergroup and Lie Superalgebra}

The supergroup $\text{OSp}(1|2,\mathbb{R})$ is defined as the subgroup of $\text{GL}(1|2,\mathbb{R})$ matrices
\begin{equation}
\label{ospdef}
g = \left[\begin{array}{cc|c}
a & b & \alpha \\
c & d & \gamma \\ \hline
\beta & \delta & e
\end{array}\right],
\end{equation}
consisting of five bosonic variables $a$, $b$, $c$, $d$, $e$ and four fermionic (Grassmann) variables $\alpha$, $\beta$, $\gamma$, $\delta$, that preserve the orthosymplectic form $\Omega$: $g^{\text{st}}\Omega g = \Omega$.  Explicitly,
\begin{equation}
\label{ospcond}
\left[\begin{array}{cc|c} 
a & c & -\beta \\
b & d & -\delta \\
\hline
\alpha & \gamma & e
\end{array} \right]
\left[\begin{array}{cc|c} 
0 & -1 & 0 \\
1 & 0 & 0 \\
\hline
0 & 0 & 1
\end{array} \right]
\left[\begin{array}{cc|c} 
a & b & \alpha \\
c & d & \gamma \\
\hline
\beta & \delta & e
\end{array} \right]
=
\left[\begin{array}{cc|c} 
0 & -1 & 0 \\
1 & 0 & 0 \\
\hline
0 & 0 & 1
\end{array} \right].
\end{equation}
The operation $g^{\text{st}}$ is the supertranspose that flips the sign of one block of fermionic variables to ensure the property $(g_1 g_2)^{\text{st}} = g_2^{\text{st}}g_1^{\text{st}}$, which implies that this subset defines a subgroup.\footnote{We deal only with even supermatrices, for which we define the supertranspose as
\begin{equation}
\label{supert}
\left[\begin{array}{c|c}
A & B \\ \hline
C & D
\end{array}\right]^\text{st}
= \left[\begin{array}{c|c}
A^T & -C^T \\ \hline
B^T & D^T
\end{array}\right].
\end{equation}
Other conventions also exist in the literature. The Berezinian is invariant under the supertranspose.}

The group OSp$(1|2,\mathbb{R})$ has the noncompact bosonic subgroup Sp$(2,\mathbb{R})$ $\simeq$ SU$(1,1)$ $\simeq$ SL$(2,\mathbb{R})$, parametrized by $a,b,c,d \in \mathbb{R}$ satisfying $ad-bc=1$, which distinguishes it from OSp$(1|2)$ with compact bosonic subgroup Sp$(2)$ (the latter group is discussed in, e.g., \cite{cmp/1103908695, Mikhaylov:2014aoa} and sometimes denoted by UOSp$(1|2)$). In the following, we sometimes abuse notation and simply denote the noncompact group of interest in this work by OSp$(1|2)$ as well.

The condition \eqref{ospcond} translates into
\begin{alignat}{2}
ad - bc - \delta\beta &= 1, \qquad & e^2 + 2\gamma\alpha &= 1, \nonumber \\
c\alpha - a\gamma - \beta e &= 0, \qquad & d\alpha - b\gamma - \delta e &= 0. \label{rela}
\end{alignat}
These relations can be conveniently solved into
\begin{align}
\alpha = \pm(a\delta - b \beta), \qquad \gamma &= \pm(c \delta - d \beta), \qquad e = \pm(1+ \beta \delta), \label{rela2} \\
ad-bc &= 1 +\delta \beta, \label{rela2last}
\end{align}
where one has a choice of sign that must be consistent across all relations \eqref{rela2}. Taking the Berezinian of the constraint \eqref{ospcond} immediately leads to $\operatorname{Ber} g = \pm 1$, which is precisely \eqref{rela2last} given the identities \eqref{rela2}. We will hence refer to \eqref{rela2last} somewhat loosely as the determinant condition.

We end up with the OSp$(1|2,\mathbb{R})$ supermatrices
\begin{equation}
g = \left[\begin{array}{cc|c} 
a & b & a\delta - b \beta \\
c & d & c\delta - d \beta \\
\hline
\beta & \delta & 1+\beta \delta
\end{array} \right], \qquad g = \left[\begin{array}{cc|c} 
a & b & -(a\delta - b \beta) \\
c & d & -(c\delta - d \beta) \\
\hline
\beta & \delta & -(1+\beta \delta)
\end{array} \right],
\label{firstgeneralparam}
\end{equation}
satisfying $ad-bc = 1 + \delta \beta$, and distinguishable by the sign of the Berezinian $\pm 1$.  These two components are related to each other by applying the elementary matrix $(-)^F$:
\begin{equation}
(-)^F = \left[\begin{array}{cc|c} 
1 & 0& 0 \\
0 & 1 & 0 \\
\hline
0 & 0 & -1
\end{array} \right].
\end{equation}
One can further quotient by the $\mathbb{Z}_2$ subgroup generated by the matrix $-I_3$, which again has Berezinian $-1$, and thereby identify both components. This restricts us to the projective supergroup denoted by $\posp = \osp/\mathbb{Z}_2$ in \cite{Stanford:2019vob}. For convenience, we may simply choose the top sign in \eqref{rela2}.

Thus an arbitrary group element is specified by $3|2$ independent parameters.  For later reference, we write down the inverse group element:
\begin{equation}
g^{-1}
= \left[\begin{array}{cc|c}
d & -b & -\delta \\
-c & a & \beta \\ \hline
\gamma & -\alpha & e
\end{array}\right],
\label{ginvstuff}
\end{equation}
which takes a simple form thanks to the OSp constraints.

The Cartan-Weyl generators of the algebra in the above defining representation are
\begin{gather}
\label{defrep}
H = \left[\begin{array}{cc|c} 
1/2 & 0 & 0 \\
0 & -1/2 & 0 \\
\hline
0 & 0 & 0
\end{array} \right], \quad E^- = \left[\begin{array}{cc|c} 
0 & 0 & 0 \\
1 & 0 & 0 \\
\hline
0 & 0 & 0
\end{array} \right], \quad E^+ = \left[\begin{array}{cc|c} 
0 & 1 & 0 \\
0 & 0 & 0 \\
\hline
0 & 0 & 0
\end{array} \right], \\[5 pt]
F^- = \left[\begin{array}{cc|c} 
0 & 0 & 0 \\
0 & 0 & -1/2 \\
\hline
1/2 & 0 & 0
\end{array} \right], \quad F^+ = \left[\begin{array}{cc|c} 
0 & 0 & 1/2 \\
0 & 0 & 0 \\
\hline
0 & 1/2 & 0
\end{array} \right],
\end{gather}
which can be readily verified to satisfy the $\mathfrak{osp}(1|2)$ algebra:
\begin{alignat}{2}
[H, E^\pm] &= \pm E^\pm, \quad & [E^+, E^-] &= 2H, \nonumber \\
[H, F^\pm] &= \pm\frac{1}{2}F^\pm, \quad & [E^\pm, F^\mp] &= -F^\pm, \label{ospline2} \\
\{F^+, F^-\} &= \frac{1}{2}H, \quad & \{F^\pm, F^\pm\} &= \pm\frac{1}{2}E^\pm. \nonumber
\end{alignat}
The corresponding Gauss-Euler representation of the first group element in \eqref{firstgeneralparam} is
\begin{align}
g(\phi, \gamma_{\m}, \gamma_{\+} | \theta_{\m}, \theta_{\+}) &= e^{2\theta_{\m} F^-}e^{\gamma_{\m} E^-}e^{2\phi H}e^{\gamma_{\+}E^+}e^{2\theta_{\+}F^+} \\
&= \left[\begin{array}{cc|c}
e^\phi & \gamma_{\+}e^\phi & e^\phi\theta_{\+} \\
\gamma_{\m} e^\phi & e^{-\phi} + \gamma_{\m}\gamma_{\+}e^\phi - \theta_{\m}\theta_{\+} & \gamma_{\m} e^\phi \theta_{\+} - \theta_{\m} \\ \hline
e^\phi\theta_{\m} & \gamma_{\+}e^\phi\theta_{\m} + \theta_{\+} & 1 + e^\phi\theta_{\m}\theta_{\+}
\end{array}\right],
\end{align}
which satisfies the relations \eqref{rela}.

\subsection{Finite-Dimensional Representations} 
\label{app:finiterep}

The spin-$j$ representation of $\mathfrak{osp}(1|2)$ has dimension $4j + 1$ (or $2j + 1|2j$), and it decomposes under the even part $\mathfrak{sl}(2, \mathbb{R})$ as $R_j^{\text{OSp}} = R_j^{\text{SL}} \oplus R_{j-1/2}^{\text{SL}}$ for $j > 0$ \cite{Frappat:1996pb}.  The summands in the direct sum decomposition of the tensor product of two irreps $j_1$ and $j_2$ range from $|j_1 - j_2|$ to $j_1 + j_2$ in half-integer steps due to the fermionic raising and lowering operators.  One can derive the generators of the spin-$j$ representation via the Clebsch-Gordan decomposition,\footnote{Along with the fact that the generators of the tensor product of two representations $R$ and $R'$ of $\mathfrak{g}$ are $T_{R\otimes R'}^a = T_R^a\otimes I_{R'} + I_R\otimes T_{R'}^a$.} but it is easier to simply postulate that
\begin{align}
H &= \left[\begin{array}{c|c}
\operatorname{diag}(j, \ldots, -j) & \mathbf{0}_{(2j + 1)\times 2j} \\ \hline
\mathbf{0}_{2j\times (2j + 1)} & \operatorname{diag}(j - \frac{1}{2}, \ldots, -j + \frac{1}{2})
\end{array}\right], \\
E^- &= \left[\begin{array}{c|c}
\begin{array}{cc} \mathbf{0}_{1\times 2j} & 0 \\ \operatorname{diag}(1, \ldots, 2j) & \mathbf{0}_{2j\times 1} \end{array} & \mathbf{0}_{(2j + 1)\times 2j} \\ \hline
\mathbf{0}_{2j\times (2j + 1)} & \begin{array}{cc} \mathbf{0}_{1\times (2j - 1)} & 0 \\ \operatorname{diag}(1, \ldots, 2j - 1) & \mathbf{0}_{(2j - 1)\times 1} \end{array}
\end{array}\right], \\
E^+ &= \left[\begin{array}{c|c}
\begin{array}{cc} \mathbf{0}_{2j\times 1} & \operatorname{diag}(2j, \ldots, 1) \\ 0 & \mathbf{0}_{1\times 2j} \end{array} & \mathbf{0}_{(2j + 1)\times 2j} \\ \hline
\mathbf{0}_{2j\times (2j + 1)} & \begin{array}{cc} \mathbf{0}_{(2j - 1)\times 1} & \operatorname{diag}(2j - 1, \ldots, 1) \\ 0 & \mathbf{0}_{1\times (2j - 1)} \end{array}
\end{array}\right], \\
F^- &= \left[\begin{array}{c|c}
\mathbf{0}_{(2j + 1)\times (2j + 1)} & \begin{array}{c} \mathbf{0}_{1\times 2j} \\ -\frac{1}{2}\operatorname{diag}(1, \ldots, 2j) \end{array} \\ \hline
\begin{array}{cc} \frac{1}{2}I_{2j} & \mathbf{0}_{2j\times 1} \end{array} & \mathbf{0}_{2j\times 2j}
\end{array}\right], \\
F^+ &= \left[\begin{array}{c|c}
\mathbf{0}_{(2j + 1)\times (2j + 1)} & \begin{array}{c} \frac{1}{2}\operatorname{diag}(2j, \ldots, 1) \\ \mathbf{0}_{1\times 2j} \end{array} \\ \hline
\begin{array}{cc} \mathbf{0}_{2j\times 1} & \frac{1}{2}I_{2j} \end{array} & \mathbf{0}_{2j\times 2j}
\end{array}\right],
\end{align}
where $I_n$ denotes the $n\times n$ identity matrix (the fermionic generators are fixed by the stated form of the bosonic generators up to an irrelevant normalization).  One can check that the algebra \eqref{ospline} is satisfied. This representation is equivalent to the one presented in \cite{Corwin:1974fi}. As for SL$(2,\mathbb{R})$, these finite representations are not unitary (aside from the trivial one).

In the spin-$j$ representation, we write
\begin{equation}
|j\rangle = \left[\begin{array}{c} 1 \\ 0 \\ \vdots \\ 0 \\ \hline 0 \\ \vdots \\ 0 \end{array}\right], \ldots, |{-j}\rangle = \left[\begin{array}{c} 0 \\ \vdots \\ 0 \\ 1 \\ \hline 0 \\ \vdots \\ 0 \end{array}\right], |{\textstyle j - \frac{1}{2}}\rangle = \left[\begin{array}{c} 0 \\ \vdots \\ 0 \\ \hline 1 \\ 0 \\ \vdots \\ 0 \end{array}\right], \ldots, |{\textstyle -j + \frac{1}{2}}\rangle = \left[\begin{array}{c} 0 \\ \vdots \\ 0 \\ \hline 0 \\ \vdots \\ 0 \\ 1 \end{array}\right].
\end{equation}
We then have the actions
\begin{align}
H|m\rangle &= m|m\rangle, \\
E^-|m\rangle &= \begin{cases} (j - m + 1)|m - 1\rangle & j - m\in \mathbb{Z}, \\ (j - m + \frac{1}{2})|m - 1\rangle & j - m\in \mathbb{Z} + \frac{1}{2}, \end{cases} \\
E^+|m\rangle &= \begin{cases} (j + m + 1)|m + 1\rangle & j - m\in \mathbb{Z}, \\ (j + m + \frac{1}{2})|m + 1\rangle & j - m\in \mathbb{Z} + \frac{1}{2}, \end{cases} \\
F^-|m\rangle &= \begin{cases} \frac{1}{2}|m - \frac{1}{2}\rangle & j - m\in \mathbb{Z}, \\ -\frac{1}{2}(j - m + \frac{1}{2})|m - \frac{1}{2}\rangle & j - m\in \mathbb{Z} + \frac{1}{2}, \end{cases} \\
F^+|m\rangle &= \begin{cases} \frac{1}{2}|m + \frac{1}{2}\rangle & j - m\in \mathbb{Z}, \\ \frac{1}{2}(j + m + \frac{1}{2})|m + \frac{1}{2}\rangle & j - m\in \mathbb{Z} + \frac{1}{2}, \end{cases}
\end{align}
as used in the computation of \eqref{fiducialspinjosp}.

\subsection{Casimir and sCasimir}

An important aspect of Lie superalgebras is the existence of elements in the universal enveloping algebra that commute or anticommute with all generators. They play an important role in classifying representations of the algebra.

Elements that commute with all generators span the \emph{centre} of the universal enveloping algebra. For OSp$(1|2)$, whose universal enveloping algebra we denote by U$(\mathfrak{osp}(1|2))$, there is a single such element: the quadratic Casimir
\begin{equation}
\label{CCas}
\mathcal{C} = H^2 + \frac{1}{2}( E^+E^- + E^- E^+) - ( F^+F^- - F^-F^+ ),
\end{equation}
which can indeed be checked to commute with all generators $H, E^\pm , F^\pm$ using \eqref{ospline2}.

Elements that commute with all bosonic generators and anticommute with all fermionic generators span the \emph{scentre} of the universal enveloping algebra \cite{Arnaudon:1996qe}. For OSp$(1|2)$, this is the \emph{sCasimir} operator. It is given by the expression:
\begin{align}
\label{sCas}
\mathcal{Q} = F^+F^- - F^- F^+ + \frac{1}{8} = \frac{1}{2}H - 2 F^- F^+ + \frac{1}{8}.
\end{align}
It has the important property that it squares to the Casimir:
\begin{equation}
\label{sCasCas}
\mathcal{Q}^2 - \frac{1}{64} = \frac{1}{4}\mathcal{C} = \frac{1}{4}H^2 + \frac{1}{8}( E^+E^- + E^- E^+) - \frac{1}{4}( F^+F^- - F^-F^+ ).
\end{equation}
For example, in the defining $j=1/2$ representation \eqref{defrep}, it is given explicitly by
\begin{equation}
\mathcal{Q} = \frac{3}{8}\left[\begin{array}{cc|c} 
1 & 0& 0 \\
0 & 1 & 0 \\
\hline
0 & 0 & -1
\end{array} \right]
\end{equation}
and is proportional to the matrix $(-)^F$ transforming between the two connected components of OSp$(1|2)$.

More generally, for the finite representations, the sCasimir operator is proportional to the fermion number $(-)^F$ and is given explicitly by
\begin{equation}
\label{Qs}
\mathcal{Q} = \left( \frac{j}{2} + \frac{1}{8}\right) \left[\begin{array}{c|c} 
I_{2j+1} & 0 \\
\hline
0 & -I_{2j} \\
\end{array} \right].
\end{equation}
From \eqref{sCasCas}, one finds the Casimir $\mathcal{C}$ to be proportional to the identity matrix:
\begin{equation}
\mathcal{C} = j (j+1/2) \left[\begin{array}{c|c} 
I_{2j+1} & 0 \\
\hline
0 & I_{2j} \\
\end{array} \right],
\end{equation}
as required by the generalization of Schur's lemma to supergroups, which states that all elements in the centre are proportional to the identity in an irreducible representation.

We can get a handle on the possible form of any element $M$ in the scentre using Schur's superlemma \cite{Kac:1977em}, as follows.\footnote{We make the assumption that the field over which we work is algebraically complete.}  The supermatrix $M^2$ commutes with the entire group, and we find by Schur's superlemma that
\begin{equation}
M^2 = z \left[\begin{array}{c|c} 
I & 0\\
\hline
0 & I
\end{array} \right]
\end{equation}
for some supernumber $z$.  Since the square root of a supernumber has only a $\pm$ sign ambiguity, we find that $M$ has to be proportional to a diagonal supermatrix with only $\pm 1$ on the diagonal.  We can go further if we assume that upon restricting to the maximal bosonic subgroup, the representation falls into two irreducible components (as happens in our case).\footnote{Any supergroup has a bosonic subgroup obtained by ignoring the fermionic part. For a finite representation, the resulting subgroup is of the form:
\begin{equation}
\left[\begin{array}{c|c} 
A_\text{body}(g) & 0\\
\hline
0 &D_\text{body}(g)
\end{array} \right].
\end{equation}
In the specific case of OSp$(1|2)$, the resulting bosonic representations $A_\text{body}(g)$ and $D_\text{body}(g)$ are irreducible as seen in the explicit construction above, and we have the branching rule
\begin{equation}
\label{branch}
R_{j, \, \text{supergroup}} \to R_{j, \, \text{bos}} \oplus R_{j-1/2, \, \text{bos}}.
\end{equation}}
Then by the ordinary Schur's lemma applied twice to the relation
\begin{equation}
\left[\begin{array}{c|c} 
A_\text{body}(g) & 0\\
\hline
0 &D_\text{body}(g)
\end{array} \right]M_\text{body} = M_\text{body} \left[\begin{array}{c|c} 
A_\text{body}(g) & 0\\
\hline
0 &D_\text{body}(g)
\end{array} \right],
\end{equation}
and by the fact that one can deduce the sign of $\pm \sqrt{z}$ unambiguously by knowing its body's, we end up with 
\begin{equation}
M = \lambda \left[\begin{array}{c|c} 
I & 0\\
\hline
0 & I
\end{array} \right] \text{ or } M = \lambda \left[\begin{array}{c|c} 
I & 0\\
\hline
0 & -I
\end{array} \right].
\end{equation}
Demanding anticommutativity with the fermionic generators of the group then shows that
\begin{equation}
M = \lambda \left[\begin{array}{c|c} 
I & 0\\
\hline
0 & -I
\end{array} \right]
\end{equation}
is the only possibility. This is indeed what we see explicitly in \eqref{Qs}.

\subsection{Principal Series Representations}
\label{app:ps}

Next to the finite-dimensional representations, the continuous representations play an important role in the harmonic analysis of OSp$(1|2,\mathbb{R})$. We construct them in this section.

\subsubsection{Invitation}

Our goal in this section is to construct the analogue of the principal series representations for OSp$(1|2,\mathbb{R})$. We take inspiration from the case of SL$(2,\mathbb{R})$:
\begin{equation}
\label{pssl}
(g\circ f)(x) = \operatorname{sgn}(bx+d)^{\epsilon}|bx + d|^{2j}f\left(\frac{ax + c}{bx + d}\right),
\end{equation}
where on the right-hand side, the group acts projectively by its transpose as $X^T g$ with
\begin{equation}
g = \left[\begin{array}{cc}
a & b  \\
c & d \end{array} \right], \quad
X = \left[\begin{array}{c} x \\ z \end{array}\right].
\end{equation}
The transpose ensures that the group action composes as required for a representation.  The carrier space is $L^2(\mathbb{R})$.

Let us now contemplate the supergroup case. The supertranspose we use was defined earlier in \eqref{supert}. In particular, it is not an involution, but rather has order four. Therefore, to imitate the construction for $\sltr$, we write
\begin{equation}
g = \left[\begin{array}{cc|c}
a & b & \alpha \\
c & d & \gamma \\ \hline
\beta & \delta & e
\end{array}\right], \quad
X = \left[\begin{array}{c} x \\ z \\ \hline \vartheta \end{array}\right]
\end{equation}
and let
\begin{equation}
X\mapsto g^{\text{st}^3}X, \quad g^{\text{st}^3}
= \left[\begin{array}{cc|c}
a & c & \beta \\
b & d & \delta \\ \hline
-\alpha & -\gamma & e
\end{array}\right].
\end{equation}
Indeed, this action can be written equivalently as\footnote{The supertranspose acts on \emph{column} vectors as follows:
\begin{equation}
\left[\begin{array}{c} \text{bos} \\ \hline \text{fer} \end{array}\right]^\text{st} = \left[\begin{array}{c|c} \text{bos} & -\text{fer} \end{array}\right], \quad \left[\begin{array}{c} \text{fer} \\ \hline \text{bos} \end{array}\right]^\text{st} = \left[\begin{array}{c|c} \text{fer} & \text{bos} \end{array}\right].
\end{equation}}
\begin{equation}
X^\text{st}\mapsto X^\text{st}g.
\end{equation}
Armed with these considerations, one is tempted to propose the following group action for OSp$(1|2,\mathbb{R})$ on $L^2(\mathbb{R}^{1|1})$:
\begin{equation}
\label{guess}
(g\circ f)(x, \vartheta) \stackrel{?}{\equiv} \operatorname{sgn}(bx+d+ \delta\vartheta)^{\epsilon}|bx + d + \delta\vartheta|^{2j}f\left(\frac{ax + c + \beta\vartheta}{bx + d + \delta\vartheta}, -\frac{\alpha x + \gamma - e\vartheta}{bx + d + \delta\vartheta}\right).
\end{equation}
However, this guess is not quite correct. Whereas the representation thus constructed is irreducible (as we show in Appendix \ref{app:irre}), it is not unitary. A more well-substantiated approach is based on the method of parabolic induction, which will allow us in the end to write down a corrected version of \eqref{guess}.

\subsubsection{Parabolic Induction}

It is a well-known fact in the mathematical literature that the so-called noncompact picture \eqref{pssl} for defining a principal series representation of SL$(2,\mathbb{R})$ has an equivalent induced picture, where one constructs the representation induced by a parabolic subgroup \cite{knapptrapa}. Since this idea is not as familiar to physicists, we first describe it in concrete terms. We then generalize it to OSp$(1|2,\mathbb{R})$, and in particular, show that the construction \eqref{guess} (properly adjusted) also deserves the name of principal series representation.

An induced representation is determined as follows.  Fix a Lie group $G$ and a subgroup $H \subset G$.  Let $D_\lambda(h)$ be a representation of the subgroup $H$ on a Hilbert space $V$.  For a continuous function $f: G \to V$, the equality
\begin{equation}
\label{consgen}
f(g h) = D_\lambda(h)^{-1} f(g)
\end{equation}
is consistent in the sense that $f(g h_1 h_2) = D_\lambda(h_2)^{-1} D_\lambda(h_1)^{-1}f(g) = D_\lambda(h_1h_2)^{-1}f(g)$. Then an action of the form
\begin{equation}
\label{act}
(g \cdot f) (g_0) \equiv f(g^{-1}g_0)
\end{equation}
on the reduced class of functions $f$ satisfying \eqref{consgen} automatically defines a representation of $G$. This is called an \emph{induced representation}. The representation one finds in this way will automatically be contained within the left regular representation of the group (to be discussed in Section \ref{app:regu} below).

Here, we specialize the above procedure to the case where we induce from a product of abelian subgroups. Principal series representations are defined as induced by a parabolic subgroup $P = MAN$ of $G$, where $M = Z_K(A)$ is the centralizer of $A$ in the maximal compact subgroup $K$ of $G$, $A$ is the abelian subgroup of positive diagonal matrices, and $N$ is the unipotent subgroup of upper triangular matrices with $1$ on the diagonal. Picking representations of $M$ and $A$, we can construct a representation of $G$.

\paragraph{SL$(2,\mathbb{R})$}

We focus first on SL$(2,\mathbb{R})$. We restrict to functions on the group $f:G \to \mathbb{C}$ respecting the constraint
\begin{equation}
\label{cons}
f(g man) = \sigma(m)^{-1} a^{-i\lambda-1} f(g).
\end{equation}
The factors appearing in \eqref{cons} are as follows:
\begin{itemize}
\item We denote by $\sigma$ the character of the finite group $M = \{\pm \mathbf{1}_{2\times 2}\}$: $\sigma(m) = (1,1)$ or $(1,-1)$.  We distinguish between the two representations by labeling the trivial and nontrivial representations by $\epsilon=0,1$, respectively.
\item The abelian diagonal subgroup $A$ has 1d irreps, which for a group element $\operatorname{diag}(a,a^{-1})$ can be parametrized as $e^{i\lambda \log a} = a^{i\lambda}$, since $\log a$ is the generator and $\lambda \in \mathbb{R}$ is the representation label. Due to the $i$ in the exponent, we will be inducing from a unitary representation of $A$.  The shift $-i\lambda \to -i\lambda -1$ is given by $\rho = \frac{1}{2}\sum_{i \in \Delta^+}\alpha_i$, half the sum of all positive roots. This is a normalization effect wherein the Haar measure on the group $dg$ gets decomposed into two parts $d\bar{n}$ and $d(man)$, with a nontrivial Jacobian which is absorbed into the transformation of the function $f$ by including the square root of the \emph{modular function} for the subgroup $P$:
\begin{equation}
\Delta(man)^{1/2} = |\det(\operatorname{Ad}_{\mathfrak{g}/\mathfrak{man}}(man))|^{1/2} = a^{\rho}.
\end{equation}
This deformation is necessary to induce a unitary representation of $G$.\footnote{In other words, we actually want to use \emph{normalized induction} to induce unitary representations from unitary representations, whereas the less sophisticated procedure that we have described in general terms is known as \emph{unnormalized induction}.} For SL$(2,\mathbb{R})$, $\rho=1$.
\item Finally, one picks the trivial representation $\mathbf{1}$ for the unipotent subgroup $N$.
\end{itemize}
A representation is then defined by the action \eqref{act} within the function space \eqref{cons}. The difficulty in making this action explicit is encoded precisely in the nontrivial structure of \eqref{cons}. For SL$(2,\mathbb{R})$, this restriction can be solved explicitly by using the identity
\begin{align}
\label{iden}
\left(\begin{array}{cc} 
d & -b \\
-c & a \\
\end{array} \right) \left(\begin{array}{cc} 
1 & 0 \\
-x & 1 \\
\end{array} \right) = \left(\begin{array}{cc} 
1 & 0 \\
-\frac{ax+c}{bx+d} & 1 \\
\end{array} \right) \left(\begin{array}{cc} 
bx+d & -b \\
0 & (bx+d)^{-1} \\
\end{array} \right),
\end{align}
which we write as $g^{-1} \bar{n} = \bar{n}' m'a'n'$ where the last three matrices are found via the following expansion:
\begin{align}
\label{expab}
&\left(\begin{array}{cc}
bx+d & -b \\
0 & (bx+d)^{-1} \\
\end{array} \right) \\
&= \left(\begin{array}{cc} 
\operatorname{sgn}(bx+d) & 0 \\
0 & \operatorname{sgn}(bx+d) \\
\end{array} \right) \left(\begin{array}{cc} 
|bx+d| & 0 \\
0 & |bx+d|^{-1} \\
\end{array} \right) \left(\begin{array}{cc} 
1 & -b (bx+d)^{-1} \\
0 & 1 \\
\end{array} \right). \nonumber
\end{align}
Namely, we decompose an arbitrary group element $g = \bar{n} m a n$ into an element $\bar{n}$ of the lower triangular unipotent subgroup $\bar{N}$ and an element of the parabolic subgroup $P=MAN$. Then by \eqref{cons}, it suffices to consider $g = \bar{n}$, and we restrict the function to this subgroup. Using the parametrization
\begin{equation}
\bar{n} = \left(\begin{array}{cc} 
1 & 0 \\
-x & 1 \\
\end{array} \right)
\end{equation}
and the identity \eqref{iden}, we can rewrite \eqref{act} more suggestively as
\begin{equation}
(g \cdot f) (\bar{n}) \equiv f(g^{-1}\bar{n}) =  f(\bar{n}' m'a'n') = \sigma(m')^{-1}\left|a'\right|^{-i\lambda-1} f(\bar{n}').
\end{equation}
Plugging in \eqref{iden} and setting $f(\bar{n}) \equiv f(x)$, this becomes
\begin{equation}
(g \cdot f) (x)  = \operatorname{sgn}(bx+d)^{\epsilon}|bx+d|^{-i\lambda-1} f \left(\frac{ax+c}{bx+d} \right),
\end{equation}
which matches our earlier definition \eqref{pssl} in the noncompact picture. This shows that the noncompact picture and induced picture give the same outcome, a well-known statement in the representation theory of SL$(2,\mathbb{R})$.

\paragraph{OSp$(1|2,\mathbb{R})$}

Now let's generalize this construction to OSp$(1|2,\mathbb{R})$. The upper and lower ``triangular'' matrices of interest are, in this case:
\begin{equation}
N = \left(\begin{array}{cc|c} 
1 & x & \vartheta \\
0 & 1 & 0 \\
\hline
0 & \vartheta & 1
\end{array} \right), \qquad \bar{N} = \left(\begin{array}{cc|c} 
1 & 0 & 0 \\
-x & 1 & \vartheta \\
\hline
-\vartheta & 0 & 1
\end{array} \right).
\end{equation}
The abelian subgroup with positive entries is parametrized by
\begin{equation}
A = \left(\begin{array}{cc|c} 
a & 0 & 0 \\
0 & a^{-1} & 0 \\
\hline
0 & 0 & 1
\end{array} \right).
\end{equation}
Finally, $M$ is the Klein four-group $\mathbb{Z}_2 \times \mathbb{Z}_2$ consisting of four elements $\operatorname{diag}(\pm 1, \pm 1, \pm' 1)$, where the first two signs are aligned. This finite group has four irreps, whose characters take the form $\sigma(M) = \zeta^{\epsilon}{\zeta'}^{\epsilon'}$ if one parametrizes $M = \text{diag}(\zeta, \zeta, \zeta')$.\footnote{The character table of $M$ is:
\begin{equation}
\begin{array}{c|cccc}
\sigma & \zeta=\zeta'=1 & \zeta=-1, \, \zeta'=1 & \zeta=1, \, \zeta'= -1 & \zeta = \zeta' = -1 \\
\hline
\text{id} & 1 & 1 & 1 & 1 \\
\epsilon=1, \, \epsilon'=0 & 1& -1 & 1 & -1 \\
\epsilon=0, \, \epsilon'=1 & 1 & 1 & -1 & -1 \\
\epsilon=1, \, \epsilon'=1 & 1 & -1 & -1 & 1
\end{array}
\end{equation}
} We can then write an arbitrary $\osp$ matrix as $g = \bar{N} MAN $, which is just our Gauss-Euler parametrization. Within $\posp = \osp/\mathbb{Z}_2$, the matrix $M$ parametrizes the two connected com\-po\-nents of $\osp$.

In this case, the following identity holds:
\begin{align}
\label{idenosp}
&\left(\begin{array}{cc|c} 
d & -b & -\delta\\
-c & a & \beta \\
\hline
\gamma & - \alpha & e
\end{array} \right) \left(\begin{array}{cc|c} 
1 & 0 & 0 \\
-x & 1 & \vartheta \\
\hline
-\vartheta & 0 & 1
\end{array} \right) \\ 
&= \left(\begin{array}{cc|c} 
1 & 0 & 0 \\
-\frac{ax + c + \beta \vartheta}{bx + d + \delta \vartheta} & 1 & - \frac{\alpha x + \gamma - e \vartheta}{bx + d + \delta \vartheta}\\
\hline
\frac{\alpha x + \gamma - e \vartheta}{bx + d + \delta \vartheta} & 0 & 1 
\end{array} \right) \left(\begin{array}{cc|c} 
bx + d +\delta \vartheta & -b & - b\vartheta - \delta \\
0 & (bx + d + \delta \vartheta)^{-1} & 0 \\
\hline
0 & \operatorname{sgn}(e) (- b\vartheta - \delta) & \operatorname{sgn}(e)
\end{array} \right), \nonumber
\end{align}
where each matrix belongs to $\osp$. We can again read this identity as a decomposition of $g^{-1}\bar{n} = \bar{n}' m'a'n'$ into an element of the lower triangular subgroup $\bar{N}$ and a remainder in $P=MAN$:
\begin{align}
&\left(\begin{array}{cc|c} 
bx+d +\delta \vartheta & -b & - b\vartheta - \delta \\
0 & (bx+d + \delta \vartheta)^{-1} & 0 \\
\hline
0 & \operatorname{sgn}(e) (- b\vartheta - \delta) & \operatorname{sgn}(e)
\end{array} \right) \nonumber \\
&= \arraycolsep=4pt \left(\begin{array}{cc|c} 
\operatorname{sgn}(bx+d +\delta \vartheta) & 0 & 0 \\
0 & \operatorname{sgn}(bx+d +\delta \vartheta) & 0 \\
\hline
0 & 0 & \operatorname{sgn}(e)
\end{array} \right) \left(\begin{array}{cc|c} 
|bx+d +\delta \vartheta| & 0 & 0 \\
0 & |bx+d + \delta \vartheta|^{-1} & 0 \\
\hline
0 & 0 & 1
\end{array} \right) \nonumber \\
&\phantom{==} \times \left(\begin{array}{cc|c} 
1 & -\frac{b}{bx+d +\delta \vartheta} & \frac{- b\vartheta - \delta}{bx+d +\delta \vartheta} \\
0 & 1 & 0 \\
\hline
0 & \frac{- b\vartheta - \delta}{bx+d +\delta \vartheta} & 1
\end{array} \right).
\end{align}
To induce a unitary representation, we again need to include a half-density of the form $a^{\rho}$ in the definition of the transformed function. This factor originates from a change of variables on the group manifold.  Therefore, for the supergroup case, we need to replace it by a super-Jacobian, which is a Berezinian in which \emph{only} the bosonic subtransformation $A$ is taken with an absolute value \cite{Stanford:2019vob}:
\begin{equation}
\operatorname{sdet}' = \operatorname{sgn}(\det A)\operatorname{sdet}.
\label{sdetprime}
\end{equation}
The precise super-Jacobian can be written as a product:
\begin{equation}
\label{prodjac}
\Delta(man) = \operatorname{sdet}'(\operatorname{Ad}_{\mathfrak{g}/\mathfrak{man}}(man)) = \operatorname{sdet}'(\operatorname{Ad}_{\mathfrak{g}/\mathfrak{man}}(m)) \operatorname{sdet}'(\operatorname{Ad}_{\mathfrak{g}/\mathfrak{man}}(a)) \operatorname{sdet}'(\operatorname{Ad}_{\mathfrak{g}/\mathfrak{man}}(n)).
\end{equation}
Analogously to the bosonic case, we have
\begin{equation}
\operatorname{sdet}'(\operatorname{Ad}_{\mathfrak{g}/\mathfrak{man}}(a)) = a^{2 (\rho_B-\rho_F)}, \qquad \operatorname{sdet}'(\operatorname{Ad}_{\mathfrak{g}/\mathfrak{man}}(n)) = 1,
\end{equation}
where $\rho_B = \frac{1}{2} \sum_{i \in \Delta_B^+} \alpha_i$ and $\rho_F = \frac{1}{2} \sum_{i \in \Delta_F^+} \alpha_i$. For $\osp$, $\rho_B=1$ and $\rho_F=1/2$.\footnote{We note that for OSp$(2|2)$, $\rho_B=\rho_F$ and there is no shift by the Weyl vector. This corresponds to the non-renormalization theorems starting with $\mathcal{N}=2$ supersymmetry.} However, the factor associated with the adjoint action of $m$ in \eqref{prodjac} is not trivial in this case. Let's work it out explicitly. The $1|1$-dimensional vector space $\mathfrak{g}/\mathfrak{man}$ is spanned by the generators $E^-$ and $F^-$. The group $M$ is four-dimensional: $m \in \{I_{2|1},-I_{2|1}, (-)^F,-(-)^F\}$.  We easily derive the adjoint action of each of these elements on the vector space:
\begin{alignat}{2}
I_{2|1} E^- I_{2|1} &= E^-, \qquad & I_{2|1} F^- I_{2|1} &= F^-, \\
(-)^F E^- (-)^F &= E^-, \qquad & (-)^F F^- (-)^F &= - F^-,
\end{alignat}
from which we read off that
\begin{equation}
\label{mdet}
\operatorname{sdet}'(\operatorname{Ad}_{\mathfrak{g}/\mathfrak{man}}(m))  = \begin{cases} 1 & m = \pm I_{2|1}, \\ -1 & m = \pm (-)^F,
\end{cases}
\end{equation}
and hence 
\begin{equation}
\Delta(man) = \pm a^{2 (\rho_B-\rho_F)},
\end{equation}
where the choice of sign depends on the element $m$ as in \eqref{mdet}.

These considerations finally lead to the definition
\begin{equation}
\label{paraend}
\boxed{
\begin{aligned}
(g \cdot f) (x,\vartheta) &= \operatorname{sgn}(e)^{\epsilon'-1/2} \operatorname{sgn}(bx+d+\delta\vartheta)^{\epsilon-1/2} \\
&\phantom{==} \times |bx+d+\delta\vartheta|^{-i\lambda - 1/2}f\left(\frac{ax+c+\beta\vartheta}{bx+d+\delta\vartheta}, -\frac{\alpha x + \gamma -e \vartheta}{bx+d + \delta \vartheta} \right).
\end{aligned}}
\end{equation}
From now on, we set $k\equiv - \lambda$, and we define the quantity $2j \equiv ik-1/2$ with $j$ being the spin representation label, yielding the Casimir $\mathcal{C} = j(j + 1/2) = -(k^2/4 + 1/16)$ for the principal series irreps.

We see that the principal series representations carry discrete sign labels $\epsilon, \epsilon'$.  Within the quotient $\osp/\mathbb{Z}_2$, the only distinguishable cases are $\epsilon=\epsilon'$ and $\epsilon \neq \epsilon'$. The representations for which one picks the trivial representation of $M$ (where $\epsilon=\epsilon'=0$) are called spherical principal series representations. When we restrict to the subsupersemigroup later on, there will be no distinction between these four cases anymore, and we can remove all of the sign factors by hand. This formula hence implies that the continuous representations of $\osp$ that emerge from the carrier space construction (\eqref{guess}, when suitably tweaked) deserve to be called principal series representations, just as in the bosonic case. Moreover, we learn that the carrier space consists of functions on a space of dimension equal to that of $\bar{N}$, a fact that would seem to be important for generalization to higher supersymmetry.

The resulting representation matrices are unitary with respect to the standard inner product on $L^2(\bar{N})$. In detail, this means
\begin{equation}
(f,g) = \begin{cases} \int_{\mathbb{R}} dx\, f^*(x) g(x) & \text{(bosonic case)}, \\ \int_{\mathbb{R}} dx\int d\vartheta\, f^*(x,\vartheta)g(x,\vartheta) & \text{(supersymmetric case)}. \end{cases}
\label{innerprod}
\end{equation}
We next demonstrate this very explicitly.

\subsubsection{Unitarity}
\label{app:unitarity}

We now verify by explicit calculation that the principal series representations defined by \eqref{paraend} are unitary, as guaranteed by the above more abstract construction.

With the measure on $\mathbb{R}^{1|1}$ in \eqref{innerprod}, a representation matrix element is constructed by performing the following operation:
\begin{equation}
\langle F|g|G\rangle \equiv \int dx\, d\vartheta\, F(x, \vartheta)^\ast (g \cdot G)(x, \vartheta).
\end{equation}
Plugging in \eqref{paraend}, we can write
\begin{align}
\int dx\, d\vartheta\, F(x, \vartheta)^\ast \left[\frac{|bx + d + \delta\vartheta|^{2j}}{\operatorname{sgn}(e)^{1/2} \operatorname{sgn}(bx+d+\delta\vartheta)^{1/2}}G\left(\frac{ax + c + \beta\vartheta}{bx + d + \delta\vartheta}, -\frac{\alpha x + \gamma - e\vartheta}{bx + d + \delta\vartheta}\right)\right].
\end{align}
We will show that the resulting representation matrix element is unitary iff $2j=ik-1/2$ by proving the identity
\begin{equation}
\int dx\, d\vartheta\, F(x, \vartheta)^\ast (g \cdot G)(x, \vartheta) = \int dx\, d\vartheta\, (g^{-1}\cdot F)(x, \vartheta)^\ast G(x, \vartheta), \quad j=-\frac{1}{4}+\frac{ik}{2},
\end{equation}
thus identifying the adjoint action $\dagger$ of $g$ with the inverse. To begin, we make the change of variables
\begin{align}
x &= \frac{dx' - c + \gamma\vartheta'}{-bx' + a - \alpha\vartheta'} = \frac{dx' - c}{-bx' + a} + \operatorname{sgn}(e) \frac{\delta x' - \beta}{(-bx' + a)^2}\vartheta', \\
\vartheta &= \frac{\delta x' - \beta + e\vartheta'}{-bx' + a - \alpha\vartheta'} = \frac{\delta x' - \beta}{-bx' + a} + \operatorname{sgn}(e) \frac{\vartheta'}{-bx' + a}.
\end{align}
This gives
\begin{equation}
bx + d + \delta\vartheta = \frac{1}{-bx' + a - \alpha\vartheta'}.
\end{equation}
To perform the change of variables in the Berezin integral, we write
\begin{align}
\left[\begin{array}{cc} A & B \\ C & D \end{array}\right] &\equiv \left[\begin{array}{cc} \frac{\partial x}{\partial x'} & \frac{\partial\vartheta}{\partial x'} \\[5 pt] \frac{\partial x}{\partial\vartheta'} & \frac{\partial\vartheta}{\partial\vartheta'} \end{array}\right] = \left[\begin{array}{cc} \frac{ad - bc}{(-bx' + a)^2} + \frac{\alpha + \operatorname{sgn}(e)b(\delta x' - \beta)}{(-bx' + a)^3}\vartheta' & \operatorname{sgn}(e) \frac{\alpha + b\vartheta'}{(-bx' + a)^2} \\ \operatorname{sgn}(e) \frac{-\delta x' + \beta}{(-bx' + a)^2} & \operatorname{sgn}(e)\frac{1}{-bx' + a} \end{array}\right],
\end{align}
and then compute the superdeterminant $\operatorname{sdet}'$:\footnote{We derive the supersymmetric Jacobian in this manner, rather than from the pullback of some differential form (i.e., from the exterior derivative of Grassmann variables), because the volume measure on a supermanifold is not a differential form (see Appendix A of \cite{Stanford:2019vob}).}
\begin{equation}
(A - BD^{-1}C)D^{-1} = \frac{ \operatorname{sgn}(e)}{-bx' + a - \alpha\vartheta'} \,\, \implies \,\, dx\, d\vartheta = dx'\, d\vartheta'\frac{\operatorname{sgn}(e) }{{-bx' + a - \alpha\vartheta'}},
\end{equation}
where we used $\operatorname{sdet}' = \operatorname{sgn}(\det A)\operatorname{sdet}$ and $\operatorname{sgn}(\det A) = \operatorname{sgn}(ad-bc)$. This gives the final result:
\begin{align}
&\scalebox{0.95}{$\displaystyle \int dx\, d\vartheta\, F(x, \vartheta)^\ast \left[\frac{|bx + d + \delta\vartheta|^{2j}}{{\operatorname{sgn}(e)^{1/2} \operatorname{sgn}(bx+d+\delta\vartheta)^{1/2}}}G\left(\frac{ax + c + \beta\vartheta}{bx + d + \delta\vartheta}, -\frac{\alpha x + \gamma - e\vartheta}{bx + d + \delta\vartheta}\right)\right]$} \\
&= \scalebox{0.95}{$\displaystyle \int dx'\, d\vartheta'\left[\frac{|{-bx' + a - \alpha\vartheta'}|^{-2j^\ast - 1}}{\operatorname{sgn}(e)^{1/2} \operatorname{sgn}(-bx' + a - \alpha\vartheta')^{1/2}}F\left(\frac{dx' - c + \gamma\vartheta'}{-bx' + a - \alpha\vartheta'}, \frac{\delta x' - \beta + e\vartheta'}{-bx' + a - \alpha\vartheta'}\right)\right]^\ast G(x', \vartheta')$}. \nonumber
\end{align}
If $2j = - 2j ^* -1$, or $j = -1/4 + ik/2$ for $k\in \mathbb{R}$, then we recognize the resulting expression as acting with the inverse group element $g^{-1}$ on $F$ (referring to the components of $g^{-1}$ in \eqref{ginvstuff}).  Hence the adjoint action is by $g^{-1}$, as desired, and the representation matrix constructed above is unitary.

\subsubsection{Infinitesimal Level: Lie Superalgebra}
\label{app:infinitesimal}

We now work out the infinitesimal action of the group and explicitly construct the resulting Lie superalgebra. Using the following parametrizations of one-parameter subgroups,
\begin{gather}
e^{2\phi H} = \left[\begin{array}{cc|c}
e^\phi & 0 & 0 \\
0 & e^{-\phi} & 0 \\ \hline
0 & 0 & 1
\end{array}\right], \quad
e^{\gamma_{\m} E^-} = \left[\begin{array}{cc|c}
1 & 0 & 0 \\
\gamma_{\m} & 1 & 0 \\ \hline
0 & 0 & 1
\end{array}\right], \quad
e^{\gamma_{\+}E^+} = \left[\begin{array}{cc|c}
1 & \gamma_{\+} & 0 \\
0 & 1 & 0 \\ \hline
0 & 0 & 1
\end{array}\right], \\[5 pt]
e^{2\theta_{\m} F^-} = \left[\begin{array}{cc|c}
1 & 0 & 0 \\
0 & 1 & -\theta_{\m} \\ \hline
\theta_{\m} & 0 & 1
\end{array}\right], \quad
e^{2\theta_{\+}F^+} = \left[\begin{array}{cc|c}
1 & 0 & \theta_{\+} \\
0 & 1 & 0 \\ \hline
0 & \theta_{\+} & 1
\end{array}\right],
\label{exponentiatedF}
\end{gather}
we read off the corresponding group actions from \eqref{paraend}:
\begin{align}
(e^{2\phi H}\circ f)(x, \vartheta) &= e^{-2j\phi}f(e^{2\phi}x, e^\phi\vartheta), \label{dilatation} \\
(e^{\gamma_{\m} E^-}\circ f)(x, \vartheta) &= f(x + \gamma_{\m}, \vartheta), \\
(e^{\gamma_{\+}E^+}\circ f)(x, \vartheta) &= \textstyle \operatorname{sgn}(\gamma_{\+}x + 1)^{\epsilon-1/2}|\gamma_{\+}x + 1|^{2j}f\left(\frac{x}{\gamma_{\+}x + 1}, \frac{\vartheta}{\gamma_{\+}x + 1}\right), \\
(e^{2\theta_{\m} F^-}\circ f)(x, \vartheta) &= f(x + \theta_{\m}\vartheta, \vartheta + \theta_{\m}), \\
(e^{2\theta_{\+}F^+}\circ f)(x, \vartheta) &= \textstyle \operatorname{sgn}(1 + \theta_{\+}\vartheta)^{\epsilon-1/2}|1 + \theta_{\+}\vartheta|^{2j}f\left(\frac{x}{1 + \theta_{\+}\vartheta}, \frac{\vartheta - \theta_{\+}x}{1 + \theta_{\+}\vartheta}\right).
\end{align}
Note that $\operatorname{sgn}(1 + \theta_{\+}\vartheta) = 1$, although we have left it explicit above.  We then derive the infinitesimal representations of the generators:\footnote{The signs and absolute values are not relevant at the infinitesimal level. However, they do mean that simply exponentiating the infinitesimal action would get these wrong.}
\begin{align}
\label{oppborelweil}
E^- &= \partial_x, \nonumber \vphantom{\frac{}{2}} \\
F^- &= \frac{1}{2}(\partial_\vartheta + \vartheta\partial_x), \nonumber \\
H &= x\partial_x + \frac{1}{2}\vartheta\partial_\vartheta - j, \\
E^+ &= -x^2\partial_x - x\vartheta\partial_\vartheta + 2jx, \nonumber \vphantom{\frac{1}{}} \\
F^+ &= -\frac{1}{2}x\partial_\vartheta - \frac{1}{2}x\vartheta\partial_x + j\vartheta. \nonumber 
\end{align}
These obey the commutation relations
\begin{alignat}{2}
[H, E^\pm] &= \pm E^\pm, \quad & [E^+, E^-] &= 2H, \nonumber \\
[H, F^\pm] &= \pm\frac{1}{2}F^\pm, \quad & [E^\pm, F^\mp] &= -F^\pm, \label{osplinerev} \\
\{F^+, F^-\} &= -\frac{1}{2}H, \quad & \{F^\pm, F^\pm\} &= \mp \frac{1}{2}E^\pm, \nonumber
\end{alignat}
which amount to almost the same algebra as the $\mathfrak{osp}(1|2)$ algebra \eqref{ospline}, except that the corresponding anticommutation relations have the opposite sign as in \eqref{ospline}:
\begin{equation}
\{F^+, F^-\} = -\frac{1}{2}H, \quad \{F^\pm, F^\pm\} = \mp\frac{1}{2}E^\pm.
\label{badsigns}
\end{equation}
These sign discrepancies come down to the fact that the bosonic generators $F^\pm$ that exponentiate to the group elements in \eqref{exponentiatedF} are represented by fermionic differential operators. Indeed, when the generators associated to fermionic group parameters (which are bosonic matrices) are represented as fermionic differential operators, the anticommutation relations of the operators must be opposite to those of the matrices, because the operators anticommute with the fermionic parameters while the matrices do not.

One can map our infinitesimal algebra \eqref{osplinerev} to the Lie superalgebra \eqref{ospline} by taking
\begin{equation}
(H, E^-, E^+, F^-, F^+)\to (-H, E^+, E^-, F^+, F^-).
\end{equation}
This reverses the signs of anticommutators while preserving the commutators.  We will hence call our infinitesimal superalgebra \eqref{osplinerev} the \emph{opposite} Lie superalgebra.\footnote{Compare to the notion of the ``opposite'' superalgebra in \cite{Stanford:2019vob}, which is obtained by taking
\begin{equation}
(H, E^-, E^+, F^-, F^+)\to (H, -E^+, -E^-, F^+, F^-).
\end{equation}
This reverses the signs of commutators while preserving the anticommutators.}

We call the set of differential operators \eqref{oppborelweil} the Borel-Weil realization of (opposite) $\mathfrak{osp}(1|2)$.  In this Borel-Weil realization, one readily computes the sCasimir:\footnote{Notice the slight change in this operator compared to \eqref{sCas} to accommodate the opposite Lie superalgebra.}
\begin{align}
\label{ssCas}
\mathcal{Q} =  F^-F^+-F^+F^-+\frac{1}{8} = \left(\frac{j}{2} + \frac{1}{8}\right)(1-2\vartheta \partial_\vartheta ) = \left(\frac{j}{2} + \frac{1}{8}\right) (-)^F.
\end{align}
Using \eqref{sCasCas}, we also obtain the Casimir operator
\begin{equation}
\mathcal{C} = 4 \left(\frac{j}{2} + \frac{1}{8}\right)^2 - \frac{1}{16} = j(j+1/2),
\end{equation}
which is diagonal in this basis. Notice that $\mathcal{C}$ is proportional to the identity operator, and $\mathcal{Q}$ is proportional to $(-)^F$. This has to be true for finite irreps by Schur's lemma, but it is true for the principal series representations as well, as we see here.

When restricting \eqref{oppborelweil} to the bosonic subgroup, we obtain the direct sum of a spin-$j$ Borel-Weil realization of SL$(2,\mathbb{R})$ (with $H, E^\pm$) and another Borel-Weil realization with spin $j-1/2$. The former part comes from acting on a purely bosonic function $f(x)$, and the latter from acting on a purely fermionic function $\vartheta f(x)$:
\begin{equation}
\label{decops}
R_j^{\text{OSp}} = R_j^{\text{SL}} \oplus R_{j-1/2}^{\text{SL}}.
\end{equation}
However, whereas both SL$(2,\mathbb{R})$ representations appearing here are irreducible, they are not unitary since $j$ is restricted by the left-hand side to take the value $j=-1/4+ik/2$, which is incompatible with the unitarity constraint for the SL$(2,\mathbb{R})$ spin label. We will see this feature very explicitly later on when we compute the characters of this representation in Section \ref{app:char}.

Fixing $j = -1/4 + ik/2$, all of the bosonic generators are antihermitian with respect to the measure $dx\, d\vartheta$:
\begin{equation}
H^\dag = -H, \qquad (E^-)^\dag = -E^-, \qquad (E^+)^\dag = -E^+.
\end{equation}
This follows straightforwardly from integrating by parts, while neglecting boundary terms (so that, e.g., $\partial_\vartheta^\dag = \partial_\vartheta$ and $\partial_x^\dag = -\partial_x$).  The fermionic generators $F^\pm$ are not antihermitian because they square to antihermitian operators.  However, accounting for the Grassmann statistics of the group parameters, all generators are antihermitian in the appropriate sense (as they must be, since the representation is unitary).

Finally, we mention that both the algebra and its opposite, which differ only in the anticommutation relations
\begin{alignat}{2}
\{F_\text{bos}^+, F_\text{bos}^-\} &= \frac{1}{2}H, \quad & \{F_\text{bos}^\pm, F_\text{bos}^\pm\} &= \pm\frac{1}{2}E^\pm, \\
\{F_\text{fer}^+, F_\text{fer}^-\} &= -\frac{1}{2}H, \quad & \{F_\text{fer}^\pm, F_\text{fer}^\pm\} &= \mp\frac{1}{2}E^\pm,
\end{alignat}
can be summarized by rewriting the $\mathfrak{osp}(1|2)$ algebra entirely in terms of commutators:
\begin{alignat}{2}
[H, E^\pm] &= \pm E^\pm, \quad & [E^+, E^-] &= 2H, \\
[H, \xi F^\pm] &= \pm\frac{1}{2}\xi F^\pm, \quad & [E^\pm, \xi F^\mp] &= -\xi F^\pm, \\
[\xi' F^+, \xi F^-] &= \frac{1}{2}\xi'\xi H, \quad & [\xi' F^\pm, \xi F^\pm] &= \pm\frac{1}{2}\xi'\xi E^\pm,
\end{alignat}
where we have introduced anticommuting parameters $\xi, \xi'$.  These relations hold regardless of whether $F = F_\text{bos}$ or $F = F_\text{fer}$.\footnote{The difference between the $F_\text{bos}$ generators \eqref{badborelweil} and the $F_\text{fer}$ generators \eqref{oppborelweil} is similar to that between the representations of the supercharges $Q$ and the supercovariant derivatives $D$ on superspace: the two resulting SUSY algebras have minus sign differences in the anticommutators.}

\subsubsection{Super-Fourier and Super-Mellin Transforms}
\label{app:bases}

To facilitate computations, and to make contact with specific representation matrices, we next discuss suitable bases of functions on the superline $\mathbb{R}^{1|1}$ that diagonalize some of the generators. These will define invertible integral transformations: the super-Fourier transform, an inverted cousin thereof, and the super-Mellin transform.

\paragraph{Super-Fourier Transform} The modes 
\begin{equation}
\label{sft}
\psi_{k,\alpha}(x,\vartheta) = \frac{1}{\sqrt{2\pi}}e^{ikx}e^{\alpha \vartheta} =  \frac{1}{\sqrt{2\pi}}(e^{ikx}+ \alpha \vartheta e^{ikx}),
\end{equation}
where $\alpha$ is an \emph{imaginary} Grassmann number in the sense that $\alpha^* = - \alpha$, form a basis for the functions on the superline $\mathbb{R}^{1|1}$:
\begin{align}
\int_{-\infty}^{+\infty} dx\, d\vartheta \, \psi_{k,\alpha}^*(x,\vartheta) \psi_{k',\alpha'}(x,\vartheta) &= \delta(k-k')\delta(\alpha-\alpha'), \\
\int_{-\infty}^{+\infty} dk\, d\alpha \, \psi_{k,\alpha}^*(x,\vartheta) \psi_{k,\alpha}(x',\vartheta') &= \delta(x-x')\delta(\vartheta'-\vartheta).
\end{align}
They are simultaneous eigenfunctions of the commuting operators
\begin{equation}
\hat{E}^- = \partial_x, \qquad \partial_\vartheta.
\end{equation}
A function on the superline can be uniquely expanded in this basis as
\begin{equation}
f(x,\vartheta) = \int dk\, d\alpha\, C_{k,\alpha} e^{ikx} e^{\alpha \vartheta},
\end{equation}
with coefficients $C_{k,\alpha} = C_{\B}(k) + \alpha C_{\T}(k)$ determined by the bosonic Fourier transform for each component:
\begin{equation}
f_{\B}(x) = \int dk\, C_{\T}(k) e^{ikx}, \qquad f_{\T}(x) = \int dk\, C_{\B}(k) e^{ikx}, \qquad f(x,\vartheta) = f_{\B}(x) + \vartheta f_{\T}(x).
\end{equation}
This basis \eqref{sft} corresponds to the superanalogue of the Fourier transform, and we will call it the \emph{super-Fourier transform}. This integral transform in superspace and its variants have been studied in the mathematics literature \cite{De_Bie_2008}.

\paragraph{Inverted Super-Fourier Transform} Quite analogously, one can find simultaneous ei\-gen\-functions of the commuting operators
\begin{equation}
\hat{E}^+ = -x^2\partial_x - x\vartheta \partial_\vartheta + 2jx , \qquad x\partial_\vartheta,
\end{equation}
as
\begin{equation}
\label{invsft}
\psi^j_{\lambda,\alpha}(x,\vartheta) = \frac{1}{\sqrt{2\pi}}\frac{|x|^{ik}}{\sqrt{x}}e^{i\lambda/x}e^{\alpha \frac{\vartheta}{x}}=  \frac{1}{\sqrt{2\pi}}\left( \frac{|x|^{ik}}{\sqrt{x}}e^{i\lambda/x}+ \alpha \vartheta \frac{|x|^{ik}}{x^{3/2}}e^{i\lambda/x} \right),
\end{equation}
satisfying the same orthonormality and completeness relations:
\begin{align}
\int_{-\infty}^{+\infty} dx\, d\vartheta \, \psi_{k,\alpha}^*(x,\vartheta) \psi_{k',\alpha'}(x,\vartheta) &= \delta(k-k')\delta(\alpha-\alpha'), \\
\int_{-\infty}^{+\infty} dk\, d\alpha \, \psi_{k,\alpha}^*(x,\vartheta) \psi_{k,\alpha}(x',\vartheta') &= \delta(x-x')\delta(\vartheta'-\vartheta).
\end{align}
These modes \eqref{invsft} are somewhat like inverted Fourier modes, and we will call the corresponding integral transform the \emph{inverted super-Fourier transform}.

\paragraph{Super-Mellin Transform} Finally, there is a basis associated to diagonalizing the hyperbolic generator $H$: the commuting operators
\begin{equation}
\hat{H} = x\partial_x + \frac{1}{2}\vartheta \partial_\vartheta - j, \qquad x^{1/2} \partial_\vartheta
\end{equation}
are diagonalized by the wavefunctions
\begin{equation} 
\label{melmode}
\psi_{s,\alpha}(x,\vartheta) = \frac{1}{\sqrt{2\pi}} x^{is-1/4} e^{\alpha x^{-1/2} \vartheta} = \frac{1}{\sqrt{2\pi}}(x^{is-1/4} + \alpha \vartheta x^{is-3/4}),
\end{equation}
with $\hat{H}$ eigenvalue $i(s - k/2)$ ($s\in \mathbb{R}$, by antihermiticity of $\hat{H}$ on the super half-line $\mathbb{R}^{+1|1}$ $(x>0,\vartheta)$).  The Grassmann variable $\alpha$ is the eigenvalue of the fermionic operator $x^{1/2} \partial_\vartheta$. This operator commutes with the superspace scaling operator $H$ because $x^{1/2} \partial_\vartheta$ is scale-invariant: $[\hat{H}, x^{1/2} \partial_\vartheta]=0$, as can be checked explicitly.

These functions transform the super half-line $(x>0,\vartheta)$ into a new pair of coordinates $(s,\alpha)$, and form an orthonormal basis with orthonormality and completeness relations:
\begin{align}
\langle \psi_{s,\alpha}|\psi_{s',\alpha'}\rangle &= \int_0^{+\infty} dx\, d\vartheta \, \psi_{s,\alpha}^*(x,\vartheta) \psi_{s',\alpha'}(x,\vartheta) = \delta(s-s')\delta(\alpha - \alpha'), \\
\langle x,\vartheta|x', \vartheta'\rangle &= \int_{-\infty}^{+\infty} ds\, d\alpha \, \psi_{s,\alpha}^*(x,\vartheta) \psi_{s,\alpha}(x',\vartheta') = \delta(x-x')\delta(\vartheta'-\vartheta).
\end{align}
This means that any function $f(x,\vartheta)$ on the super half-line can be uniquely expanded in this basis as
\begin{equation}
f(x,\vartheta) = \int ds\, d\alpha \, C_{s,\alpha} \psi_{s,\alpha}(x,\vartheta),
\end{equation}
with explicit expansion coefficients
\begin{equation}
C_{s,\alpha} = \int dx\, d\vartheta \, \psi_{s,\alpha}^*(x,\vartheta) f(x,\vartheta).
\end{equation}
The integral transform in superspace defined by the modes \eqref{melmode} is a superanalogue of the Mellin transform, and we will call it the \emph{super-Mellin transform}. To find a basis on the full superline $\mathbb{R}^{1|1}$, one needs a pair of such modes. This is identical to the description of Rindler modes to describe physics in the right wedge of the Minkowski plane.

For all three integral transforms defined above, we have taken $\alpha$ to satisfy the conjugation property $\alpha^*=-\alpha$. This was done so as to retain our convention that complex conjugation preserves the order of Grassmann numbers.\footnote{One could alternatively extract an ``$i$'' from $\alpha$ and take it be real. This would require absorbing an extra $i$ into the measure $d\alpha$, which we choose not to do.}

The above bases correspond to diagonalizing a single bosonic generator, augmented with a fermionic partner that is not in the algebra. This fermionic partner is unique in the following sense. Choosing a particular bosonic generator $D$ to diagonalize, one can prove that there is (up to a prefactor) only a single operator $\mathcal{O}$ that has the two properties:
\begin{align}
&\text{$\mathcal{O}$ is fermionic, i.e., $\mathcal{O}^2 = 0$}, \\
&\text{$\mathcal{O}$ and $D$ are simultaneously diagonalizable: $[\mathcal{O}, D] = 0$}.
\end{align}

\subsubsection{Discrete Representations Revisited: Monomial Realization}
\label{monomialreps}

Just as for SL$(2,\mathbb{R})$, one can realize the discrete representations on the carrier space $\mathbb{R}^{1|1}$ in terms of monomials. This complements earlier accounts of this representation \cite{Hughes:1981, Backhouse:1982}.

We work in a basis that diagonalizes $H = x\partial_x + \frac{1}{2} \vartheta \partial_\vartheta-j$, and we call its eigenvalue the weight of the state. Then the OSp$(1|2)$ algebra has raising (lowering) operators $F^\pm$ and $E^\pm$ that raise (lower) the weight by $1/2$ and $1$, respectively. If the representation has a lowest-weight state $\psi_{\text{LW},j}(x,\vartheta)$, then it satisfies $(\partial_\vartheta + \vartheta \partial_x)\psi_{\text{LW},j}(x,\vartheta) =0$, leading to
\begin{equation}
\psi_{\text{LW},j}(x,\vartheta) = 1, \qquad H = -j.
\end{equation}
Likewise, a highest-weight state (if present in the representation) satisfies $(x\partial_\vartheta + x \vartheta \partial_x - 2 j \vartheta)\psi_{\text{LW},j}(x,\vartheta) = 0$:
\begin{equation}
\psi_{\text{HW},j}(x,\vartheta) = x^{2j}, \qquad H = +j.
\end{equation}
If both of these states are present, then one needs $2j \in \mathbb{N}$, as one readily sees by consecutive applications of the raising and lowering operators. The representation becomes finite,  and consists of the monomials 
\begin{equation}
\{1, \vartheta, x, \vartheta x, x^2, \ldots, x^{2j}\}.
\end{equation}
For $2j \notin \mathbb{N}$ ($j$ can be negative), the representation is unbounded either from above (lowest-weight) or from below (highest-weight). For the group OSp$(1|2)$, there is a further restriction to $2j \in - \mathbb{N}$ for these representations that is not visible at the level of our current treatment. For its universal cover $\widetilde{\text{OSp}}(1|2)$, this further discretization is not present, and we are not sensitive here to this difference.

If either such state is present in the representation space, then all states are simultaneous eigenfunctions of $(-)^F= 1-2 \vartheta \partial_\vartheta$, where $\vartheta \partial_\vartheta$ measures the $\mathbb{Z}_2$ grading of the representation space. This is because the lowest- or highest-weight state is an eigenstate of $(-)^F$ with eigenvalue $1$, and applying fermionic raising or lowering operators flips the eigenvalue of $(-)^F$. This hence automatically holds both for finite-dimensional representations and for lowest/highest-weight discrete representations. This leads to a natural decomposition of the representation in terms of SL$(2,\mathbb{R})$ representations \cite{Hughes:1981,Backhouse:1982}:
\begin{equation}
\label{deco}
R_j^{\text{OSp}} = R_j^{\text{SL}} \oplus R_{j-1/2}^{\text{SL}},
\end{equation}
with $\mathbb{Z}_2$-grading $0$ and $1$, respectively. For the principal series representations, there are no lowest- or highest-weight states in the representation space, and hence the states need not be eigenstates of $(-)^F$. As examples, we refer to the super-Fourier and super-Mellin bases constructed in Section \ref{app:bases}.

Next, we define the adjoint wavefunctions for the finite representations, assuming the same inner product as the one used for the principal series representations. The adjoint wavefunctions are obtained by demanding orthonormality as:
\begin{equation}
\langle m|m'\rangle = \int dx\, d\vartheta\, \langle m|x, \vartheta\rangle\langle x, \vartheta|m'\rangle = \delta_{mm'}.
\end{equation}
Denoting the highest-weight state of the spin-$j$ representation by $|\text{h.w.}\rangle_j = |j\rangle$, whose wavefunction is annihilated by $F^+$ (and hence $E^+$), we fix the normalization by setting
\begin{equation}
\langle x, \vartheta|\text{h.w.}\rangle_j = x^{2j}.
\end{equation}
Applying $E^-$ consecutively, we construct the (normalized) monomials:
\begin{equation}
\langle x, \vartheta|m\rangle = \frac{(2j)!}{(j - m)!(j + m)!}x^{j+m}, \quad j - m\in \mathbb{Z}.
\end{equation}
This fixes the wavefunctions of the dual states to be
\begin{equation}
\langle m|x, \vartheta\rangle = \frac{(j - m)!}{(2j)!}(-\partial_x)^{j+m}\vartheta\delta(x), \quad j - m\in \mathbb{Z},
\end{equation}
and in particular, we obtain the (normalized) lowest-weight state
\begin{equation}
\langle \text{l.w.}|x, \vartheta\rangle_j = \langle -j|x, \vartheta\rangle = \vartheta\delta(x) = \delta(x, \vartheta).
\end{equation}
The (normalized) fermionic states in the representation are constructed analogously:
\begin{equation}
\langle x, \vartheta|m\rangle = \frac{(2j)!}{(j - m - \frac{1}{2})!(j + m - \frac{1}{2})!}\vartheta x^{j + m - \frac{1}{2}}, \quad j - m\in {\textstyle \mathbb{Z} + \frac{1}{2}}.
\end{equation}
This fixes the wavefunctions of the dual states to be
\begin{equation}
\langle m|x, \vartheta\rangle = \frac{(j - m - \frac{1}{2})!}{(2j)!}(-\partial_x)^{j + m - \frac{1}{2}}\delta(x), \quad j - m\in {\textstyle \mathbb{Z} + \frac{1}{2}}.
\end{equation}
Compare the action of the generators on these states to those in Appendix \ref{app:finiterep}.\footnote{The Borel-Weil realization that furnishes a representation of the conventional (rather than the opposite) $\mathfrak{osp}(1|2)$ superalgebra \eqref{ospline} has instead
\begin{equation}
\hat{F}^- = -\frac{1}{2}(\partial_\vartheta - \vartheta\partial_x), \qquad \hat{F}^+ = \frac{1}{2}x\partial_\vartheta - \frac{1}{2}x\vartheta\partial_x + j\vartheta,
\label{badborelweil}
\end{equation}
with the bosonic generators the same as in \eqref{borelweil}.}

This way of describing the finite-dimensional representations allows a direct generalization to the infinite-dimensional lowest-weight representations \cite{Hikida:2018eih}. As in the bosonic case, we formally take $j\to -j$ and replace the factorials by appropriate Pochhammer symbols.

\subsection{Left Regular Representation of \texorpdfstring{$\text{OSp}(1|2, \mathbb{R})$}{OSp(1|2, R)}}
\label{app:regu}

The most basic representation of any group is the regular representation. For a Lie (super)group, the left regular representation is defined by the following group action on the group itself:
\begin{equation}
f(g_0) \xrightarrow{g} f(g^{-1}g_0),
\end{equation}
with $f:G \to \mathbb{C}$. The left regular realization can be studied infinitesimally for each one-parameter subgroup as:
\begin{equation}
\hat{L}_i f(g_0) \equiv \frac{d}{d\epsilon}\left. f( e^{-\epsilon X_i} g_0) \right|_{\epsilon =0},
\end{equation}
where the generators are now realized as first-order differential operators satisfying
\begin{equation}
\label{defreg}
\hat{L}_i g = -X_i g.
\end{equation}
This definition can be worked out explicitly once a suitable coordinatization of the group element $g$ is chosen. We choose to work in Gauss-Euler coordinates:
\begin{equation}
g = e^{2\theta_{\m} F^-} e^{\gamma_{\m} E^-} e^{2\phi H} e^{\gamma_{\+} E^+} e^{2\theta_{\+}F^+}.
\end{equation}
Using the exponentiated commutator identities
\begin{gather}
[H,e^{\gamma_{\+} E^+}] = \gamma_{\+} E^+ e^{\gamma_{\+}E^+}, \qquad [e^{\gamma_{\m} E^-},H] = \gamma_{\m} E^- e^{\gamma_{\m} E^-}, \nonumber \\
[H,e^{2\theta_{\+}F^+}] = \theta_{\+}F^+ e^{2\theta_{\+}F^+}, \qquad [e^{2\theta_{\m} F^-},H] = \theta_{\m} F^- e^{2\theta_{\m} F^-}, \nonumber \\
[F^-,e^{\gamma_{\+} E^+}] = \gamma_{\+}F^+ e^{\gamma_{\+}E^+}, \qquad  [e^{\gamma_{\m} E^-},F^+] = - \gamma_{\m} F^- e^{\gamma_{\m} E^-}, \nonumber \\
e^{-\phi} F^- e^{2\phi H} = e^{2\phi H} F^-, \qquad e^{-\phi}e^{2\phi H}F^+ = F^+ e^{2\phi H}, \nonumber \\
[e^{\gamma_{\m} E^-},E^+] = e^{\gamma_{\m} E^-}(-2\gamma_{\m} H + \gamma_{\m}^2 E^-),
\end{gather}
we obtain the left regular representation of the algebra:\footnote{There exists an analogous right regular representation, where care has to be taken in the supergroup case that those differential operators act from the right \cite{cmp/1103908695}.}
\begin{align}
\label{left1}
\hat{L}_{F^-} &= -\frac{1}{2}\left(\partial_{\theta_{\m}} - \theta_{\m} \partial_{\gamma_{\m}}\right), \\
\hat{L}_{E^-} &= -\partial_{\gamma_{\m}}, \\
\hat{L}_H &= -\frac{1}{2} \partial_\phi + \gamma_{\m} \partial_{\gamma_{\m}} + \frac{1}{2} \theta_{\m} \partial_{\theta_{\m}}, \\
\hat{L}_{F^+} &= -\frac{1}{2}e^{-\phi}\left(\partial_{\theta_{\+}} +  \theta_{\+} \partial_{\gamma_{\+}} \right) - \frac{1}{2}\gamma_{\m}\left(\partial_{\theta_{\m}} - \theta_{\m} \partial_{\gamma_{\m}}\right) - \frac{1}{2}\theta_{\m} \partial_\phi, \\
\label{left5}
\hat{L}_{E^+} &= -e^{-2\phi} \partial_{\gamma_{\+}} -  \gamma_{\m} \partial_\phi + \gamma_{\m}^2 \partial_{\gamma_{\m}} + \gamma_{\m} \theta_{\m} \partial_{\theta_{\m}} + e^{-\phi} \theta_{\m} \left(\partial_{\theta_{\+}} + \theta_{\+} \partial_{\gamma_{\+}}\right).
\end{align}
Using \eqref{defreg}, one can check that these differential operators satisfy the algebra:
\begin{alignat}{2}
\left[\hat{L}_i,\hat{L}_j\right] g &= - \left[X_i,X_j\right] g = f_{ijk} \hat{L}_k g, \qquad && i,j \text{ not both odd}, \\
\left\{\hat{L}_i,\hat{L}_j \right\} g &= \left\{X_i,X_j\right\} g = -f_{ijk} \hat{L}_k g, \qquad && i,j \text{ both odd},
\end{alignat}
which has flipped signs for all anticommutators. We hence again see the appearance of the \emph{opposite} algebra, just as for the Borel-Weil realization on the superline $\mathbb{R}^{1|1}$.

In deriving these relations, we assumed the abstract generators $X_i$ satisfy the algebra \eqref{ospline}. However, for the Borel-Weil realization in terms of differential operators on $\mathbb{R}^{1|1}$, we work with the opposite algebra where the fermionic generators are Grassmann-valued. Working instead with this assumption, we can redo the above calculation to find precisely the same set of generators \eqref{left1}--\eqref{left5}.

We can construct the sCasimir operator in this realization as:
\begin{align}
\hat{\mathcal{Q}} &= \hat{L}_{F^-} \hat{L}_{F^+} - \hat{L}_{F^+} \hat{L}_{F^-} + \frac{1}{8} \\
&= \frac{1}{2}e^{-\phi}\left( \partial_{\theta_{\m}} - \theta_{\m} \partial_{\gamma_{\m}} \right) \left( \partial_{\theta_{\+}} + \theta_{\+} \partial_{\gamma_{\+}}\right) - \frac{1}{2} \theta_{\m} \partial_{\theta_{\m}} \partial_\phi - \frac{1}{4} \theta_{\m} \partial_{\theta_{\m}} + \frac{1}{4} \partial_\phi + \frac{1}{8}.
\end{align}
From the definition \eqref{defreg}, this operator has the property:
\begin{equation}
\hat{\mathcal{Q}} g = \mathcal{Q} g, \quad \mathcal{Q} = F^+ F^- - F^- F^+ + \frac{1}{8},
\end{equation}
in the abstract basis of operators satisfying \eqref{ospline}. For Grassmann-valued generators $F^\pm$, on the other hand (satisfying the opposite algebra), we find instead:
\begin{equation}
\hat{\mathcal{Q}} g = \mathcal{Q} g, \quad \mathcal{Q} = F^- F^+ - F^+ F^- + \frac{1}{8},
\end{equation}
where the right-hand side is the sCasimir operator in, e.g., the Borel-Weil realization \eqref{ssCas}. All of this can be viewed as consistency requirements on our calculations.

The sCasimir operator squares to the Casimir differential operator:
\begin{equation}
\hat{\mathcal{C}} = \frac{1}{4} \partial_\phi^2 + \frac{1}{4} \partial_\phi + e^{-2\phi} \partial_{\gamma_{\m}} \partial_{\gamma_{\+}} - \frac{1}{2} e^{-\phi}\left( \partial_{\theta_{\m}} + \theta_{\m} \partial_{\gamma_{\m}} \right) \left( \partial_{\theta_{\+}} + \theta_{\+} \partial_{\gamma_{\+}}\right).
\end{equation}

The above relations imply that generic representation matrix elements $\langle \psi_-|g|\psi_+\rangle$ for any bra and ket states satisfy the set of coupled differential equations:
\begin{align}
\hat{\mathcal{Q}}\langle \psi_-|g|\psi_+\rangle &= \langle \psi_-|\mathcal{Q}g|\psi_+\rangle, \nonumber \\
\hat{\mathcal{Q}} \langle \psi_-|\mathcal{Q}g|\psi_+\rangle &= \left(\frac{j}{2} + \frac{1}{8} \right)^2\langle \psi_-|g|\psi_+\rangle, \label{coupleige}
\end{align}
relating the two matrix elements $\langle \psi_-|g|\psi_+\rangle$  and $\langle \psi_-|\mathcal{Q}g|\psi_+\rangle$. On the left-hand side, the operator $\hat{\mathcal{Q}}$ is a differential operator acting on the supergroup coordinates $\phi,\gamma_{\m},\gamma_{\+},\theta_{\m},\theta_{\+}$ hidden in the representation matrix element:
\begin{equation}
\langle \psi_-|g|\psi_+\rangle = \langle \psi_-|e^{2\theta_{\m} F^-} e^{\gamma_{\m} E^-} e^{2\phi H} e^{ \gamma_{\+} E^+} e^{2 \theta_{\+} F^+}|\psi_+\rangle.
\end{equation}
On the right-hand side, the quantity $\mathcal{Q}$ is evaluated in either the discrete representations or in the Borel-Weil realization \eqref{ssCas}. In both cases, its square is proportional to the identity matrix, leading to \eqref{coupleige}.

The equations \eqref{coupleige} can be decoupled by combining them into the Casimir eigenvalue equation:
\begin{equation}
\hat{\mathcal{C}}\langle \psi_-|g|\psi_+\rangle =  j(j+1/2)\langle \psi_-|g|\psi_+\rangle.
\end{equation}
These results then show that (irreducible) representation matrix elements solve the Casimir eigenvalue equation. The coupled differential system \eqref{coupleige} shows that in a supergroup with a nontrivial sCasimir operator, irrep matrix elements come in pairs related by acting with this supercharge. Recognizing this structure is important when attempting further generalizations such as $q$-deformation, as will be presented elsewhere \cite{Fan:2021bwt}.

\subsection{Harmonic Analysis}
\label{app:hasu}

We now solve the Casimir eigenvalue equation explicitly. To find all Casimir eigenfunctions, following \cite{Hikida:2007sz}, we diagonalize $\partial_{\gamma_{\m}} = i\nu$ and $\partial_{\gamma_{\+}} = i\lambda$ as well as 
\begin{align}
\Xi = \left( \partial_{\theta_{\m}} + i\theta_{\m} \nu \right) \left(\partial_{\theta_ {\+}} + i\theta_{\+} \lambda \right) = \xi,
\end{align}
which has a four-dimensional space of eigenvectors, split into doubly degenerate eigenspaces for each choice of sign of $\xi$. They are spanned by the following eigenstates:
\begin{align}
\label{eigenspac}
\xi &= +\sqrt{\nu\lambda}: \qquad 1-\sqrt{\nu\lambda}\theta_{\m} \theta_{\+},  \quad \sqrt{\nu} \theta_{\m} - i\sqrt{\lambda} \theta_{\+}, \nonumber \\
\xi &= -\sqrt{\nu\lambda}: \qquad 1+\sqrt{\nu\lambda}\theta_{\m} \theta_{\+},  \quad \sqrt{\nu} \theta_{\m} + i\sqrt{\lambda} \theta_{\+},
\end{align}
where we have chosen to describe the eigenspaces with eigenvectors that are either bosonic or fermionic.

This transforms the Casimir operator into
\begin{equation}
\hat{\mathcal{C}} = \frac{1}{4} \partial_\phi^2 + \frac{1}{4} \partial_\phi - e^{-2\phi}\nu\lambda \mp \frac{1}{2} e^{-\phi}\sqrt{\nu\lambda}.
\end{equation}
Setting $f(\phi) = e^{-\phi/2}g(\phi)$, the Casimir equation can be reformulated as a Schr\"odinger equation
\begin{equation}
\left(-\frac{1}{4} \partial_\phi^2 + \left(\nu\lambda e^{-2\phi} \pm \frac{1}{2} \sqrt{\nu\lambda}e^{-\phi} \right)\right) g(\phi) = \frac{k^2}{4} g(\phi)
\end{equation}
describing a nonrelativistic quantum particle of energy $k^2/4$ in a Morse-like potential
\begin{equation}
V(\phi) = \nu\lambda e^{-2\phi} \pm \frac{1}{2} \sqrt{\nu\lambda}e^{-\phi}.
\end{equation}
We draw these potentials in Figure \ref{MorsePanel} below. We can shift $\phi \to \phi +\log 4 \sqrt{\left|\nu\lambda\right|}$ to map the potential into a canonical form:
\begin{equation}
\label{morse}
V(\phi) = \frac{1}{16}\left(\operatorname{sgn}(\nu\lambda) e^{-2\phi} \pm 2\sqrt{\operatorname{sgn}(\nu\lambda)} e^{-\phi} \right).
\end{equation}
The resulting wavefunctions depend on the relative signs of the various terms. For $\nu\lambda > 0$, we get the delta-normalizable eigenfunctions:
\begin{alignat}{2}
\nu\lambda > 0, \quad \xi &= +\sqrt{\nu\lambda}: \qquad & g(\phi) &= ie^{-\phi/2}\left( K_{\frac{1}{2}+ik}\left(2\sqrt{\nu\lambda}e^{-\phi}\right) - K_{-\frac{1}{2}+ik}\left(2\sqrt{\nu\lambda}e^{-\phi}\right) \right), \nonumber \\
\nu\lambda > 0, \quad \xi &= -\sqrt{\nu\lambda}: \qquad & g(\phi) &= e^{-\phi/2}\left( K_{\frac{1}{2}+ik}\left(2\sqrt{\nu\lambda}e^{-\phi}\right) + K_{-\frac{1}{2}+ik}\left(2\sqrt{\nu\lambda}e^{-\phi}\right) \right). 
\end{alignat}
When $\xi <0$, the Morse potential has a small potential well whose height in the energy variable $k^2/4$ ranges over $\left(-\frac{1}{16},0\right)$. No bound states exist in this small potential well.\footnote{The Morse potential in quantum mechanics was introduced as a model for bound states of diatomic molecules. It is amusing that here, we are in the opposite regime where no bound states exist.}\textsuperscript{,}\footnote{When $\nu\lambda >0$ and $\xi <0$, the most generic solution to the differential equation takes the form:
\begin{align}
&C_1 e^{-\phi/2}\left( K_{\frac{1}{2}+ik}\left(2\sqrt{\nu\lambda}e^{-\phi}\right) + K_{-\frac{1}{2}+ik}\left(2\sqrt{\nu\lambda}e^{-\phi}\right) \right) \nonumber \\
&\phantom{==} + C_2e^{-\phi/2}\left( I_{\frac{1}{2}+ik}\left(2\sqrt{\nu\lambda}e^{-\phi}\right) - I_{-\frac{1}{2}+ik}\left(2\sqrt{\nu\lambda}e^{-\phi}\right) \right),
\end{align}
where we look for solutions that are both damped at $\phi \to \pm \infty$ and have value of $k=0 \to i/2$ in order for the energy variable $k^2/4$ to lie in the range $\left(-\frac{1}{16},0\right)$. Since the BesselI diverges when its argument $\to \infty$ and the BesselK when its argument $\to 0$, both functions have to independently become damped. Using the asymptotics of BesselI and the series expansion of BesselK, we see that this cannot happen unless $C_1=C_2=0$. Hence no bound states exist in the small potential well.

Alternatively, the Morse potential eigenvalue problem
\begin{equation}
(-\partial_x^2 + \lambda^2 ( e^{-2x}-2e^{-x})) \psi(x) = \epsilon_n \psi(x)
\end{equation}
has known bound-state solutions for $n = 0, 1, \ldots, \lfloor\lambda-1/2\rfloor$. From \eqref{morse}, we see that we are in the limiting case $\lambda=1/2$, and one can check that the only $(n=0)$ candidate bound-state wavefunction in this limit is indeed not normalizable.}

For the case $\nu\lambda <0$, the Morse potential becomes complex. Delta-normalizable eigenfunctions can still be constructed and take the form:
\begin{alignat}{2}
\nu\lambda < 0, \quad \xi &= +\sqrt{\nu\lambda}: \qquad & g(\phi) &= e^{-\phi/2}\left( iJ_{\frac{1}{2}+ik}\left(2\sqrt{-\nu\lambda}e^{-\phi}\right) - J_{-\frac{1}{2}+ik}\left(2\sqrt{-\nu\lambda}e^{-\phi}\right) \right), \nonumber \\
\nu\lambda < 0, \quad \xi &= -\sqrt{\nu\lambda}: \qquad & g(\phi) &= e^{-\phi/2}\left( iJ_{\frac{1}{2}+ik}\left(2\sqrt{-\nu\lambda}e^{-\phi}\right) + J_{-\frac{1}{2}+ik}\left(2\sqrt{-\nu\lambda}e^{-\phi}\right) \right). 
\end{alignat}
All of these wavefunctions have Casimir eigenvalue $\frac{1}{16} + \frac{k^2}{4}$ and fall into the principal series (continuous) representations; they have positive energies $k^2/4$ in the Schr\"odinger problem (Figure \ref{MorsePanel}).

\begin{figure}[ht] 
\centering
\includegraphics[width=0.75\textwidth]{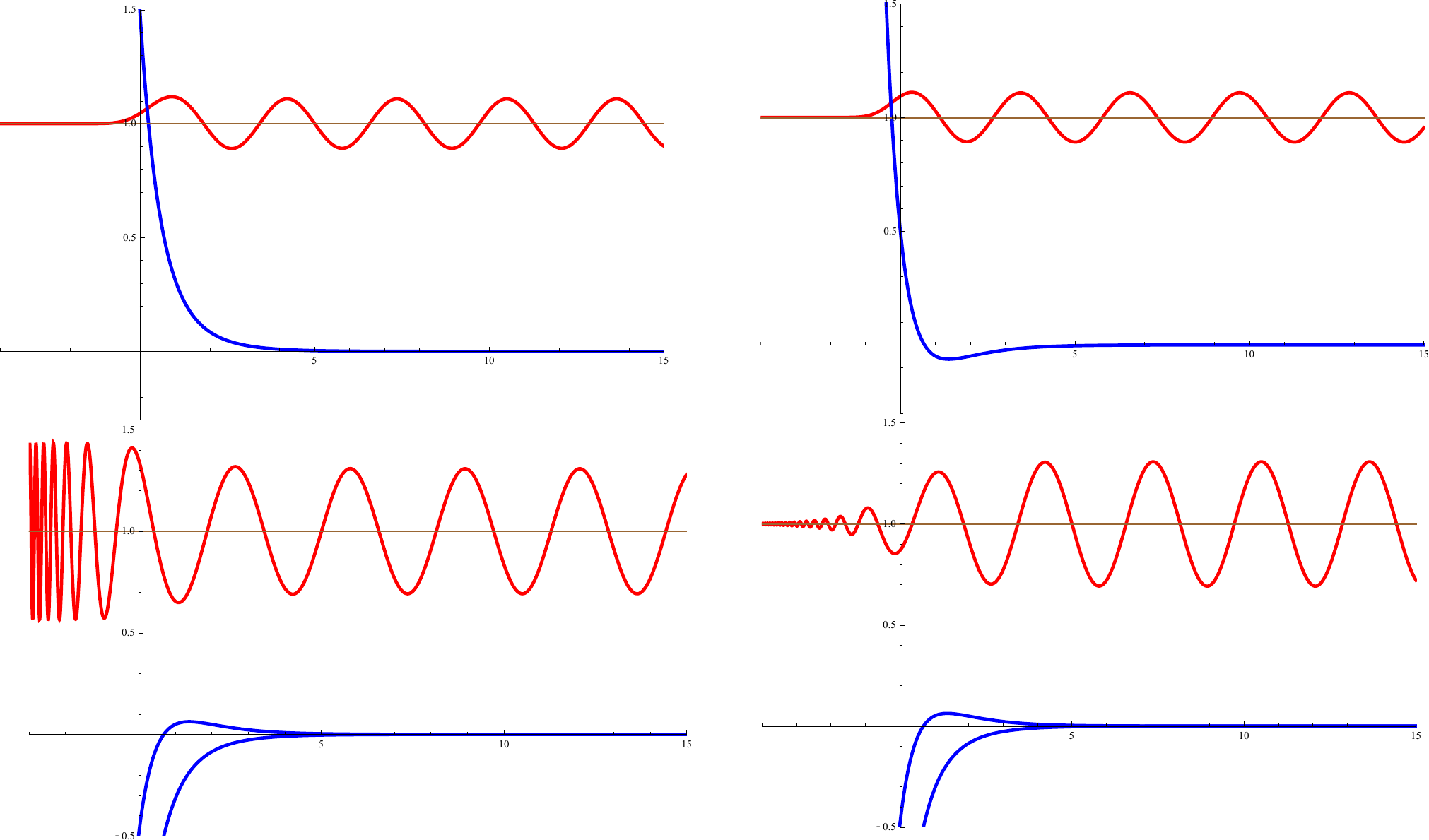}
\caption{Blue: Morse potential $V(\phi)$. Red: wavefunction solution for $k=2$ with energy value $k^2/4=1$ marked by the brown line. Top left: $\nu\lambda >0, \xi >0$. Top right: $\nu\lambda >0, \xi <0$. Bottom left: $\nu \lambda <0, \xi >0$. Bottom right: $\nu \lambda <0, \xi <0$. When $\nu\lambda <0$, the Morse potential is complex-valued. We can get intuition by realizing that in the asymptotic regions $\phi \to \pm \infty$, the potential becomes real. We hence draw the ``envelope'' of the Morse potential instead, obtained by replacing the imaginary coefficient $\sqrt{\nu\lambda}$ by $\pm \sqrt{-\nu\lambda}$. For illustrative purposes, we draw $\mathfrak{Re}(g(\phi))$ ($\mathfrak{Im}(g(\phi))$ is qualitatively similar). Notice the unbounded increase of oscillation frequency of these modes as $\phi \to - \infty$, as we would expect from the profile of the potential.}
\label{MorsePanel}
\end{figure}

\begin{figure}[ht] 
\centering
\includegraphics[width=0.45\textwidth]{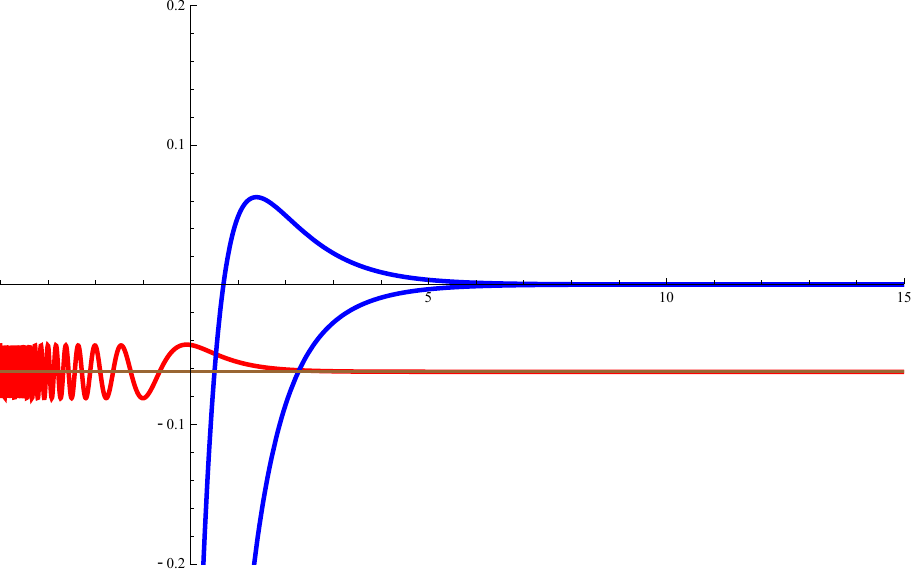}
\caption{Blue: Morse potential. Red: discrete state wavefunction solution $\mathfrak{Re}(g(\phi))$ for $\nu\lambda <0$ and with the energy variable $k^2/4=-1/16$.}
\label{MorseDiscrete}
\end{figure}

For $\nu\lambda <0$, there exist eigenfunctions that are damped for positive $\phi$. These take the form:
\begin{equation}
\label{discrep}
g(\phi) = e^{-\phi/2}\left( iJ_{2j+1}\left(2\sqrt{-\nu\lambda}e^{-\phi}\right) - \text{sgn}(\xi) J_{2j}\left(2\sqrt{-\nu\lambda}e^{-\phi}\right) \right), 
\end{equation}
depending on the sign of $\xi$. These modes form a continuum and correspond to the lowest-weight representation matrices of the universal cover of OSp$(1|2,\mathbb{R})$. They have Casimir eigenvalue $j(j+1/2)$ for $j\in \mathbb{R}^+$, and appear at negative energies $-j(j+1/2)-1/16$ in the associated Schr\"odinger problem (Figure \ref{MorseDiscrete}).

Just as for SL$(2,\mathbb{R})$ discussed in Appendix \ref{app:ha}, the actual discretization happens when demanding single-valuedness as a function of the complex variable $z=e^{-\phi}$, restricting to $2j \in -\mathbb{N}$ and singling out the correct representations of OSp$(1|2,\mathbb{R})$ (instead of its universal cover). It would be interesting to construct these explicitly from the representation theory perspective, but we do not pursue this problem here.

For $\nu\lambda >0$, we can summarize the four linearly independent Casimir eigenfunctions as:
\begin{align}
\left(1-\sqrt{\nu\lambda}\theta_{\m}\theta_{\+}\right)e^{i\nu\gamma_{\m}}e^{i\lambda\gamma_{\+}}e^{-\phi} &\left( K_{\frac{1}{2}+ik}\left(2\sqrt{\nu\lambda}e^{-\phi}\right) - K_{-\frac{1}{2}+ik}\left(2\sqrt{\nu\lambda}e^{-\phi}\right) \right), \nonumber \\
\left(\sqrt{\nu} \theta_{\m} - i\sqrt{\lambda} \theta_{\+}\right)e^{i\nu\gamma_{\m}}e^{i\lambda\gamma_{\+}}e^{-\phi} &\left( K_{\frac{1}{2}+ik}\left(2\sqrt{\nu\lambda}e^{-\phi}\right) - K_{-\frac{1}{2}+ik}\left(2\sqrt{\nu\lambda}e^{-\phi}\right) \right), \nonumber \\
\left(1+\sqrt{\nu\lambda}\theta_{\m}\theta_{\+}\right)e^{i\nu\gamma_{\m}}e^{i\lambda\gamma_{\+}}e^{-\phi} &\left( K_{\frac{1}{2}+ik}\left(2\sqrt{\nu\lambda}e^{-\phi}\right) + K_{-\frac{1}{2}+ik}\left(2\sqrt{\nu\lambda}e^{-\phi}\right) \right), \nonumber \\
\left(\sqrt{\nu} \theta_{\m} + i\sqrt{\lambda} \theta_{\+}\right)e^{i\nu\gamma_{\m}}e^{i\lambda\gamma_{\+}}e^{-\phi} &\left( K_{\frac{1}{2}+ik}\left(2\sqrt{\nu\lambda}e^{-\phi}\right) + K_{-\frac{1}{2}+ik}\left(2\sqrt{\nu\lambda}e^{-\phi}\right) \right).
\end{align}
Adding and subtracting, we can equivalently write the following set of linearly independent eigenfunctions:
\begin{align}
&e^{i\nu\gamma_{\m}}e^{i\lambda\gamma_{\+}}e^{-\phi}\left( K_{\frac{1}{2}+ik}\left(2\sqrt{\nu\lambda}e^{-\phi}\right) + \sqrt{\nu\lambda}\theta_{\m}\theta_{\+} K_{-\frac{1}{2}+ik}\left(2\sqrt{\nu\lambda}e^{-\phi}\right) \right), \nonumber \\
&e^{i\nu\gamma_{\m}}e^{i\lambda\gamma_{\+}}e^{-\phi}\left(  \sqrt{\nu\lambda}\theta_{\m}\theta_{\+} K_{\frac{1}{2}+ik}\left(2\sqrt{\nu\lambda}e^{-\phi}\right) +  K_{-\frac{1}{2}+ik}\left(2\sqrt{\nu\lambda}e^{-\phi}\right) \right), \nonumber \\
&e^{i\nu\gamma_{\m}}e^{i\lambda\gamma_{\+}}e^{-\phi}\left( \sqrt{\nu} \theta_{\m} K_{\frac{1}{2}+ik}\left(2\sqrt{\nu\lambda}e^{-\phi}\right) + i\sqrt{\lambda} \theta_{\+} K_{-\frac{1}{2}+ik}\left(2\sqrt{\nu\lambda}e^{-\phi}\right) \right),  \nonumber \\
&e^{i\nu\gamma_{\m}}e^{i\lambda\gamma_{\+}}e^{-\phi}\left( i\sqrt{\lambda} \theta_{\+} K_{\frac{1}{2}+ik}\left(2\sqrt{\nu\lambda}e^{-\phi}\right) +  \theta_{\m} K_{-\frac{1}{2}+ik}\left(2\sqrt{\nu\lambda}e^{-\phi}\right) \right), \label{Casfull}
\end{align}
which we will reproduce by an explicit calculation of the representation matrix element in Section \ref{app:whit} below.

For the subsemigroup OSp$^+(1|2,\mathbb{R})$, we need to set $\lambda \to i\lambda$ and $\nu \to -i \nu$ with $\nu,\lambda >0$. This leads to the Casimir eigenfunctions:
\begin{align}
\left(1-\sqrt{\nu\lambda}\theta_{\m}\theta_{\+}\right)e^{\nu \gamma_{\m}}e^{-\lambda \gamma_{\+}}e^{-\phi} &\left( K_{\frac{1}{2}+ik}\left(2\sqrt{\nu\lambda}e^{-\phi}\right) - K_{-\frac{1}{2}+ik}\left(2\sqrt{\nu\lambda}e^{-\phi}\right) \right), \nonumber \\*
\left(\sqrt{\nu} \theta_{\m} + \sqrt{\lambda} \theta_{\+}\right)e^{\nu \gamma_{\m}}e^{-\lambda \gamma_{\+}}e^{-\phi} &\left( K_{\frac{1}{2}+ik}\left(2\sqrt{\nu\lambda}e^{-\phi}\right) - K_{-\frac{1}{2}+ik}\left(2\sqrt{\nu\lambda}e^{-\phi}\right) \right), \nonumber\\
\left(1+\sqrt{\nu\lambda}\theta_{\m}\theta_{\+}\right)e^{\nu \gamma_{\m}}e^{-\lambda \gamma_{\+}}e^{-\phi} &\left( K_{\frac{1}{2}+ik}\left(2\sqrt{\nu\lambda}e^{-\phi}\right) + K_{-\frac{1}{2}+ik}\left(2\sqrt{\nu\lambda}e^{-\phi}\right) \right), \nonumber \\
\left(\sqrt{\nu} \theta_{\m} - \sqrt{\lambda} \theta_{\+}\right)e^{\nu \gamma_{\m}}e^{-\lambda \gamma_{\+}}e^{-\phi} &\left( K_{\frac{1}{2}+ik}\left(2\sqrt{\nu\lambda}e^{-\phi}\right) + K_{-\frac{1}{2}+ik}\left(2\sqrt{\nu\lambda}e^{-\phi}\right) \right). \label{Cassubsemi}
\end{align}
Within this subsector, we are only probing the top two potentials in Figure \ref{MorsePanel}. In particular, the discrete representations do not appear in the class of Casimir eigenfunctions relevant for the subsemigroup. This is one way of arguing that only the principal series representations appear in the conjectured Plancherel decomposition of OSp$^+(1|2,\mathbb{R})$ in \eqref{PlOsp}.

\subsection{Characters}
\label{app:char}

In this section, we compute the characters for all of the irreps of OSp$(1|2,\mathbb{R})$ discussed up to this point. Our main interest is the principal series representations, since these characters are used in particular to glue surfaces together in super-Teichm\"uller space. For completeness and consistency, we also discuss the characters for the discrete highest-weight and the finite-dimensional representations.

\subsubsection{Principal Series Character}

We can compute the character in the principal series representations by writing the Borel-Weil realization of the algebra in terms of a kernel $K(x,\vartheta| y, \vartheta')$ as
\begin{equation}
f(x,\vartheta) = \int dy\, d\vartheta'\, K(x,\vartheta| y, \vartheta') f(y, \vartheta'),
\end{equation}
where\footnote{A fermionic delta function works in much the same way as a bosonic one. In particular, we have
\begin{equation}
\int d\vartheta\, \delta(\vartheta - f(\alpha)) F(\vartheta) = F(f(\alpha))
\end{equation}
for any fermionic quantity $f(\alpha)$. The proof proceeds by writing $F(\vartheta) = F_0 + \vartheta F_1$, using $\delta (\vartheta -f(\alpha)) = \vartheta - f(\alpha)$, and taking care of the minus sign obtained by pulling $f(\alpha)$ through the measure.}
\begin{equation}
\label{repco}
K(x,\vartheta| y, \vartheta') = \scalebox{0.98}{$\displaystyle \frac{\left|bx+d+\delta\vartheta\right|^{2j}}{\operatorname{sgn}(e)^{1/2}\operatorname{sgn}(bx+d+\delta\vartheta)^{1/2}} \delta\left(\frac{ax+c +\beta \vartheta}{bx+d +\delta \vartheta} - y\right)\delta\left(\frac{-\alpha x - \gamma + e \vartheta}{b x + d + \delta\vartheta } - \vartheta'\right)$}.
\end{equation}
This corresponds to working in a coordinate basis on the carrier space $L^2(\mathbb{R}^{1|1})$:
\begin{align}
f_{x_1,\vartheta_1}(x,\vartheta) \equiv \langle x ,\vartheta|x_1,\vartheta_1\rangle = \delta(x-x_1)\delta(\vartheta-\vartheta_1),
\end{align}
with orthonormality and completeness relations:
\begin{align}
\int dx\, d\vartheta\, f_{x_1,\vartheta_1}(x,\vartheta)^* f_{x_2,\vartheta_2}(x,\vartheta) &= \delta(x_1 - x_2)\delta(\vartheta_1-\vartheta_2), \\
\int dx_1\, d\vartheta_1\, f_{x_1,\vartheta_1}(x,\vartheta)^* f_{x_1,\vartheta_1}(y,\vartheta') &= \delta(x - y) \delta(\vartheta-\vartheta').
\end{align}
In the coordinate basis, the representation matrices have the simple form of \eqref{repco}:
\begin{equation}
R^k_{x|\vartheta,y|\vartheta'}(g) = K(x,\vartheta| y, \vartheta').
\end{equation}
The character $\chi_j(g)$ in representation $j$ is then computed as
\begin{align}
\label{chardef}
\chi_j(g) &\equiv \int dx\, d\vartheta\, K(x,\vartheta| x, \vartheta) \\
&= \scalebox{0.98}{$\displaystyle \int dx\, d\vartheta\, \frac{\left|bx+d+\delta\vartheta\right|^{2j}}{{\operatorname{sgn}(e)^{1/2}\operatorname{sgn}(bx+d+\delta\vartheta)^{1/2}}} \delta\left(\frac{ax+c +\beta \vartheta}{bx+d +\delta \vartheta}-x\right)\delta\left(\frac{-\alpha x - \gamma + e \vartheta}{b x + d + \delta\vartheta} - \vartheta \right)$}. \nonumber
\end{align}
The superspace integral evaluates to all of the fixed points of the supergroup action on the superline.

To simplify the calculation, we use the fact that the character is a class function. For a group element of hyperbolic conjugacy class, we can hence consider
\begin{equation}
g = \left[\begin{array}{cc|c} 
e^{\phi} & 0 & 0 \\
\epsilon & e^{-\phi} & 0 \\
\hline
0 & 0 & \pm 1
\end{array} \right],
\end{equation}
without loss of generality.  We focus first on the R sector ($+$). The parameter $\epsilon$ serves as a regulator since the number of fixed points jumps at $\epsilon=0$. We assume $\phi >0$ in the following. For this particular group element, we get:
\begin{equation}
\label{charint}
\chi_j(g) = \int dx\, d\vartheta\, |\epsilon x+e^{-\phi}|^{2j} \delta\left(\frac{e^{\phi}x}{\epsilon x + e^{-\phi}}-x\right)\delta\left(\frac{\vartheta}{\epsilon x + e^{-\phi}} - \vartheta \right).
\end{equation}
The bosonic delta function evaluates to
\begin{equation}
\delta\left(\frac{e^{\phi}x}{\epsilon x + e^{-\phi}}-x\right) = \frac{\delta(x)}{e^{2\phi}-1} + \frac{\delta(x-\frac{e^{\phi}- e^{-\phi}}{\epsilon})}{1-e^{-2\phi}},
\end{equation}
giving two fixed points at $x=0$ and $x=+\infty$, respectively. The fermionic delta function gives:
\begin{equation}
\delta\left(\frac{\vartheta}{\epsilon x + e^{-\phi}} - \vartheta \right) = \frac{1-\epsilon x -e^{-\phi}}{\epsilon x + e^{-\phi}} \vartheta.
\end{equation}
Doing the integrals in \eqref{charint} then gives:
\begin{align}
\label{chir}
\chi_j^{\R}(g) =  \frac{(e^{-\phi})^{2j}(e^{\phi}-1)}{e^{2\phi}-1} + \frac{(e^{\phi})^{2j}(e^{-\phi}-1)}{1- e^{-2\phi}}
= i\frac{\sin (k \phi)}{ \cosh (\phi /2)},
\end{align}
where we used $j=-1/4 + ik/2$ in the last line. An analogous computation for the NS sector gives instead:
\begin{equation}
\label{chins}
\chi_j^{\NS}(g) = i\frac{\cos (k \phi)}{\sinh (\phi/2)}.
\end{equation}
Notice that this character can be decomposed into SL$(2,\mathbb{R})$ continuous irrep characters as:
\begin{equation}
\frac{\cosh ((4j+1) \phi/2)}{\sinh (\phi/2)} = \frac{\cosh ((2j+1) \phi)}{\sinh \phi} + \frac{\cosh (2j \phi)}{\sinh \phi}.
\end{equation}
However, this cannot be interpreted as a sum of \emph{unitary} SL$(2,\mathbb{R})$ principal series characters since the left-hand side requires $j=-1/4 + ik/2$, whereas the first and second terms on the right-hand side require $\mathfrak{Re}(j)=-1/2$ and $\mathfrak{Re}(j)=0$, respectively. This is an example of the statement made earlier that the principal series representations cannot be obtained as a direct sum of those of the SL$(2,\mathbb{R})$ subgroup \cite{Backhouse:1982}.

For a group element in the elliptic conjugacy class, $g$ is equivalent to an element:
\begin{equation}
g = \left[\begin{array}{cc|c} 
\cos\theta & \sin\theta & 0 \\
-\sin\theta & \cos\theta & 0 \\
\hline
0 & 0 & \pm 1
\end{array} \right].
\end{equation}
Then for the bosonic delta function in the definition \eqref{chardef}, the evaluation boils down to the SL$(2,\mathbb{R})$ calculation. The resulting character vanishes since there are no fixed points on the real line $\mathbb{R}$ of an elliptic SL$(2,\mathbb{R})$ element. However, to make contact with elliptic defects in JT (super)gravity, one needs a formal analytic continuation of the hyperbolic defects, by letting $\phi \to i\phi$. We comment on this in the main text. Finally, the parabolic conjugacy class is of lower dimensionality and will not be important for the coming discussions.

The characters should satisfy an orthonormality relation:
\begin{equation}
\int d\mu(t)\, \chi_j(t) \chi_{j'}(t)^* = \delta(j-j')
\end{equation}
for some measure $d\mu(t) = d\mu(\phi)$ on the space of hyperbolic conjugacy class elements, and where we can restrict to the hyperbolic conjugacy class elements since the elliptic characters vanish and the parabolic characters are of measure zero. Since the supergroup at hand falls apart into two connected components, this equality boils down to two explicit relations
\begin{align}
\label{charotho}
\int_{-\infty}^{+\infty} d\mu(\phi)\, \chi_j^{\NS}(\phi)\chi_{j'}^{\NS}(\phi)^* = \delta(j-j'), \\
\int_{-\infty}^{+\infty} d\mu(\phi)\, \chi_j^{\R}(\phi)\chi_{j'}^{\R}(\phi)^* = \delta(j-j'),
\end{align}
where we insert the characters \eqref{chins} and \eqref{chir}. In order to prove this relation, we will need the correct measure on the space of conjugacy class elements $d\mu(\phi) $, which follows from the superanalogue of the Weyl integration formula. We turn to this next, and come back to the relation \eqref{charotho} and its interpretation further on.

\subsubsection{Interlude: Weyl Integration Formula for Compact Supergroups}

We first review the proof of the Weyl integration formula in a physicist's fashion, and then generalize it to compact Lie supergroups.

Recall that every group element $g$ is conjugate to an element in a maximal torus (conjugacy theorem): 
\begin{equation}
\label{decos}
g = c t c^{-1}, \quad c \in G/T, \quad t \in T,
\end{equation}
where one considers the left coset $gT$. This writing is not unique, with the ambiguity being parametrized by the Weyl group:
\begin{equation}
W(T) = \frac{N(T)}{T} = \left\{x\in G/T \,|\, x t x^{-1} \in T, \forall t \in T\right\}.
\end{equation}
To elaborate, suppose we have two ways of writing $g = c_1 t_1 c_1^{-1} = c_2 t_2 c_2^{-1}$. Then $t_1 = c_1^{-1} c_2 t_2 c_2^{-1} c_1$. Set $w \equiv c_2^{-1} c_1$. If $w \in T$, then $c_1 \sim c_2$ and the two ways of writing $g$ are equivalent in the decomposition of $G$ into $G/T \times T$. If $w \in G/T$, then the above equality implies $w \in N(T)/T = W(T)$, the Weyl group. This is hence the only ambiguity in the decomposition \eqref{decos}.

Since $W(T)$ is a finite group, this just implies that one has a $|W|$-fold covering of the group $G$:
\begin{equation}
\int_G dg\, f(g) = \frac{1}{|W|}\int_{G/T \times T} d(ctc^{-1})\, f(ctc^{-1}).
\end{equation}
Next, we need to perform the change of variables \eqref{decos} explicitly and track the Jacobian in the transformation. The Maurer-Cartan one-form can be written out explicitly as:
\begin{align}
g^{-1}dg &= c t^{-1} c^{-1}dc \,t c^{-1} - dc \, c^{-1} + c t^{-1} dt \,c^{-1} \\
&= \operatorname{Ad}(c)\left[\operatorname{Ad}(t^{-1})-1\right]c^{-1} dc + \operatorname{Ad}(c) t^{-1} dt.
\end{align}
The metric is written as
\begin{equation}
ds^2 = \Tr(g^{-1} dg \otimes g^{-1} dg).
\end{equation}
Writing $g^{-1}dg = \sum_{i,j}J_{ij}X^i\, dx^j$, we can rewrite it as:
\begin{equation}
ds^2 = \Tr(g^{-1} dg \otimes g^{-1} dg) = J_{ij}J^{i}{}_k\, dx^j dx^k,
\end{equation}
where indices are raised with the Cartan-Killing metric $h^{i\ell} = \Tr(X^i X^\ell)$. This immediately leads to the Haar measure $\det J \bigwedge_i dx^i$.

In our case, since we have $t^{-1}dt = \sum_{i, j} t_{ij}T^i\, dx^j$ and $c^{-1} dc = \sum_{i, j} c_{ij} E^i\, dy^j$ and the Cartan generators are orthogonal to the other generators (with respect to the Cartan-Killing metric), the metric is block diagonal. We can extract the Jacobian from this transformation explicitly as:
\begin{equation}
J = \det\operatorname{Ad}(c)_\mathfrak{g} \det( \operatorname{Ad}(t^{-1})-\mathbf{1})_{\mathfrak{g}/\mathfrak{t}},
\end{equation}
where
\begin{equation}
\det \operatorname{Ad}(c)_\mathfrak{g} = 1
\end{equation}
for a unimodular group. The remaining determinant is evaluated explicitly as:
\begin{equation}
\det( \operatorname{Ad}(t^{-1})-\mathbf{1})_{\mathfrak{g}/\mathfrak{t}} = \prod_{\alpha} (e^{\alpha(t)}-1) = \prod_{\alpha > 0}-4 \left|\sinh \frac{\alpha(t)}{2}\right|^2 = (-)^{\#\text{roots}/2}\left|\Delta(t)\right|^2,
\end{equation}
by exponentiating the algebra to get $\operatorname{Ad}(t^{-1}) X^\alpha = t^{-1} X^\alpha t = e^{-\alpha(t)} X^\alpha$ and by the symmetry $\alpha \leftrightarrow - \alpha$ of the root space. We end up with the Weyl integration formula:
\begin{equation}
\int_G dg\, f(g) = \frac{1}{|W|}\int_T dt \left|\Delta(t)\right|^2 \left( \int_{G/T} dc\, f(ctc^{-1}) \right).
\end{equation}

For a Lie supergroup, the only difference is that some of the coordinates are Grassmann numbers, and hence the Jacobian in the above coordinate transformation gets replaced by a super-Jacobian, which is a Berezinian with an absolute value sign included for the bosonic subtransformation:
\begin{equation}
\Delta_{\R}(t) \equiv \operatorname{sdet}'( \operatorname{Ad}(t^{-1})-\mathbf{1})_{\mathfrak{g}/\mathfrak{t}} = \frac{\prod_{\alpha \in \Delta_B} |e^{\alpha(t)}-1|}{\prod_{\alpha \in \Delta_F} (e^{\alpha(t)}-1)} = \frac{\prod_{\alpha \in \Delta_B^+} 4 \sinh^2 \alpha_B(t)}{\prod_{\alpha \in \Delta_F^+} -4 \sinh^2 \alpha_F(t)},
\end{equation}
where we used the common notation $\Delta_B$ and $\Delta_F$ for the bosonic and fermionic roots, respectively, and the $+$ superscript indicates a restriction to the positive roots of said statistics.

\begin{figure}[!htb]
\centering
\includegraphics[width=0.4\textwidth]{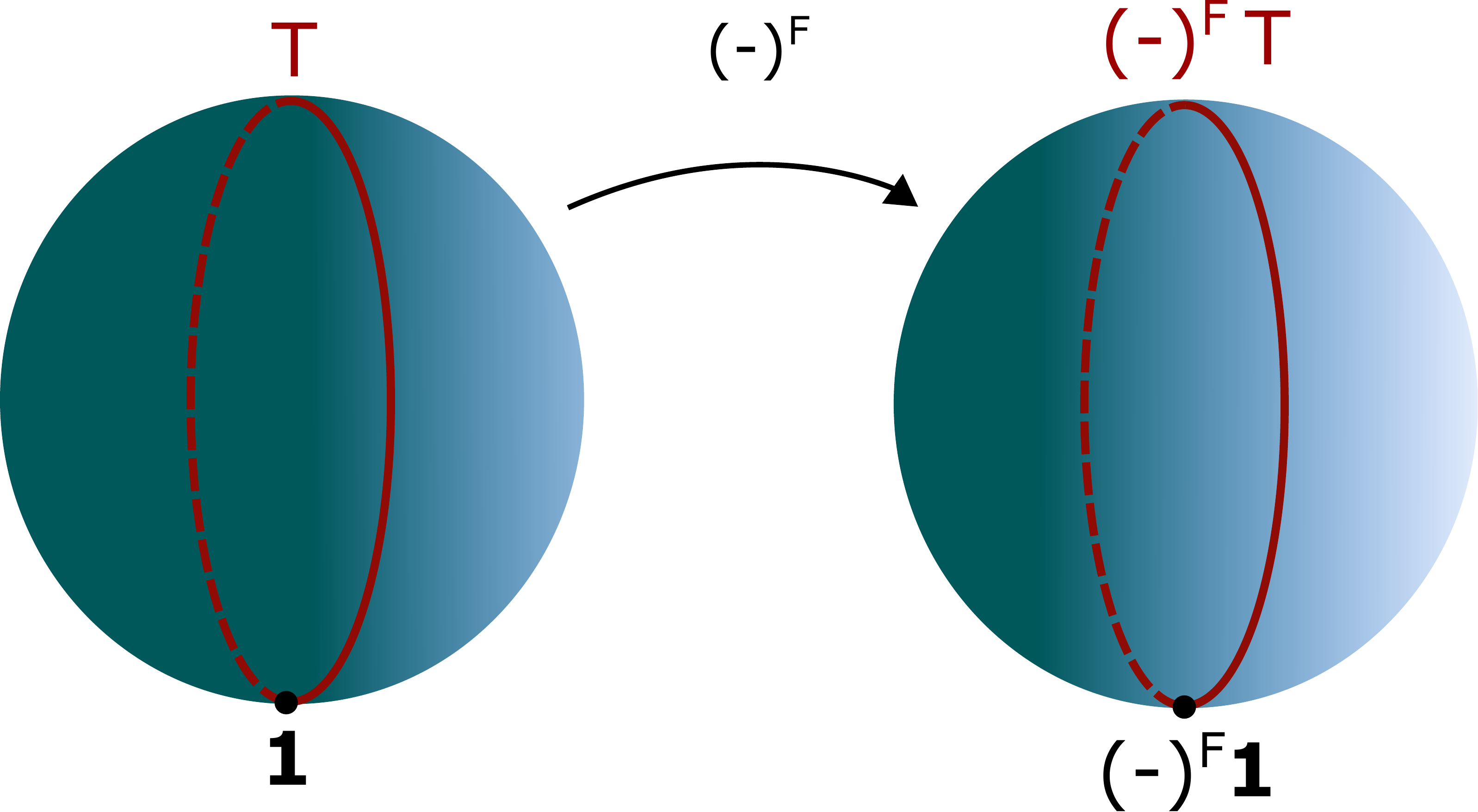}
\caption{Schematic of the topology of the compact supergroup discussed here, with two connected components related by acting with $(-)^F$. The full maximal torus $= \left\{T,(-)^F T\right\}$.}
\label{topospgroup}
\end{figure}

In the case that the Lie supergroup is disconnected, one has to perform a calculation as above for each connected component. For the specific case where there exists a sCasimir operator that distinguishes the connected components from one another (which is the case for, e.g., OSp$(1|2n)$), we can work out an explicit formula. For the component of the Lie supergroup connected to the element $(-)^F$ (satisfying $\operatorname{sdet} (-)^F = -1$ and $((-)^F)^2 = \mathbf{1}$), we instead parametrize:
\begin{equation}
g = c (-)^F t c^{-1},
\end{equation}
in terms of which the preceding argument goes through identically with $t \to (-)^F t$ and $t^{-1} (-)^F d(-)^F t = t^{-1} dt$. The only difference appears in the end:\footnote{We use the fact that there are an even number of fermionic roots.}
\begin{equation}
\Delta_{\NS}(t) \equiv \operatorname{sdet}'( \operatorname{Ad}(t^{-1}(-)^F{})-\mathbf{1})_{\mathfrak{g}/\mathfrak{t}} = \frac{\prod_{\alpha \in \Delta_B} |e^{\alpha(t)}-1|}{\prod_{\alpha \in \Delta_F} (e^{\alpha(t)}+1)} = \frac{\prod_{\alpha \in \Delta_B^+} 4 \sinh^2 \alpha_B(t)}{\prod_{\alpha \in \Delta_F^+} 4 \cosh^2 \alpha_F(t)},
\end{equation}
because $\operatorname{Ad}(t^{-1}(-)^F) X^\alpha = t^{-1}(-)^F X^\alpha (-)^F t = (-)^{\epsilon(\alpha) }t^{-1} X^\alpha t = (-)^{\epsilon(\alpha) } e^{-\alpha(t)} X^\alpha$, where $\epsilon(\alpha)$ denotes the $\mathbb{Z}_2$ grading of the root $\alpha$. This is because $(-)^F$ commutes with all bosonic generators but anticommutes with all fermionic ones. We end up with the supergroup Weyl integration formula:
\begin{equation}
\boxed{
\int_G dg\, f(g) = \frac{1}{|W_{\R}|}\int_T dt \int_{G/T} dc\, \Delta_{\R}(t) f(ctc^{-1}) + \frac{1}{|W_{\NS}|}\int_T dt \int_{G/T} dc\, \Delta_{\NS}(t) f(c(-)^Ftc^{-1}). }
\end{equation}
A comment on the Weyl groups is in order. Assuming the Cartan subalgebra is the same as the one from the bosonic subalgebra (which happens for, e.g., a basic superalgebra \cite{Frappat:1996pb}), the element $(-)^F$ commutes with the component $T$ of the maximal torus connected to the identity. This means the full maximal torus contains two connected components that are related by multiplying by $(-)^F$. The above Weyl group is then computed with respect to this full maximal torus. In this case, one has $W_{\R} = W_{\NS}$, but this is not necessarily true in the more generic case.  See Figure \ref{topospgroup}.

\subsubsection{Principal Series Character Revisited}

For a noncompact supergroup, the decomposition \eqref{decos} is further complicated by the fact that not every element is conjugate to an element within the same maximal torus. This corresponds to the different types of conjugacy classes: elliptic, parabolic, and hyperbolic for OSp$(1|2,\mathbb{R})$. This situation basically entails a further summation over these three types of classes to allow for a decomposition of any group element:
\begin{align}
\int_G dg\, f(g) &= \sum_{\text{type $i$}}\bigg[\frac{1}{|W_{i,\R}|}\int_{T_i} dt_i \int_{G/T_i} dc\, \Delta_{\R}(t) f(ct_ic^{-1}) \nonumber \\
&\hspace{2 cm} + \frac{1}{|W_{i,\NS}|}\int_{T_i} dt_i \int_{G/T_i} dc\, \Delta_{\NS}(t) f(c(-)^Ft_ic^{-1})\bigg].
\end{align}
This corresponds to the supergroup analogue of the Harish-Chandra formula for reductive (possibly noncompact) Lie groups. While we have no proof for the general case, for OSp$(1|2,\mathbb{R})$, we have explicit knowledge of the conjugacy classes and the above formula manifestly holds for the three types elliptic, parabolic, and hyperbolic.

For OSp$(1|2,\mathbb{R})$, the characters are zero for elliptic holonomy. Since the parabolic conjugacy class is of measure zero, we focus on the hyperbolic class. In this case, we have two bosonic roots and two fermionic roots for which:
\begin{equation}
e^{\alpha_B(t)} = e^{\pm 2 \phi}, \quad e^{\alpha_F(t)} = e^{\pm \phi},
\end{equation}
which is found by exponentiating the algebra relations:
\begin{equation}
[H,E^\pm] = \pm E^\pm, \quad [H,F^\pm] = \pm \frac{1}{2} F^\pm. 
\end{equation}
Hence we find:
\begin{align}
\Delta_{\NS}(t) &= \frac{(e^{2\phi}-1)(1-e^{-2\phi})}{(e^{\phi}+1)(e^{-\phi}+1)} = 4\sinh^2 (\phi/2), \\
\Delta_{\R}(t) &= \frac{(e^{2\phi}-1)(e^{-2\phi}-1)}{(e^{\phi}-1)(e^{-\phi}-1)} = 4\cosh^2 (\phi/2).
\end{align}
Since the Weyl supergroup for a fixed component (R or NS) has two elements in this case (just as for SL$(2,\mathbb{R})$),\footnote{The normalizer is given by:
\begin{equation}
N(T \cup (-)^F T) = \left\{\left[\begin{array}{cc|c} 
e^{\phi} & 0 & 0 \\
0 & e^{-\phi} & 0 \\
\hline
0 & 0 & \pm \epsilon
\end{array} \right], \quad \left[\begin{array}{cc|c} 
0 & -e^{\phi} & 0 \\
e^{-\phi} & 0 & 0 \\
\hline
0 & 0 & \pm \epsilon
\end{array} \right], \quad \phi \in \mathbb{R}, \quad \epsilon = \pm 1\right\}.
\end{equation}
Modding out by the maximal torus, we get the Weyl group:
\begin{equation}
W(T \cup (-)^F T) = \left\{\left[\begin{array}{cc|c} 
1 & 0 & 0 \\
0 & 1 & 0 \\
\hline
0 & 0 & 1
\end{array} \right], \quad \left[\begin{array}{cc|c} 
0 & -1 & 0 \\
1 & 0 & 0 \\
\hline
0 & 0 & 1
\end{array} \right]\right\}.
\end{equation}} we can indeed verify the orthonormality relations:
\begin{align}
\label{charotho2}
\frac{1}{|W_{\NS}|}\int d\phi\, 4\sinh^2 (\phi/2) \chi_j^{\NS}(\phi)\chi_{j'}^{\NS}(\phi)^* &= 2\int d \phi \cos(k\phi)\cos(k'\phi) = 2\pi \delta(k-k'), \\
\frac{1}{|W_{\R}|}\int d\phi\, 4\cosh^2 (\phi/2) \chi_j^{\R}(\phi)\chi_{j'}^{\R}(\phi)^* &= 2\int d \phi \sin(k\phi)\sin(k'\phi) = 2\pi \delta(k-k').
\end{align}
Since $\phi$ is interpretable as the geodesic length, this means that a complete set of states for cutting open a surface is obtained by summing over all geodesic lengths. This calculation is performed in the main text in Section \ref{s:defglue}.

\subsubsection{Discrete Representations and Relations}

The characters of the finite representations can be calculated easily and are given by:\footnote{We use the same notation for these characters as in the book \cite{VK}.}
\begin{align}
\chi_{j}^{0\NS}(g) &= \operatorname{STr} e^{2 \phi H} = \sum_{n=-2j}^{+2j} e^{n\phi} = \frac{\sinh (4j+1) \frac{\phi}{2}}{\sinh \frac{\phi}{2}} = \frac{\sinh (2j+1) \phi}{\sinh \phi} + \frac{\sinh 2j \phi}{\sinh \phi}, \\
\chi_{j}^{0\R}(g) &= \operatorname{STr}[(-)^F e^{2 \phi H}] = \sum_{n=-2j}^{+2j} (-)^{n-2j} e^{n\phi} = \frac{\cosh (4j+1) \frac{\phi}{2}}{\cosh \frac{\phi}{2}} =  \frac{\sinh (2j+1) \phi}{\sinh \phi} - \frac{\sinh 2j \phi}{\sinh \phi}, \nonumber
\end{align}
for hyperbolic holonomy parametrized by $\phi$. In the last equalities, we made explicit the decomposition $R_j^{\text{OSp}} = R_j^{\text{SL}} \oplus R_{j-1/2}^{\text{SL}}$ of the representation in terms of SL$(2,\mathbb{R})$ finite representations. Note that $\chi_{j}^{0\NS}(\mathbf{1}) = \dim R$ and that $\chi_{j}^{0\R}(\mathbf{1}) = \#B - \#F$ is the Witten index counting the number of bosonic states minus the number of fermionic states.

In the highest-weight discrete representations, the characters are given by:
\begin{align}
\chi_{j}^{+\NS}(g) &= \operatorname{STr} e^{2 \phi H} = \sum_{n=-\infty}^{-2j} e^{n\phi} = \frac{e^{-(4j-1) \phi/2}}{2\sinh \frac{\phi}{2}}, \\*
\chi_{j}^{+\R}(g) &= \operatorname{STr}[(-)^F e^{2 \phi H}] = \sum_{n=-\infty}^{-2j} (-)^{n-2j} e^{n\phi} = \frac{e^{-(4j-1) \phi/2}}{2\cosh \frac{\phi}{2}}.
\end{align}
If we analytically continue the principal series representation characters \eqref{chins} and \eqref{chir} to $j \in \mathbb{N}/2$, we get the equalities:
\begin{align}
\chi_{j}^{\NS}(g) &= \chi_{j+1/2}^{+\NS}(g) + \chi_{j+1/2}^{-\NS}(g) + \chi_{j}^{0\NS}(g), \\
\chi_{j}^{\R}(g) &= -\chi_{j+1/2}^{+\R}(g) - \chi_{j+1/2}^{-\R}(g) + \chi_{j}^{0\R}(g),
\end{align}
corresponding to decomposing the representation into its irreducible representation content.\footnote{Notice the signs appearing for the R sector. This can happen when rearranging the supertrace of a larger matrix into supertraces of its blocks.} Adding up these equations and dividing by two, we get the well-known SL$(2,\mathbb{R})$ relation for the character \cite{VK}:
\begin{align}
\chi_{j}^{\text{SL}}(g) &= \chi_{j+1}^{+\text{SL}}(g) + \chi_{j+1}^{-\text{SL}}(g) + \chi_{j}^{0\text{SL}}(g),
\end{align}
corresponding to the insertion of the projection operator $\frac{1}{2}(\mathbf{1} + (-)^F)$ onto bosonic states. This corresponds to the decomposition of the Borel-Weil realization at $j \in \mathbb{N}/2$ as $R_j^{\text{OSp}} = R_j^{\text{SL}} \oplus R_{j-1/2}^{\text{SL}}$, and the diagonalization of $(-)^F$.

\subsection{Whittaker Function and Plancherel Measure}
\label{app:whit}

In this section, we explicitly compute the mixed parabolic representation matrix element. This serves both as an example, and to introduce a relatively convenient basis in which to compute the Plancherel measure on the principal series representations of $\text{OSp}(1|2, \mathbb{R})$.

The Plancherel measure for SL$(2,\mathbb{R})$ can be found by computing the orthogonality relation of the representation matrix elements:
\begin{equation}
\int dg \, R^{k}_{\nu\lambda}(g)^*R^{k'}_{\nu'\lambda'}(g) = \frac{\delta(k-k')\delta(\nu-\nu')\delta(\lambda-\lambda')}{\rho(k)},
\end{equation}
where $\rho(k) = k \tanh \pi k$. This calculation was performed explicitly in the mixed parabolic basis in \cite{Blommaert:2018iqz}.

This result is basis-independent. For instance, in the coordinate basis on $\mathbb{R}$, the representation matrices satisfy a similar orthogonality relation:
\begin{equation}
\int dg \, R^{k}_{xy}(g)^*R^{k'}_{x'y'}(g) = \frac{\delta(k-k')\delta(x-x')\delta(y-y')}{\rho(k)}.
\end{equation}
Notice that the basis does not need to diagonalize any of the generators in order to make use of it and derive this orthogonality relation.

For $\text{OSp}(1|2, \mathbb{R})$, the representation is defined on the carrier space $L^2(\mathbb{R}^{1|1})$, and hence the coordinate representation calculation of the Plancherel measure would yield:
\begin{equation}
\int dg \, R^{k}_{x|\theta,y|\zeta}(g)^*R^{k'}_{x'|\theta',y'|\zeta'}(g) = \frac{\delta(k-k')\delta(x-x')\delta(\theta-\theta')\delta(y-y')\delta(\zeta-\zeta')}{\rho(k)},
\end{equation}
where the ``indices'' of the representation matrices are $1|1$ coordinates $x|\theta$, $y|\zeta$, etc. We have already used this coordinate representation when computing the characters in Appendix \ref{app:char}. Here, we will write down the representation matrices explicitly within a super-Fourier basis on $\mathbb{R}^{1|1}$ introduced in \ref{app:bases}, and perform the above calculation to determine $\rho(k)$.

\subsubsection{Mixed Parabolic Matrix Element}

The mixed parabolic mode eigenfunctions are given by the coordinate space expressions
\begin{align}
\langle x, \vartheta|\lambda_+, \alpha\rangle &= \frac{1}{\sqrt{2\pi}}\left(1 + \frac{\alpha\vartheta}{x}\right)\frac{|x|^{ik}}{\sqrt{x}}e^{i\lambda/x}, \label{mixedparab1} \\
\langle x, \vartheta|\nu_-, \beta\rangle &= \frac{1}{\sqrt{2\pi}}(1 - \beta\vartheta)e^{i\nu x}. \label{mixedparab2}
\end{align}
We parametrize the group element in Gauss-Euler form:
\begin{equation}
g = e^{2\theta_{\m} F^-} e^{\gamma_{\m} E^-} e^{2\phi H} e^{\gamma_{\+} E^+} e^{2\theta_{\+}F^+}.
\end{equation}
Writing $j = -1/4 + ik/2$, we have the group actions
\begin{align}
(e^{2\phi H}\circ f)(x, \vartheta) &= e^{(1/2 - ik)\phi}f(e^{2\phi}x, e^\phi\vartheta), \\
(e^{\gamma_{\+}E^+}\circ f)(x, \vartheta) &= \textstyle (\gamma_{\+}x + 1)^{-1/2}|\gamma_{\+}x + 1|^{ik}f\left(\frac{x}{\gamma_{\+}x + 1}, \frac{\vartheta}{\gamma_{\+}x + 1}\right), \\
(e^{2\theta_{\+}F^+}\circ f)(x, \vartheta) &= \textstyle |1 + \theta_{\+}\vartheta|^{-1/2 + ik}f\left(\frac{x}{1 + \theta_{\+}\vartheta}, \frac{\vartheta - \theta_{\+}x}{1 + \theta_{\+}\vartheta}\right),
\end{align}
where we have used $1 + \theta_{\+}\vartheta = |1 + \theta_{\+}\vartheta|$.  The action of the non-Cartan group elements on the states \eqref{mixedparab1} and \eqref{mixedparab2} gives:\footnote{The latter expression can be computed, e.g., as the conjugate of $\langle x, \vartheta|e^{-\gamma_{\m}E^-}e^{2\theta_{\m}(F^-)^\dag}|\nu_-, \beta\rangle$.  Crucially, when acting on functions whose top components are fermionic such as \eqref{mixedparab1} and \eqref{mixedparab2}, we have $(F^-)^\dag = -\frac{1}{2}(\partial_\vartheta + \vartheta\partial_x)$ (in contrast to the situation in Section \ref{sect:whittaker}).  This follows from the fact that on such functions,
\begin{equation}
\int dx\, d\vartheta\, f(x, \vartheta)^\ast\partial_\vartheta g(x, \vartheta) = -\int dx\, d\vartheta\, (\partial_\vartheta f(x, \vartheta))^\ast g(x, \vartheta).
\end{equation}
On the other hand, the bottom components $f(x, 0)$ are still assumed to be bosonic, so that $\vartheta$ commutes with any $f(x, \vartheta)$.}
\begin{align}
\langle x, \vartheta|e^{\gamma_{\+}E^+}e^{2\theta_{\+}F^+}|\lambda_+, \alpha\rangle &= \frac{1}{\sqrt{2\pi}}e^{i\lambda\gamma_{\+}}\left(1 + \theta_{\+}\alpha + \frac{(\alpha + i\lambda\theta_{\+})\vartheta}{x}\right)\frac{|x|^{ik}}{\sqrt{x}}e^{i\lambda/x}, \\
\langle\nu_-, \beta|e^{2\theta_{\m} F^-}e^{\gamma_{\m} E^-}|x, \vartheta\rangle &= \frac{1}{\sqrt{2\pi}}e^{i\nu\gamma_{\m}}(1 + \theta_{\m}\beta + (\beta + i\nu\theta_{\m})\vartheta)e^{-i\nu x}.
\end{align}
Inserting a resolution of the identity $\int dx\, d\vartheta\, |x, \vartheta\rangle\langle x, \vartheta|$, we therefore obtain the representation matrix element:
\begin{align}
&\langle\nu_-, \beta|g|\lambda_+, \alpha\rangle \label{repmatr} \\*
&= \frac{1}{2\pi}e^{i\nu\gamma_{\m}}e^{i\lambda\gamma_{\+}}e^{-\phi}\int dx \left((\beta + i\nu\theta_{\m})(1 + \theta_{\+}\alpha) + \frac{(1 + \theta_{\m}\beta)(\alpha + i\lambda\theta_{\+})}{x}\right)\frac{|x|^{ik}}{\sqrt{x}}e^{-i\nu e^{-\phi}x + i\lambda e^{-\phi}/x} \nonumber \\
&= \frac{2}{\pi}e^{i\nu\gamma_{\m}}e^{i\lambda\gamma_{\+}}e^{-\phi}\cosh\left(\frac{\pi k}{2}\right)\Bigg[(\beta + i\nu\theta_{\m})(1 + \theta_{\+}\alpha)e^{-i\pi/4}\left(\frac{\lambda}{\nu}\right)^{1/4 + ik/2}K_{\frac{1}{2} + ik}(2e^{-\phi}\sqrt{\nu\lambda}) \nonumber \\
&\hspace{5 cm} + (1 + \theta_{\m}\beta)(\alpha + i\lambda\theta_{\+})e^{i\pi/4}\left(\frac{\lambda}{\nu}\right)^{-1/4 + ik/2}K_{\frac{1}{2} - ik}(2e^{-\phi}\sqrt{\nu\lambda})\Bigg]. \nonumber
\end{align}
In the first equality, we have used the Cartan action \eqref{dilatation} and shifted $x\to e^{-\phi}x$.  In the second equality, we have used the integral formula
\begin{equation}
\int_0^\infty dx\, x^{2j-1}e^{\pm(i\nu e^{-\phi}x - i\lambda e^{-\phi}/x)} = 2e^{\pm i\pi j}\left(\frac{\lambda}{\nu}\right)^j K_{2j}(2e^{-\phi}\sqrt{\nu\lambda}),
\end{equation}
which holds for $\nu, \lambda > 0$ and which follows from \eqref{besselint} after analytic continuation in $\nu, \lambda$ via $(\nu, \lambda)\to e^{-\phi}(e^{\mp i\pi/2}\nu, e^{\pm i\pi/2}\lambda)$ \cite{Blommaert:2018iqz}.
This matrix element \eqref{repmatr} has four independent components when expanding in both Grassmann numbers $\alpha$ and $\beta$. Each of these four components is readily checked to be equivalent to one of the four Casimir eigenfunctions given in \eqref{Casfull} above. This indeed shows that this representation matrix element is a solution to the Casimir eigenvalue equation, as it should be by consistency.

Next, we want to evaluate the orthogonality relation
\begin{equation}
\int dg\, \langle\nu_-, \beta|g|\lambda_+, \alpha\rangle_{k_2}^\ast\langle\nu_-', \beta'|g|\lambda_+', \alpha'\rangle_{k_1},
\end{equation}
for which we require explicit knowledge of the Haar measure on OSp$(1|2,\mathbb{R})$, to which we turn next.

\subsubsection{Interlude: Haar Measure on \texorpdfstring{$\osp$}{OSp(1|2)}}
\label{app:HM}

The Haar measure on supergroups is defined analogously as for bosonic groups and can be determined in a physical manner by considering the Maurer-Cartan metric on the algebra:
\begin{equation}
ds^2 = \frac{1}{2}\operatorname{STr} (g^{-1} dg \otimes g^{-1} dg) = G_{ij}\, dx^i dx^j.
\end{equation}
Upon choosing an arbitrary parametrization of the group element $g$ in terms of coordinates $x_1,\ldots, x_n$, the volume form is determined as:
\begin{equation}
\omega = \sqrt{\operatorname{sdet} G}\, dx^1 \wedge \cdots\wedge dx^n.
\end{equation}
For the specific case of OSp$(1|2)$, this calculation can in principle be done. Writing the Maurer-Cartan one-form as
\begin{equation}
g^{-1}dg = \sum_{i,j} J_{ij} X^i\, dx^j,
\end{equation}
the resulting transformation matrix $J$ is:
\begin{equation}
J = \left[\begin{array}{ccc|cc} 
2 & -2\gamma_{\+}e^{2\phi} & 0 & -2 e^{\phi} \theta_{\+} - 2 \gamma_{\+} e^{2\phi} \theta_{\m} & 0 \\
2\gamma_{\+} & -\gamma_{\+}^2 e^{2\phi} & 1 & -\gamma_{\+}^2 e^{2\phi} \theta_{\m} -2 \gamma_{\+} e^{\phi} \theta_{\+} & -\theta_{\+} \\
0 & e^{2\phi} & 0 & e^{2\phi} \theta_{\m} & 0 \\
\hline
2\theta_{\+} & -2\gamma_{\+} e^{2\phi}\theta_{\+} & 0 & 2 \gamma_{\+}e^{\phi} + 2 \gamma_{\+} \theta_{\m} \theta_{\+} & 2 \\
0 & -2 e^{2\phi} \theta_{\+} & 0 & 2 e^{\phi} + 2 e^{2\phi} \theta_{\m} \theta_{\+} & 0
\end{array} \right]
\end{equation}
where the rows are ordered as $(H,E^+,E^-,F^+,F^-)$ and the columns as $(\phi, \gamma_{\m}, \gamma_{\+}, \theta_{\m},\theta_{\+})$, with Berezinian
\begin{equation}
\label{berJ}
\operatorname{sdet} J = \frac{\det A}{\det(D- C A^{-1}B)} = \frac{-2e^{2\phi}}{-4e^{\phi}} =\frac{1}{2}e^{\phi}.
\end{equation}
The matrix $J$ is roughly a square root of the metric, since $G_{ij} = \frac{1}{2} J_{ki}h^{k\ell}J_{\ell j} = \frac{1}{2} J_{ki}J^{k}{}_j$ where indices are raised with the Killing form:
\begin{align}
h^{k\ell} \equiv \operatorname{STr}(X^k X^\ell) = \operatorname{diag}\left(\frac{1}{2}, \left[\begin{array}{cc} 0 & 1 \\ 1 & 0 \end{array} \right], \left[\begin{array}{cc} 0 & \frac{1}{2} \\ \frac{1}{2} & 0 \end{array} \right] \right),
\end{align}
with $\operatorname{sdet} h^{k\ell} = 2$. Hence $\sqrt{\operatorname{sdet} G} = \operatorname{sdet} J$, as given by \eqref{berJ} above.

However, there is a more sophisticated mathematical argument that requires virtually no calculation once one proves several auxiliary results first. It goes as follows. Consider the supergroup GL$(p|q,\mathbb{R})$. The Haar measure for this supergroup is given by:\footnote{The wedge product notation is slightly formal here. We will adhere to the notation of \cite{Witten:2012bg} below.}
\begin{equation}
\label{Haargen}
\omega = \frac{1}{(\operatorname{sdet} x)^{p+q}} \bigwedge_{i,j} dx_{ij},
\end{equation}
in terms of the $(p+q)^2$ entries $x_{ij}$ of the matrix.\footnote{The proof mimics the bosonic proof for GL$(n,\mathbb{R})$. We identify the set of matrices with a vector space $\smash{\mathbb{R}^{p^2+q^2|2pq}}$, where each element of the matrix becomes a separate component of the vector. The ordering of this identification is chosen to go ``down'' each column before proceeding with the next column on the right. With this identification, the left action of the group $g \to g_0 g$ gets mapped into a corresponding left action on the vector space $x \to \mathbf{g}_0 x$, where the matrix $\mathbf{g}_0$ is now a $(p+q)^2 \times (p+q)^2$ dimensional supermatrix that is block diagonal with $(p+q)$ copies of the original $g_0$. This directly leads to the left invariance of the measure \eqref{Haargen}. Right invariance is checked analogously.}

Restricting to subgroups of GL$(p|q,\mathbb{R})$ is done by imposing a suitable matrix delta function such as $\bm{\delta} (g^{\text{st}} \Omega g - \Omega)$. This matrix delta function generically contains several copies of the same component delta function, which we define to be omitted. The Haar measure on the subgroup is then proportional to
\begin{equation}
\label{Haarred}
\omega = \frac{\bm{\delta} (g^{\text{st}} \Omega g - \Omega)}{(\operatorname{sdet} x)^{p+q}} \bigwedge_{i,j} dx_{ij}.
\end{equation}
Indeed, one readily checks left/right invariance under the subgroup directly. 

For the specific case of OSp$(1|2)$, this delta-function constraint reduces to a product of two bosonic and two fermionic delta functions:
\begin{align}
\delta(ad - bc -\delta \beta -1) \delta(e^2 +2 \gamma \alpha - 1)\delta(c\alpha - a \gamma  - \beta e) \delta(d\alpha - b \gamma - \delta e).
\end{align}
Using these delta functions to evaluate the $d, e, \gamma, \delta$ integrals respectively, we pick up the additional factor
\begin{equation}
\frac{1}{a}\frac{1}{2e} (a+\beta \alpha) = \frac{1}{2},
\end{equation}
where we made use of the relation $e=1-\alpha\beta/a$. The Haar measure becomes:\footnote{We have used the notation of \cite{Witten:2012bg} to denote the superspace version of a top differential form.}
\begin{equation}
\omega = \frac{1}{2} \left[da \,  db\, dc \,|\, d\alpha \, d\beta\right].
\end{equation}
If one is interested in the parametrization in Gauss-Euler coordinates, then we perform the coordinate transformation
\begin{equation}
a= e^{\phi}, \quad b= \gamma_{\+} e^{\phi}, \quad c= \gamma_{\m} e^{\phi}, \quad \alpha = e^{\phi} \theta_{\+}, \quad \beta = e^{\phi} \theta_{\m}
\end{equation}
to map $(e^\phi, \gamma_{\m}, \gamma_{\+}, \theta_{\m}, \theta_{\+})$ to $(a,b,c,\alpha,\beta)$ and find Berezinian 1. So we get the Haar measure in the Gauss-Euler parametrization as:
\begin{equation}
\label{Haar}
\boxed{ \omega = \frac{1}{2} e^{\phi}\left[d\phi\, d\gamma_{\m}\, d\gamma_{\+} \,|\, d\theta_{\m}\, d\theta_{\+}\right], }
\end{equation}
matching the above explicit computation.

\subsubsection{Orthogonality and Plancherel Measure}

With this Haar measure \eqref{Haar}, we can proceed with the orthogonality calculation:
\begin{equation}
\int dg\, \langle\nu_-, \beta|g|\lambda_+, \alpha\rangle_{k_2}^\ast\langle\nu_-', \beta'|g|\lambda_+', \alpha'\rangle_{k_1}.
\end{equation}
The evaluation of this integral is quite involved. In the process, we will have need of the following relations:
\begin{align}
\label{orth1}
\int_0^{+\infty}dx \left(K_{\frac{1}{2}+ip}(x)K_{\frac{1}{2}+ip'}(x) + K_{\frac{1}{2}-ip}(x) K_{\frac{1}{2}-ip'}(x) \right) &= \frac{\pi^2}{\cosh \pi p} \delta(p+p'), \\
\label{orth2}
\int_0^{+\infty}dx \left(K_{\frac{1}{2}+ip}(x)K_{\frac{1}{2}-ip'}(x) + K_{\frac{1}{2}-ip}(x) K_{\frac{1}{2}+ip'}(x) \right) &= \frac{\pi^2}{\cosh \pi p} \delta(p-p').
\end{align}
After inserting the representation matrices \eqref{repmatr}, we first evaluate the $\gamma_{\m}$ and $\gamma_{\+}$ integrals, yielding a factor of
\begin{equation}
4\pi^2\delta(\nu - \nu')\delta(\lambda - \lambda').
\end{equation}
Next, we get four terms by pairwise multiplication of the two terms in \eqref{repmatr} for each matrix element.  Doing the $\theta_{\m}$ and $\theta_{\+}$ integrals projects onto the $\theta_{\m} \theta_{\+}$ component of the integrand, leaving the following remaining part of the integral to be evaluated:
\begin{equation}
\frac{2}{\pi^2}\cosh\left(\frac{\pi k_1}{2}\right)\cosh\left(\frac{\pi k_2}{2}\right)\int d\phi\, e^{-\phi}(I_\text{diag} + I_\text{cross})
\end{equation}
where (setting $x = 2e^{-\phi}\sqrt{\nu\lambda}$) the contributions from the diagonal terms and the cross terms arising from the multiplication are\footnote{Note that complex conjugation acts on the fermionic parts of the matrix elements as if $\alpha, \beta$ are \emph{imaginary} Grassmann numbers, and in order-preserving fashion.  The order-preserving convention was important for our proof of the unitarity of the principal series representations in Appendix \ref{app:unitarity}, since conjugation should leave a super-M\"obius transformation invariant.}
\begin{align}
I_\text{diag} &= -i\nu(\alpha - \alpha')(\beta - \beta')\left(\frac{\lambda}{\nu}\right)^{\frac{1}{2} + \frac{i(k_1 - k_2)}{2}}K_{\frac{1}{2} + ik_1}(x)K_{\frac{1}{2} - ik_2}(x) \nonumber \\
&\phantom{==} - i\lambda(\alpha - \alpha')(\beta - \beta')\left(\frac{\lambda}{\nu}\right)^{-\frac{1}{2} + \frac{i(k_1 - k_2)}{2}}K_{\frac{1}{2} - ik_1}(x)K_{\frac{1}{2} + ik_2}(x), \\
I_\text{cross} &= i(i\nu\alpha\alpha' + i\lambda\beta\beta' + \nu\lambda - \alpha\alpha'\beta\beta')\left(\frac{\lambda}{\nu}\right)^{\frac{i(k_1 - k_2)}{2}}K_{\frac{1}{2} + ik_1}(x)K_{\frac{1}{2} + ik_2}(x) \nonumber \\
&\phantom{==} - i(-i\nu\alpha\alpha' - i\lambda\beta\beta' - \nu\lambda + \alpha\alpha'\beta\beta')\left(\frac{\lambda}{\nu}\right)^{\frac{i(k_1 - k_2)}{2}}K_{\frac{1}{2} - ik_1}(x)K_{\frac{1}{2} - ik_2}(x).
\end{align}
Using \eqref{orth1}, we can see that the $I_\text{cross}$ contribution is proportional to $\delta(k_1+k_2)$, which vanishes under our assumption $k_i > 0$.

Finally, using \eqref{orth2}, we deduce the orthogonality relation of the representation matrices:
\begin{align}
&\int dg\, \langle\nu_-, \beta|g|\lambda_+, \alpha\rangle_{k_2}^\ast\langle\nu_-', \beta'|g|\lambda_+', \alpha'\rangle_{k_1} \nonumber \\*
&= -8i\delta(\alpha - \alpha')\delta(\beta - \beta')\delta(\nu - \nu')\delta(\lambda - \lambda')\cosh\left(\frac{\pi k_1}{2}\right)\cosh\left(\frac{\pi k_2}{2}\right)\left(\frac{\lambda}{\nu}\right)^{\frac{i(k_1 - k_2)}{2}} \nonumber \\*
&\phantom{==} \times \int_{-\infty}^\infty d\phi\, e^{-\phi}\sqrt{\nu\lambda}\left[K_{\frac{1}{2} + ik_1}(x)K_{\frac{1}{2} - ik_2}(x) + K_{\frac{1}{2} - ik_1}(x)K_{\frac{1}{2} + ik_2}(x)\right] \\
&= -4\pi^2 i\delta(\alpha - \alpha')\delta(\beta - \beta')\delta(\nu - \nu')\delta(\lambda - \lambda')\cosh^2\left(\frac{\pi k_1}{2}\right)\frac{\delta(k_1 - k_2)}{\cosh(\pi k_1)},
\end{align}
from which we read off the Plancherel measure
\begin{equation}
\label{N1P}
\boxed{\rho(k) = \frac{1}{4\pi^2}\frac{\cosh(\pi k)}{\cosh^2\left(\frac{\pi k}{2}\right)} = \frac{1}{2\pi^2}\frac{\cosh(\pi k)}{1 + \cosh(\pi k)}.}
\end{equation}
This expression holds for the spherical principal series representations (where $\epsilon = \epsilon'=0$). We can readily generalize it to the other principal series representations, getting:
\begin{equation}
\rho(k) = \frac{1}{2\pi^2}\frac{\cosh(\pi k)}{1 + (-)^{\epsilon}\cosh(\pi k)}.
\end{equation}
The change is caused by dividing $\cosh(\pi k)$ by $\sinh^2\left(\frac{\pi k}{2}\right)$ instead of $\cosh^2\left(\frac{\pi k}{2}\right)$ when $\epsilon = -1$.

\subsubsection{Global Structure of the \texorpdfstring{$\osp$}{OSp(1|2)} Group Manifold}

Finally, we must address the subtlety that the $\osp$ (super)group (super)manifold consists of multiple patches, and that the Gauss decomposition \eqref{gaussparam} we have used covers only one of them (this is not an issue for $\ospp$).  It is helpful to recall how this works in the case of ordinary linear groups.  In general, $\text{GL}(N, \mathbb{C})$ is covered by $N!$ patches (or cells) where each patch has a Gauss decomposition of the form $g = LDU\omega$ with $L$ and $U$ lower and upper triangular unidiagonal matrices, $D$ a diagonal matrix, and $\omega$ a permutation matrix.  The LDU decomposition for subgroups of $\text{GL}(N, \mathbb{C})$ is induced by this one.  Our parametrization \eqref{gausssltr} for $g\in \sltr$ assumes $\omega = 1$, and has the further limitation that it requires positive diagonal entries for $D = e^{2\phi H}$ in \eqref{gausssltr}.  To cover the entire $\sltr$ group manifold, we must therefore allow $\omega$ to range over all \emph{signed} permutation matrices in the group, leading to four patches given by taking $g\mapsto g\omega$ with $\omega\in \{1, -1, \mathbf{s}, -\mathbf{s}\}$ and $\mathbf{s}\equiv \left[\begin{smallmatrix} 0 & 1 \\ -1 & 0 \end{smallmatrix}\right]$ \cite{Forgacs:1989ac}.

By analogy with the case of $\sltr$, we postulate that the various patches of $\osp$ are given by multiplying the Gauss parametrization in \eqref{gaussparam} on the right by
\begin{equation}
\Omega\equiv \pm \left[\begin{array}{c|c} \omega & 0 \\ \hline 0 & 1 \end{array}\right], \quad \omega\in \left\{\left[\begin{array}{cc} 1 & 0 \\ 0 & 1 \end{array}\right], \left[\begin{array}{cc} -1 & 0 \\ 0 & -1 \end{array}\right], \left[\begin{array}{cc} 0 & 1 \\ -1 & 0 \end{array}\right], \left[\begin{array}{cc} 0 & -1 \\ 1 & 0 \end{array}\right]\right\},
\label{osppatches}
\end{equation}
which comprise the only signed permutation matrices in $\osp$.  This would imply that eight patches are required to cover the $\osp$ group manifold.  This group consists of two connected components, one containing the identity and one containing minus the identity, which are distinguished by the sign of the Berezinian.  The projective group $\posp$ would contain only four patches, and its resulting measure would be half that of the full group.\footnote{In $\sltr$, the $\mathbb{Z}_2$ quotient to $\text{PSL}(2, \mathbb{R})$ simply folds the single connected component in half, whereas in $\osp$, the $\mathbb{Z}_2$ quotient to $\posp$ identifies the two connected components.  The difference is that in $\sltr$, $\pm I$ are continuously connected, while in $\osp$, they are not.}

With this setup in place, we now consider the results for the inner product in different patches of $\osp$.  The overall sign of $\Omega$ cancels out of the computation, so we focus on the four independent choices of $g$.  In addition to the original ($\omega = 1$) patch
\begin{equation}
g = \left[\begin{array}{cc|c}
e^\phi & \gamma_{\+}e^\phi & e^\phi \theta_{\+} \\
\gamma_{\m} e^\phi & e^{-\phi} + \gamma_{\m}\gamma_{\+}e^\phi - \theta_{\m}\theta_{\+} & \gamma_{\m} e^\phi \theta_{\+} - \theta_{\m} \\ \hline
e^\phi\theta_{\m} & \gamma_{\+}e^\phi\theta_{\m} + \theta_{\+} & 1 + e^\phi\theta_{\m}\theta_{\+}
\end{array}\right]
\end{equation}
(reproduced from \eqref{gaussparamdefining}), we have the $\omega = -1$ patch
\begin{equation}
g\left[\begin{array}{cc|c}
-1 & 0 & 0 \\
0 & -1 & 0 \\ \hline
0 & 0 & 1
\end{array}\right]
=
\left[\begin{array}{cc|c}
-e^\phi & -\gamma_{\+}e^\phi & e^\phi \theta_{\+} \\
-\gamma_{\m} e^\phi & -e^{-\phi} - \gamma_{\m}\gamma_{\+}e^\phi + \theta_{\m}\theta_{\+} & \gamma_{\m} e^\phi \theta_{\+} - \theta_{\m} \\ \hline
-e^\phi\theta_{\m} & -\gamma_{\+}e^\phi\theta_{\m} - \theta_{\+} & 1 + e^\phi\theta_{\m}\theta_{\+}
\end{array}\right],
\end{equation}
which is given by taking
\begin{equation}
e^\phi\to -e^\phi, \quad \theta_{\+}\to -\theta_{\+}
\end{equation}
in $g$, the $\omega = \mathbf{s}$ patch
\begin{equation}
g\left[\begin{array}{cc|c}
0 & 1 & 0 \\
-1 & 0 & 0 \\ \hline
0 & 0 & 1
\end{array}\right]
=
\left[\begin{array}{cc|c}
-\gamma_{\+}e^\phi & e^\phi & e^\phi \theta_{\+} \\
-e^{-\phi} - \gamma_{\m}\gamma_{\+}e^\phi + \theta_{\m}\theta_{\+} & \gamma_{\m} e^\phi & \gamma_{\m} e^\phi \theta_{\+} - \theta_{\m} \\ \hline
-\gamma_{\+}e^\phi\theta_{\m} - \theta_{\+} & e^\phi\theta_{\m} & 1 + e^\phi\theta_{\m}\theta_{\+}
\end{array}\right],
\end{equation}
which is given by taking
\begin{equation}
e^\phi\to -\gamma_{\+}e^\phi, \quad \gamma_{\m}\to \gamma_{\m} + \frac{e^{-2\phi}}{\gamma_{\+}} - \frac{e^{-\phi}}{\gamma_{\+}}\theta_{\m}\theta_{\+}, \quad \gamma_{\+}\to -\frac{1}{\gamma_{\+}}, \quad \theta_{\m}\to \theta_{\m} + \frac{e^{-\phi}}{\gamma_{\+}}\theta_{\+}, \quad \theta_{\+}\to -\frac{\theta_{\+}}{\gamma_{\+}}
\end{equation}
in $g$, and the $\omega = -\mathbf{s}$ patch
\begin{equation}
g\left[\begin{array}{cc|c}
0 & -1 & 0 \\
1 & 0 & 0 \\ \hline
0 & 0 & 1
\end{array}\right]
=
\left[\begin{array}{cc|c}
\gamma_{\+}e^\phi & -e^\phi & e^\phi \theta_{\+} \\
e^{-\phi} + \gamma_{\m}\gamma_{\+}e^\phi - \theta_{\m}\theta_{\+} & -\gamma_{\m} e^\phi & \gamma_{\m} e^\phi \theta_{\+} - \theta_{\m} \\ \hline
\gamma_{\+}e^\phi\theta_{\m} + \theta_{\+} & -e^\phi\theta_{\m} & 1 + e^\phi\theta_{\m}\theta_{\+}
\end{array}\right],
\end{equation}
which is given by taking
\begin{equation}
e^\phi\to \gamma_{\+}e^\phi, \quad \gamma_{\m}\to \gamma_{\m} + \frac{e^{-2\phi}}{\gamma_{\+}} - \frac{e^{-\phi}}{\gamma_{\+}}\theta_{\m}\theta_{\+}, \quad \gamma_{\+}\to -\frac{1}{\gamma_{\+}}, \quad \theta_{\m}\to \theta_{\m} + \frac{e^{-\phi}}{\gamma_{\+}}\theta_{\+}, \quad \theta_{\+}\to \frac{\theta_{\+}}{\gamma_{\+}}
\end{equation}
in $g$.  We write
\begin{equation}
\langle\psi|\psi'\rangle|_\omega\equiv \int d(g\Omega)\, \langle\nu_-, \beta|g\Omega|\lambda_+, \alpha\rangle_{k_2}^\ast\langle\nu_-', \beta'|g\Omega|\lambda_+', \alpha'\rangle_{k_1}.
\end{equation}
Let us first derive $\langle\psi|\psi'\rangle|_{-1}$ by taking $e^\phi\to -e^\phi$ and $\theta_{\+}\to -\theta_{\+}$ in our earlier computation of $\langle\psi|\psi'\rangle|_1 = \langle\psi|\psi'\rangle$.  The effect of taking $e^\phi\to -e^\phi$ is that instead of using the transformation rule
\begin{equation}
(e^{2\phi H}\circ f)(x, \vartheta) = (e^{-\phi})^{-1/2}|e^{-\phi}|^{ik}f(e^{2\phi}x, e^\phi\vartheta)
\end{equation}
to compute each matrix element (wavefunction), we should use the action
\begin{equation}
(e^{2\phi H}\circ f)(x, \vartheta) = (-e^{-\phi})^{-1/2}|{-e^{-\phi}}|^{ik}f(e^{2\phi}x, e^\phi\vartheta) = -ie^{(1/2 - ik)\phi}f(e^{2\phi}x, e^\phi\vartheta).
\end{equation}
Thus each matrix element is modified by an overall phase of $-i$, but these phases cancel after conjugation and leave no imprint in the inner product.  Taking $\theta_{\+}\to -\theta_{\+}$ flips the sign of both the integrand and the Haar measure, again having no effect (recall that the fermionic part of the super-Jacobian \eqref{sdetprime} does not involve an absolute value).  Similar statements apply to $\langle\psi|\psi'\rangle|_{-\mathbf{s}}$ and $\langle\psi|\psi'\rangle|_{\mathbf{s}}$, which are related by right multiplication by the $\omega = -1$ element.  Finally, an explicit computation along the lines of that for $\sltr$ in Appendix G of \cite{Blommaert:2018iqz} shows that $\langle\psi|\psi'\rangle|_1 = \langle\psi|\psi'\rangle|_{\mathbf{s}}$.  Therefore, we conclude that
\begin{equation}
\langle\psi|\psi'\rangle|_1 = \langle\psi|\psi'\rangle|_{-1} = \langle\psi|\psi'\rangle|_{\mathbf{s}} = \langle\psi|\psi'\rangle|_{-\mathbf{s}}.
\label{fourcontributions}
\end{equation}
Summing over the different patches leads to an overall factor of eight in the inner product, and hence the Plancherel measure
\begin{equation}
\rho(k) = \frac{1}{16\pi^2}\frac{\cosh(\pi k)}{1 + \cosh(\pi k)}.
\end{equation}

\section{\texorpdfstring{$\text{OSp}^+(1|2, \mathbb{R})$}{OSp+(1|2, R)} Representation Theory}
\label{ospprep}

In this appendix, we collect some useful results on the subsemigroup $\text{OSp}^+(1|2, \mathbb{R})$. In particular, we prove that the restriction of the principal series representations of the full supergroup OSp$(1|2,\mathbb{R})$ to its positive subsemigroup leads to representations that are still irreducible and unitary. The irreducibility proof is given in Section \ref{app:irre}, and the unitarity proof is given in Section \ref{app:unit}. This appendix complements the description in the main text in Section \ref{sect:whittaker}, where one can find the definition of the principal series representations of the subsemigroup $\text{OSp}^+(1|2, \mathbb{R})$.

\subsection{Irreducibility}
\label{app:irre}

In this subsection, we prove irreducibility of the principal series representations. Because the only valid basis on the subsemigroup is the hyperbolic basis introduced in \ref{app:bases} in terms of the super-Mellin transform, the proof differs from the usual one that uses the elliptic basis. Hence we first redo the proof in the bosonic case for SL$(2,\mathbb{R})$ and SL$^+(2,\mathbb{R})$. This in particular proves irreducibility for the bosonic subsemigroup, a result that had not yet been shown explicitly. After that, we generalize the construction and prove irreducibility for the subsupersemigroup as well.

\subsubsection*{SL$^+(2,\mathbb{R})$ and SL$(2,\mathbb{R})$}

For the bosonic system, irreducibility has been proven in the mathematics literature (see, e.g., Section 6.4.2 of \cite{VK}) in terms of a carrier space on $S^1$ for which one has a discretized (countably infinite) set of modes $e^{i n \theta}$, $n \in \mathbb{Z}$, diagonalizing the elliptic one-parameter subgroup $\left(\begin{smallmatrix}
\cos \theta & \sin \theta  \\
-\sin \theta & \cos \theta
\end{smallmatrix}\right) $, which is the exponentiated Cartan generator $J_0$ of $SU(1,1)$. Since the representation of an abelian subgroup of the principal series representation is equivalent to its regular representation, each invariant subspace is one-dimensional and occurs with multiplicity one. The argument proceeds to show that, assuming the invariant subspace of the full group SL$(2,\mathbb{R})$ is nonempty, at least one eigenmode of the subgroup is present. By acting with the full group, one then shows that it has to contain all of them, except when $2j \in \mathbb{N}$. In the latter case, the representation decomposes into a combination of lowest, highest, and finite irreps.

To generalize this argument to the supersymmetric case, it is convenient to redo the analysis for a carrier space $\mathbb{R}$ with continuous modes diagonalizing the hyperbolic generator $H$ of SL$(2,\mathbb{R})$. To simplify matters, we immediately specialize to the subsemigroup where we take carrier space $\mathbb{R}^+$. We come back to the full group below.

The generator $H = - x \partial_x + j$ is readily diagonalized by the orthonormal Rindler modes
\begin{equation}
\psi_s(x) = \frac{1}{\sqrt{2\pi}}x^{is-1/2}
\end{equation}
forming a complete basis on $\mathbb{R}^+$, where one explicitly sees that each eigenspace indeed occurs with multiplicity one. Now assume a single such mode $\psi_s(x)$ is contained within $\mathcal{I}$. Acting on this mode with the one-parameter parabolic subgroup generated by $E^- = \partial_x$, we obtain the translated modes
\begin{equation}
\frac{1}{\sqrt{2\pi}}(x+a)^{is-1/2}, \quad a > 0,
\end{equation}
all part of the same invariant subspace. Now, taking a suitable linear combination of these modes as:\footnote{To ensure convergence, one requires a regulator $ib \to ib + \epsilon$. The integral only converges in a distributional sense when $\epsilon =0$, which is sufficient for our purposes.}
\begin{equation}
\int_0^{+\infty} da\, (x+a)^{is-1/2}a^{ib-1} =  \frac{\Gamma(ib)\Gamma(1/2-is-ib)}{\Gamma(1/2-is)} x^{is-1/2 + ib},
\end{equation}
we obtain another single such mode with $s \to s+b$. Since $b\in \mathbb{R}$, we can generate all of the basis modes and hence we span the entire space. An exception occurs when $1/2-is \in -\mathbb{N}$, in which case one might not be able to generate all basis functions. This cannot occur unless $is-1/2 = j-m$ is an integer, and in particular can only happen if $2j \in \mathbb{N}$, leading to the same conclusion as above.

The generalization to the full group SL$(2,\mathbb{R})$ is not that hard. In this case, one has a doubled hyperbolic basis, covering both the $x>0$ and the $x<0$ regions:
\begin{equation}
\psi_{s,+}(x) = \frac{1}{\sqrt{2\pi}}x^{is-1/2}, \mbox{ } x>0, \qquad \psi_{s,-}(x) = \frac{1}{\sqrt{2\pi}}(-x)^{is-1/2}, \mbox{ } x<0.
\end{equation}
These functions form a complete and orthonormal basis for any function on the entire real line, where the $x>0$ and $x<0$ regions live independent lives. The analysis for the $x>0$ region proceeds along precisely the same lines as above. The analysis for the $x<0$ region requires only some small changes. In particular, one transforms a single mode into
\begin{equation}
\frac{1}{\sqrt{2\pi}}(-x+a)^{is-1/2}, \quad a > 0,
\end{equation}
and uses
\begin{equation}
\int_0^{+\infty} da\, (-x+a)^{is-1/2}a^{ib-1} = \frac{\Gamma(ib)\Gamma(1/2-is-ib)}{\Gamma(1/2-is)} (-x)^{is-1/2 + ib},
\end{equation}
thus spanning all of the $x<0$ hyperbolic eigenmodes.

\subsubsection*{OSp$^+(1|2,\mathbb{R})$ and OSp$(1|2,\mathbb{R})$}

Now let's generalize the argument to the subsupersemigroup OSp$^+(1|2,\mathbb{R})$. One might think that the simplest route is to make use of a doubled basis starting with
\begin{equation}
\psi_{s,s'}(x, \vartheta) = \frac{1}{\sqrt{2\pi}}\left(x^{is-1/2} + \vartheta x^{is'-1/2}\right).
\end{equation}
These modes are orthonormal and complete for the bosonic and fermionic components separately, and hence form a basis for functions on the super half-line $(x,\vartheta)$. These modes, however, do not correspond to eigenmodes of any of the generators, and hence the above irreducibility argument fails.

A hyperbolic eigenmode takes the form:
\begin{equation} 
\psi_{s,\alpha}(x,\vartheta) = \frac{1}{\sqrt{2\pi}} \left(x^{is-1/4} + \alpha \vartheta x^{is-3/4}\right).
\end{equation}
Since $H$ generates a one-parameter subgroup of OSp$(1|2)$, invariant subspaces of the full group decompose into invariant subspaces of the subgroup. Since the latter is abelian, the invariant subspaces are one-dimensional. A new feature, however, is that these subspaces themselves occur more than once, with a continuous multiplicity $\int_\oplus d\alpha$. We write schematically:
\begin{equation}
\mathcal{H}_{\text{inv}} = \bigoplus_s \mathcal{H}_{\text{inv}(H)}\int_\oplus d\alpha.
\end{equation}
However, this degeneracy is intuitively only ``infinitesimal,'' and we expect that due to the peculiarities of Grassmann numbers, it will not affect the argument.

The argument proceeds along similar lines as the bosonic argument, except that we have to define what is meant precisely by linear independence in supervector spaces:
\begin{quote}
\emph{A set of supervectors $V_i$ is linearly dependent iff there exist $c_i \in \Lambda_\infty$, not all zero, such that $\sum_i c_i V_i = 0$.}
\end{quote}
Suppose that the invariant subspace of OSp$(1|2)$ contains a single hyperbolic eigenmode:
\begin{equation} 
\label{startmode}
\psi_{s,\alpha}(x,\vartheta) = \frac{1}{\sqrt{2\pi}} \left(x^{is-1/4} + \alpha \vartheta x^{is-3/4}\right).
\end{equation}
We should really understand this as an equivalence class generated by identifying linearly dependent modes, i.e.,
\begin{equation} 
\psi_{s,\alpha}(x,\vartheta) \sim c \psi_{s,\alpha}(x,\vartheta), \qquad c \in \Lambda_\infty.
\end{equation}
A specific linearly dependent (equivalent) function is then:
\begin{equation} 
\label{endmode}
\psi_{s,\alpha=0}(x,\vartheta) = \frac{1}{\sqrt{2\pi}} x^{is-1/4}.
\end{equation}
One can view the transition from \eqref{startmode} to \eqref{endmode} as coming from the idea that the soul of any function is infinitesimal, and thus irrelevant for the argument at hand. Acting with the one-parameter parabolic subgroup generated by $E^- = \partial_x$, we get
\begin{equation}
\int_0^{+\infty} da\, (x+a)^{is-1/4}a^{ib-1} = \frac{\Gamma(ib)\Gamma(1/4-is-ib)}{\Gamma(1/4-is)} x^{is-1/4 + ib},
\end{equation}
so we generate all possible bodies of all hyperbolic eigenmodes. Instead acting with the one-parameter subgroup generated by $F^-$, we obtain:
\begin{equation}
\frac{1}{\sqrt{2\pi}}(x-\delta \vartheta)^{is-1/4} = \frac{1}{\sqrt{2\pi}} x^{is-1/4} - \frac{1}{\sqrt{2\pi}} (is-1/4) \delta\vartheta x^{is-5/4},
\end{equation}
which is a linear combination of our starting mode and a new mode. Acting with $E^-$ on the new mode, we can obtain:
\begin{equation}
\int_0^{+\infty} da\, (x+a)^{is-5/4}a^{ib+1/2-1} = \frac{\Gamma(ib+1/2)\Gamma(3/4-is-ib)}{\Gamma(5/4-is)} x^{is-3/4 + ib},
\end{equation}
producing the generic soul parts of all hyperbolic eigenmodes:
\begin{equation}
 \delta\vartheta x^{is-3/4 + ib}.
\end{equation}
In particular, this shows that we are able to generate all hyperbolic eigenmodes:
\begin{equation} 
\psi_{s,\delta}(x,\vartheta) = \frac{1}{\sqrt{2\pi}} \left(x^{is-1/4} + \delta \vartheta x^{is-3/4}\right).
\end{equation}
Hence the invariant subspace spans the entire representation, making the principal series representation irreducible, as was to be shown.

The generalization to the entire supergroup OSp$(1|2,\mathbb{R})$ is again just a doubling of the argument with appropriate sign factors. The details are left implicit.

\subsection{Unitarity}
\label{app:unit}

In this subsection, we write down explicit formulas for the representation matrices using the only basis available for the subsemigroup: the hyperbolic basis introduced in Section \ref{app:bases}. This is a basis on $\mathbb{R}^{1|1}$ consisting of the super-Mellin eigenmodes. Moreover, we use these formulas to give a brute-force proof that the resulting matrix elements are unitary.

Within this hyperbolic basis, the explicit matrix elements are computed as
\begin{equation}
K^{++}_{s_1|\alpha_1, s_2|\alpha_2}(g) = \langle s_1| g |s_2\rangle = \int_{0}^{+\infty} dx\, d\vartheta\, \psi_{s_1|\alpha_1}^*(x,\vartheta)(g \cdot \psi_{s_2|\alpha_2}(x,\vartheta)).
\end{equation}
For each of the five constituent one-parameter subgroups corresponding to the generators $H$, $F^+$, $F^-$, $E^-$, and $E^+$, one obtains the respective matrix elements ($j=-\frac{1}{4}+\frac{ik}{2}$):
\begin{align}
\label{matgenexpl}
K^{++}_{s_1|\alpha_1, s_2|\alpha_2}(\phi) &= e^{2i(s_2-k/2)\phi}\delta(s_1-s_2) \delta(\alpha_1-\alpha_2), \\
K^{++}_{s_1|\alpha_1, s_2|\alpha_2}(\theta_{\+}) &= \delta(s_1-s_2)\delta(\alpha_1-\alpha_2) +\left(-\alpha_1\alpha_2-(ik-is_2-1/4)\right) \delta(s_1-s_2+i/2)\theta_{\+}, \nonumber \\
K^{++}_{s_1|\alpha_1, s_2|\alpha_2}(\theta_{\m}) &= \delta(s_1-s_2)\delta(\alpha_1-\alpha_2) + \left(\alpha_1\alpha_2-(is_2-1/4)\right) \delta (s_1-s_2-i/2) \theta_{\m}, \nonumber \\
K^{++}_{s_1|\alpha_1, s_2|\alpha_2}(\gamma_{\m}) &= \frac{1}{2\pi}\left[\alpha_1\frac{\Gamma(-is_1+1/4)}{\Gamma(-is_2+1/4)} - \alpha_2\frac{\Gamma(-is_1+3/4)}{\Gamma(-is_2+3/4)}\right]\Gamma(is_1-is_2)\gamma_{\m}^{is_2-is_1}, \nonumber \\
K^{++}_{s_1|\alpha_1, s_2|\alpha_2}(\gamma_{\+}) &= \frac{1}{2\pi}\left[\alpha_1\frac{\Gamma(is_1+1/4 - ik)}{\Gamma(is_2+1/4-ik)}-\alpha_2\frac{\Gamma(is_1+3/4 - ik)}{\Gamma(is_2+3/4-ik)} \right]\Gamma(is_2-is_1) \gamma_{\+}^{is_1-is_2}. \nonumber
\end{align}

\textbf{Unitarity.} These representation matrix elements are unitary. This is a statement that we proved before for the full group in Section \ref{app:unitarity}, but it is necessary to redo the proof for the subsemigroup. As examples, let's first consider some of the one-parameter subgroups separately. For $H$, we get simply
\begin{align}
&\int ds\, d\alpha\, K^{++}_{s_1|\alpha_1, s|\alpha}(\phi) K^{++}_{s_2|\alpha_2, s|\alpha}(\phi)^* \\*
&= \int ds\, d\alpha\, e^{2i(s_2-k/2)\phi}\delta(s_1-s) \delta(\alpha_1-\alpha) e^{-2i(s_2-k/2)\phi}\delta(s_2-s) (-) \delta(\alpha_2-\alpha) \nonumber \\
&= \delta(s_1-s_2) \delta(\alpha_2-\alpha_1). \nonumber
\end{align}
For $E^-$, we obtain
\begin{align}
&\int ds\, d\alpha\, K^{++}_{s_1|\alpha_1, s|\alpha}(\gamma_{\m}) K^{++}_{s_2|\alpha_2, s|\alpha}(\gamma_{\m})^* \\
&= \frac{1}{4\pi^2}\int ds\, d\alpha \int dx\, dy \left[-\alpha_1x^{-is-1/4}(x+\gamma_{\m})^{is-3/4} + \alpha x^{-is_1-3/4}(x+\gamma_{\m})^{is_1-1/4}\right] \nonumber \\
&{}\hspace{1cm}\times (-)\left[-\alpha_2 x^{-is-1/4}(x+\gamma_{\m})^{is-3/4} + \alpha y^{-is_2-3/4}(y+\gamma_{\m})^{is_2-1/4}\right] \nonumber \\
&= \delta(s_1-s_2) \delta(\alpha_2-\alpha_1). \nonumber
\end{align}
Finally, for $F^+$, we write
\begin{align}
&\int ds\, d\alpha\, K^{++}_{s_1|\alpha_1, s|\alpha}(\theta_{\+}) K^{++}_{s_2|\alpha_2, s|\alpha}(\theta_{\+})^* \\
&= \int ds\, d\alpha \left[ \delta(s_1-s)\delta(\alpha_1-\alpha) +\left(-\alpha_1\alpha - (ik-is-1/4)\right) \delta(s_1-s+i/2)\theta_{\+}\right] \nonumber \\
&{}\hspace{1cm}\times \left[ -\delta(s_2-s)\delta(\alpha_2-\alpha) +\left(-\alpha_2\alpha + (ik-is+1/4)\right) \delta(s_2-s-i/2)\theta_{\+}\right] \nonumber \\
&= \delta(s_1-s_2)\delta(\alpha_2-\alpha_1), \nonumber
\end{align}
where extreme care has to be exerted for relative minus signs. Notice the ordering in the fermionic delta function. The calculations for the remaining two generators can be done similarly. Since all of these individual objects satisfy the unitarity property, and since one can use the Gauss-Euler decomposition to write the full matrix element as a composition,
\begin{align}
&K^{++}_{s_1|\alpha_1, s_2|\alpha_2}(g) = \int ds_{i_1} \cdots ds_{i_4}\, d\alpha_{i_1} \cdots d\alpha_{i_4} \\
&\hspace{1cm}\times K^{++}_{s_1|\alpha_1, s_{i_1}|\alpha_{i_1}}(\theta_{\m}) K^{++}_{s_{i_1}|\alpha_{i_1}, s_{i_2}|\alpha_{i_2}}(\gamma_{\m}) K^{++}_{s_{i_2}|\alpha_{i_2}, s_{i_3}|\alpha_{i_3}}(\phi)   K^{++}_{s_{i_3}|\alpha_{i_3}, s_{i_4}|\alpha_{i_4}}(\gamma_{\+})   K^{++}_{s_{i_4}|\alpha_{i_4}, s_2|\alpha_2}(\theta_{\+}), \nonumber
\end{align}
after pairwise simplification using the unitarity of the constituents, one obtains
\begin{equation}
\boxed{\int ds\, d\alpha\, K^{++}_{s_1|\alpha_1, s|\alpha}(g) K^{++}_{s_2|\alpha_2, s|\alpha}(g)^* = \delta(s_1-s_2)\delta(\alpha_2-\alpha_1),}
\end{equation}
which proves that these matrices are unitary, much like what happens in the bosonic case \eqref{unitarity}.

\textbf{Explicit expressions.} For the full matrix element, we can do the resulting integrals explicitly. This leads to some suggestive results, as we now show. Writing $j=-1/4+ik/2$, $n=-is_2+ik$, and $m=-is_1+ik$, we find the full representation matrix for the bosonic subgroup (i.e., the case $\theta_{\m}=\theta_{\+}=0$):
\begin{equation}
R_{s_1|\alpha_1, s_2|\alpha_2}^k (g_B) = \alpha_1 \mathcal{G}^j_{m,n}(g_B) + \alpha_2 \mathcal{G}^{j-1/2}_{m,n}(g_B),
\end{equation}
where we introduced the notation
\begin{equation}
\mathcal{G}^j_{m,n}(g_B) \equiv \frac{1}{2\pi}\gamma_{\+}^{n}\gamma_{\m}^{m}\sinh^{2j}\zeta \frac{\Gamma(-j-m)\Gamma(-j+m)}{\Gamma(-2j)}{}_2F_1\left(-j-m,-j-n;-2j;-\frac{1}{\sinh^2\zeta}\right),
\end{equation}
and where $\sinh^2\zeta = \gamma_{\+}\gamma_{\m} e^{2\phi}$. The full representation matrix element is then:
\begin{align}
R_{s_1|\alpha_1, s_2|\alpha_2}^k (g) &= \alpha_1 \mathcal{G}^j_{m,n}(g_B) + \alpha_2 \mathcal{G}^{j-1/2}_{m,n}(g_B) \\
&\phantom{==} + \theta_{\m} \left((m-1/2)\mathcal{G}^j_{m-1/2,n}(g_B) - \alpha_1\alpha_2 \mathcal{G}^{j-1/2}_{m-1/2,n}(g_B) \right) \nonumber \\
&\phantom{==} + \theta_{\+} \left(-\alpha_1\alpha_2 \mathcal{G}^j_{m,n-1/2}(g_B) - (2j+n) \mathcal{G}^{j-1/2}_{m,n-1/2}(g_B) \right) \nonumber \\
&\phantom{==} + \theta_{\m}\theta_{\+} \left(-(m-1/2) \alpha_2 \mathcal{G}^j_{m-\frac{1}{2},n-\frac{1}{2}}(g_B) + (2j+n) \alpha_1 \mathcal{G}^{j-\frac{1}{2}}_{m-\frac{1}{2},n-\frac{1}{2}}(g_B) \right). \nonumber
\end{align}
Just as we noticed in the bosonic case in Appendix \ref{app:slr}, it is intriguing to note that all of these components can be interpreted as global superconformal blocks \cite{Belavin_2007,Belavin_2008,Hikida:2018eih}.

\mciteSetMidEndSepPunct{}{\ifmciteBstWouldAddEndPunct.\else\fi}{\relax}
\bibliographystyle{utphys}
{\small \bibliography{references}{}}

\end{document}